\documentclass[final,onefignum,onetabnum]{siamart190516}

\usepackage{amsmath}
\RequirePackage[numbers]{natbib}
\RequirePackage{hyperref}
\usepackage[utf8]{inputenc} 
\usepackage{amsfonts}
\usepackage{bm}
\usepackage{cleveref}
\usepackage{graphicx}
\usepackage[export]{adjustbox}
\usepackage[caption=false,font=footnotesize,indention=0cm]{subfig}
\usepackage[graphicx]{realboxes}
\usepackage{nameref}

\def\G{\mathcal{G}}
\def\V{\mathcal{V}}
\def\E{\mathcal{E}}
\def\Aintra{\bm{A}_{\mathrm{intra}}}
\def\Ainter{\bm{A}_{\mathrm{inter}}}
\def\A1{\bm{A}^{(1)}}
\def\AL{\bm{A}^{(L)}}
\def\Al{\bm{A}^{(l)}}
\def\R{\mathbb{R}}
\def\lmax{\lambda_{\mathrm{max}}}

\ifpdf
\hypersetup{
  pdftitle={Orientations and matrix function-based centralities in multiplex network analysis of urban public transport},
  pdfauthor={K. Bergermann, and M. Stoll}
}
\fi

\headers{Multiplex urban public transport networks}{K. Bergermann, and M. Stoll}

\title{Orientations and matrix function-based centralities in multiplex network analysis of urban public transport}

\author{Kai Bergermann\thanks{Chair of Scientific Computing, Technische Universit\"at Chemnitz, Department of Mathematics, 09107 Chemnitz, Germany
		(\email{kai.bergermann@math.tu-chemnitz.de}, \email{martin.stoll@math.tu-chemnitz.de}).}
	\and Martin Stoll\footnotemark[1]
}

\begin{document}

\maketitle

\begin{abstract}
We study urban public transport systems by means of multiplex networks in which stops are represented as nodes and each line is represented by a layer.
We determine and visualize public transport network orientations and compare them with street network orientations of the $36$ largest German as well as $18$ selected major European cities.
We find that German urban public transport networks are mainly oriented in a direction close to the cardinal east-west axis, which usually coincides with one of two orthogonal preferential directions of the corresponding street network.
While this behavior is present in only a subset of the considered European cities it remains true that none but one considered public transport network has a distinct north-south-like preferential orientation.
Furthermore, we study the applicability of the class of matrix function-based centrality measures, which has recently been generalized from single-layer networks to layer-coupled multiplex networks, to our more general urban multiplex framework.
Numerical experiments based on highly efficient and scalable methods from numerical linear algebra show promising results, which are in line with previous studies.
The centrality measures allow detailed insights into geometrical properties of urban systems such as the spatial distribution of major transport axes, which can not be inferred from orientation plots.
We comment on advantages over existing methodology, elaborate on the comparison of different measures and weight models, and present detailed hyper-parameter studies.
All results are illustrated by demonstrative graphical representations.
\end{abstract}



\section{Introduction}

By the year $2050$ two-thirds of the growing world population are expected to live in urban areas \cite{bolay2020urban}.
The sustainable planning and development of cities will greatly impact global economic, environmental, and social challenges, which demand interdisciplinary approaches to urban science \cite{acuto2018building}.

In recent decades, the interdisciplinary field of complex network science has provided models and methods, which today impact everyday life \cite{milgram1967small,watts1998collective,barabasi1999emergence,brin1998anatomy,page1999pagerank}.
Network modeling approaches for urban systems can be traced back almost $300$ years \cite{euler1741solutio} and the abstraction of urban areas into geometrical models continues to form the basis of modern urban science \cite{porta2006network,porta2006bnetwork,batty2008size,barthelemy2008modeling,barthelemy2011spatial,courtat2011mathematics,chan2011urban,barthelemy2013self,barthelemy2016structure,sharifi2019resilient}.
The combination of complex network models, urban science applications, and mathematical methodology from the field of numerical linear algebra is at the heart of this paper.

\begin{figure}
	\begin{center}
	\includegraphics[width=.95\textwidth]{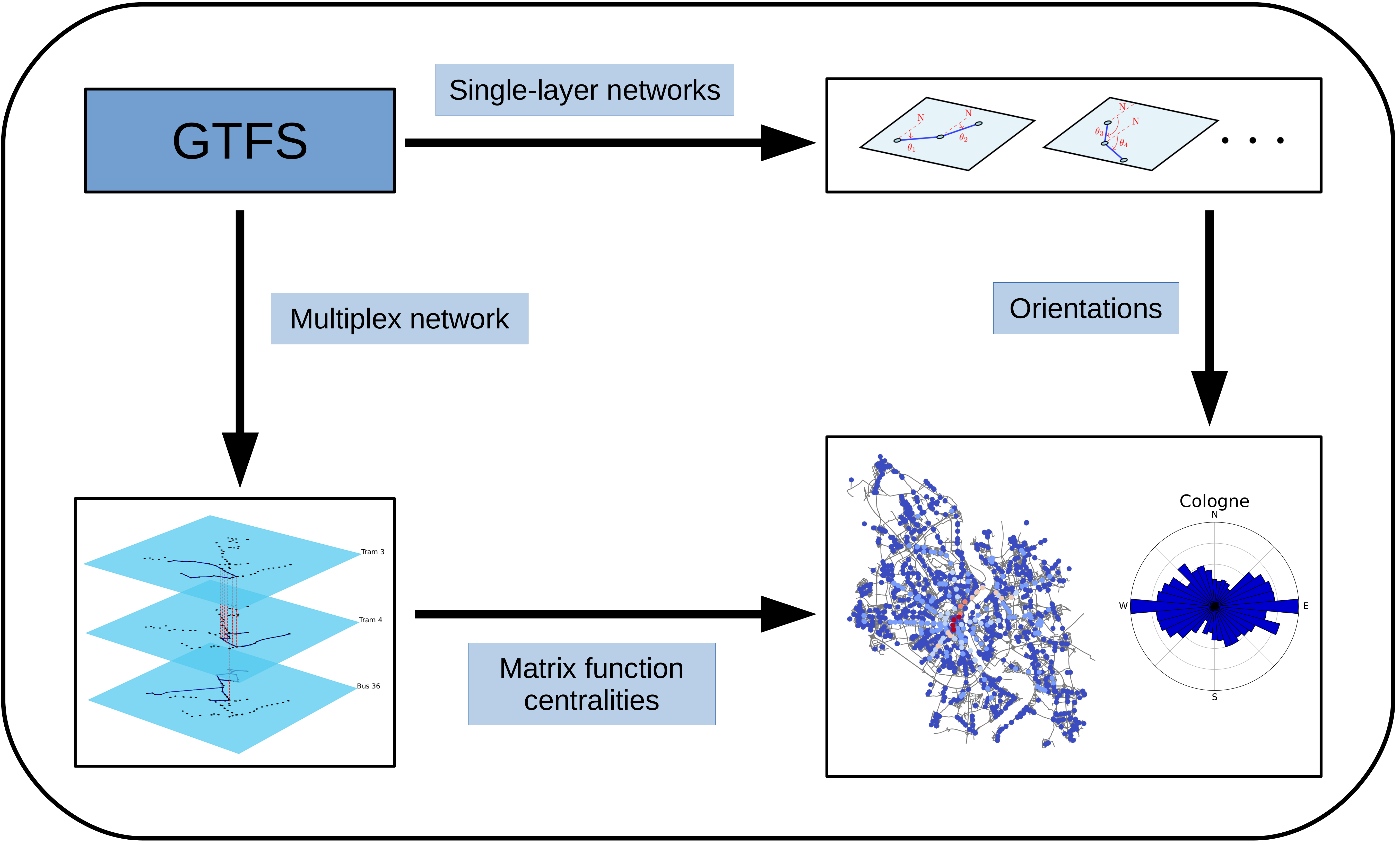}
	\caption{Schematic overview of the methods introduced in this paper.
		Both aspects of geometrical city modeling rely on GTFS timetable data, which is described in the section \nameref{sec:data}.
		The methodology for the computation of public transport orientations using sequences of single-layer networks is described in the section \nameref{sec:orientations_methodology}.
		Details on the construction of the multiplex network representation of urban public transport systems are given in the section \nameref{sec:multiplex_networks}.
		The methodology for the determination of the centrality of stops and lines of the multiplex networks using matrix function-based centrality measures is presented in the sections \nameref{sec:centralities_definition} and \nameref{sec:centralities_methods}.}\label{fig:flowchart}
	\end{center}
\end{figure}

The main contribution of this paper is the development of a general methodology for two aspects of the spatial analysis of urban public transport networks.
Exemplary results are given for major German and European cities.
\Cref{fig:flowchart} gives a schematic overview of the developed methods, which we describe further in the following paragraphs.
This paper is accompanied by publicly available python implementations of all developed methods\footnote{\url{https://github.com/KBergermann/Urban-multiplex-networks}}.

The first aspect of geometrical city modeling studied in this paper is represented by the determination of orientations of complex urban networks, which reflect different urban organization patterns ranging from top-down planning to self-organization dynamics \cite{batty2008size,barthelemy2008modeling,barthelemy2013self}.
Earlier works focused on the study of orientations of street networks \cite{courtat2011mathematics,chan2011urban,gudmundsson2013entropy,mohajeri2013evolution,boeing2019urban}.
Our contribution to this aspect of geometrical urban analysis is to provide methodology and example results of orientations of public transport systems.
We compare these to street network orientations obtained with existing methodology \cite{boeing2017osmnx,boeing2019urban} and find interesting relations between the orientations of the two modes of transportation for the largest German and several major European cities.

The second aspect studied in this paper is the identification and ranking of the most central nodes of urban public transport networks.
Complex network scientists have intensively studied a variety of different centrality measures in recent decades.
Among the most prominent examples are degree, betweenness \cite{freeman1977set}, closeness \cite{freeman1978centrality}, and versions of eigenvector centrality \cite{bonacich1987power,brin1998anatomy,kleinberg1999authoritative,page1999pagerank}.
Urban science has been one of many disciplines to pose the natural question of identifying and ranking the most important entities of complex networks \cite{crucitti2006centrality,crucitti2006bcentrality,porta2006network,scheurer2006centrality,to2015centrality,nourian2016spectral,agryzkov2019centrality,hellervik2019preferential,hong2019exploring,curado2021understanding}.

Our approach to the computation of centralities of urban public transport networks relies on two main ingredients: matrix function-based centrality measures and multiplex networks.
The class of matrix function-based centrality measures  \cite{katz1953new,estrada2005subgraph,estrada2010network,estrada2012structure,benzi2013total,benzi2020matrix} has attracted a lot of attention in recent years but has not been used in the urban science literature yet.
These measures have the interesting property to interpolate between the concepts of local degree and global eigenvector centrality \cite{benzi2015limiting}.
Furthermore, these measures require less assumptions on the graph structure than classical eigenvector centrality, which makes them applicable to a wider range of problems.

For the network modeling we rely on a special class of multilayer networks.
Research on multilayer networks exploded since $2014$ when two survey papers unified terminology and gleaned the main aspects of a field, which had before been studied across various disciplines \cite{kivela2014multilayer,boccaletti2014structure}.
Multilayer networks provide the means for building increasingly realistic models of complex systems as they allow entities to interact in different ways and on various levels.
Urban science was one discipline to witness the rise of multilayer network modeling approaches in recent years \cite{strano2015multiplex,barthelemy2016structure,alessandretti2016user,aleta2017multilayer,zheng2018understanding,curado2021understanding}.

Some recent works provide generalizations of centrality measures well-studied on single-layer graphs to the case of different multilayer architectures \cite{de2015ranking,taylor2017eigenvector,wang2017identifying,tudisco2018node,taylor2019supracentrality,wu2019tensor,taylor2021tunable,bergermann2021matrix}.
Most importantly for this paper, the class of matrix function-based centrality measures has very recently been generalized to the case of layer-coupled multiplex networks.
We adapt that methodology and study its applicability for the more general multiplex models of urban public transport networks.
Highly efficient and scalable methods from numerical linear algebra enable the rapid and stable approximation of the centralities for small- to large-scale networks.
We present various numerical results for multiplex networks with the number of physical nodes in the thousands and the number of layers in the hundrets and extensively study the influence of the different hyper-parameters involved in the model.

\section{Data}\label{sec:data}

The focus of this paper lies on the analysis of urban public transport networks.
For their realistic modeling we rely on the General Transit Feed Specification (GTFS)\footnote{\url{https://developers.google.com/transit/}}, which defines a standardized format for public transport timetables used around the globe.
For comparison and visualization purposes we also consider street networks.
Here, we rely on the python package OSMnx\footnote{\url{https://osmnx.readthedocs.io/en/stable/}} \cite{boeing2017osmnx} built on top of the open source platform OpenStreetMap\footnote{\url{https://www.openstreetmap.org/}}.
OSMnx can, amongst other things, be used to generate graph structures representing street networks according to user-specified queries and provides various routines for their analysis.
Note, that while OpenStreetMap also contains a public transport tagging feature, timetable information can not be expected to be as accurate as high-quality GTFS data \cite{osmPublicTransport}.

We consider the $36$ largest German cities by population as well as $18$ selected major European cities for which we could obtain complete GTFS data.
The three German cities Bochum, Gelsenkirchen, and Magdeburg had to be excluded from our studies as no complete GTFS data was available.

For Germany, several high quality GTFS feeds are publicly available\footnote{\url{https://gtfs.de/de/feeds/}} on a daily basis and this paper builds on the local public transport data set\footnote{\url{https://gtfs.de/de/feeds/de_nv/}}\footnote{\url{https://download.gtfs.de/germany/nv_free/latest.zip}} for April 22nd, 2021 (feed version ``light-2021-04-22'').
This data set covers over $20\,000$ public transport lines serving over $450\,000$ stops across Germany.
However, only required and conditionally required files (no optional ones) are provided\footnote{\url{https://developers.google.com/transit/gtfs/reference}}, e.g., no explicit information about route frequencies or transfer options between stops is provided.

We also obtained several individual GTFS data sets for $18$ selected major European cities\footnote{\url{https://transitfeeds.com/l/60-europe}}.
However, as data integrity and availability varies among the different feeds, no common feed date could be guaranteed for all data sets.
Instead, we downloaded the latest complete data set available as of May 21st, 2021 in each case.

For all considered cities, we use polygons representing the administrative boundaries of the cities proper (excluding suburban areas) obtained via appropriate OSMnx queries to filter for all stops within the city limits.
In GTFS terminology, all ``routes'' with at least one associated ``trip'' connecting at least two of the filtered ``stops'' are considered a valid line in the public transport network.
In addition to the stops' spatial coordinates, we process the fields ``arrival time'' and ``departure time'' of each ``trip'' from the file ``stop\_times'' to determine travel times between connected pairs of stops.

The polygons used to filter public transport stops from the GTFS data sets are also deployed in all OSMnx routines used in the context of this paper.
For our analysis, we require filtering for streets within each city's limits for two purposes:
firstly, to produce street network orientation plots and, secondly, to create and plot street networks for visualization purposes later in this paper.

The following two sections provide details on how the data is transformed into different graph structures.
While a sequence of single-layered graphs is sufficient for the determination of public transport network orientations we introduce a more sophisticated modeling approach based on multiplex networks for the determination of the most central stops and lines of the networks.

\section{Public transport network orientations}\label{sec:orientations}

Urban areas in large parts of the world have undergone rapid expansion in past decades and are expected to grow further \cite{bolay2020urban}.
Global economic, environmental, and social challenges will be crucially impacted by future urban planning \cite{acuto2018building}.
One of many important aspects of the growth of urban systems is their geometrical expansion \cite{porta2006network,porta2006bnetwork,batty2008size,barthelemy2008modeling,barthelemy2011spatial,courtat2011mathematics,chan2011urban,barthelemy2013self,barthelemy2016structure,sharifi2019resilient}.
Previous works have identified very different mechanisms for the spatial evolution of cities ranging from top-down planning to self-organization dynamics \cite{batty2008size,barthelemy2008modeling,barthelemy2013self}.
A recent work beautifully illustrates these different mechanisms by means of orientation plots of urban street networks on a global scale \cite{boeing2019urban}.

In this section, we consider orientations of both street and public transport networks of the $36$ largest German cities by population as well as $18$ selected major European cities specified in the previous section.
We describe the developed methodology in a first, and illustrate and discuss the obtained results in a second step.

\subsection{Methodology}\label{sec:orientations_methodology}

For the determination of street network orientations we use readily implemented routines\footnote{\url{https://github.com/gboeing/osmnx-examples/blob/main/notebooks/17-street-network-orientations.ipynb}} from the python package OSMnx \cite{boeing2017osmnx} in combination with appropriate search queries, which follow the methodology described in \cite{boeing2019urban}.
This methodology relies on the creation of a primal \cite{porta2006network} undirected single-layer graph in which intersections are represented by nodes and streets by straight edges between them (ignoring curvature).
The compass bearings of both directions (always including the reciprocal of any street bearing) of each street segment are recorded and added in an unweighted manner, cf.~\cite{mohajeri2013evolution} for a discussion on weighting techniques.
Afterwards, they are divided into $36$ equal-sized bins with the cardinal directions located in the middle of the associated bins and plotted as a rose diagram.

Our first contribution to the geometrical analysis of urban systems is the investigation of orientations in public transport networks.
To this end, we use GTFS data describing public transport timetables as specified in the section \nameref{sec:data} and we develop a methodology for the computation of public transport network orientations similar to that of street network orientations.

Our methodology also relies on the creation of single-layered graphs and the computation of bearings between pairs of nodes.
In our case, nodes represent stops within the city limits and edges represent connections of public transport lines between stops, which we also assume to be straight and which are also not weighted (by, e.g., segment lengths or travel times).
Deviating from the street network case, we consider these edges to be directed.
The edges' directions are determined by the sequence of stops of, in GTFS terminology, each ``trip'' of each valid ``route'' of the urban public transport network.
We realize this by successively processing all ``trips'' belonging to all valid ``routes'' as specified in the section \nameref{sec:data}.
This procedure has the effect of assigning an implicit weight to all lines proportional to their operation frequency.
An alternative approach would be to only process unique trips of all lines, which would, in the simplest case, lead to two copies of the same sequence of stops in reverse order.
Interestingly, we empirically found only small deviations in the results of the two approaches, which we will comment on in the subsection \nameref{sec:orientations_results_discussion}.
In both cases, most of the created directed graphs are chain graphs connecting only a small subset of the stops within the city limits.
We then store the directed bearings (not including reciprocals) between all pairs of connected nodes of all graphs, divide them into $36$ equal-sized bins, and visualize the results in a rose diagram as in the case of street network orientations.
Note that the computations of orientations of different public transport lines are entirely independent, which offers great potential for parallelization.

\subsection{Results and discussion}\label{sec:orientations_results_discussion}

\begin{figure}
	\begin{center}
	\includegraphics[width=.95\textwidth]{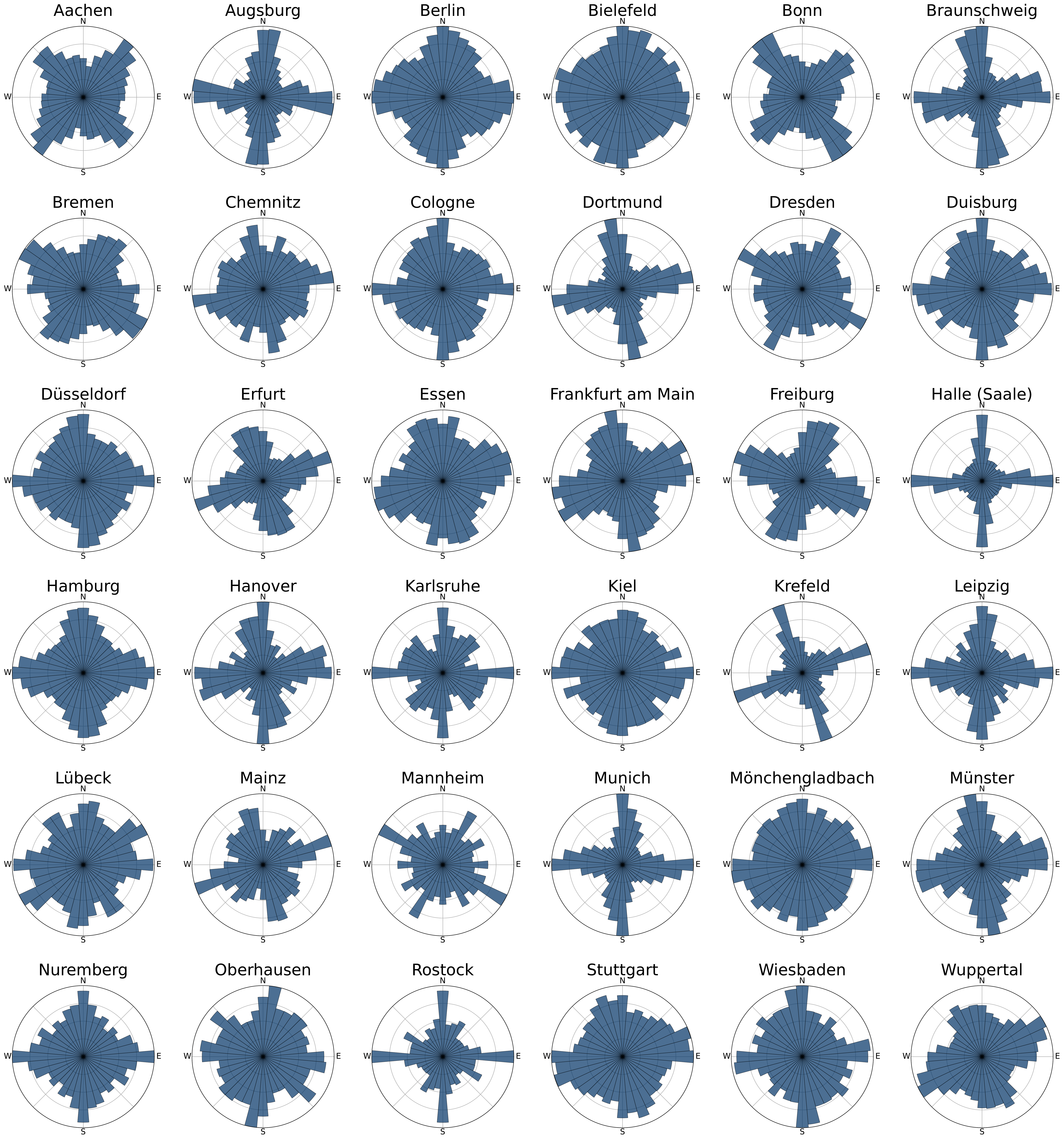}
	\caption{Street network orientations of the $36$ largest German cities with valid GTFS data as described in the section \nameref{sec:data}. The plots were created using the OSMnx python package \cite{boeing2017osmnx}.}\label{fig:orientations_streets_germany}
	\end{center}
\end{figure}

The methodology for the computation of street network orientations described in the beginning of the section \nameref{sec:orientations_methodology} leads to the results displayed in \Cref{fig:orientations_streets_germany} for the $36$ German cities under investigation.
The results show that most large German cities' street networks have two orthogonal preferential directions, which is consistent with previous findings \cite{chan2011urban}.
How strongly pronounced these directions are, however, differs significantly:
while, e.g., Halle (Saale) or Krefeld exhibit quite distinct cross-shaped patterns throughout the city area other cities like, e.g., Bielefeld or M\"onchengladbach show almost equally distributed orientations.
These differences likely reflect diverse historic city developments and urban planning approaches.
Furthermore, the preferential directions seldomly coincide with the cardinal directions, but can often be linked to geographical constraints such as rivers or mountains.
Some cities like, e.g., L\"ubeck or Rostock appear to comprise of two sets of orthogonal preferential directions prevailing in different regions of the city area.

At first glance, one might expect public transport network orientations to strongly correlate with street network orientations as buses and usually trams, which dominate the public transport system in most German cities, cf.~the layer column in \Cref{tab:germany_nodes_layers}, operate on or alongside streets.
However, as public transport stops are usually not located at street intersections but on street segments one could interpret public transport networks as dual graphs \cite{porta2006bnetwork} of street networks in which nodes represent streets and edges represent intersections and in which only a subset of street segments is equipped with public transport stops.
In the case of a line following the same street between two stops the orientations of both networks will coincide, but as soon as a public transport line takes a turn at a street intersection, the public transport bearing will point into a direction in between the two (often orthogonal) street orientations.

\begin{figure}
	\begin{center}
	\includegraphics[width=.15\textwidth]{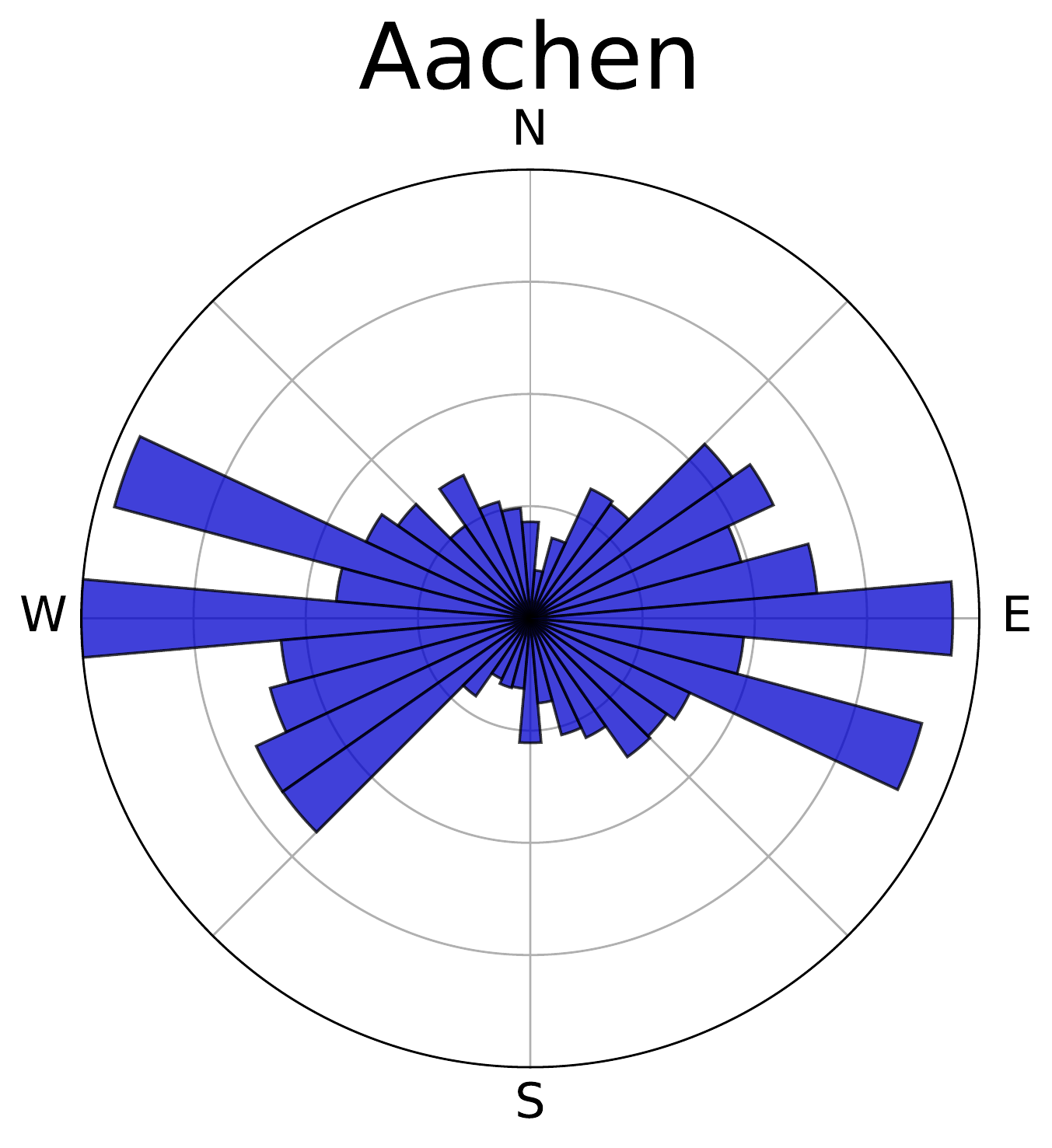}
	\includegraphics[width=.15\textwidth]{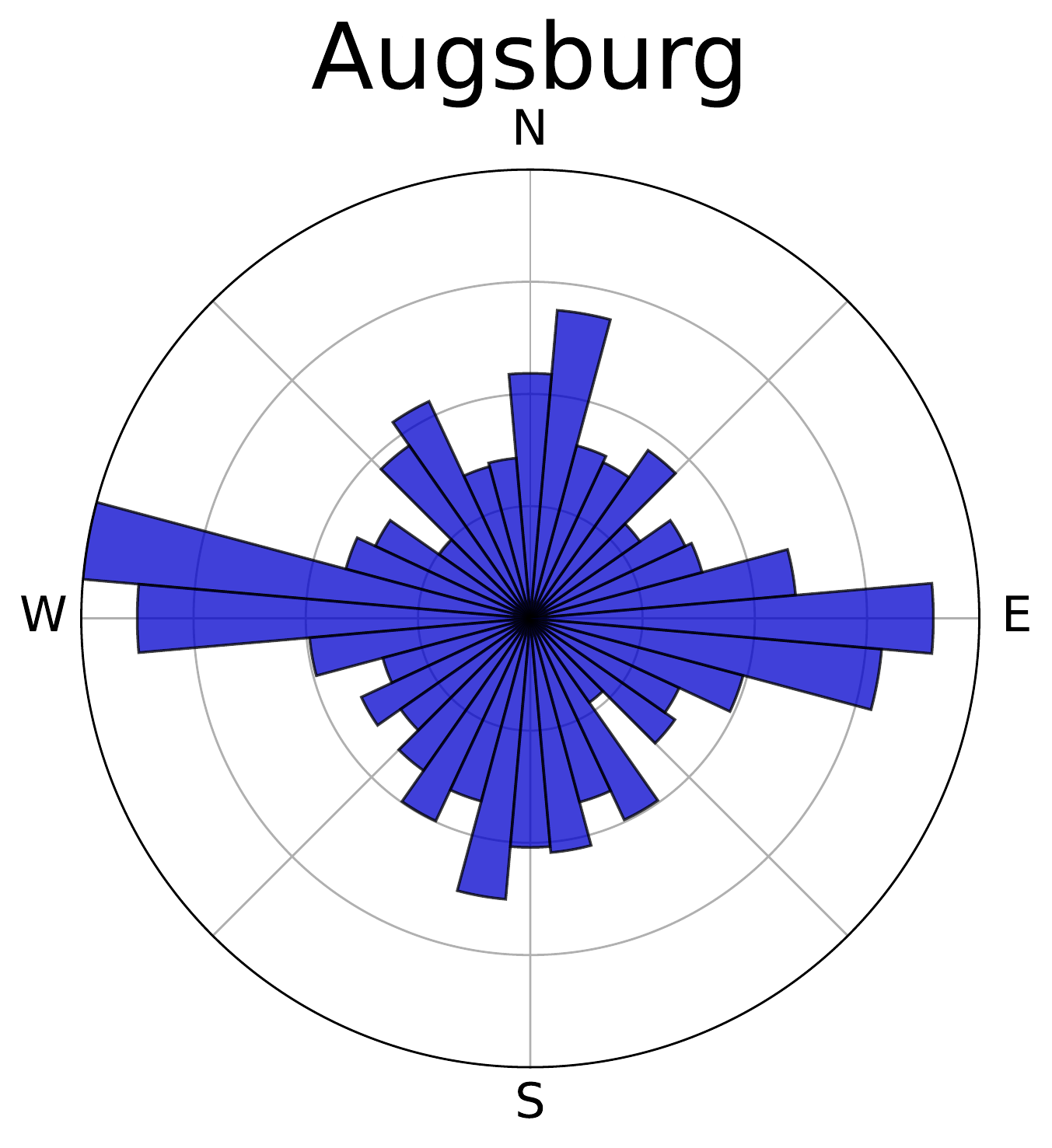}
	\includegraphics[width=.15\textwidth]{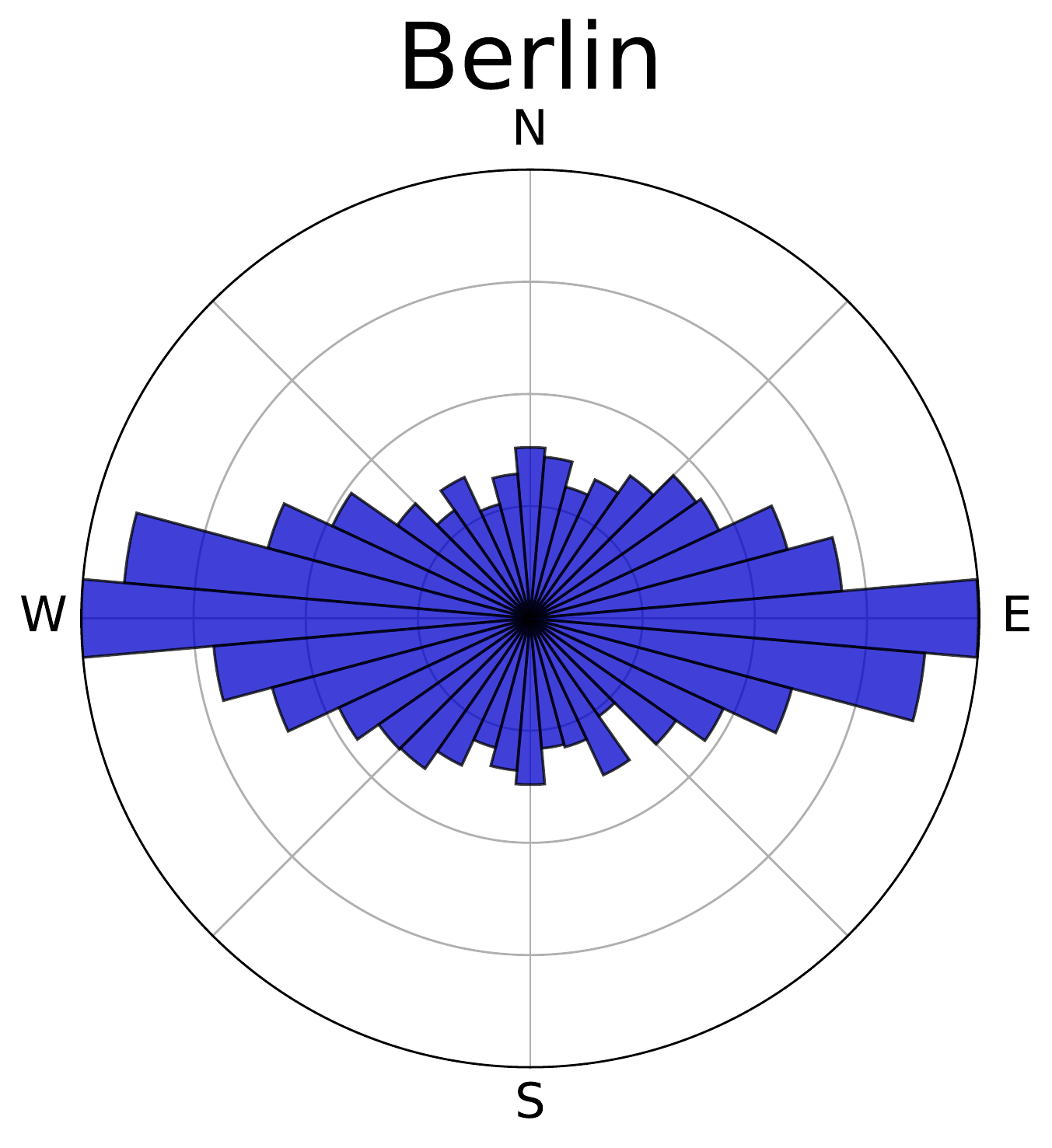}
	\includegraphics[width=.15\textwidth]{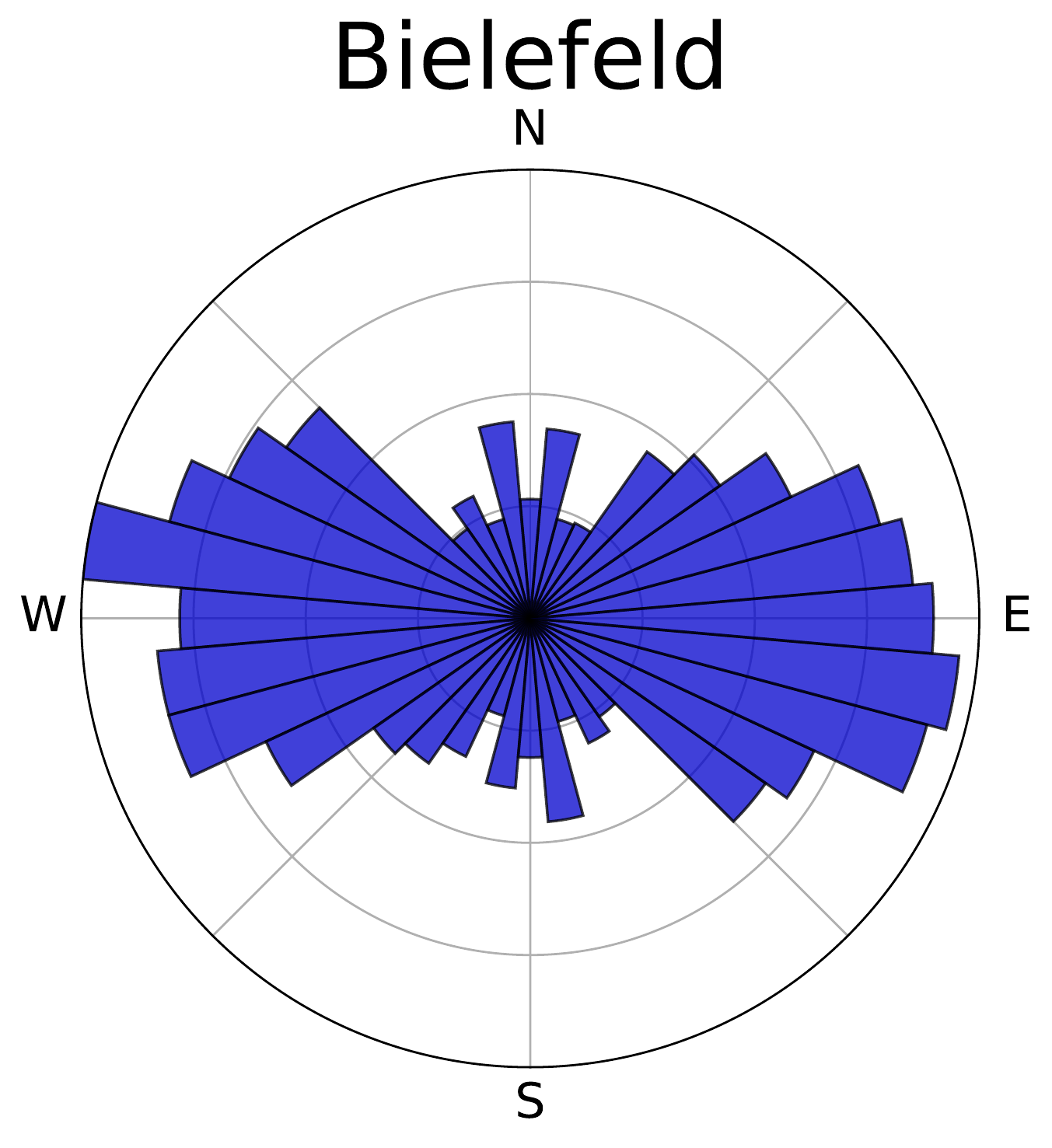}
	\includegraphics[width=.15\textwidth]{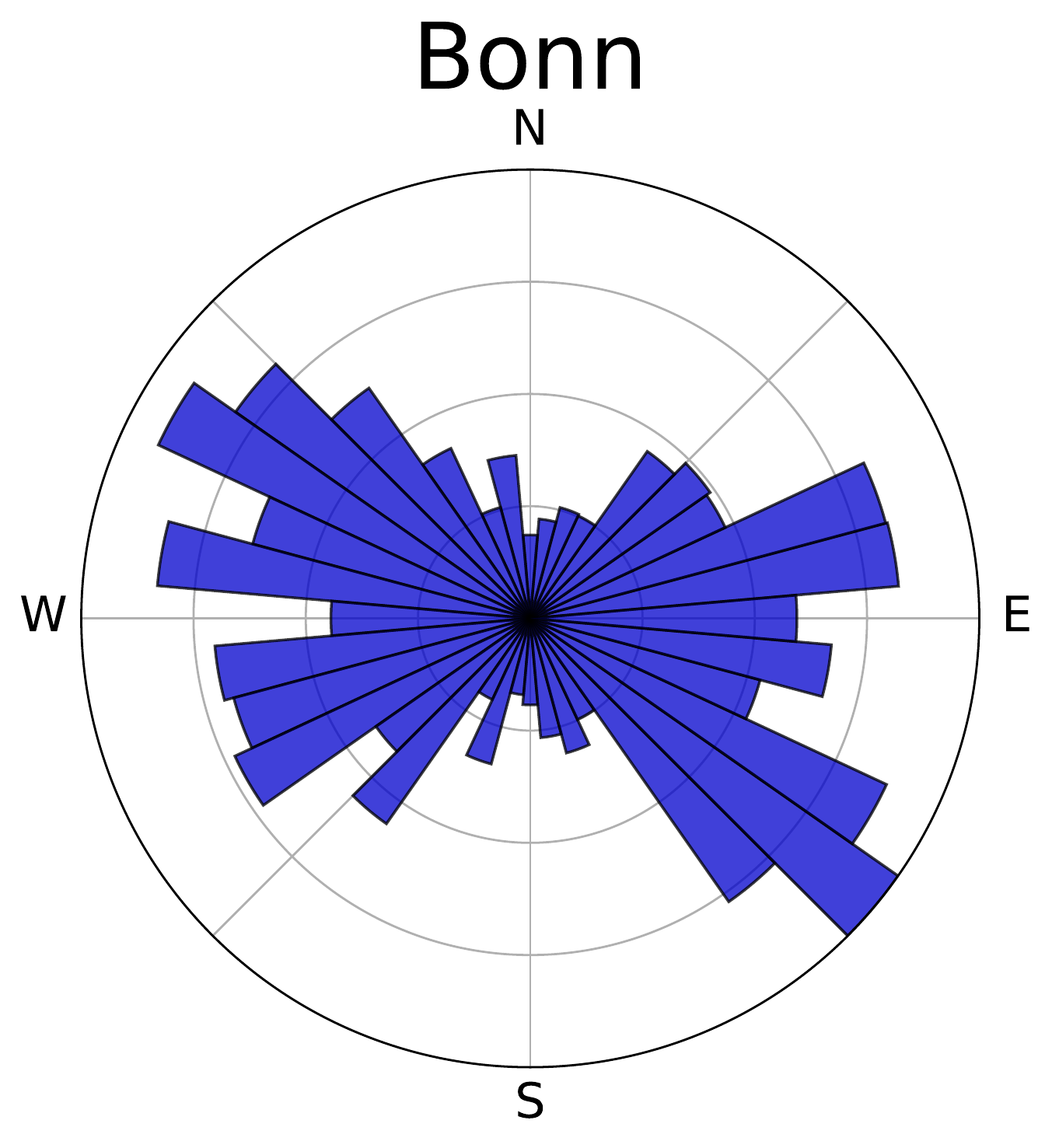}
	\includegraphics[width=.15\textwidth]{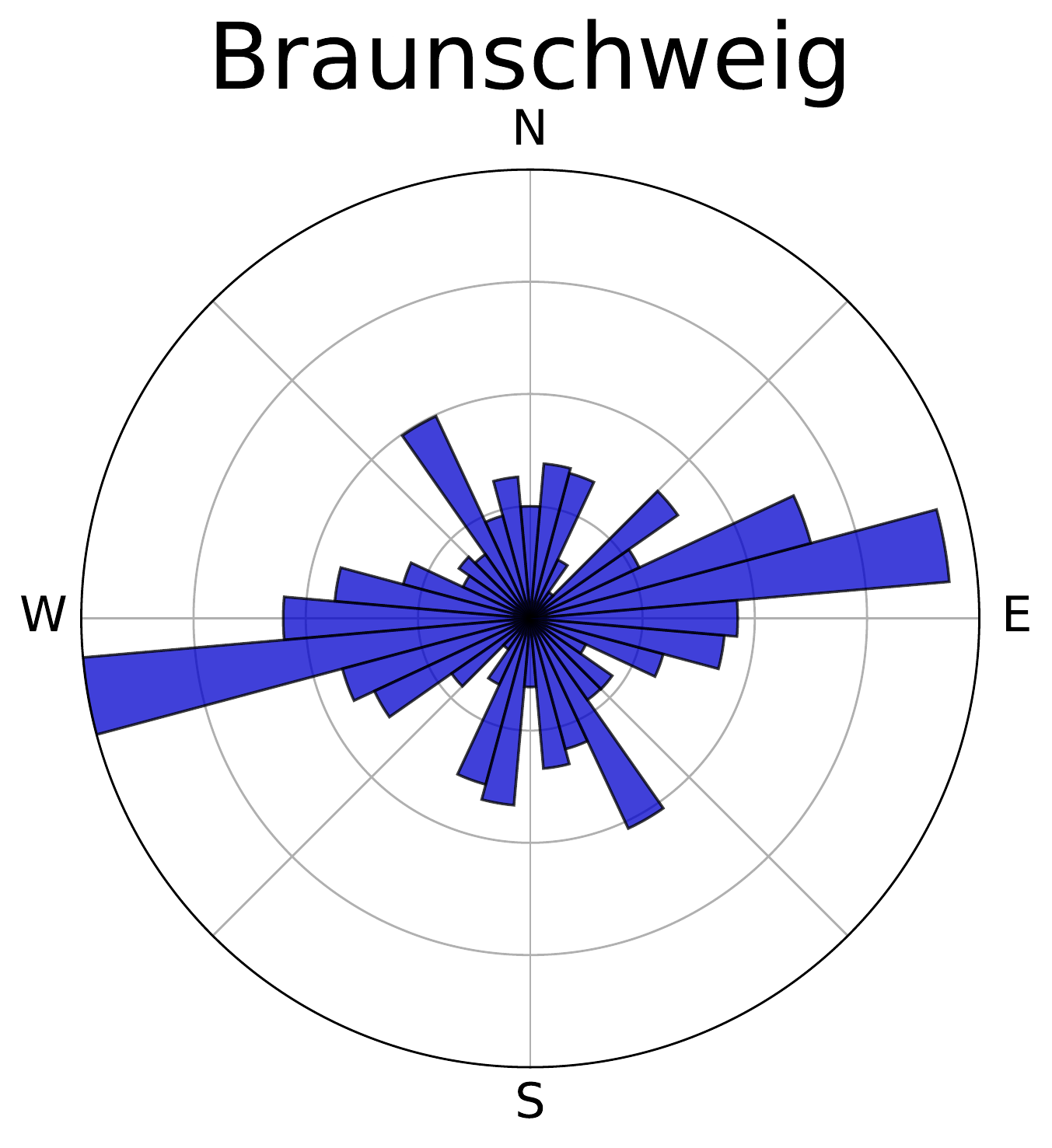}
	
	\includegraphics[width=.15\textwidth]{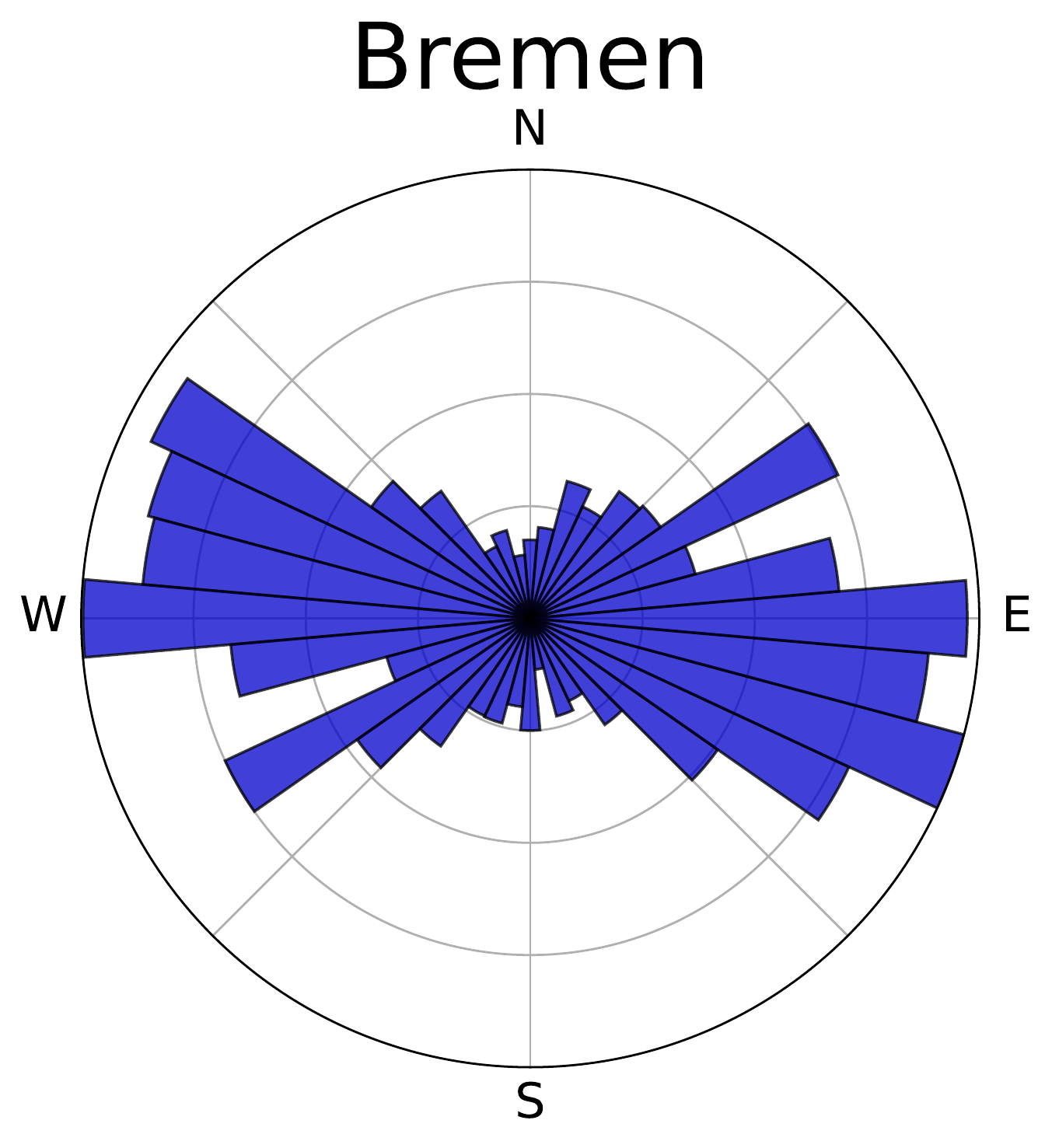}
	\includegraphics[width=.15\textwidth]{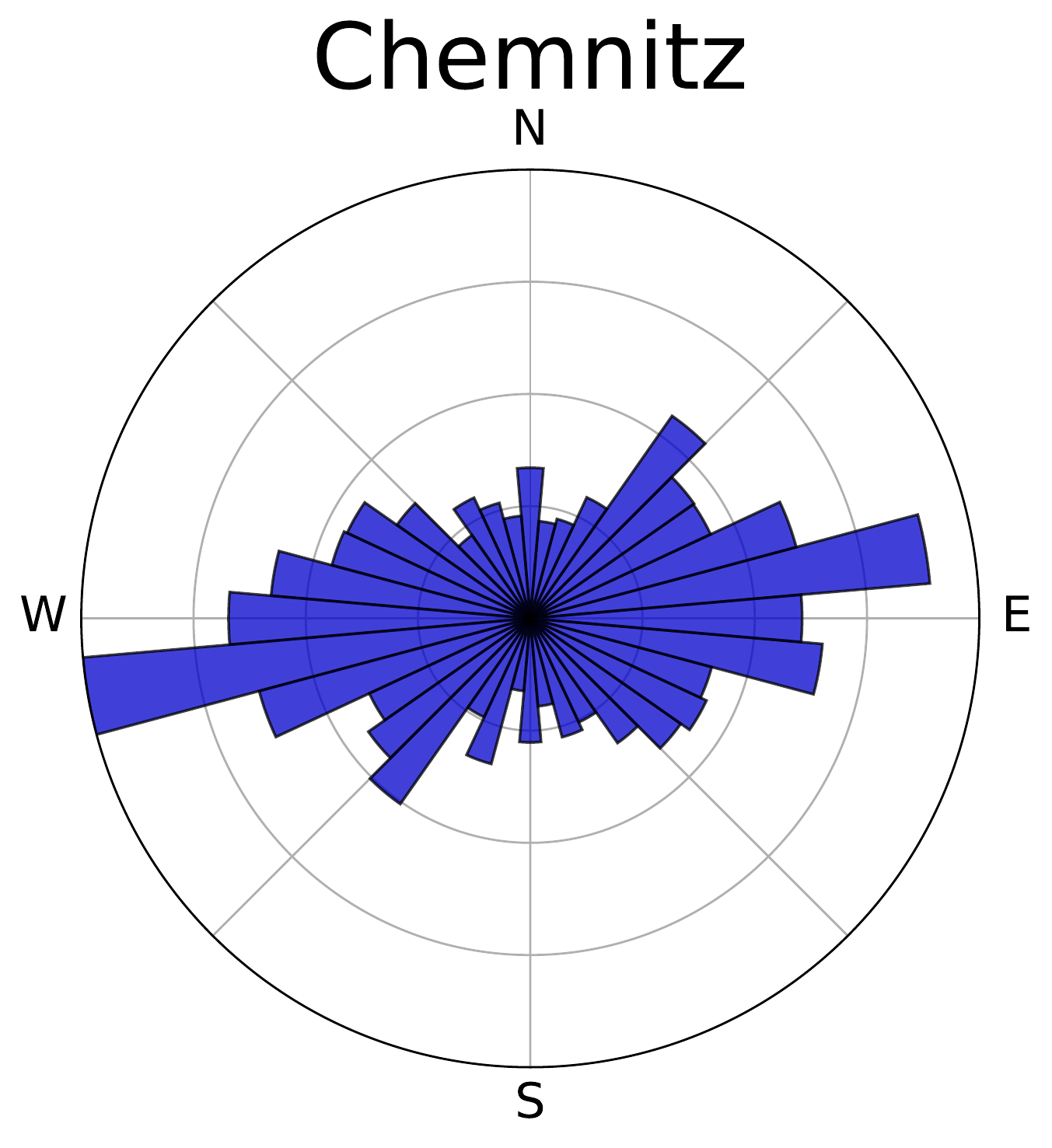}
	\includegraphics[width=.15\textwidth]{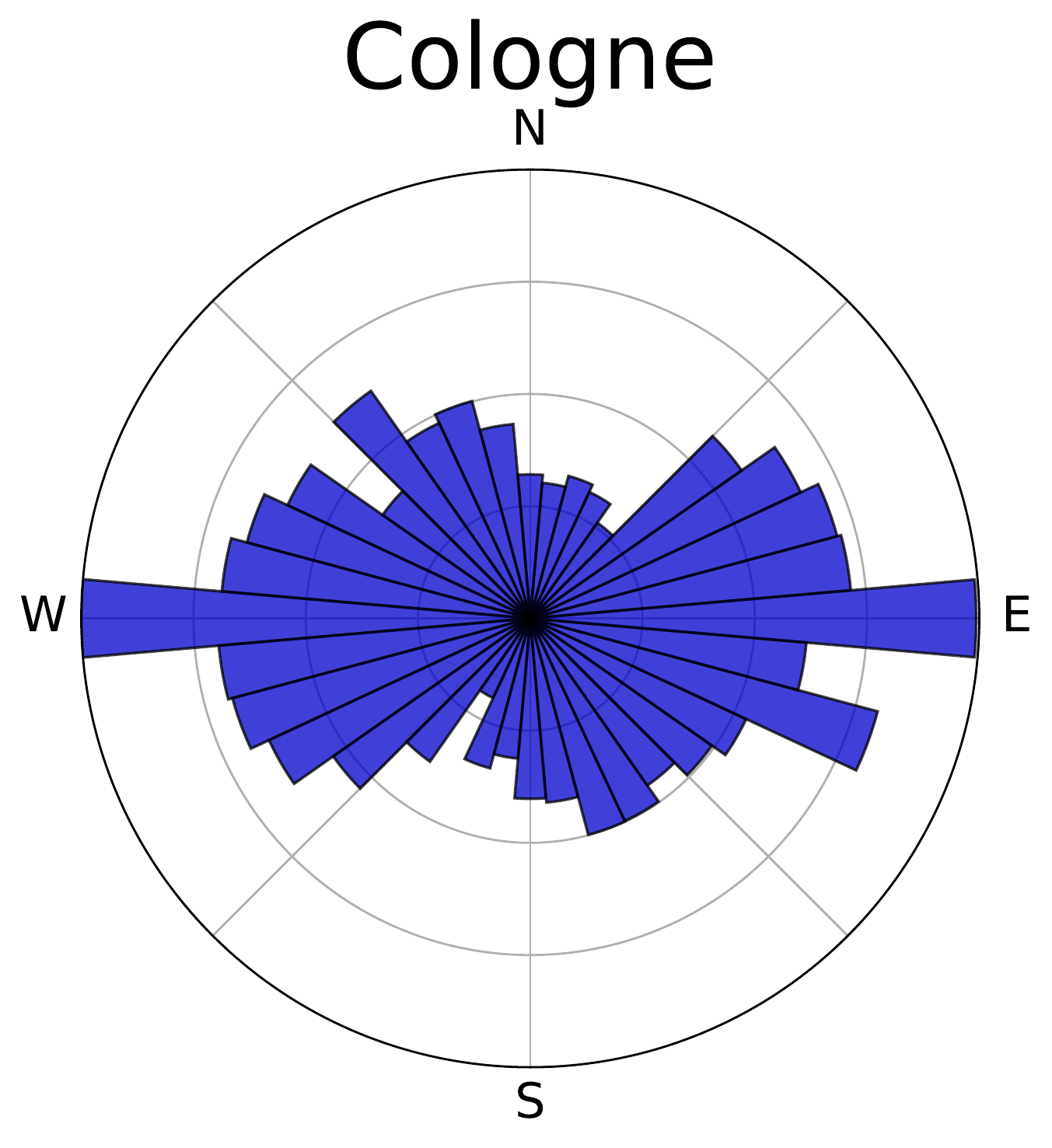}
	\includegraphics[width=.15\textwidth]{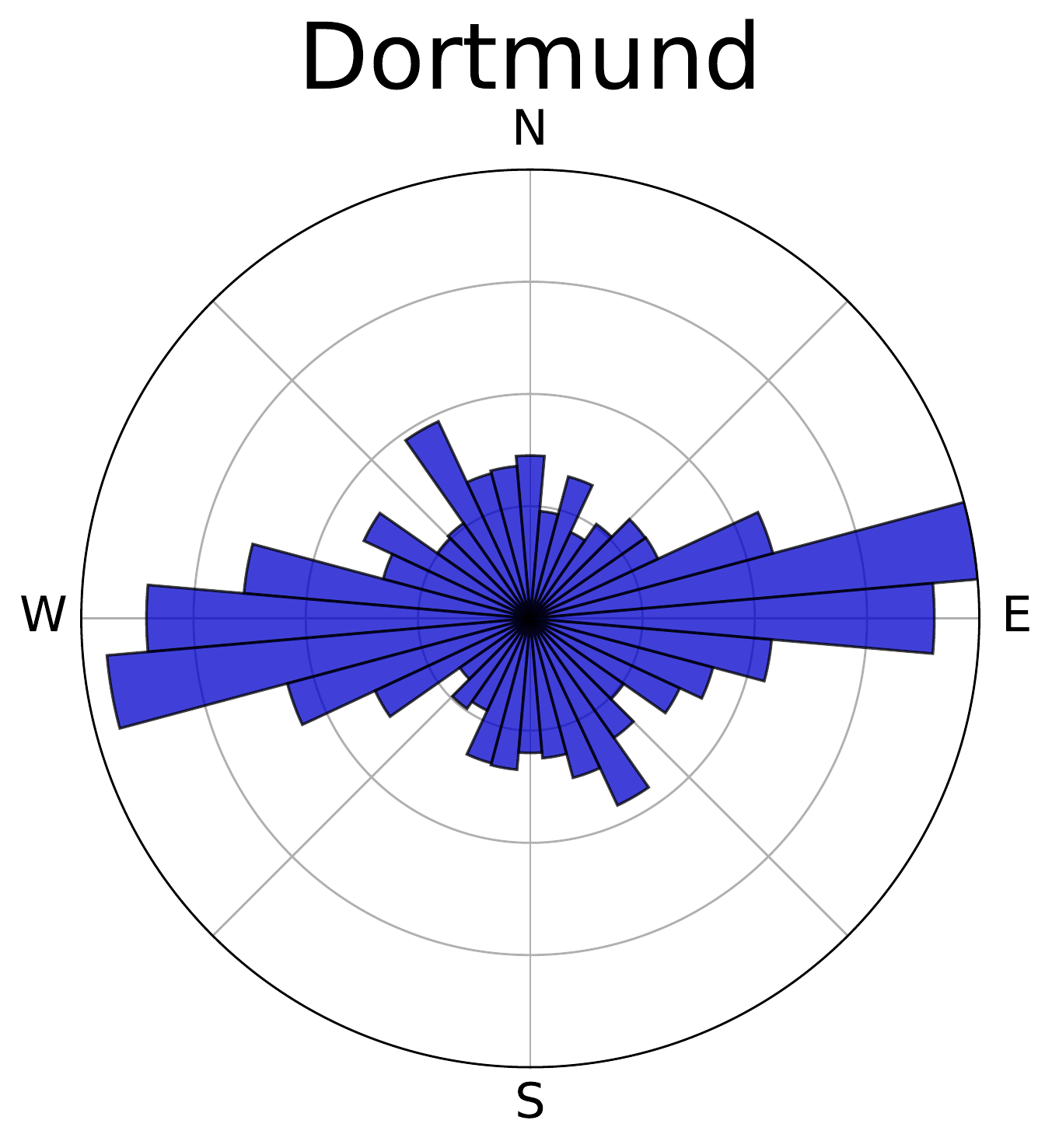}
	\includegraphics[width=.15\textwidth]{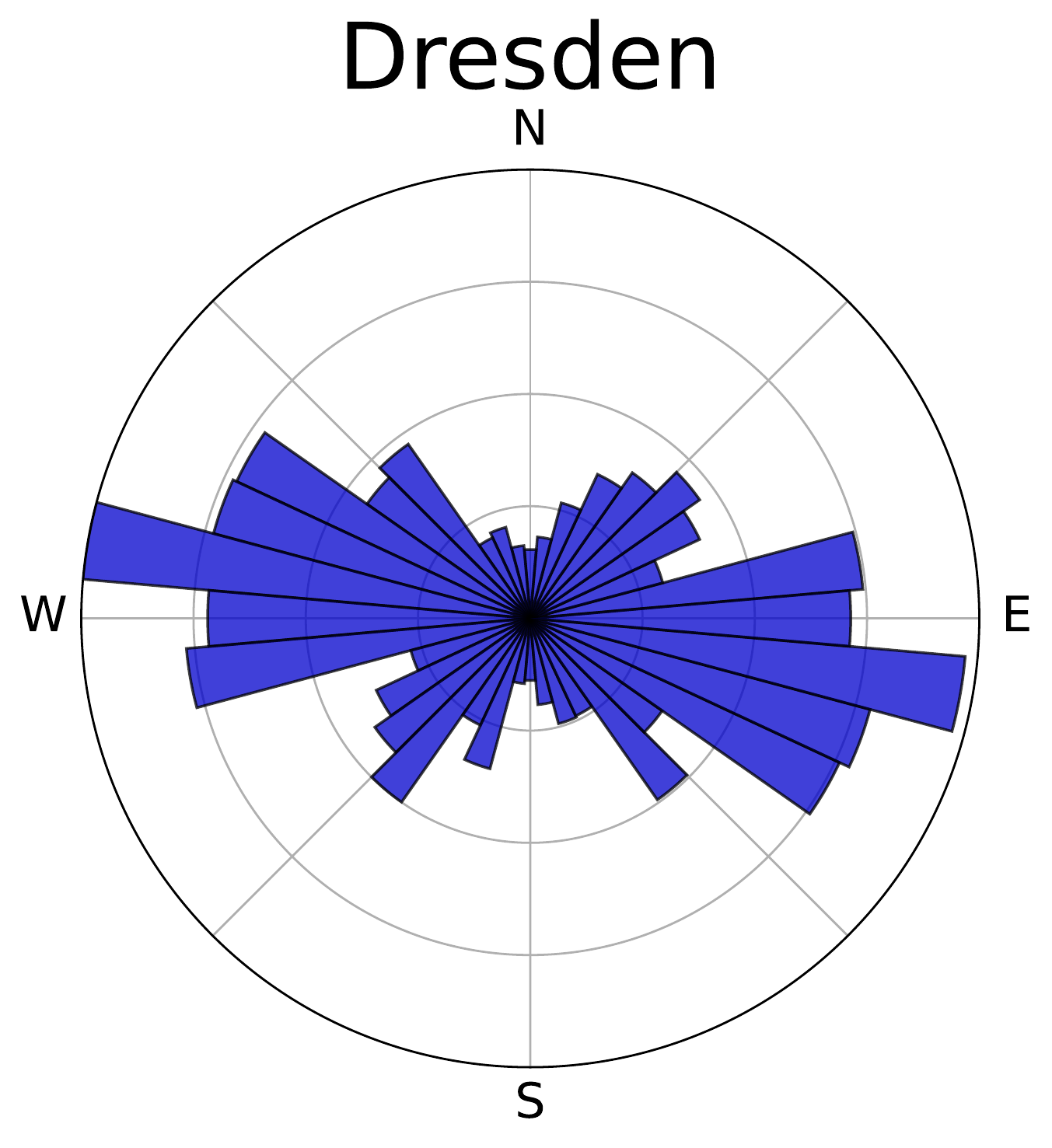}
	\includegraphics[width=.15\textwidth]{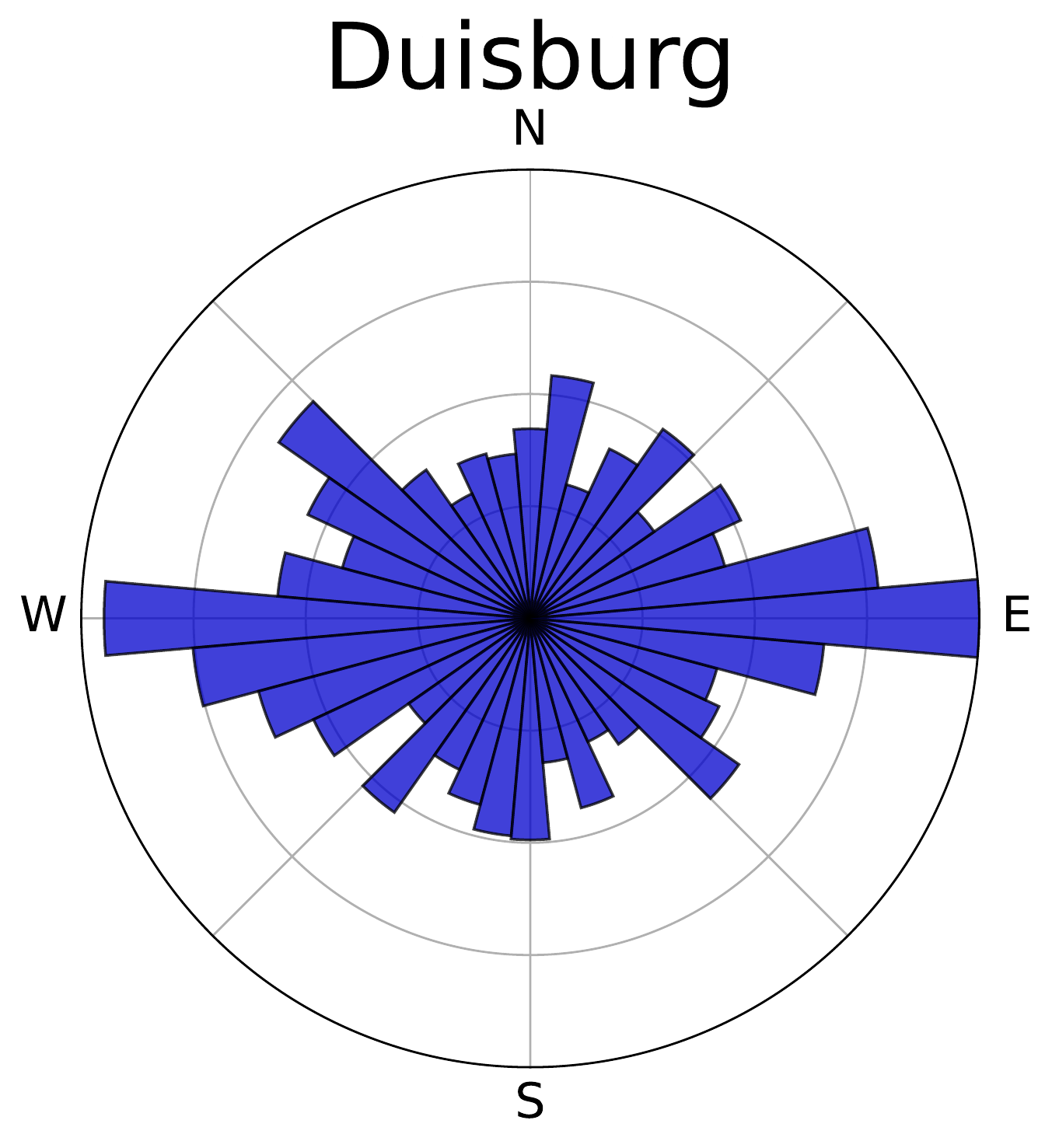}
	
	\includegraphics[width=.15\textwidth]{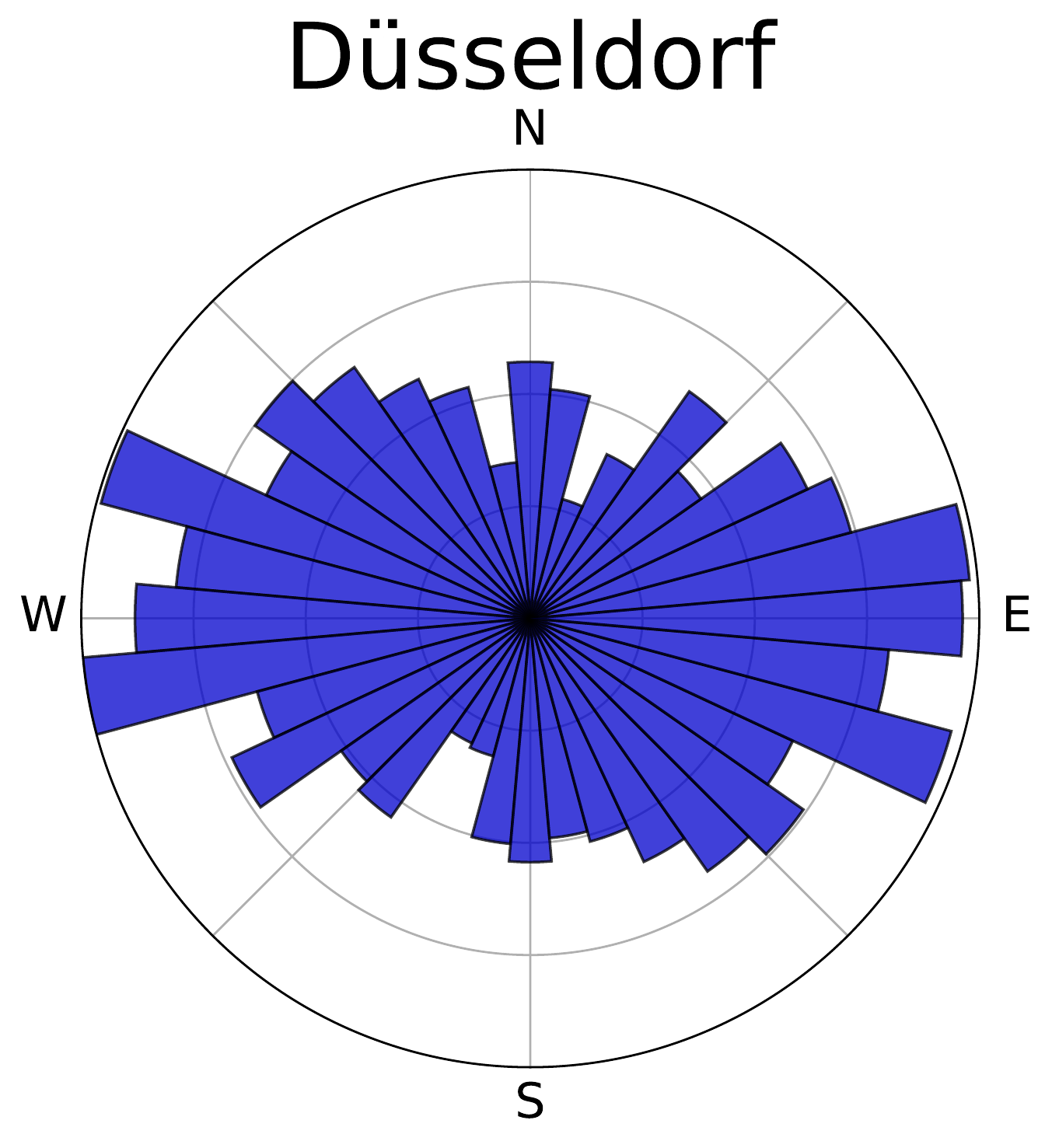}
	\includegraphics[width=.15\textwidth]{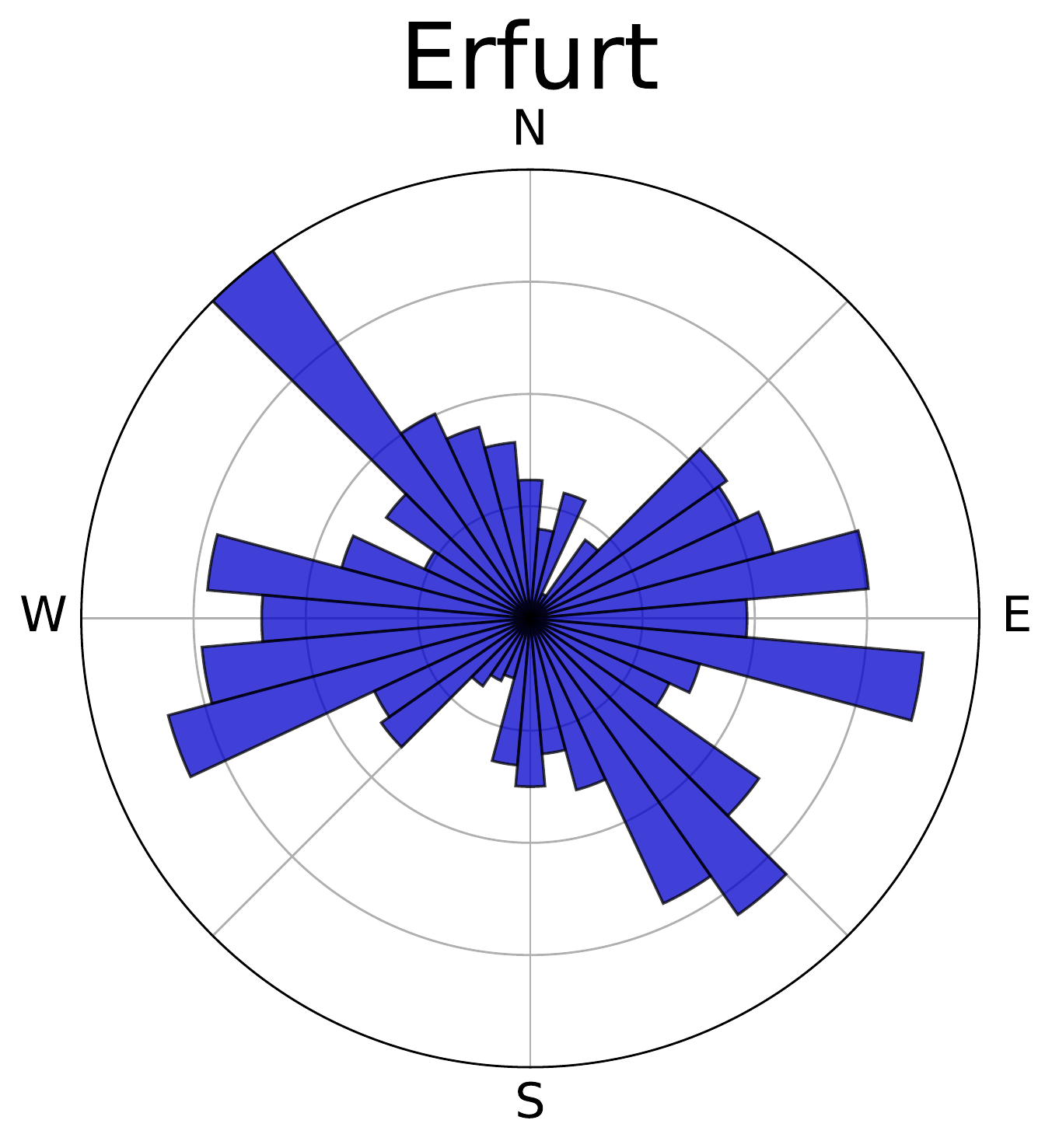}
	\includegraphics[width=.15\textwidth]{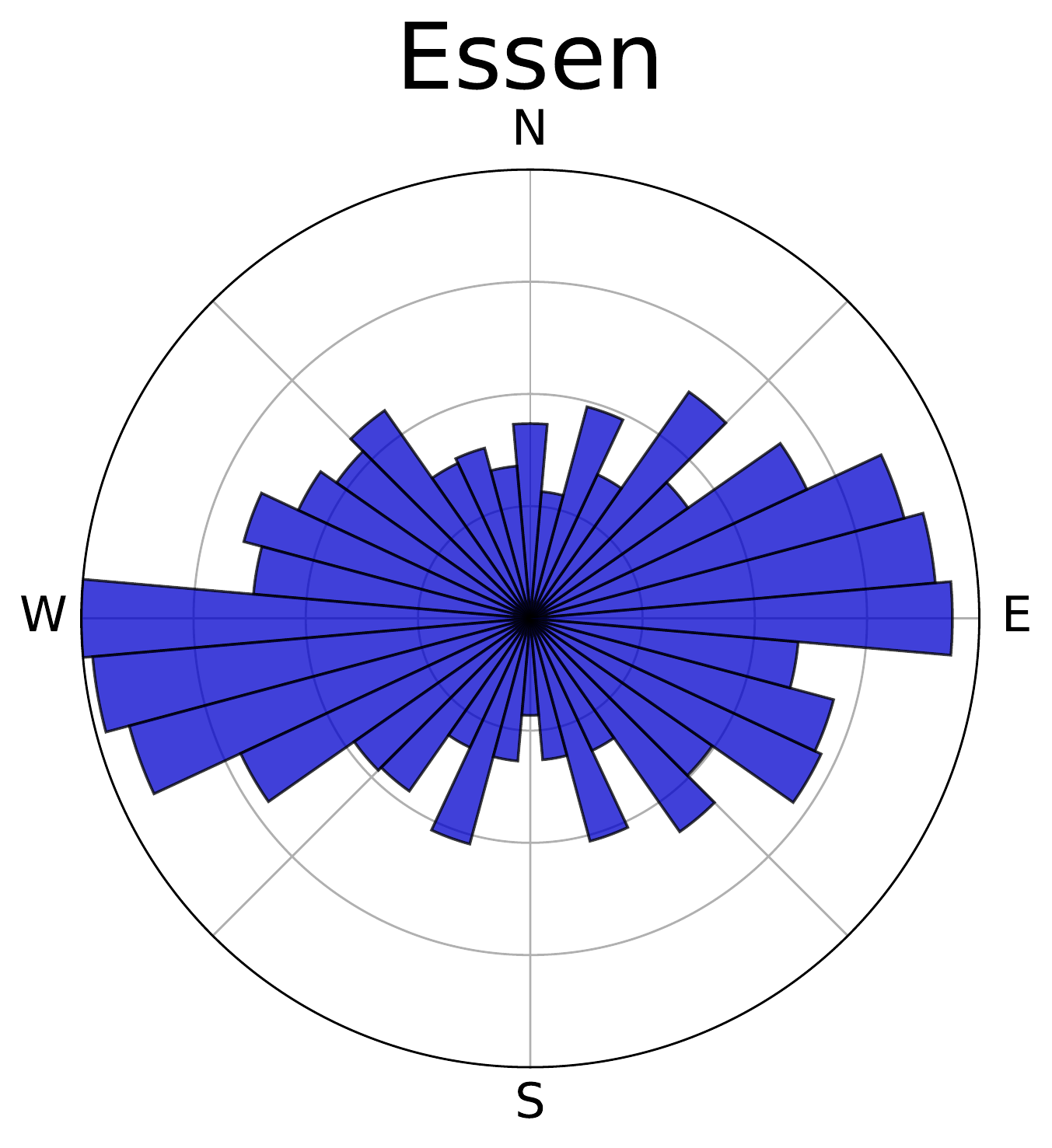}
	\includegraphics[width=.15\textwidth]{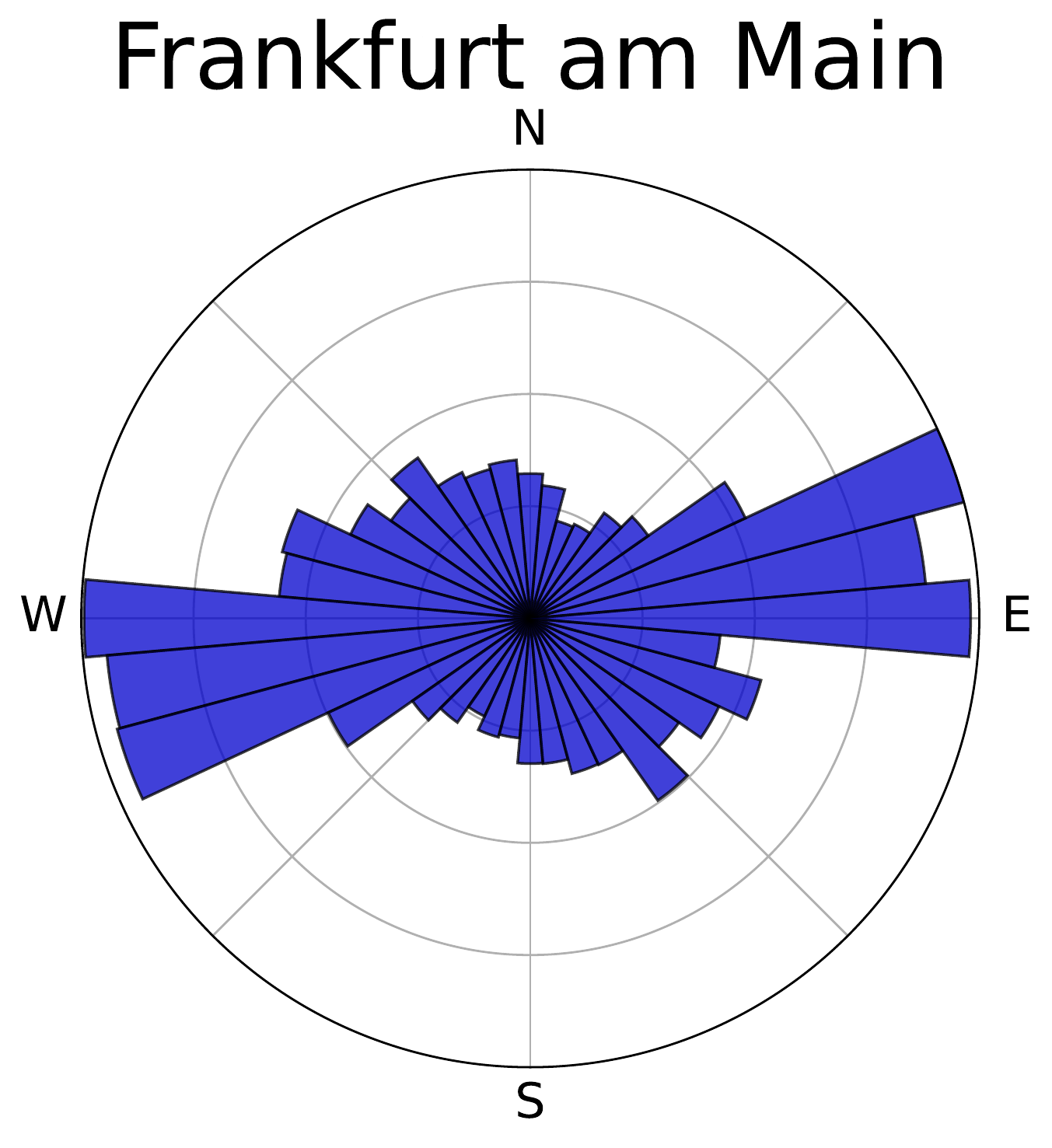}
	\includegraphics[width=.15\textwidth]{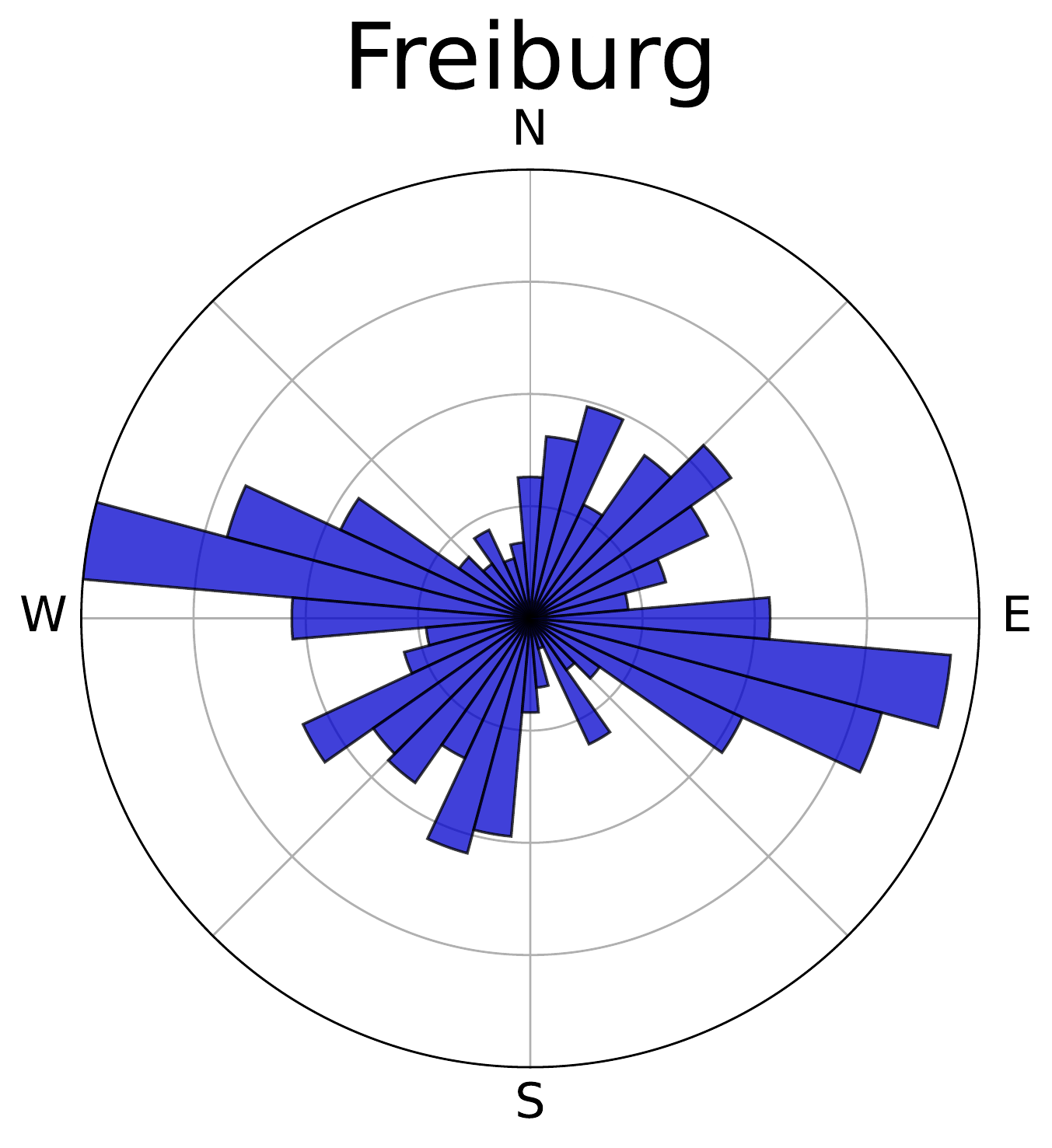}
	\includegraphics[width=.15\textwidth]{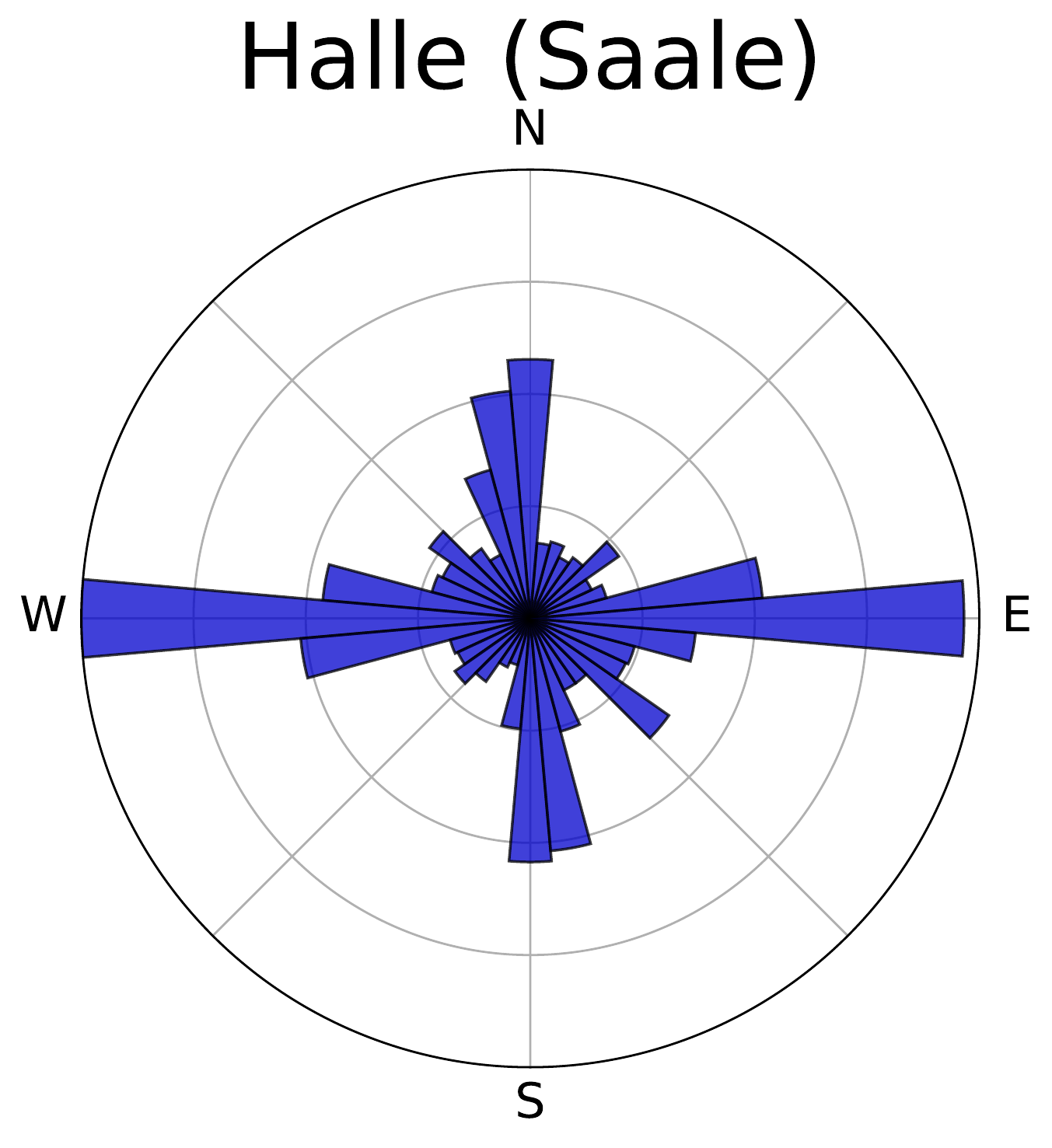}
	
	\includegraphics[width=.15\textwidth]{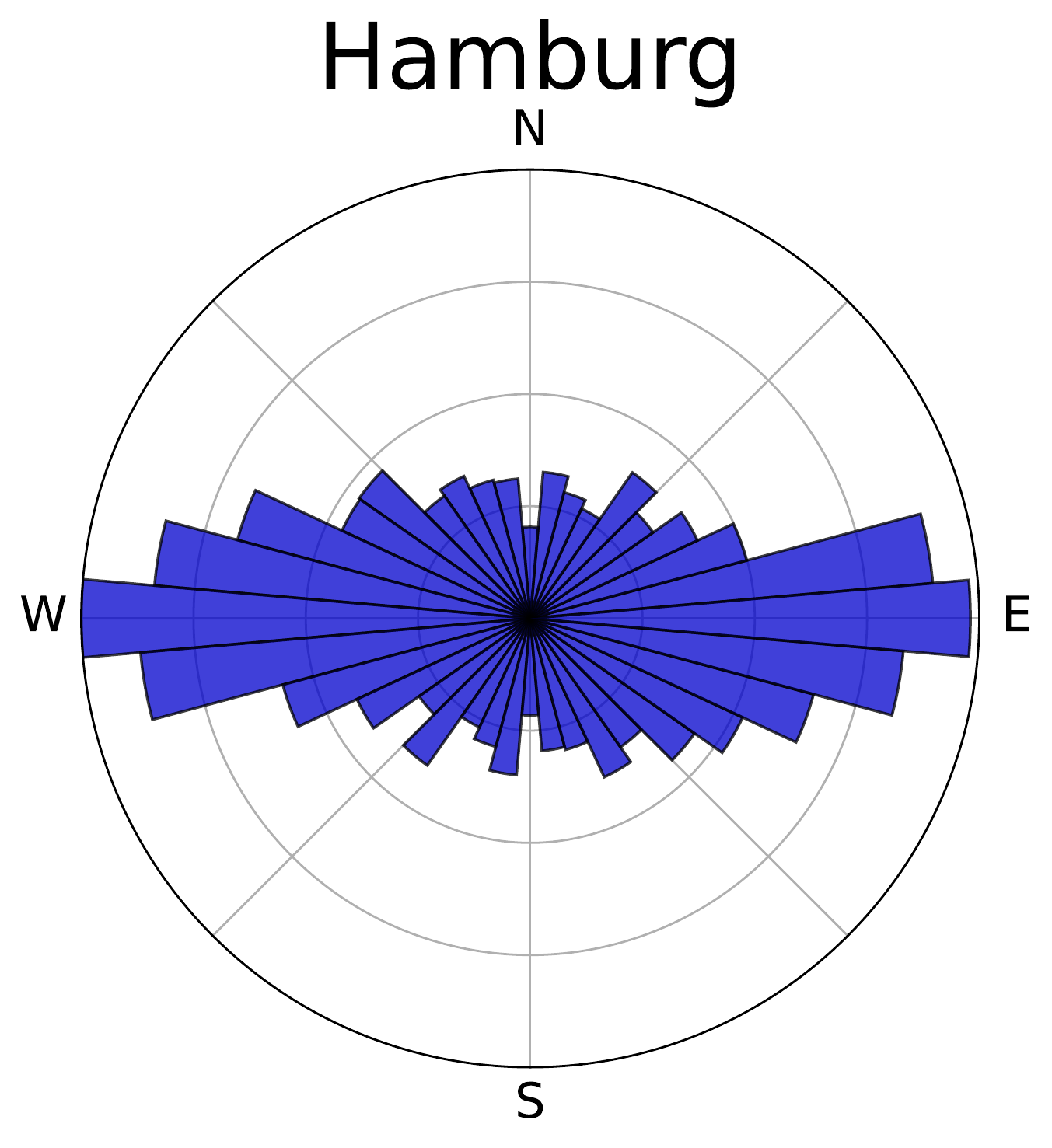}
	\includegraphics[width=.15\textwidth]{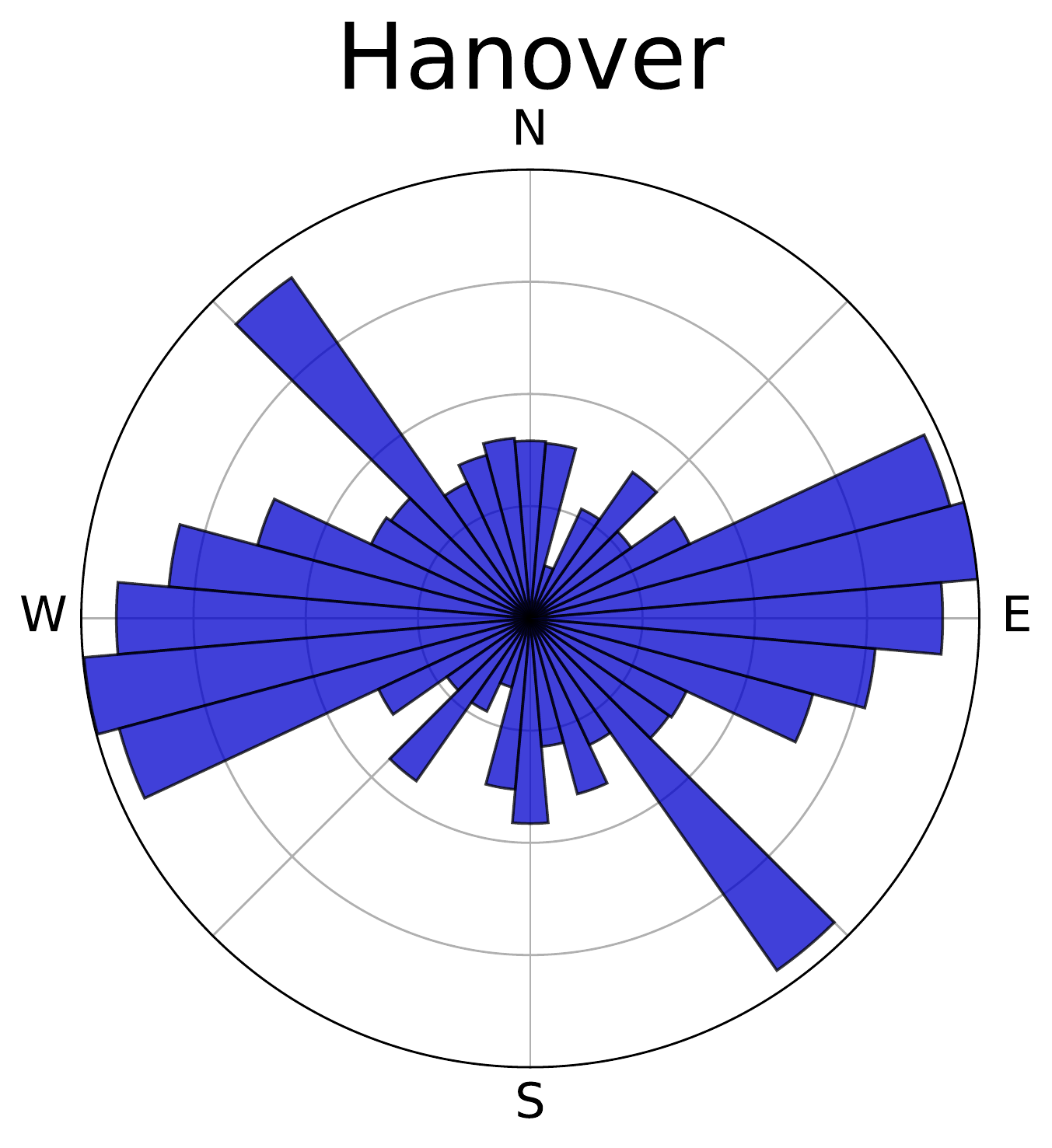}
	\includegraphics[width=.15\textwidth]{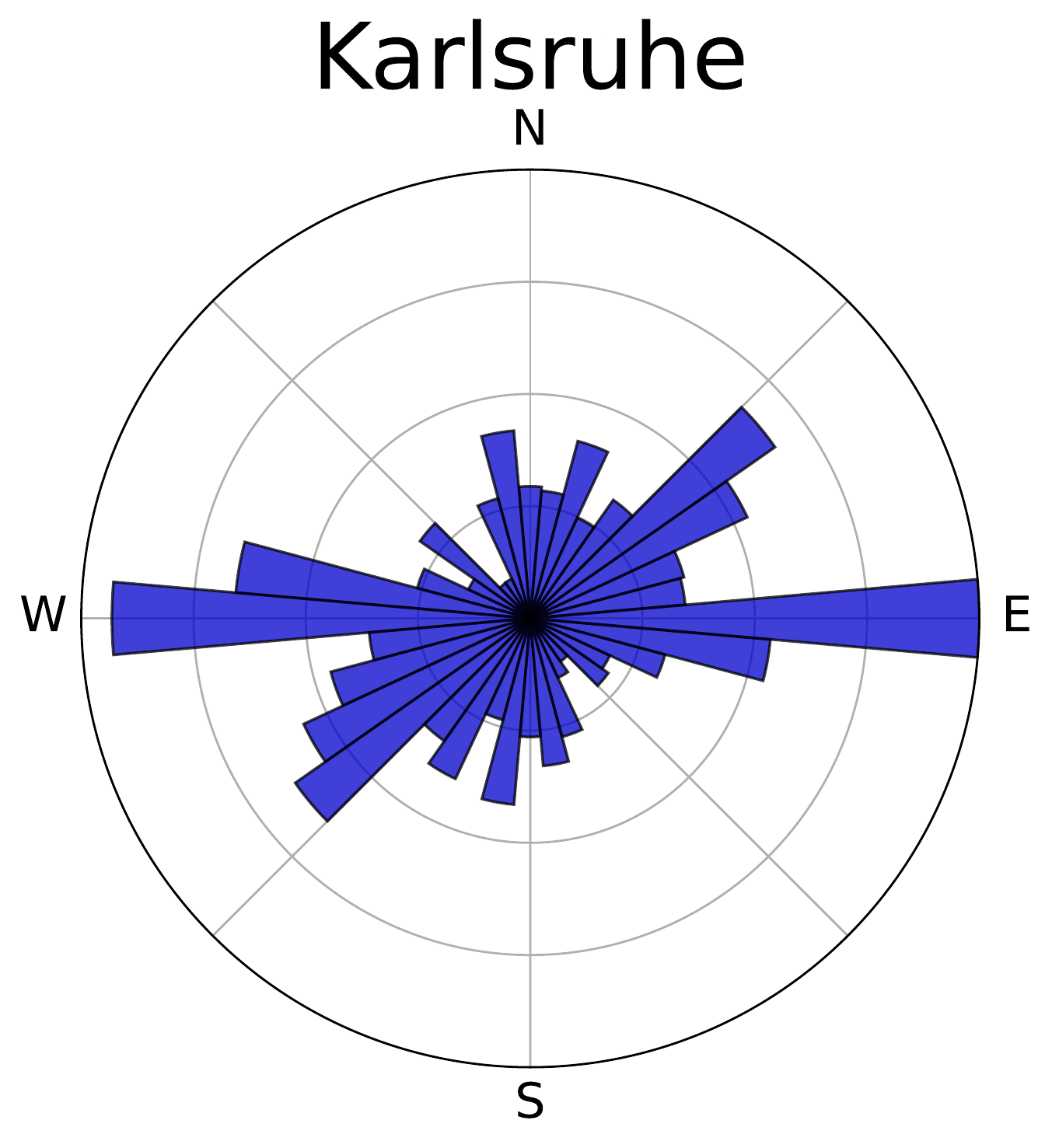}
	\includegraphics[width=.15\textwidth]{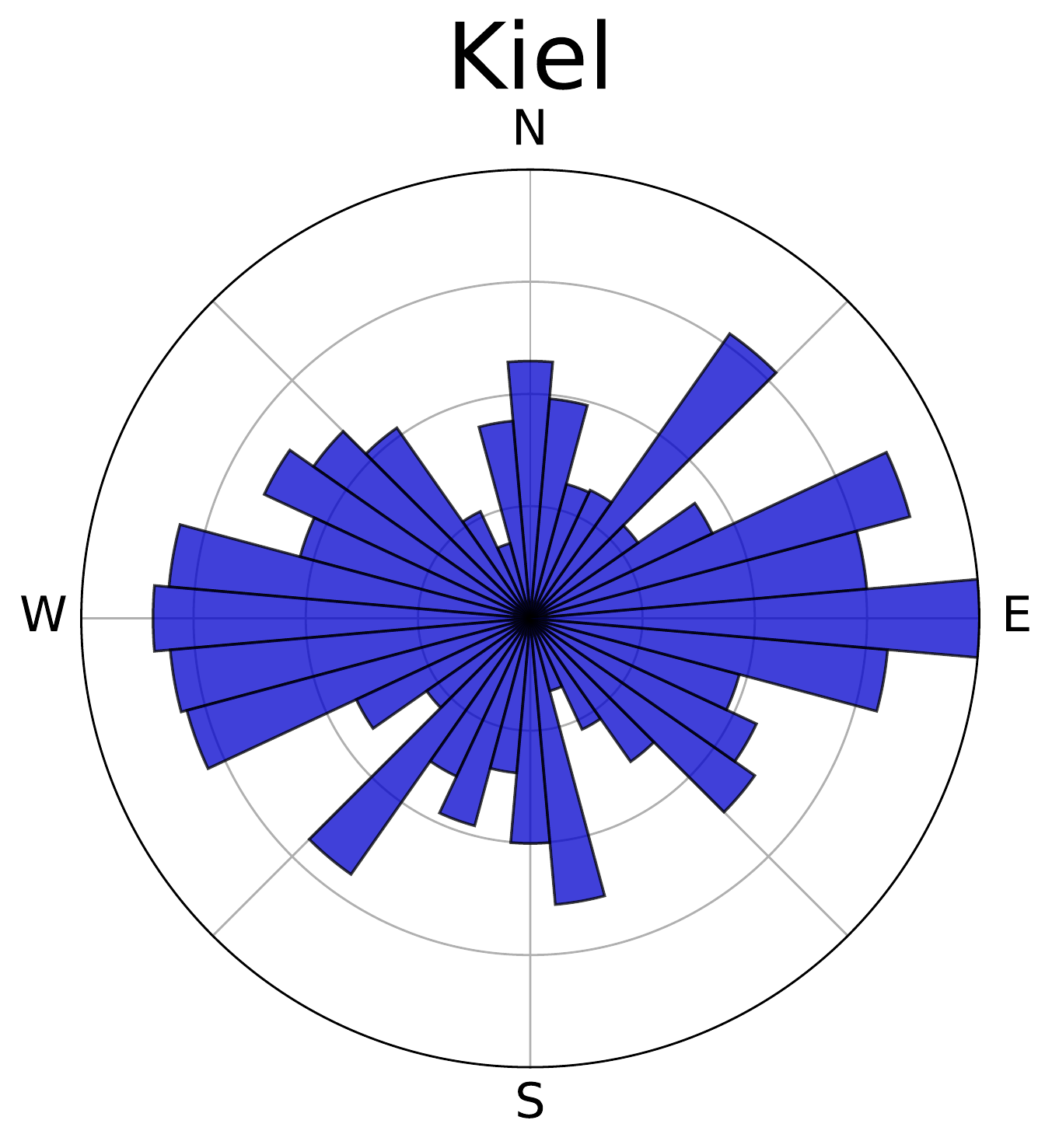}
	\includegraphics[width=.15\textwidth]{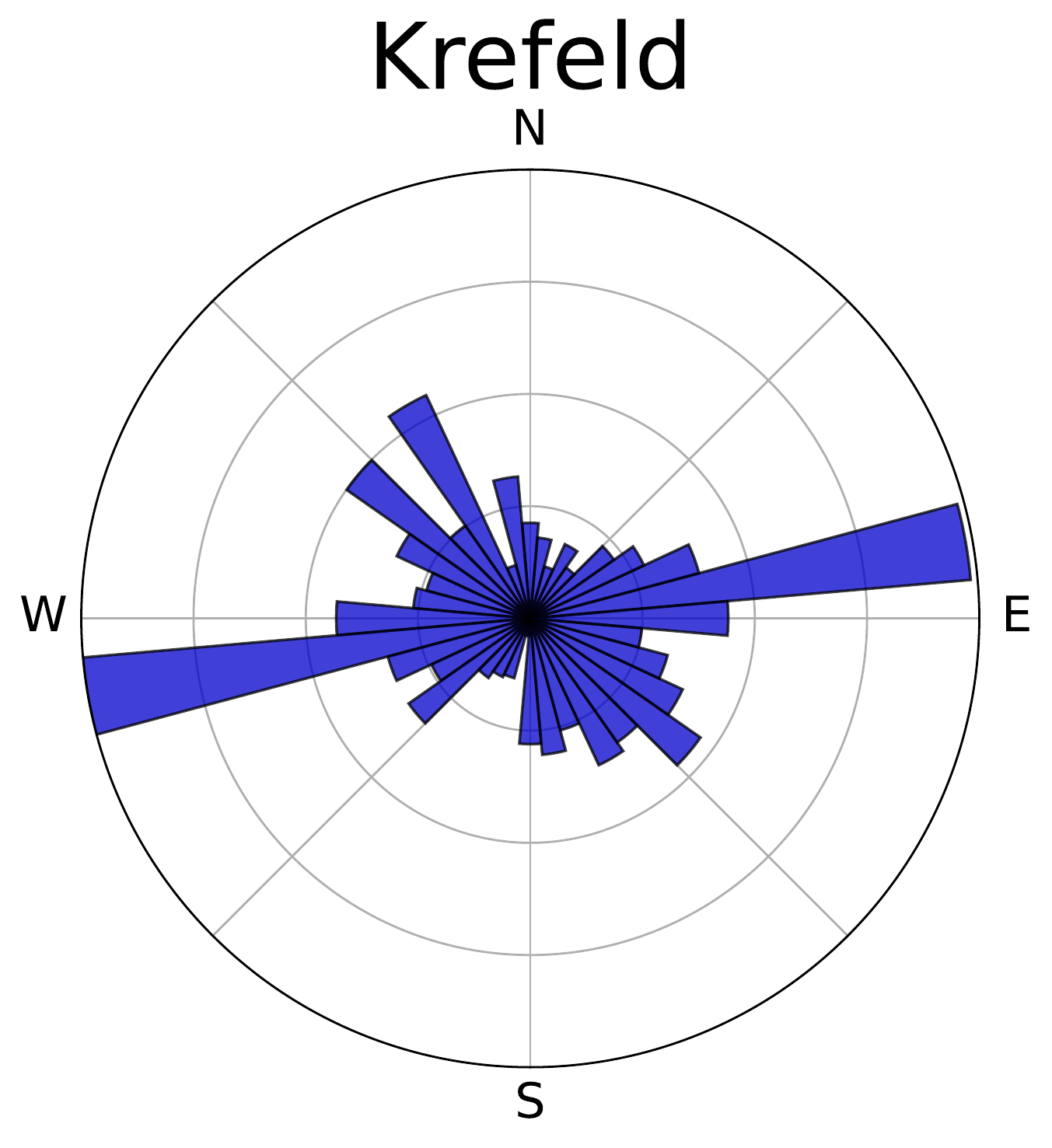}
	\includegraphics[width=.15\textwidth]{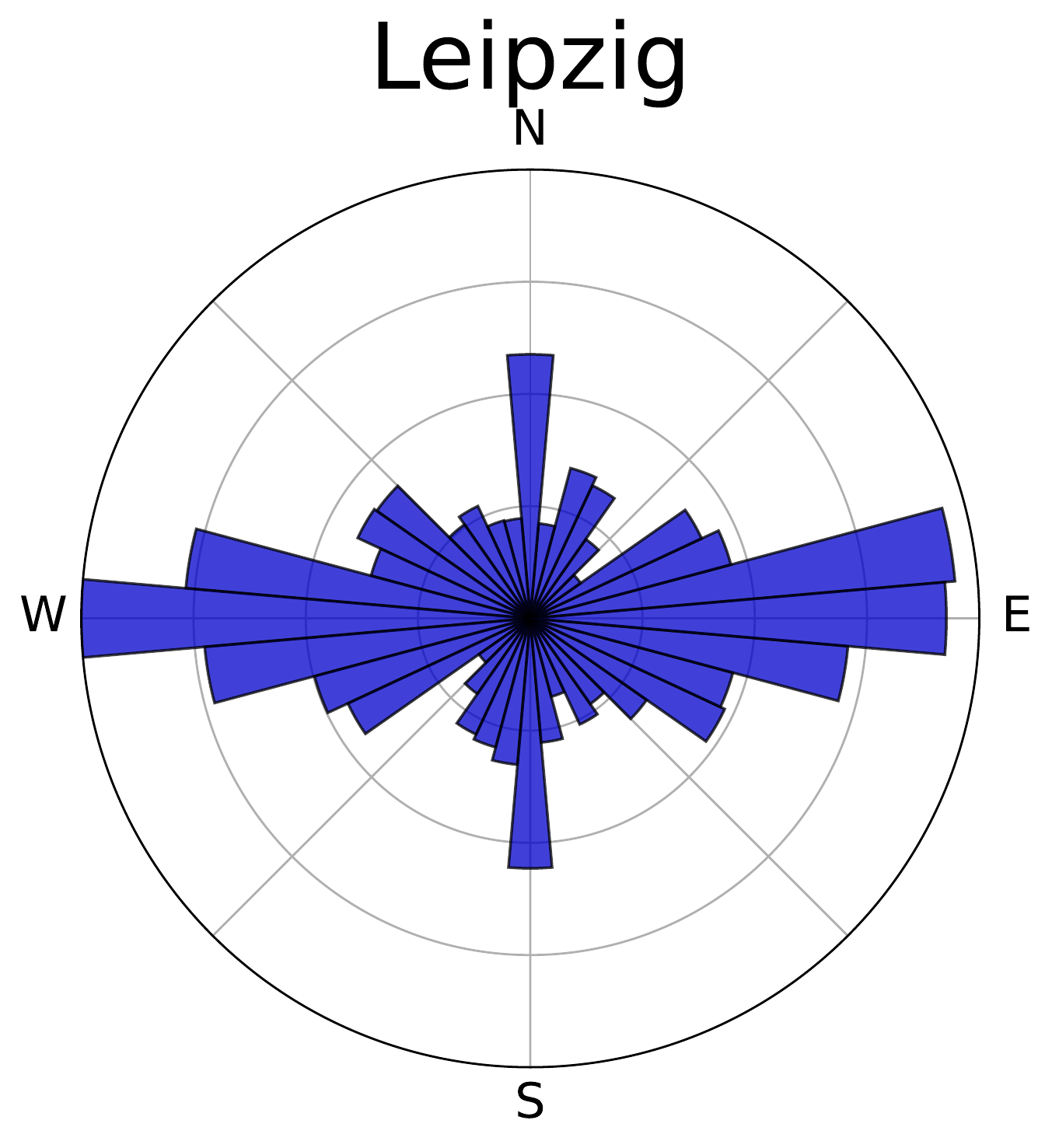}
	
	\includegraphics[width=.15\textwidth]{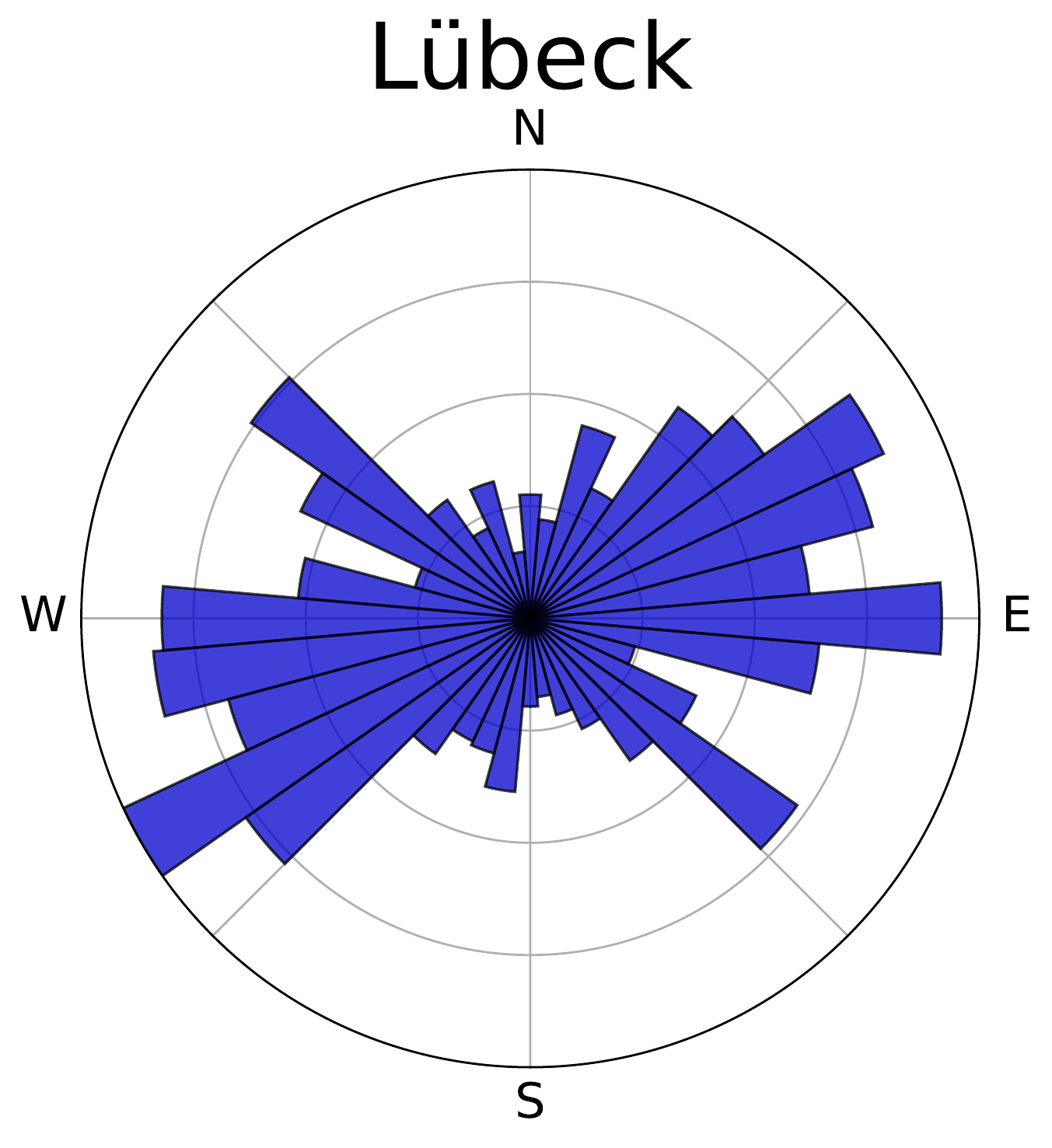}
	\includegraphics[width=.15\textwidth]{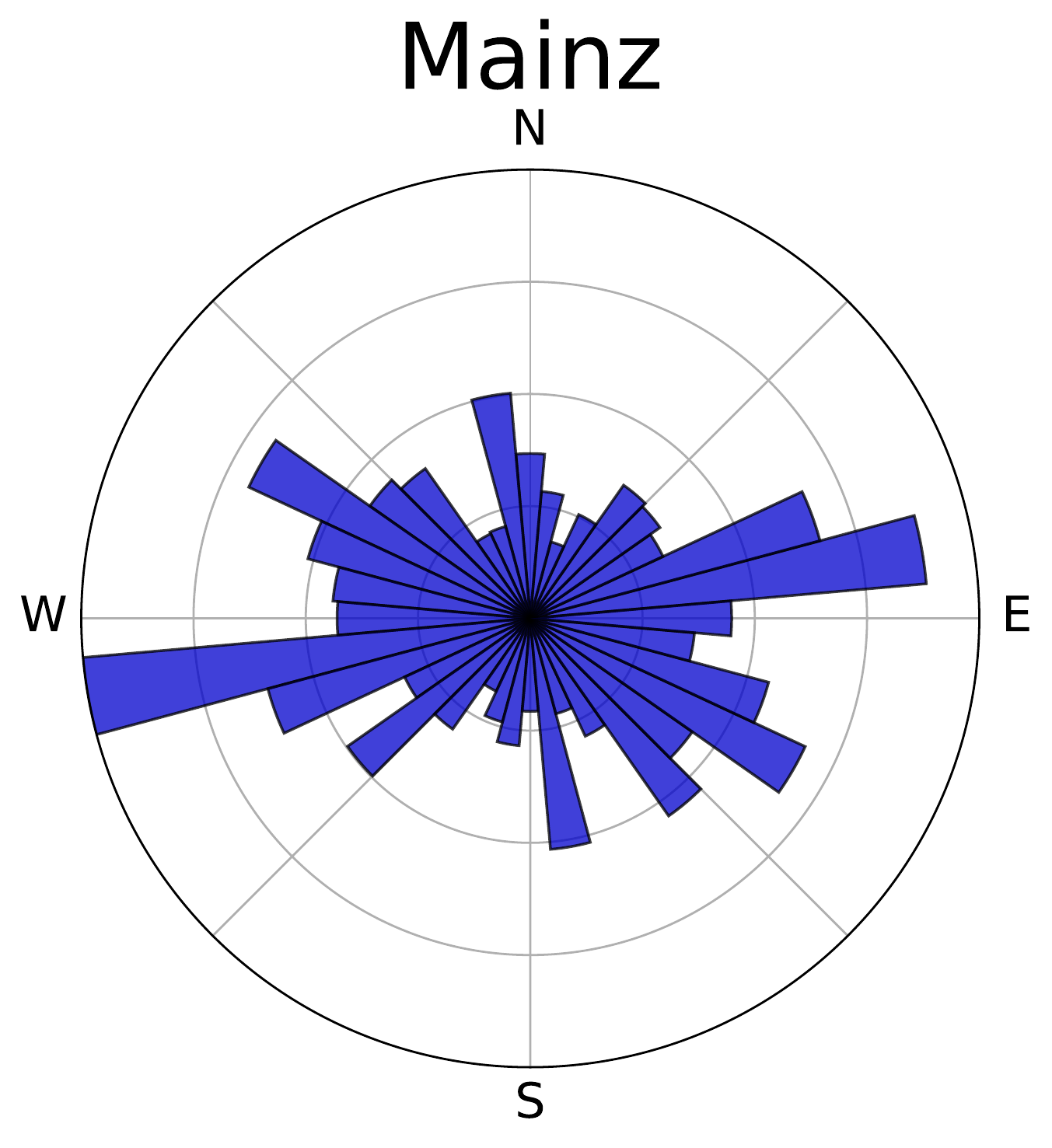}
	\includegraphics[width=.15\textwidth]{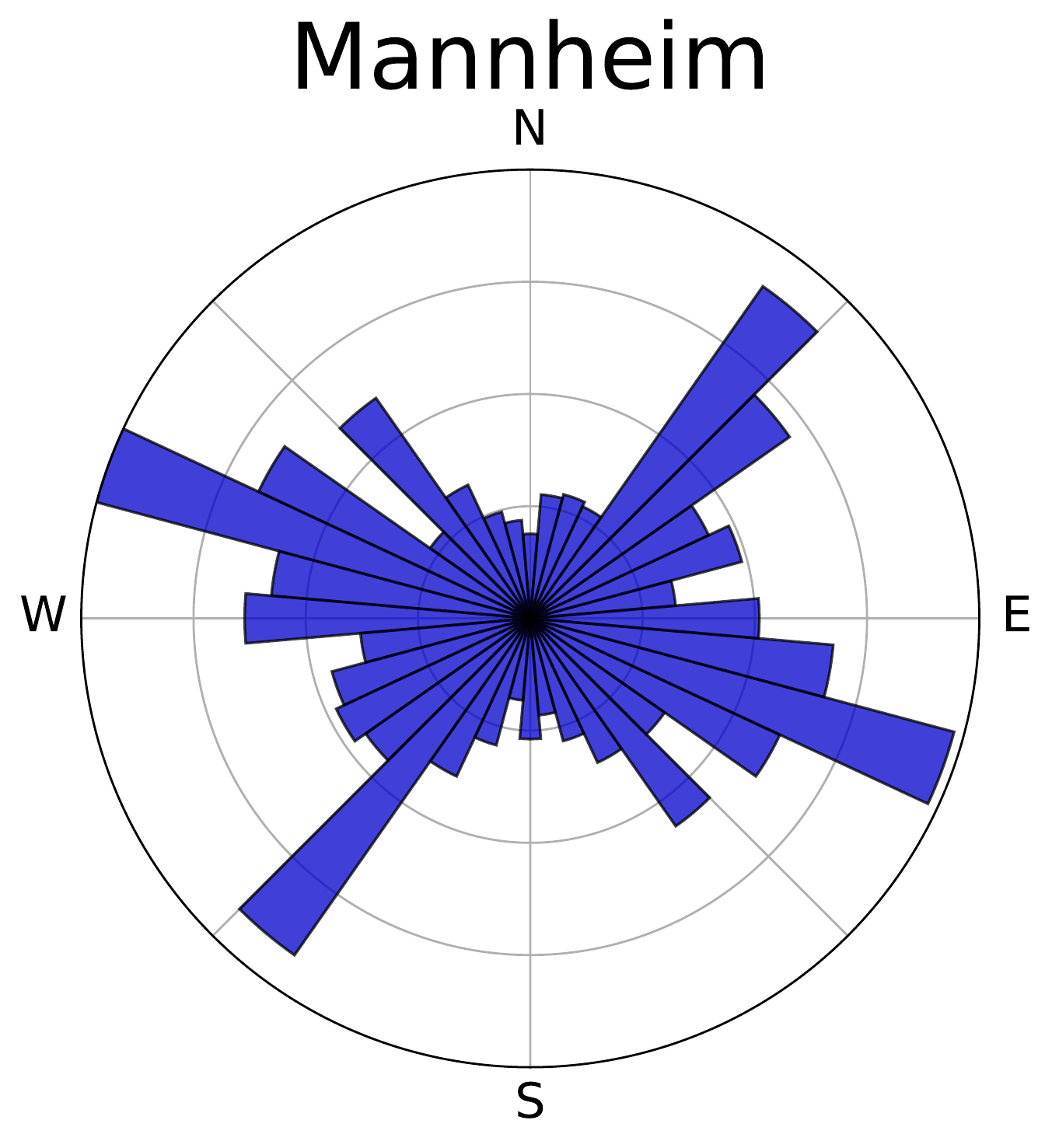}
	\includegraphics[width=.15\textwidth]{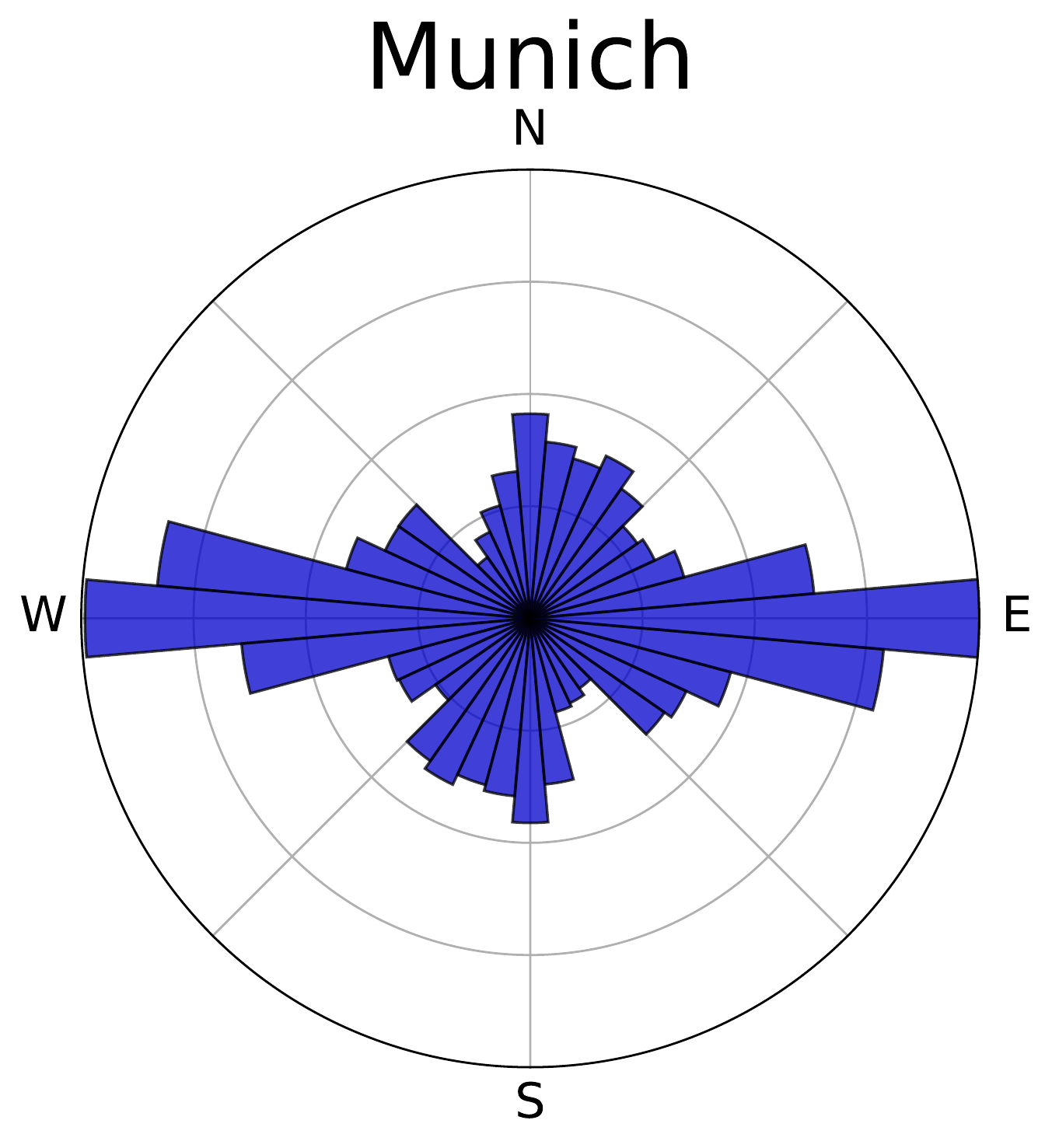}
	\includegraphics[width=.15\textwidth]{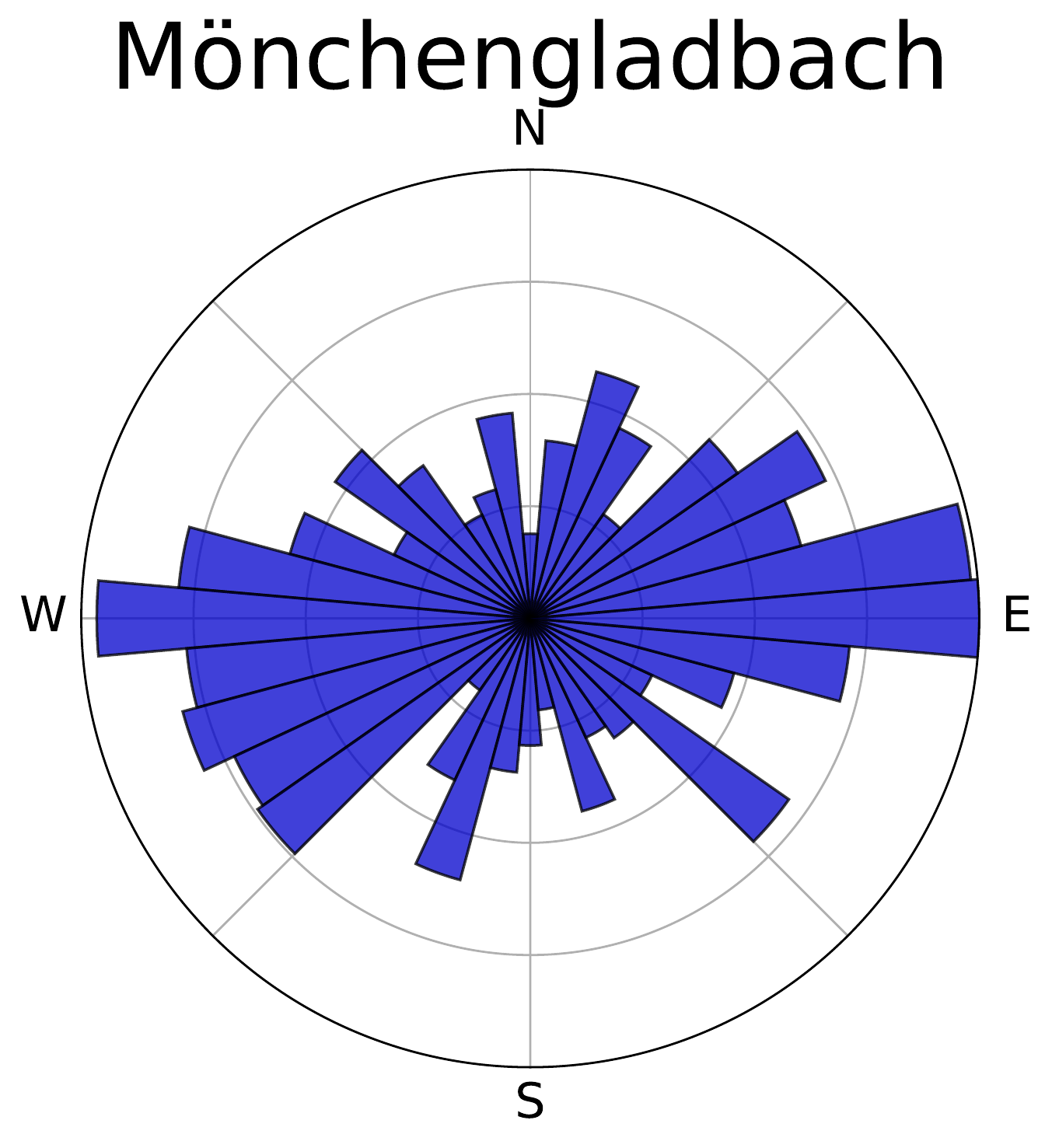}
	\includegraphics[width=.15\textwidth]{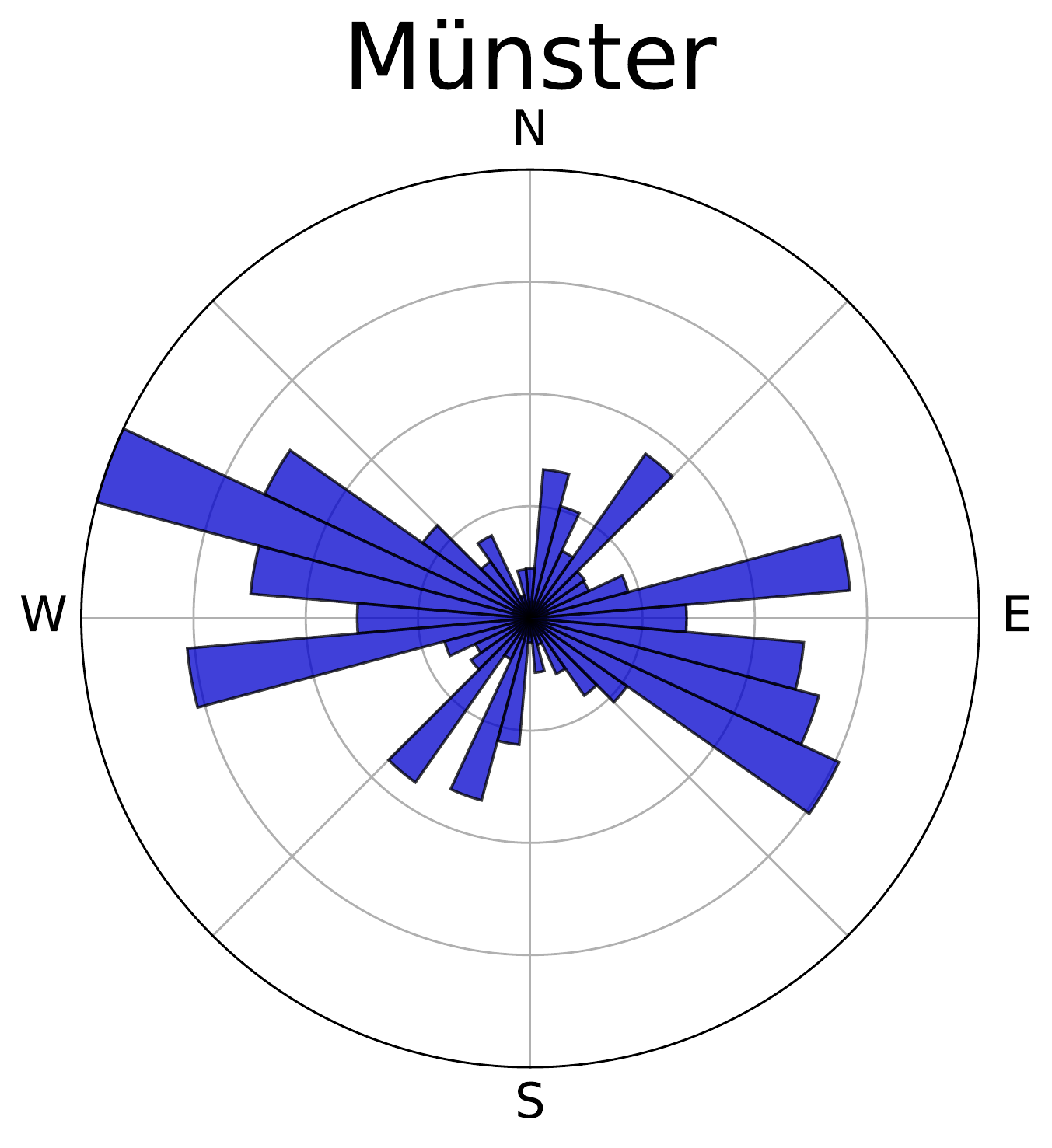}
	
	\includegraphics[width=.15\textwidth]{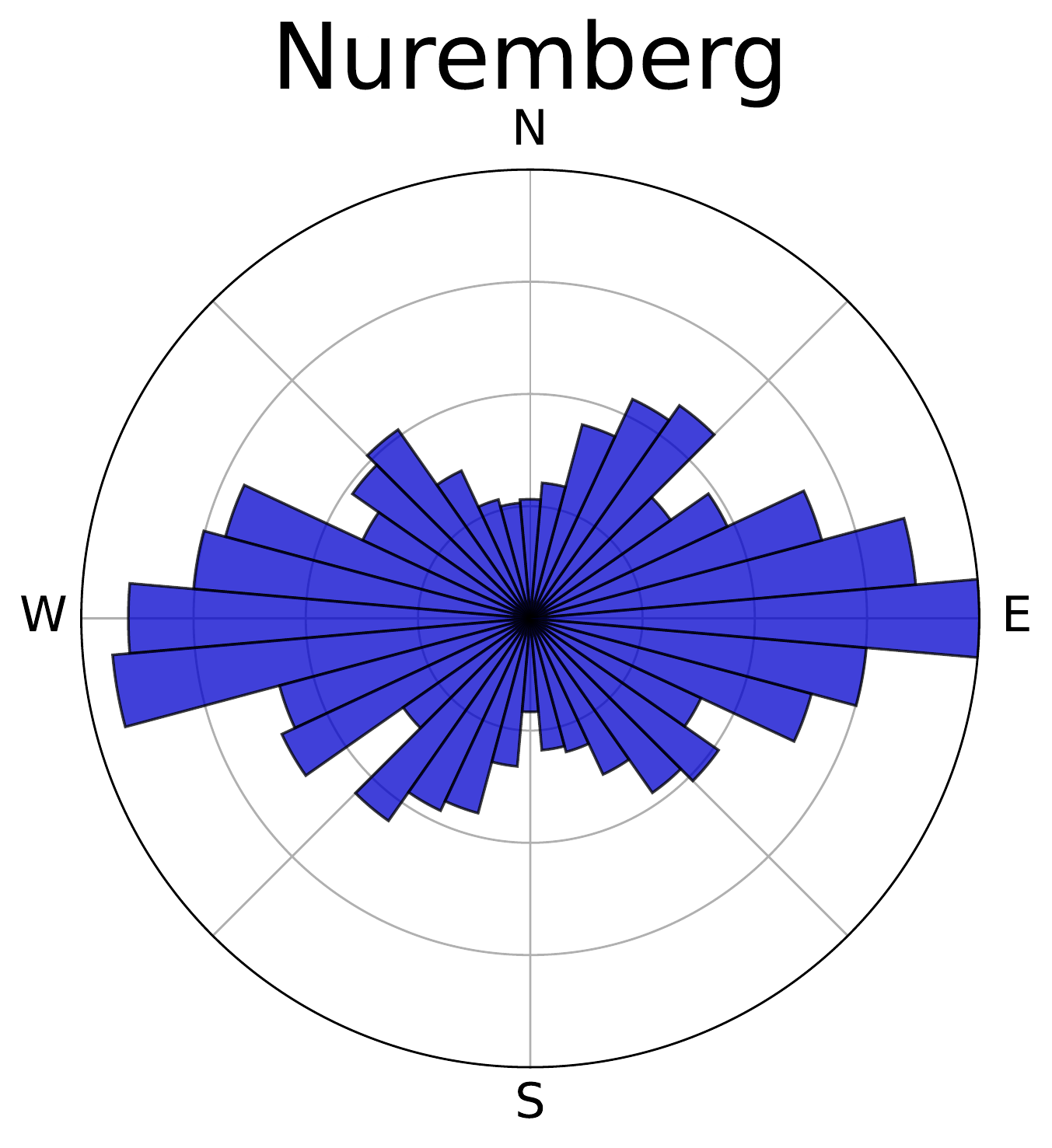}
	\includegraphics[width=.15\textwidth]{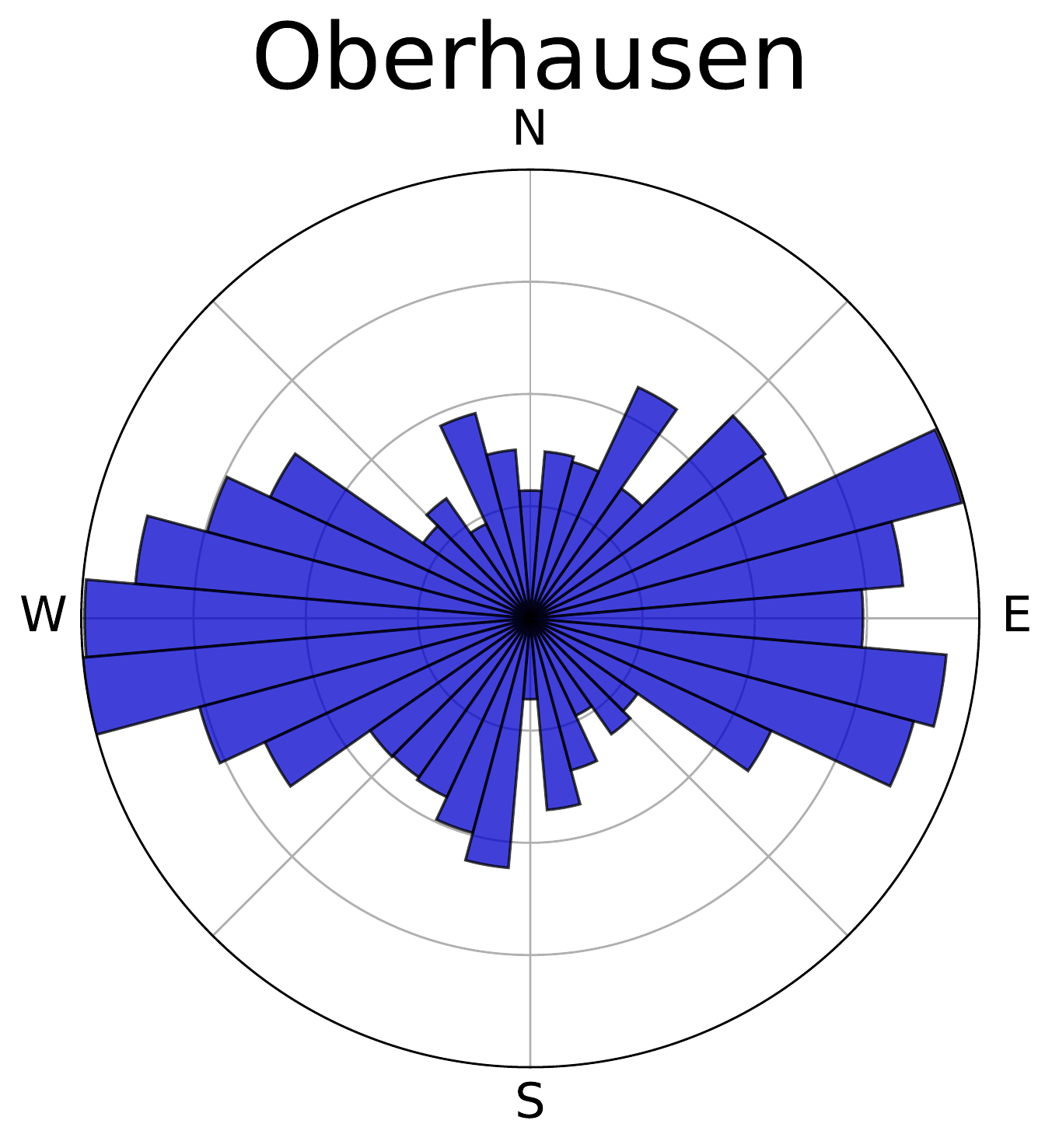}
	\includegraphics[width=.15\textwidth]{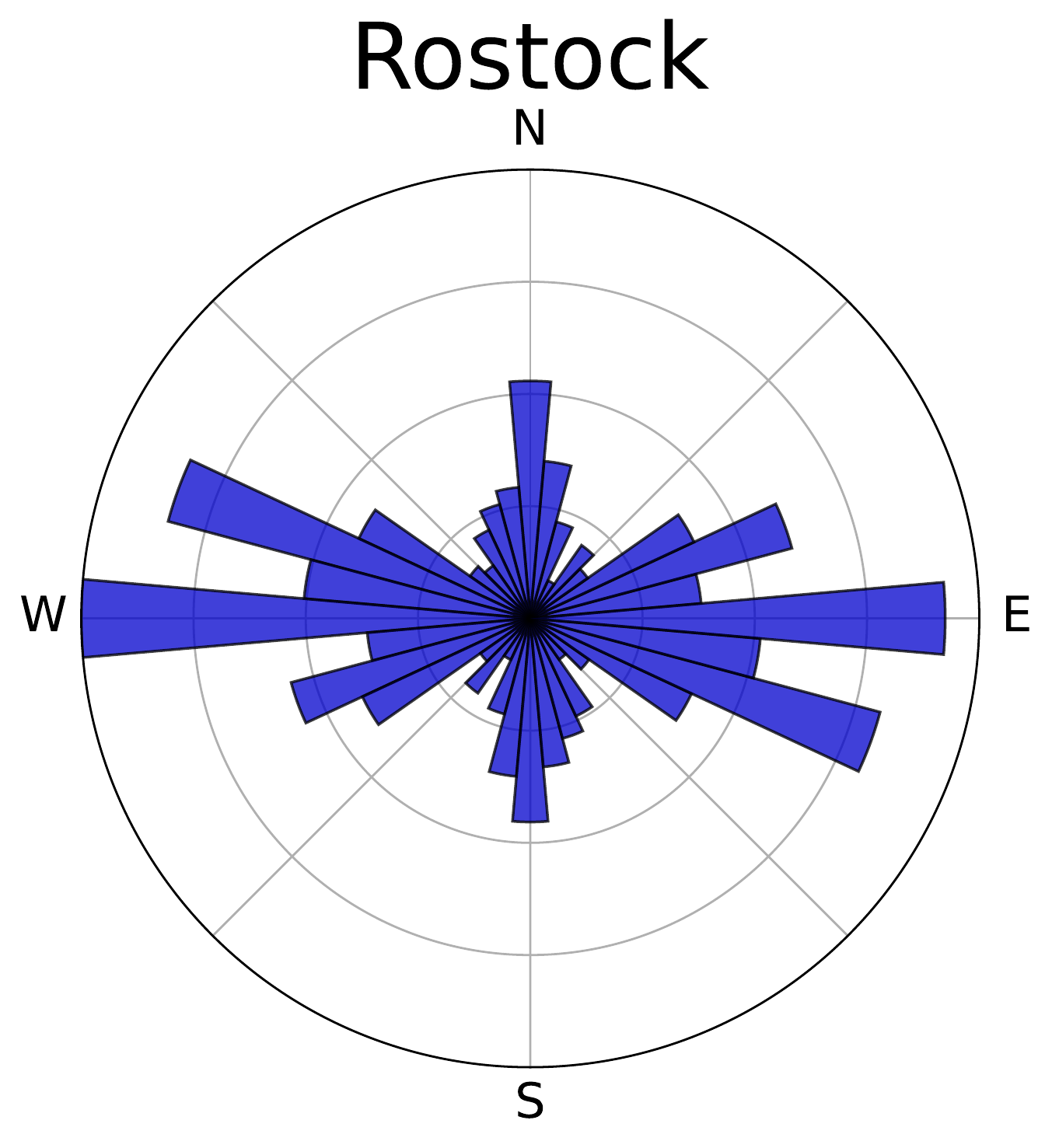}
	\includegraphics[width=.15\textwidth]{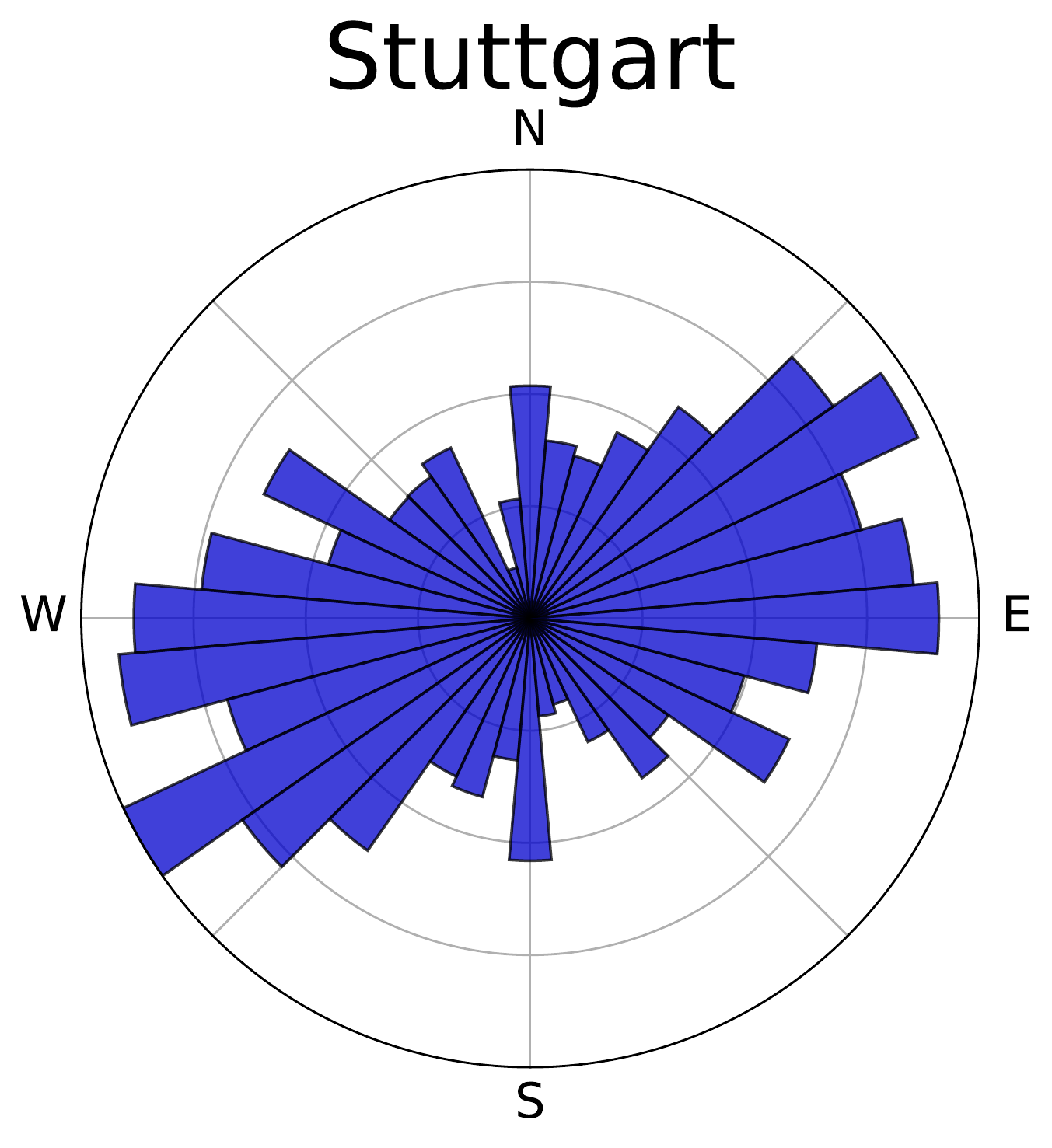}
	\includegraphics[width=.15\textwidth]{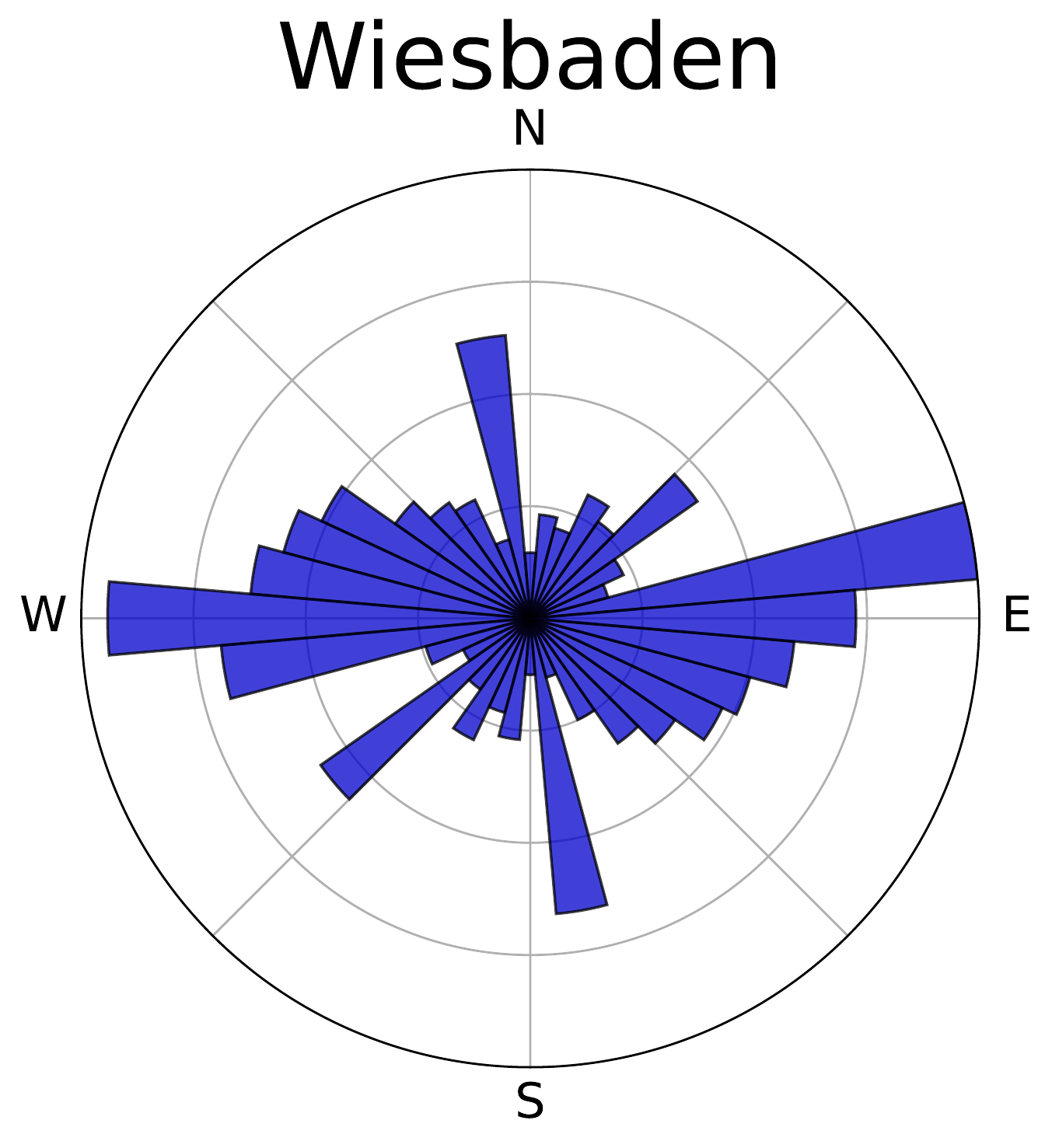}
	\includegraphics[width=.15\textwidth]{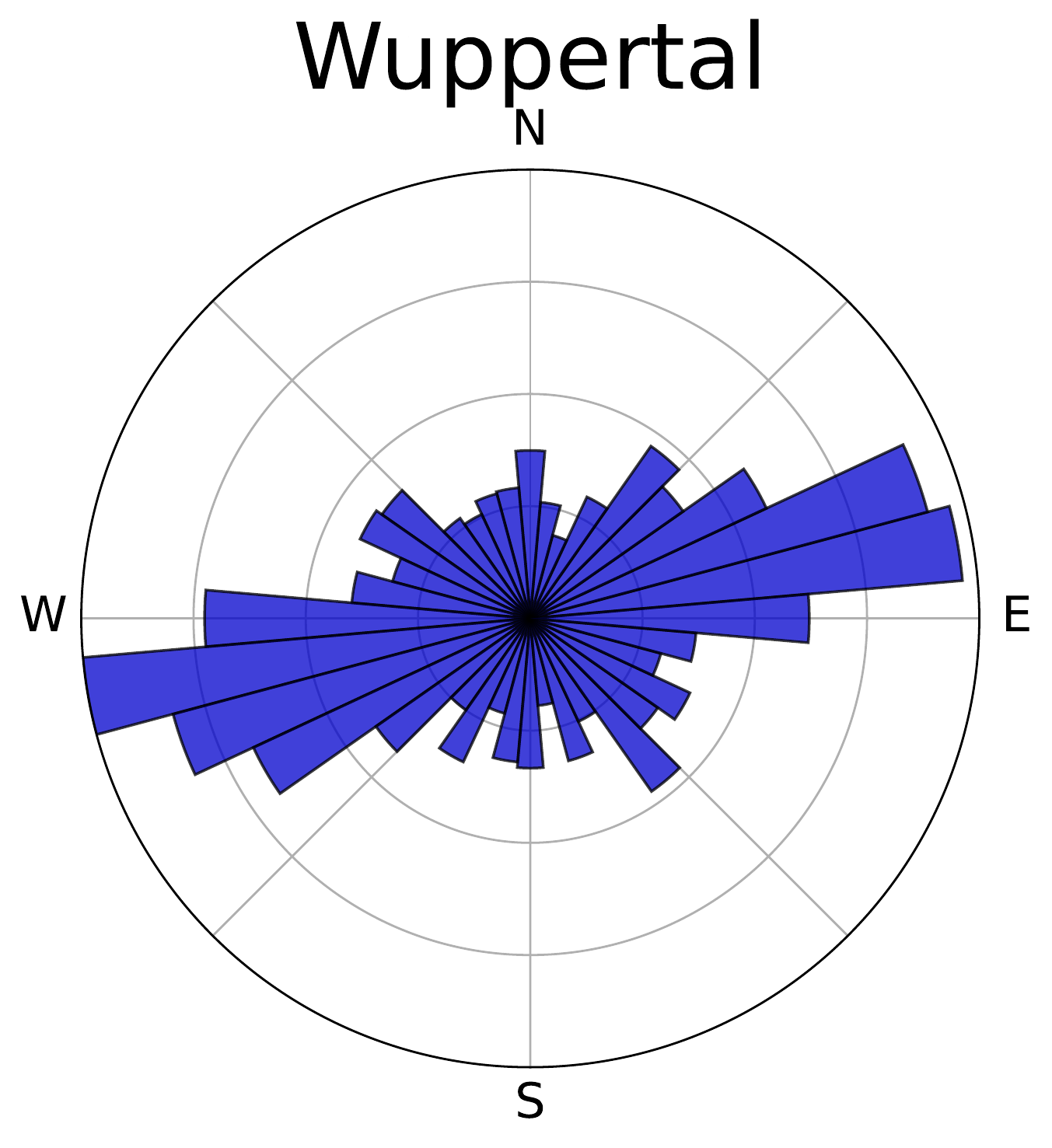}
	\caption{Public transport network orientations of the $36$ largest German cities with valid GTFS data as described in the section \nameref{sec:data}.}\label{fig:orientations_public_transport_germany}
	\end{center}
\end{figure}

\Cref{fig:orientations_public_transport_germany} shows the public transport network orientations for the $36$ considered German cities.
In contrast to the street network orientations of the same cities in \Cref{fig:orientations_streets_germany} the public transport network orientations lack clear orthogonal preferential directions for most cities.
Instead, one often observes only one blurred preferential direction consisting of several bins of the rose diagram with no or only a weakly pronounced second orthogonal direction.
Interestingly, this one blurred preferential direction often approximately coincides with the one preferential direction of the corresponding street network, which is closer to the east-west axis.
None of the considered German cities has its most pronounced public transport direction closer than $45$ degrees to the north-south axis.

\begin{table}
	\scriptsize
	\begin{tabular}{|c|c|cccc|c|c|c|c|c|}
		\hline \hline
		\textbf{City} & \textbf{Nodes} &&& \textbf{Layers} &\multicolumn{1}{c}{}&& \textbf{Node-layer} & \textbf{Non-isolated} & \textbf{Intra-layer} & \textbf{Inter-layer}\\
		&&&&&\multicolumn{1}{c}{}&&\textbf{pairs} &\textbf{node-layer pairs} & \textbf{edges} & \textbf{edges}\\ \cline{3-7}
		&& Tram & Subway & Bus & Other & \textbf{Total} &&&& \\ \hline \hline
		Aachen & $1\,113$ & $0$ & $0$ & $108$ & $0$ & $108$ & $120\,204$ & $4\,872$ & $4\,887$ & $19\,025$\\
		Augsburg & $901$ & $5$ & $0$ & $63$ & $0$ & $68$ & $61\,268$ & $1\,381$ & $1\,340$ & $942$\\
		Berlin & $7\,565$ & $23$ & $9$ & $242$ & $5$ & $279$ & $2\,110\,635$ & $13\,845$ & $14\,062$ & $11\,812$\\
		Bielefeld & $1\,561$ & $8$ & $0$ & $101$ & $0$ & $109$ & $170\,149$ & $4\,668$ & $4\,645$ & $8\,171$\\
		Bonn & $1\,264$ & $9$ & $0$ & $61$ & $0$ & $70$ & $88\,480$ & $3\,864$ & $3\,824$ & $6\,995$\\
		Braunschweig & $954$ & $7$ & $0$ & $45$ & $0$ & $52$ & $49\,608$ & $2\,029$ & $2\,008$ & $2\,188$\\ \hline
		Bremen & $1\,602$ & $22$ & $0$ & $82$ & $0$ & $104$ & $166\,608$ & $4\,256$ & $4\,186$ & $6\,707$\\
		Chemnitz & $1\,168$ & $9$ & $0$ & $70$ & $0$ & $79$ & $92\,272$ & $2\,521$ & $2\,481$ & $2\,920$\\
		Cologne & $2\,084$ & $13$ & $0$ & $92$ & $0$ & $105$ & $218\,820$ & $3\,456$ & $3\,336$ & $2\,217$\\
		Dortmund & $2\,232$ & $0$ & $8$ & $129$ & $2$ & $139$ & $310\,248$ & $5\,399$ & $5\,408$ & $7\,123$\\
		Dresden & $1\,865$ & $13$ & $0$ & $93$ & $4$ & $110$ & $205\,150$ & $4\,198$ & $4\,246$ & $5\,212$\\
		Duisburg & $1\,639$ & $2$ & $1$ & $47$ & $0$ & $50$ & $81\,950$ & $2\,186$ & $2\,140$ & $841$\\ \hline
		D\"usseldorf & $1\,878$ & $7$ & $11$ & $86$ & $0$ & $104$ & $195\,312$ & $3\,850$ & $3\,782$ & $4\,067$\\
		Erfurt & $675$ & $8$ & $0$ & $36$ & $0$ & $44$ & $29\,700$ & $1\,246$ & $1\,315$ & $1\,136$\\
		Essen & $1\,724$ & $8$ & $3$ & $80$ & $0$ & $91$ & $156\,884$ & $2\,814$ & $2\,702$ & $1\,529$\\
		Frankfurt am Main & $2\,118$ & $8$ & $9$ & $96$ & $0$ & $113$ & $239\,334$ & $3\,530$ & $3\,450$ & $2\,203$\\
		Freiburg & $608$ & $5$ & $0$ &$33$ & $0$ & $38$ & $23\,104$ & $954$ & $993$ & $569$\\
		Halle (Saale) & $824$ & $19$ & $0$ & $43$ & $0$ & $62$ & $51\,088$ & $2\,397$ & $2\,468$ & $5\,892$\\ \hline
		Hamburg & $4\,630$ & $0$ & $4$ & $241$ & $8$ & $253$ & $1\,171\,390$ & $10\,193$ & $10\,011$ & $13\,544$\\
		Hanover & $1\,188$ & $0$ & $12$ & $52$ & $0$ & $64$ & $76\,032$ & $1\,881$ & $1\,823$ & $984$\\
		Karlsruhe & $793$ & $9$ & $0$ & $44$ & $0$ & $53$ & $42\,029$ & $1\,526$ & $1\,567$ & $1\,130$\\
		Kiel & $996$ & $0$ & $0$ & $64$ & $2$ & $66$ & $65\,736$ & $2\,511$ & $2\,463$ & $3\,279$\\
		Krefeld & $799$ & $4$ & $2$ & $31$ & $0$ & $37$ & $29\,563$ & $1\,565$ & $1\,576$ & $1\,401$\\
		Leipzig & $1\,748$ & $16$ & $0$ & $64$ & $0$ & $80$ & $139\,840$ & $3\,385$ & $3\,430$ & $4\,193$\\ \hline
		L\"ubeck & $886$ & $0$ & $0$ & $35$ & $0$ & $35$ & $31\,010$ & $1\,619$ & $1\,600$ & $1\,754$\\
		Mainz & $515$ & $5$ & $0$ & $34$ & $0$ & $39$ & $20\,085$ & $1\,753$ & $1\,891$ & $4\,274$\\
		Mannheim & $805$ & $14$ & $0$ & $42$ & $2$ & $58$ & $46\,690$ & $1\,562$ & $1\,639$ & $1\,612$\\
		Munich & $2\,850$ & $24$ & $8$ & $126$ & $0$ & $158$ & $450\,300$ & $5\,258$ & $5\,165$ & $4\,070$\\
		M\"onchengladbach & $1\,196$ & $0$ & $0$ & $39$ & $0$ & $39$ & $46\,644$ & $2\,254$ & $2\,307$ & $1\,917$\\
		M\"unster & $479$ & $0$ & $0$ & $42$ & $0$ & $42$ & $20\,118$ & $1\,187$ & $1\,160$ & $1\,960$\\ \hline
		Nuremberg & $1\,430$ & $6$ & $3$ & $67$ & $0$ & $76$ & $108\,680$ & $2\,359$ & $2\,315$ & $1\,487$\\
		Oberhausen & $1\,060$ & $2$ & $0$ & $86$ & $0$ & $88$ & $93\,280$ & $3\,065$ & $2\,996$ & $8\,478$\\
		Rostock & $336$ & $6$ & $0$ & $39$ & $2$ & $47$ & $15\,792$ & $793$ & $817$ & $915$\\
		Stuttgart & $1\,333$ & $1$ & $17$ & $70$ & $2$ & $90$ & $119\,970$ & $2\,154$ & $2\,121$ & $1\,423$\\
		Wiesbaden & $1\,042$ & $0$ & $0$ & $68$ & $0$ & $68$ & $70\,856$ & $2\,814$ & $2\,857$ & $4\,920$\\
		Wuppertal & $1\,688$ & $0$ & $0$ & $175$ & $1$ & $176$ & $297\,088$ & $4\,815$ & $4\,708$ & $10\,929$\\ \hline \hline
	\end{tabular}
	\normalsize
	\caption{Multiplex network size of the $36$ largest German cities with valid GTFS data as described in the section \nameref{sec:data}.
		Nodes correspond to stops and layers correspond to public transport lines, i.e.~e.g., the public transport network of Augsburg consists of $5$ tram lines, $63$ bus lines, and a total of $68$ public transport lines.
		Each line only serves a subset of all stops of the city, which is typically small compared to the full network.
		This entails that most node-layer pairs in the individual layers and consequently in the full networks are isolated, i.e, possess no incident edge.
		Hence, the number of non-isolated node-layer pairs and the number of intra-layer edges is similar for all considered networks.}\label{tab:germany_nodes_layers}
\end{table}

Moreover, the blurred tilted east-west-like preferential directions of the public transport networks are often accompanied by a visually rather discontinuous distribution of bearings in the remaining directions.
As there is a tendency for more continuous bearing distributions for cities with many public transport lines we conjecture that part of the recorded orientations can be considered random, e.g., due to the effect induced by taking turns described above, which decouples public transport orientations from street network orientations.
Cities with more lines and consequently a larger sample size of different bearings seem to get closer to the actual, relatively smooth bearing distribution.
For instance, the bearing distribution of Mainz with only $39$ lines appears much more discontinuous than that of Hamburg with $253$ lines.

By the choice of considering directed edges in public transport networks the corresponding orientation diagrams are nonsymmetric.
However, the degree of non-symmetry in \Cref{fig:orientations_public_transport_germany} is rather small indicating that most lines serve the same sequence of stops roughly equally in both travel directions.

Comparing the two approaches of considering all ``trips'' per ``route'' against considering only unique ``trips'' the orientation plots using only unique trips are visually almost identical to those displayed in \Cref{fig:orientations_public_transport_germany} for almost all $36$ cities.
Very few cities like Freiburg and Krefeld, which both comprise of few lines contain somewhat less pronounced tilted east-west axes, implying that the respective bearings belong to highly frequented routes.

\begin{figure}
	\begin{center}
	\includegraphics[width=.95\textwidth]{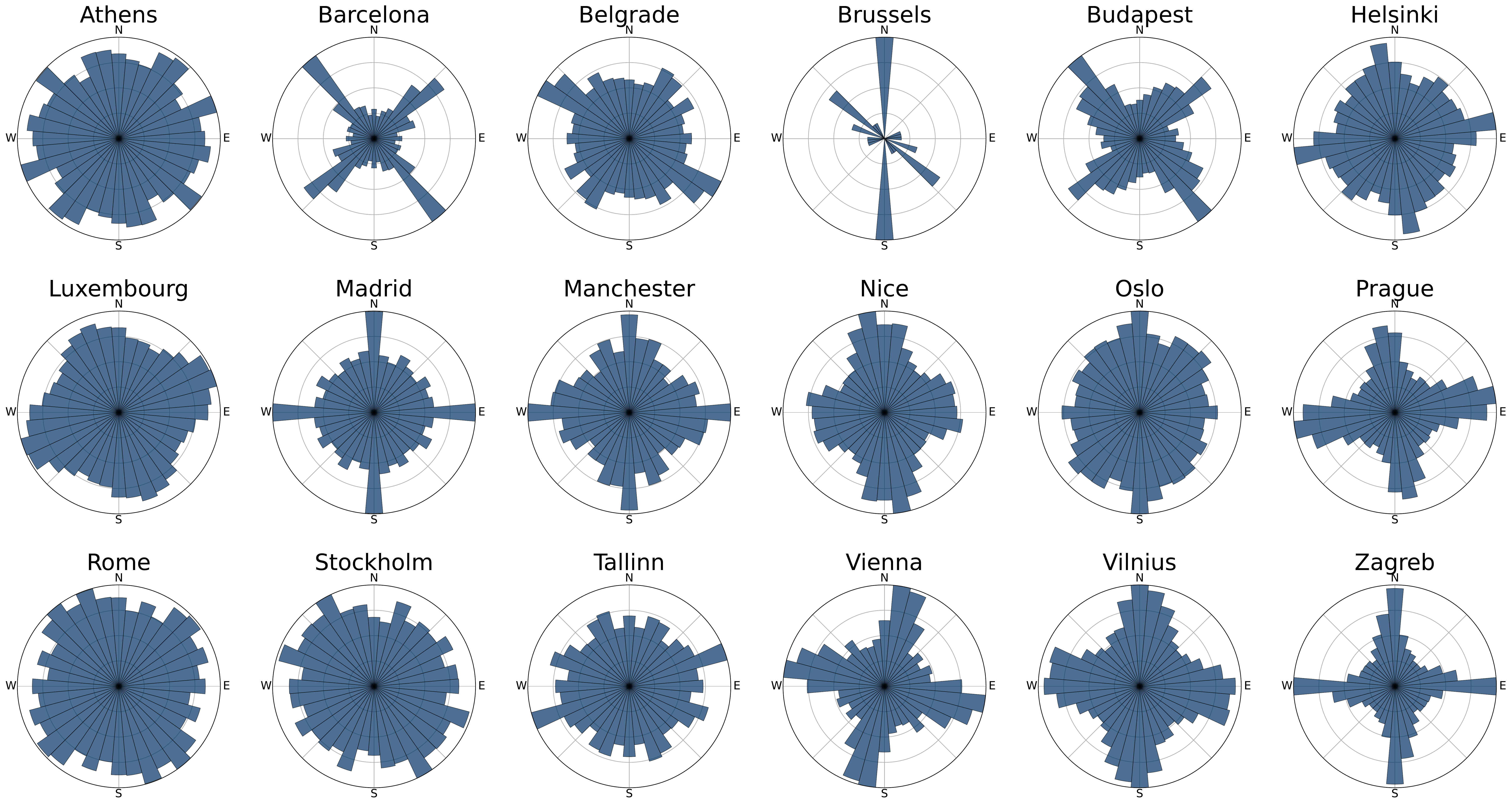}
	\caption{Street network orientations of $18$ selected European cities for which valid GTFS data could be obtained via \url{https://transitfeeds.com/l/60-europe}. The plots were created using the OSMnx python package \cite{boeing2017osmnx}.}\label{fig:orientations_streets_europe}
	\end{center}
\end{figure}

\begin{figure}
	\begin{center}
	\includegraphics[width=.15\textwidth]{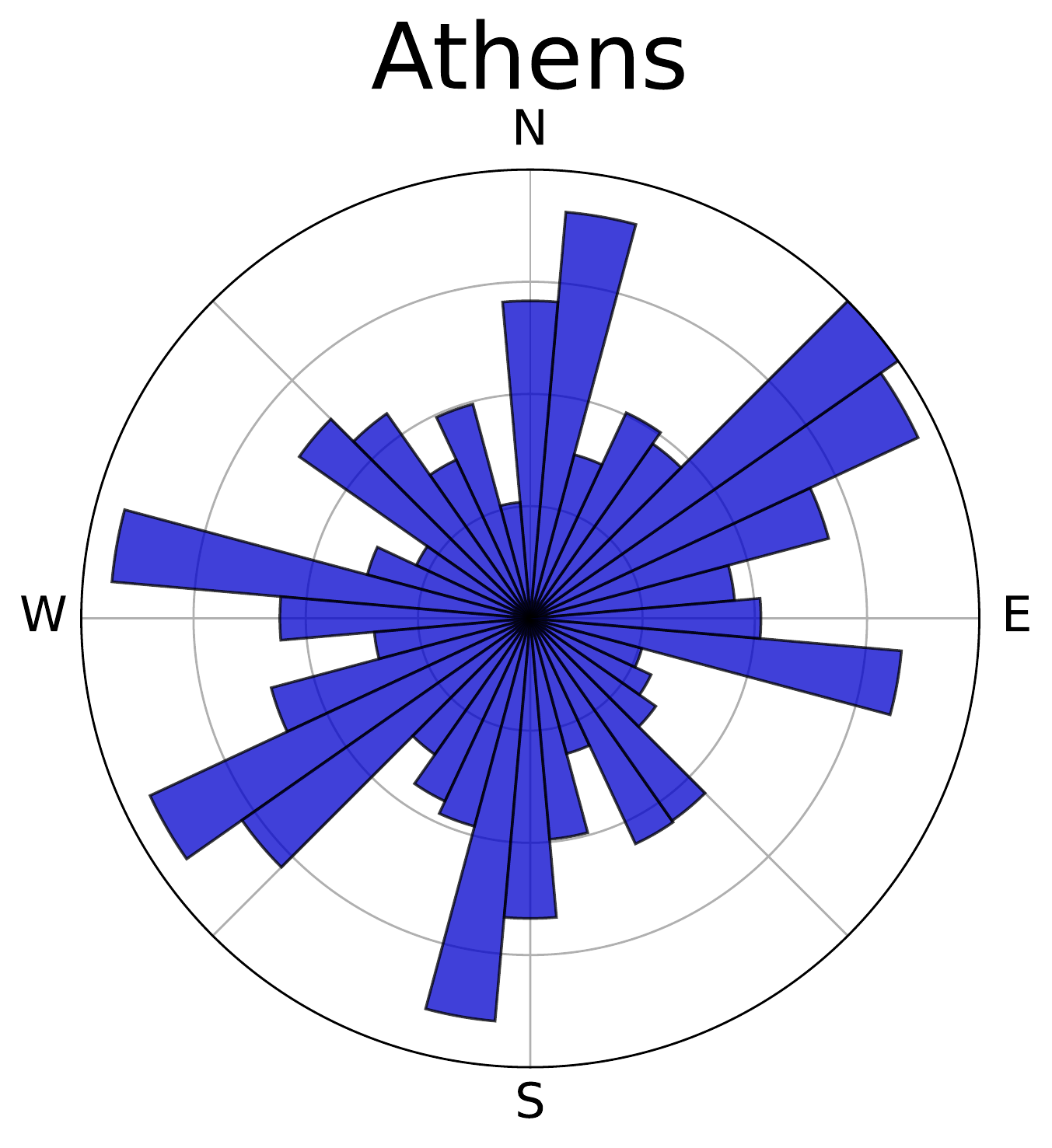}
	\includegraphics[width=.15\textwidth]{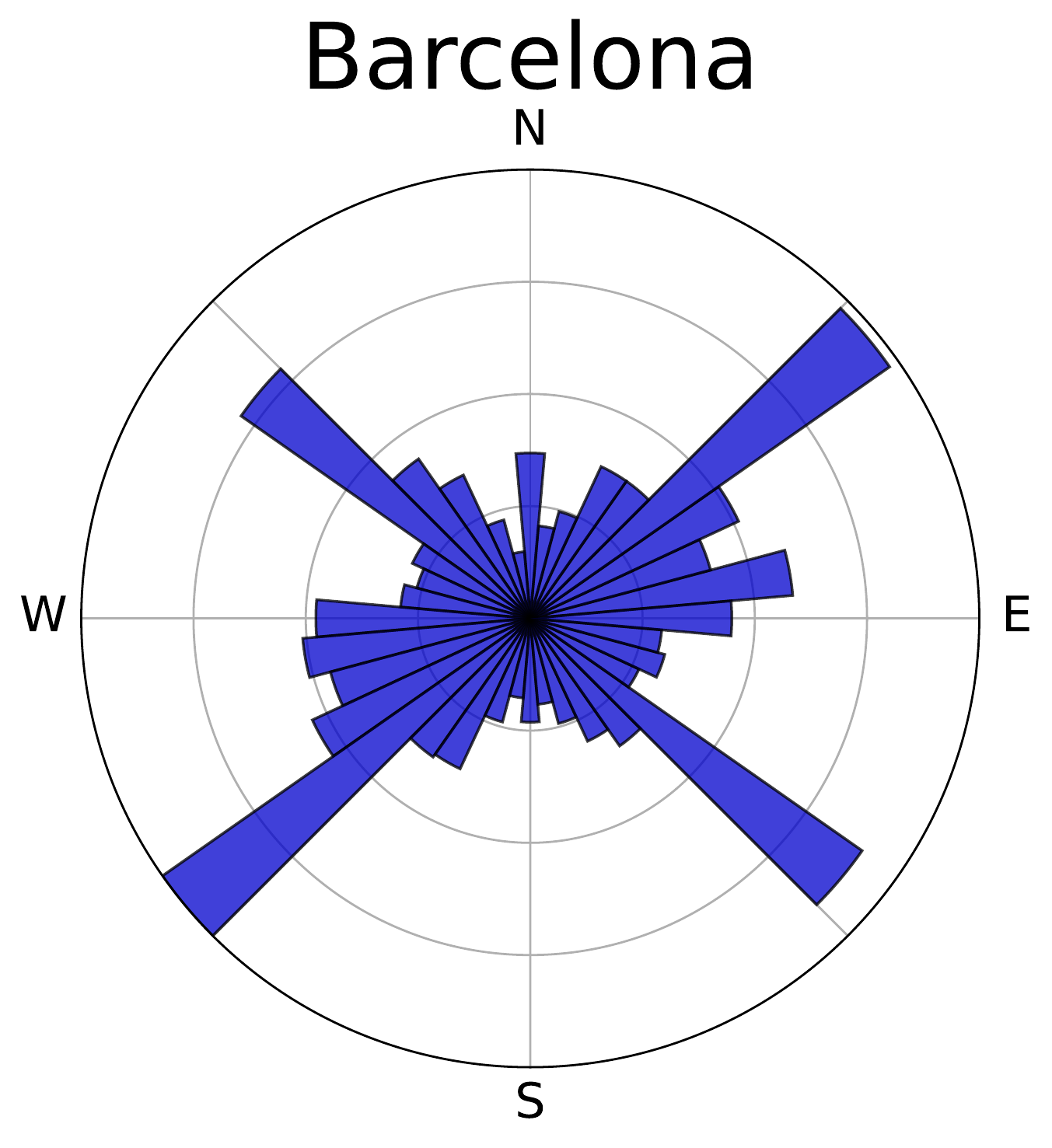}
	\includegraphics[width=.15\textwidth]{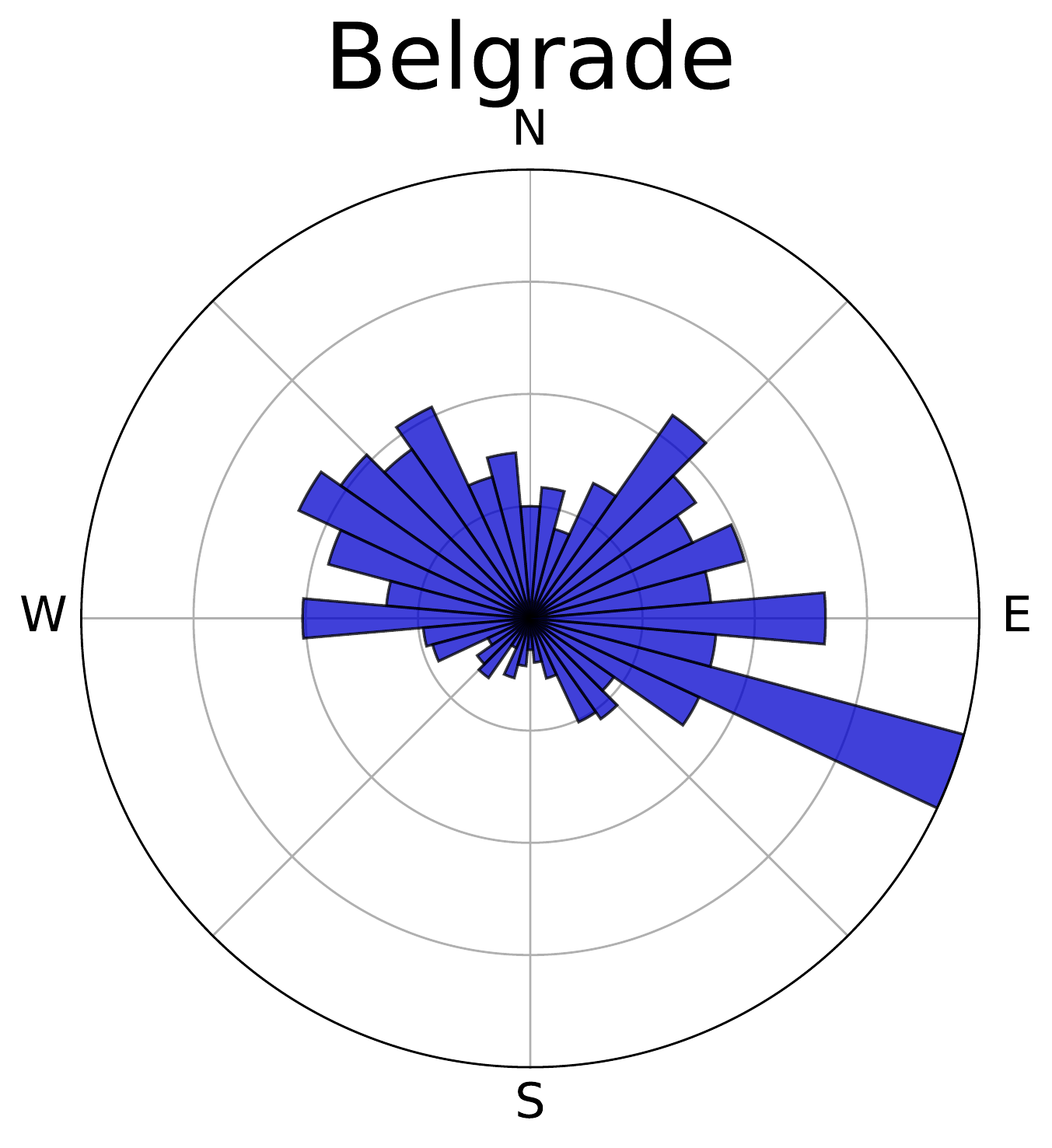}
	\includegraphics[width=.15\textwidth]{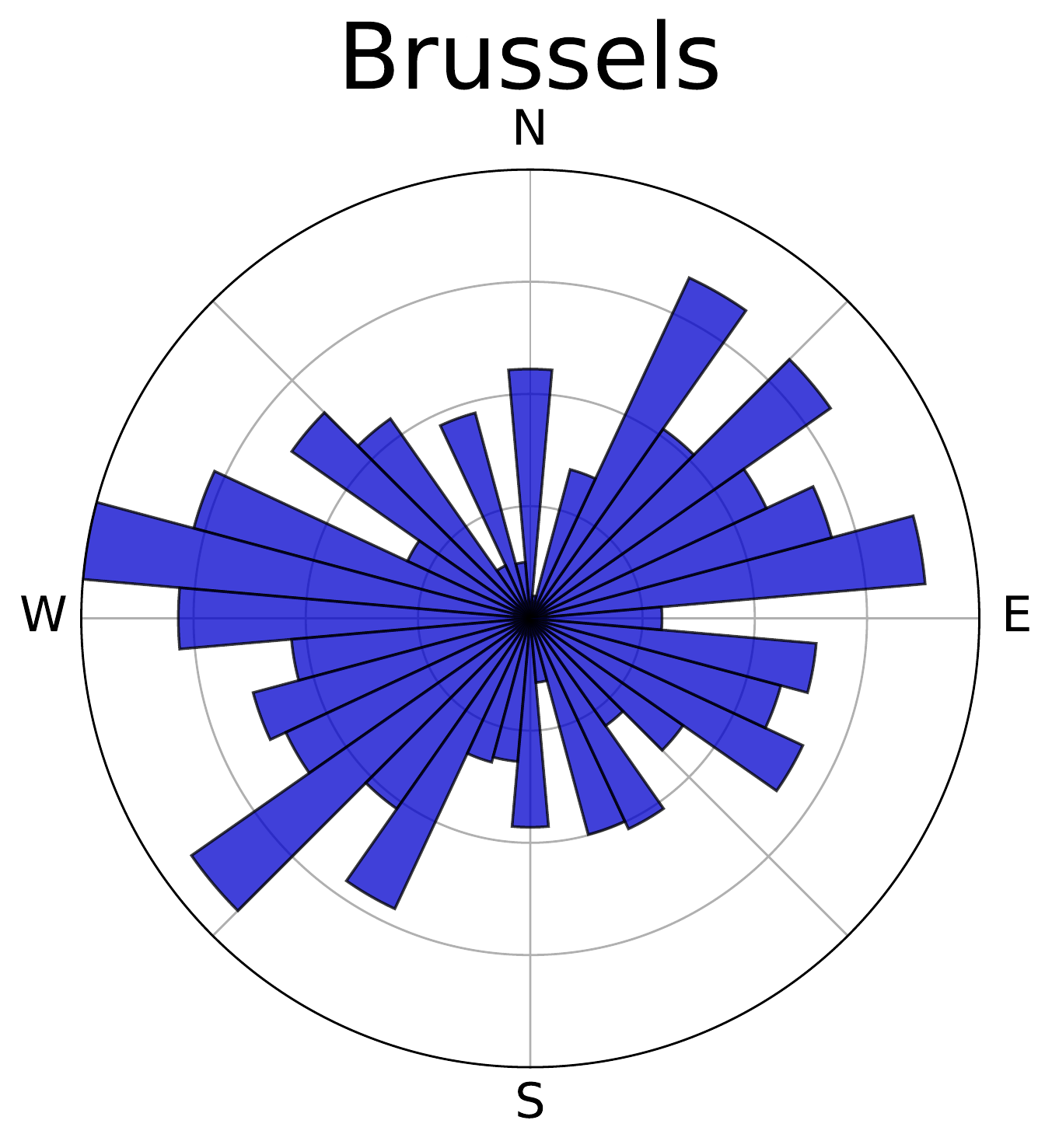}
	\includegraphics[width=.15\textwidth]{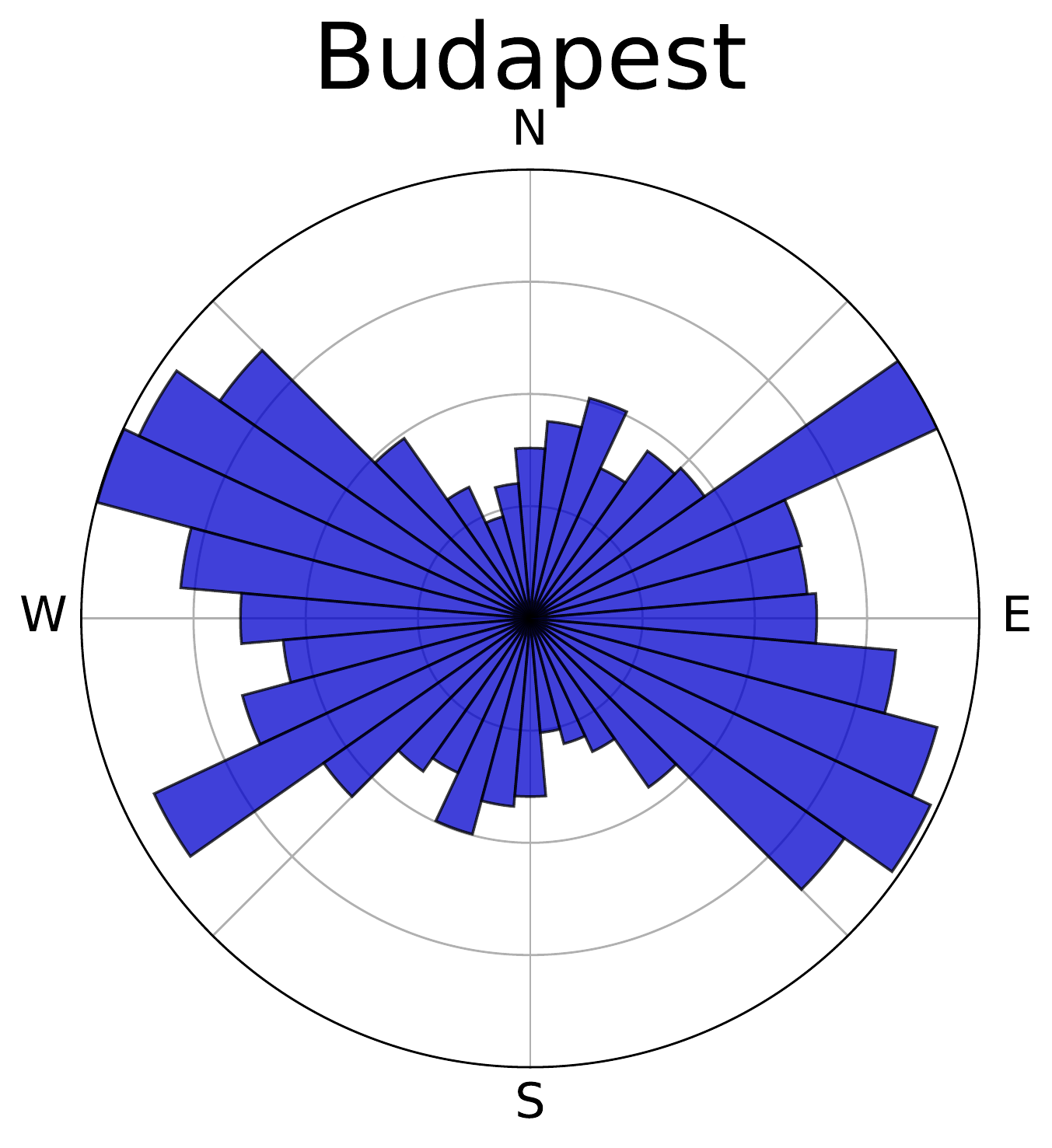}
	\includegraphics[width=.15\textwidth]{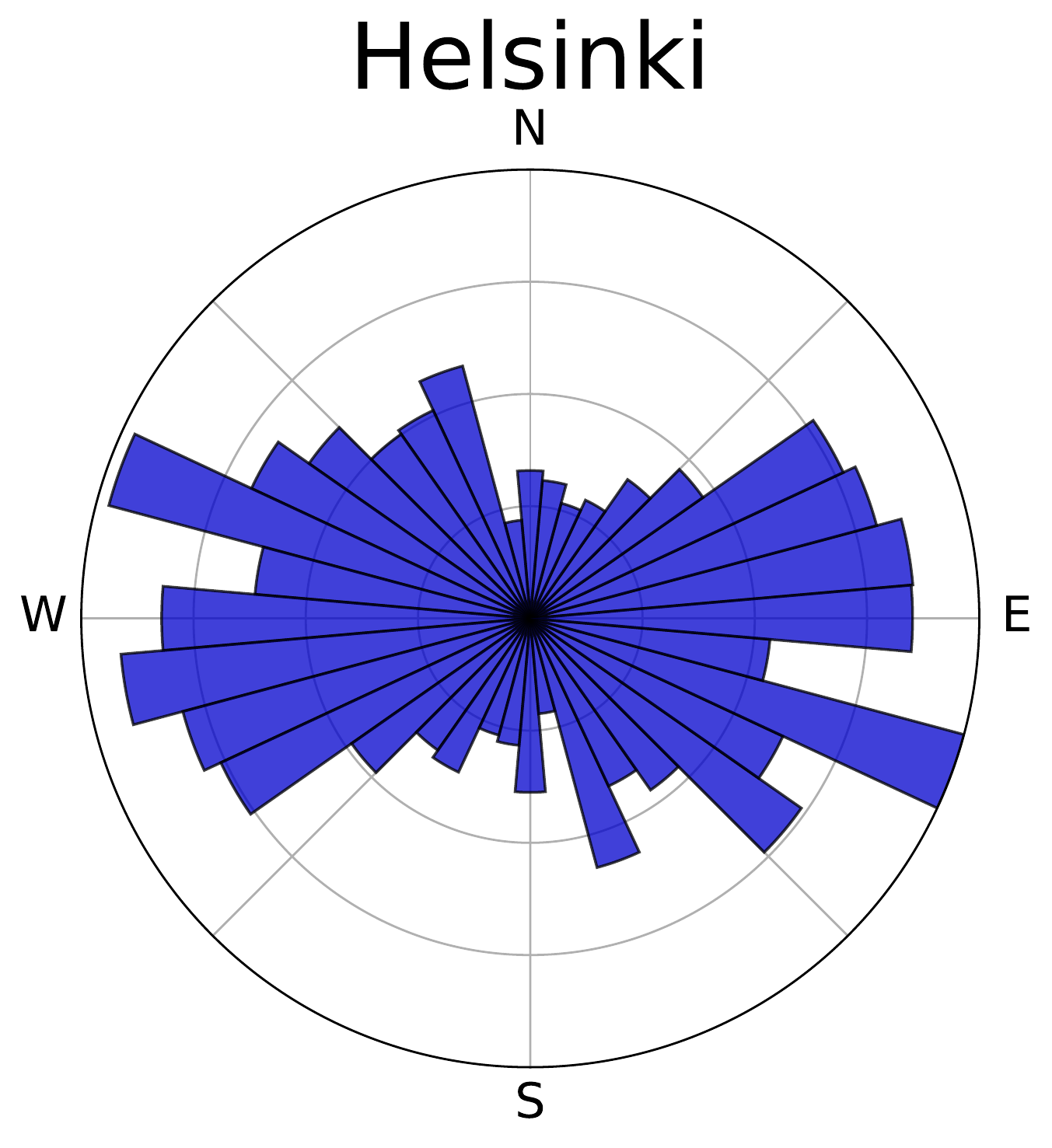}
	
	\includegraphics[width=.15\textwidth]{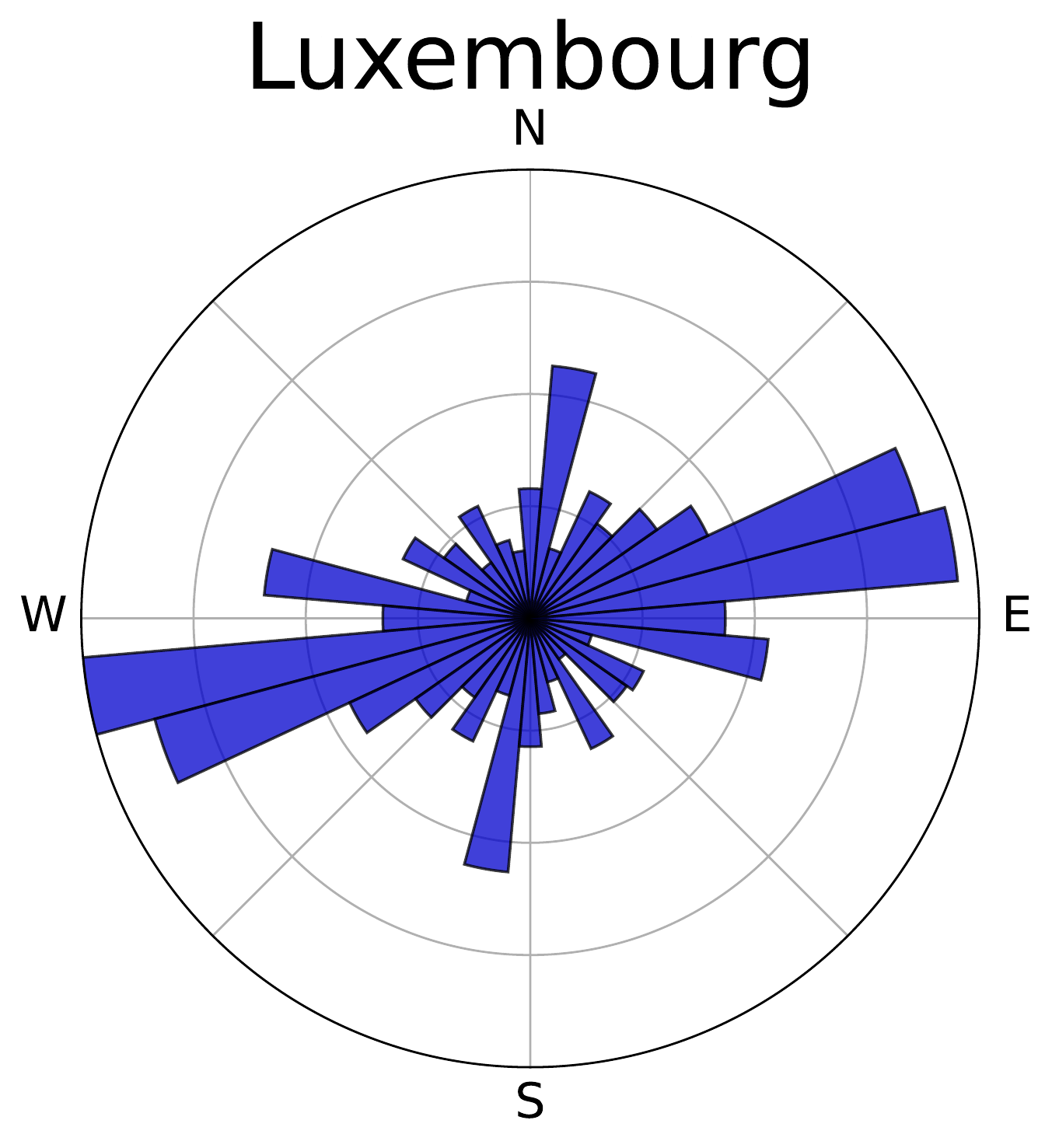}
	\includegraphics[width=.15\textwidth]{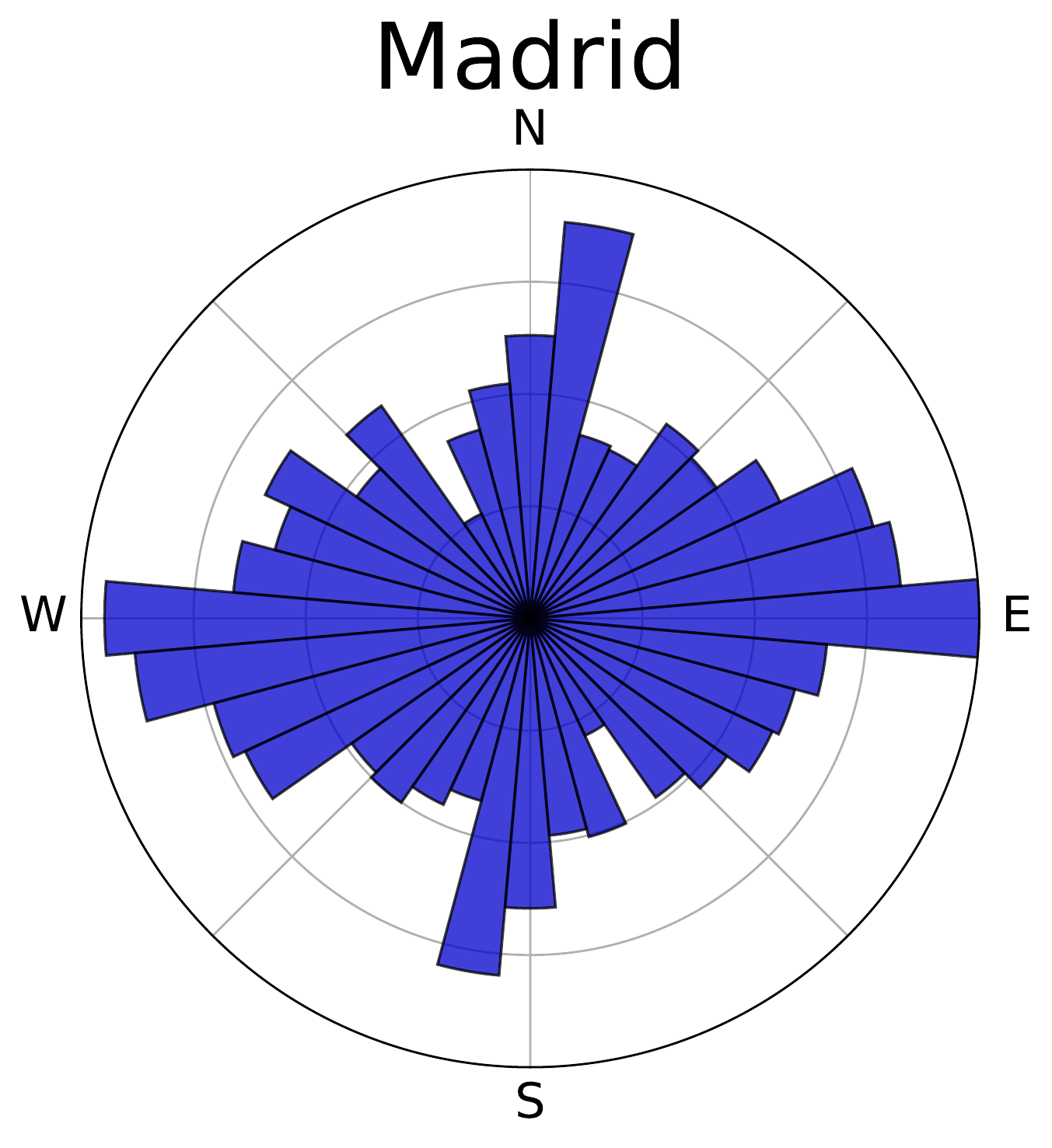}
	\includegraphics[width=.15\textwidth]{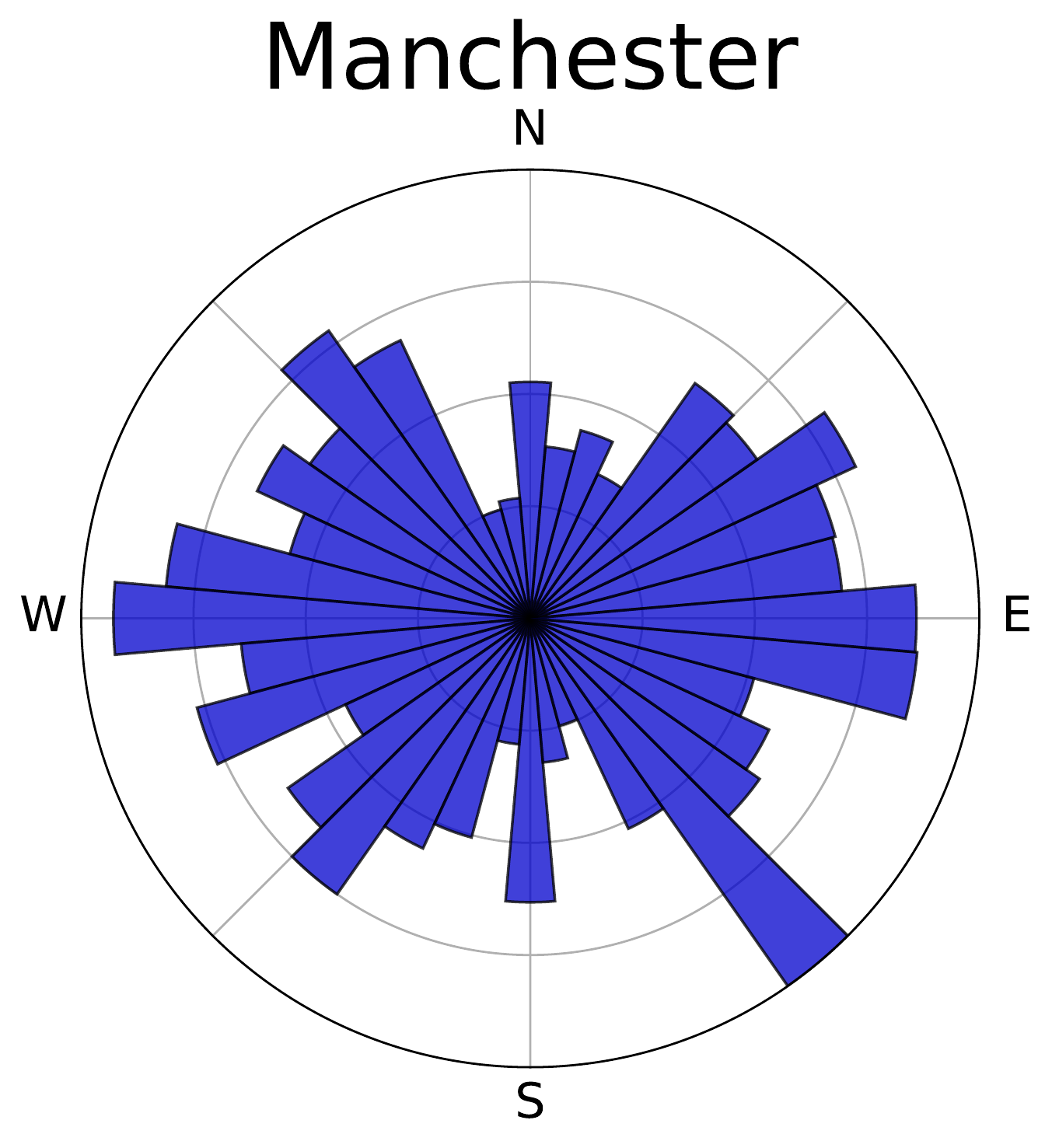}
	\includegraphics[width=.15\textwidth]{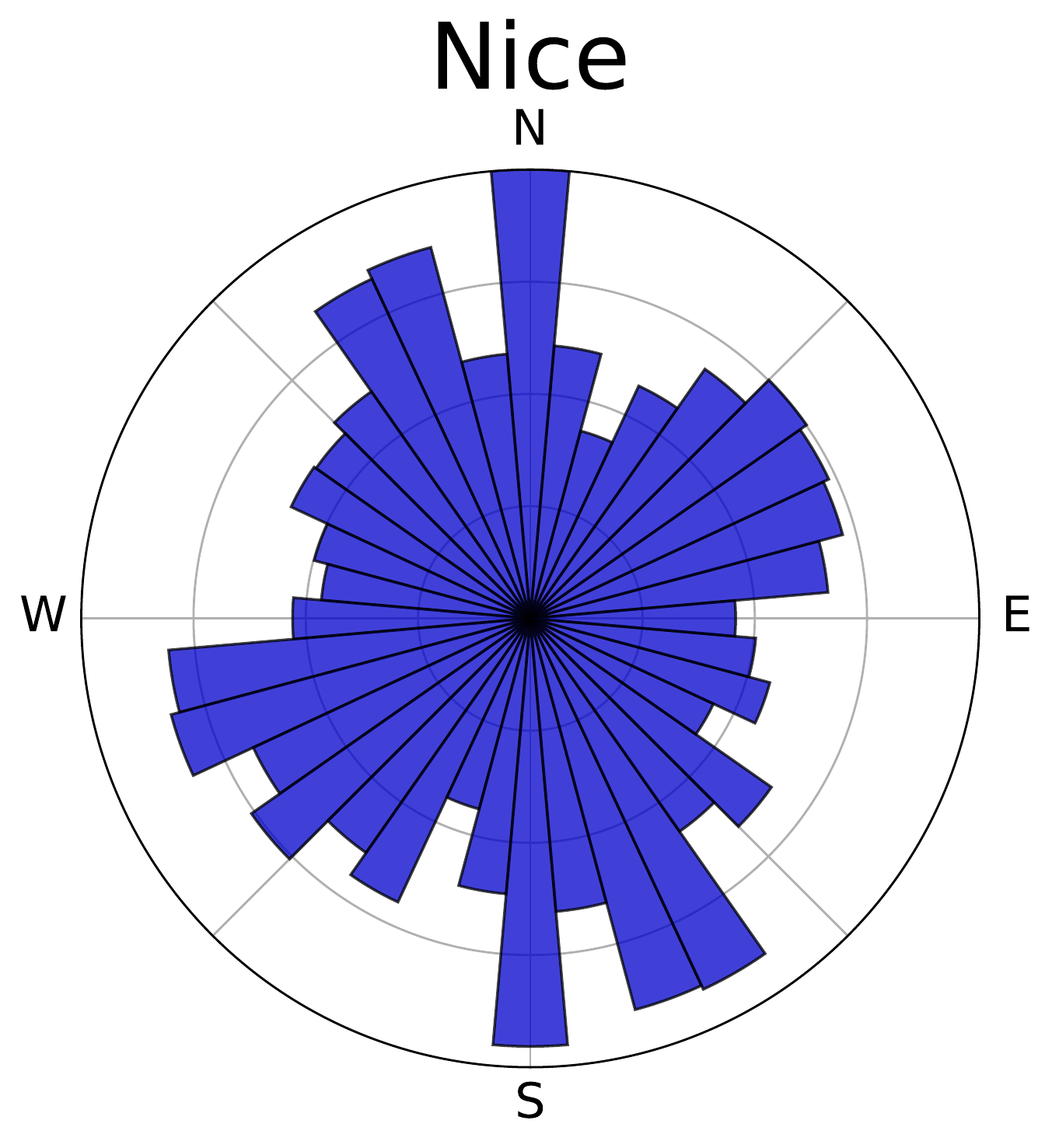}
	\includegraphics[width=.15\textwidth]{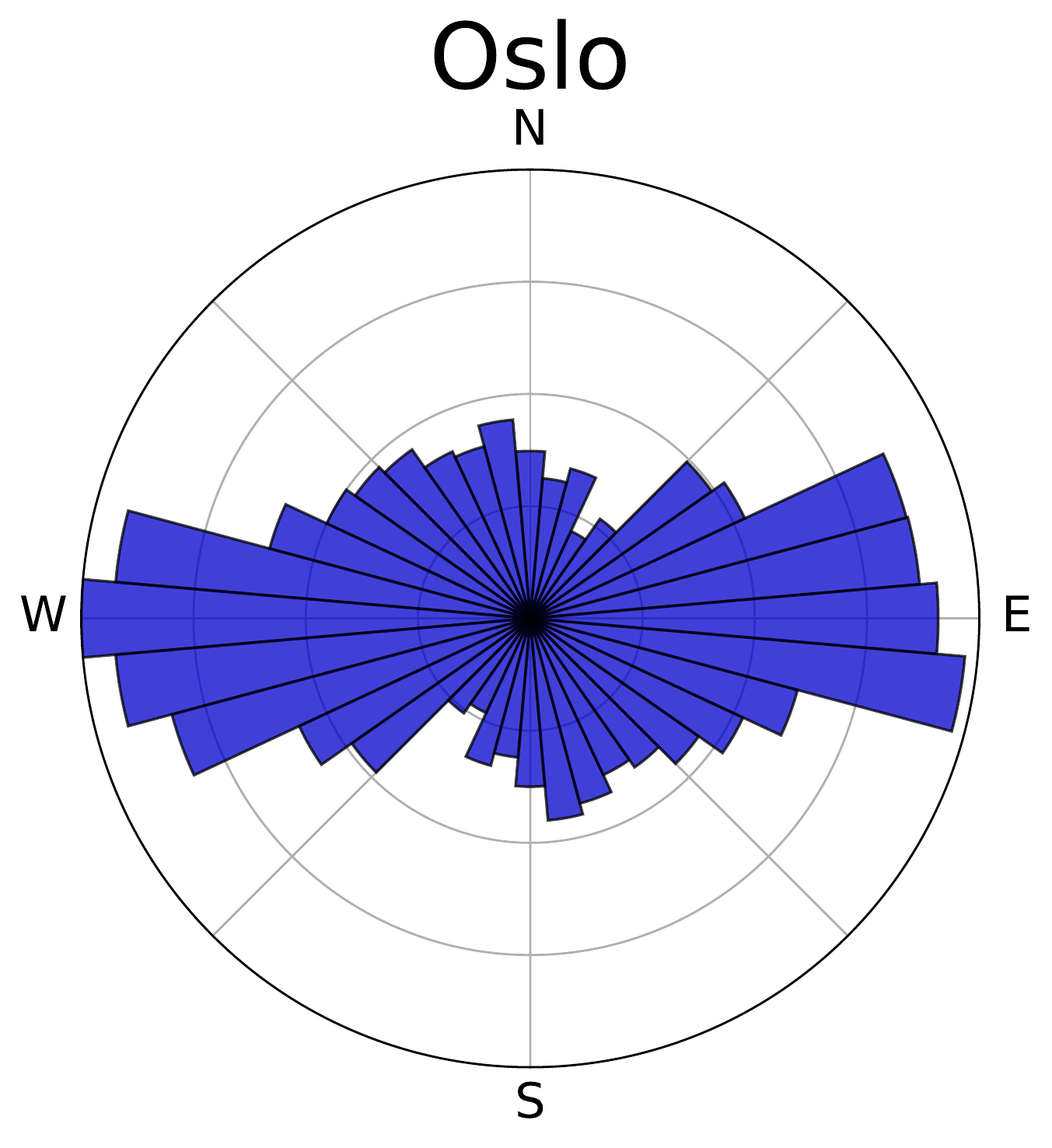}
	\includegraphics[width=.15\textwidth]{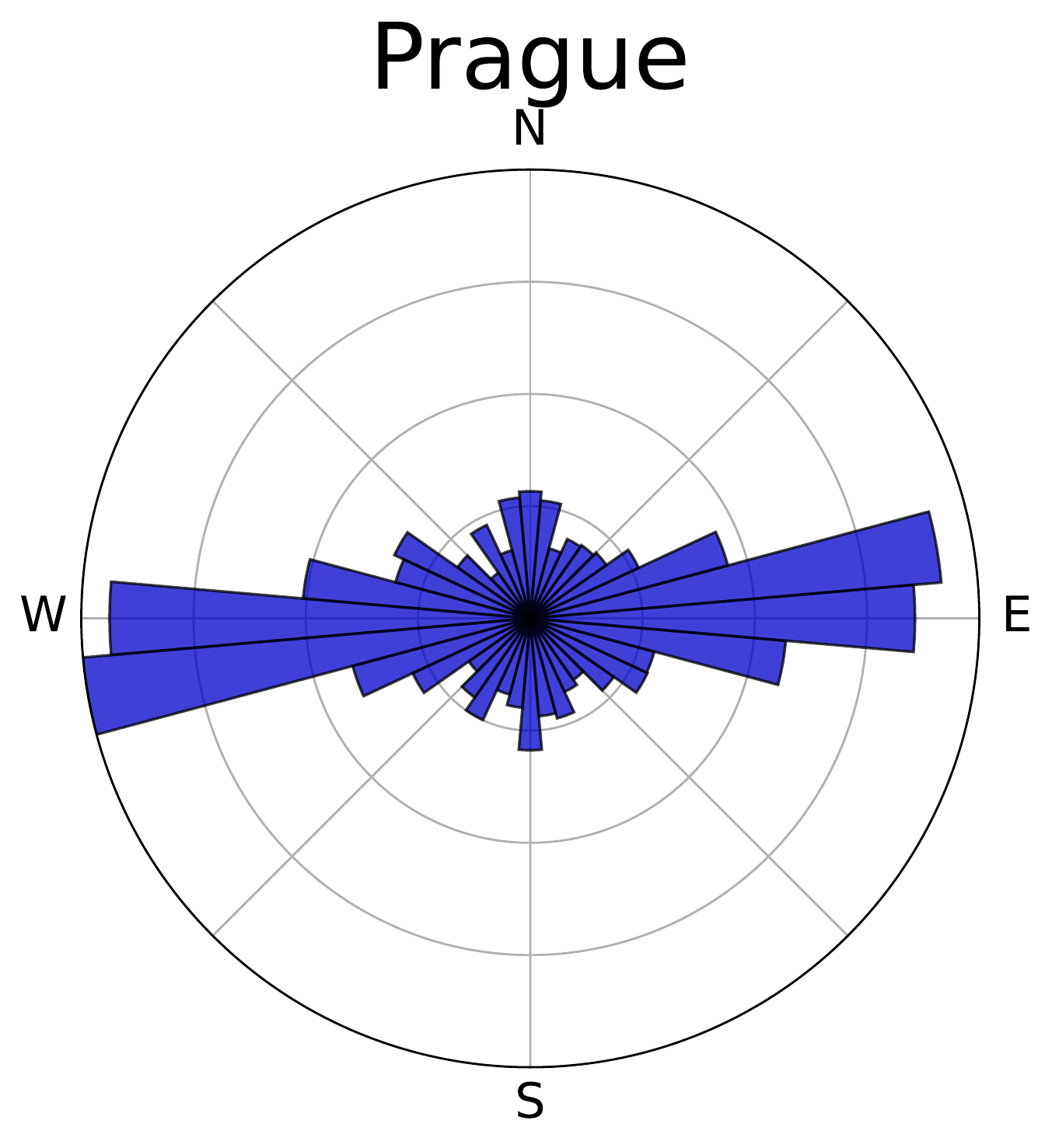}
	
	\includegraphics[width=.15\textwidth]{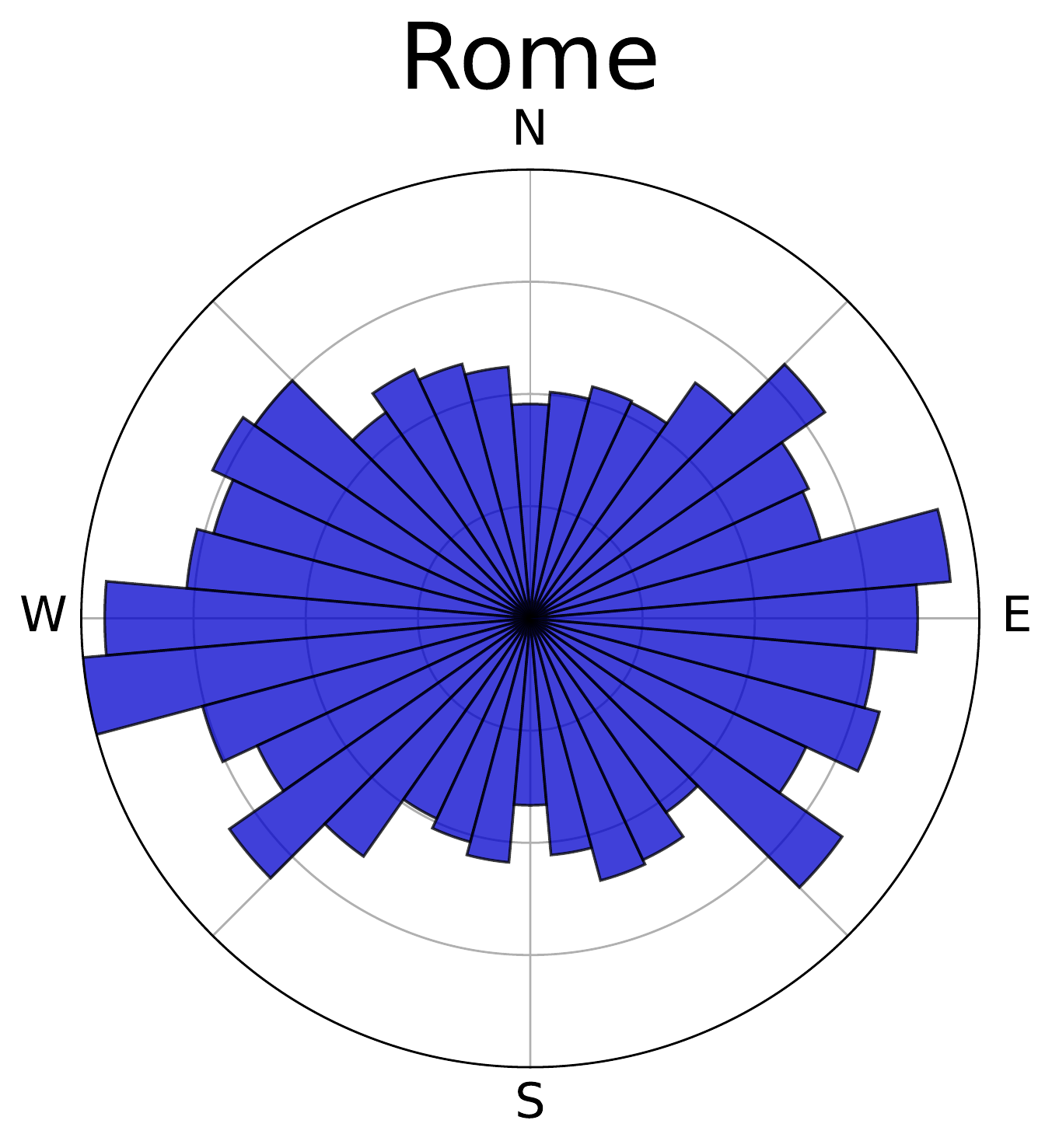}
	\includegraphics[width=.15\textwidth]{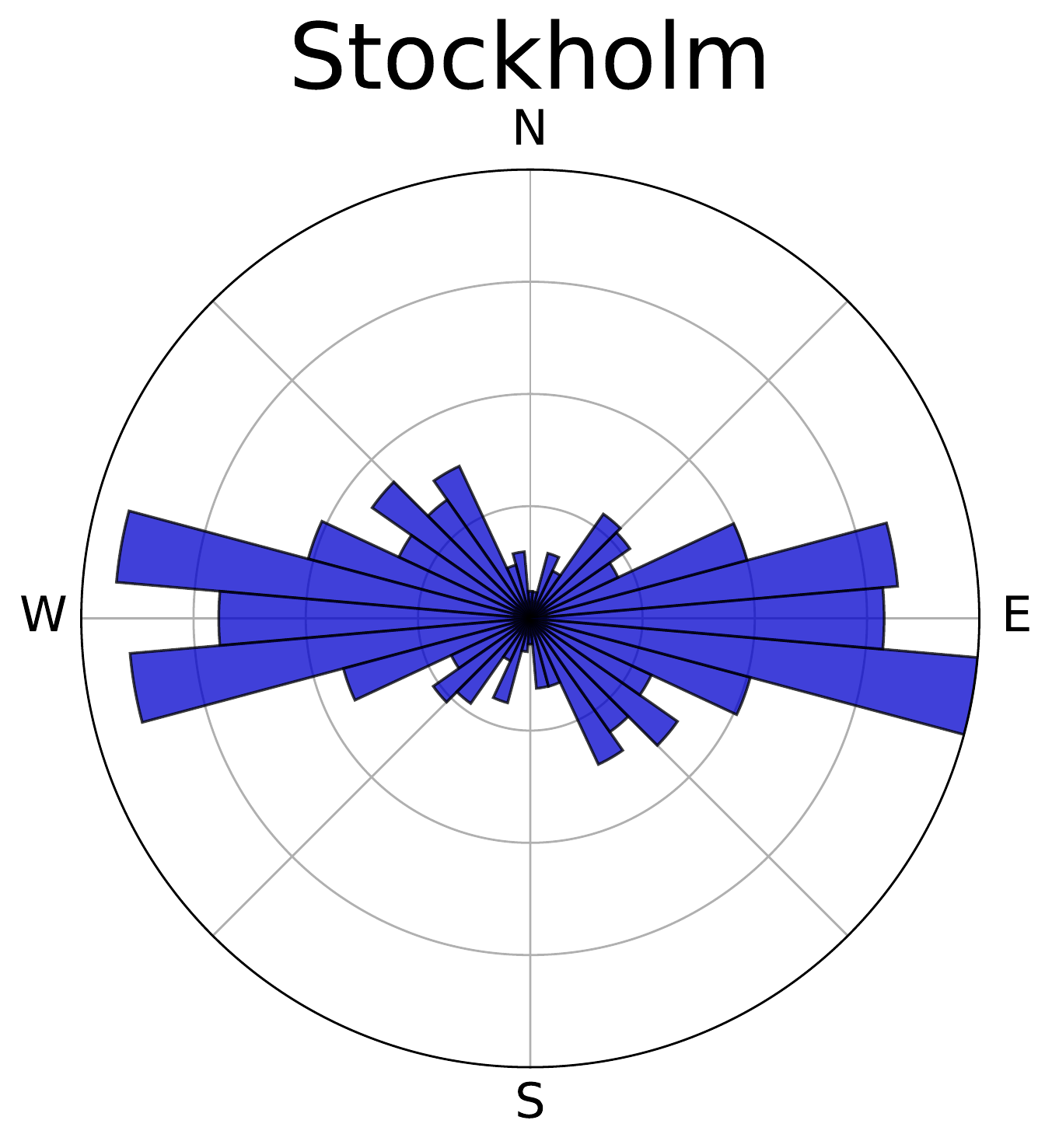}
	\includegraphics[width=.15\textwidth]{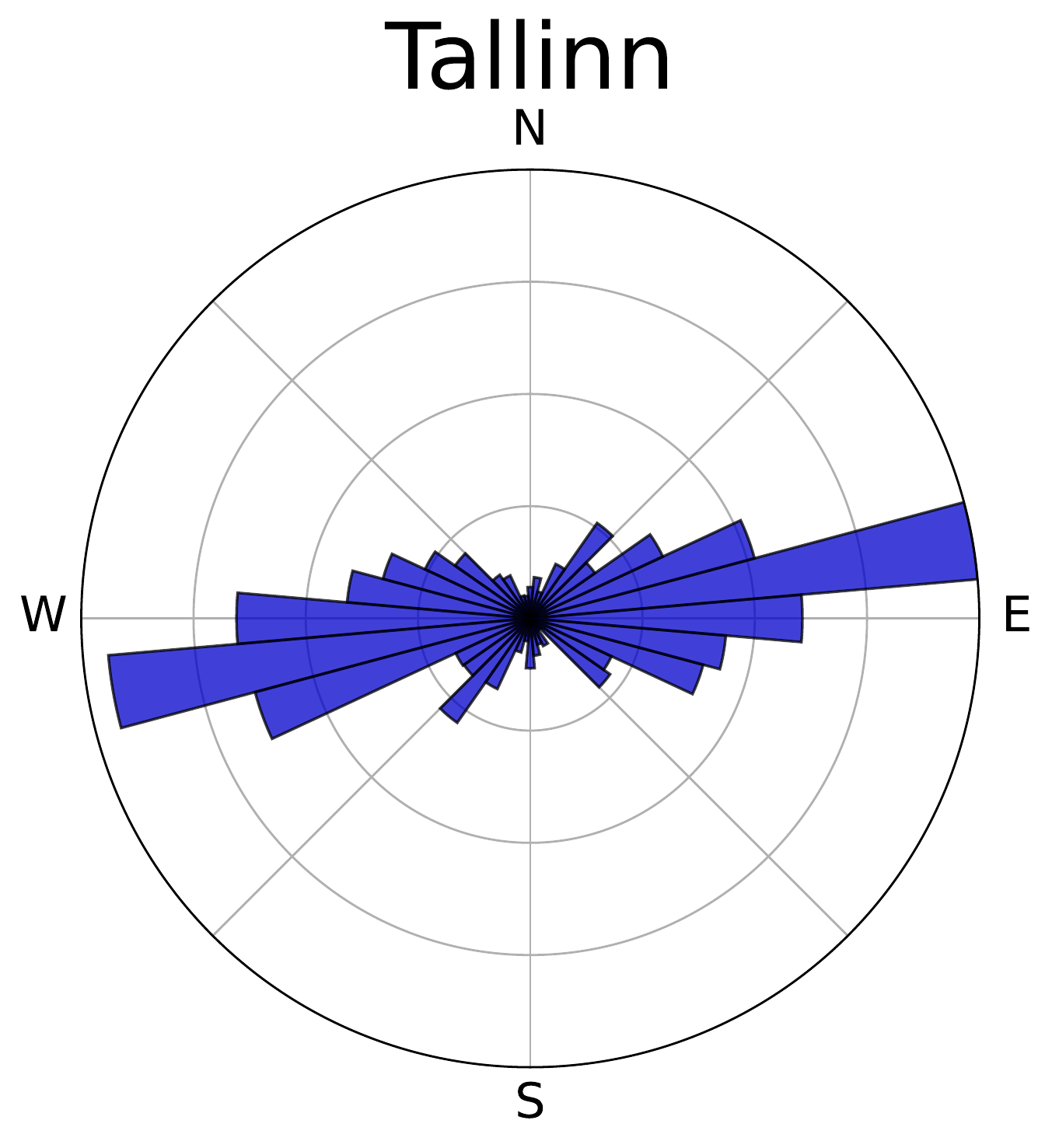}
	\includegraphics[width=.15\textwidth]{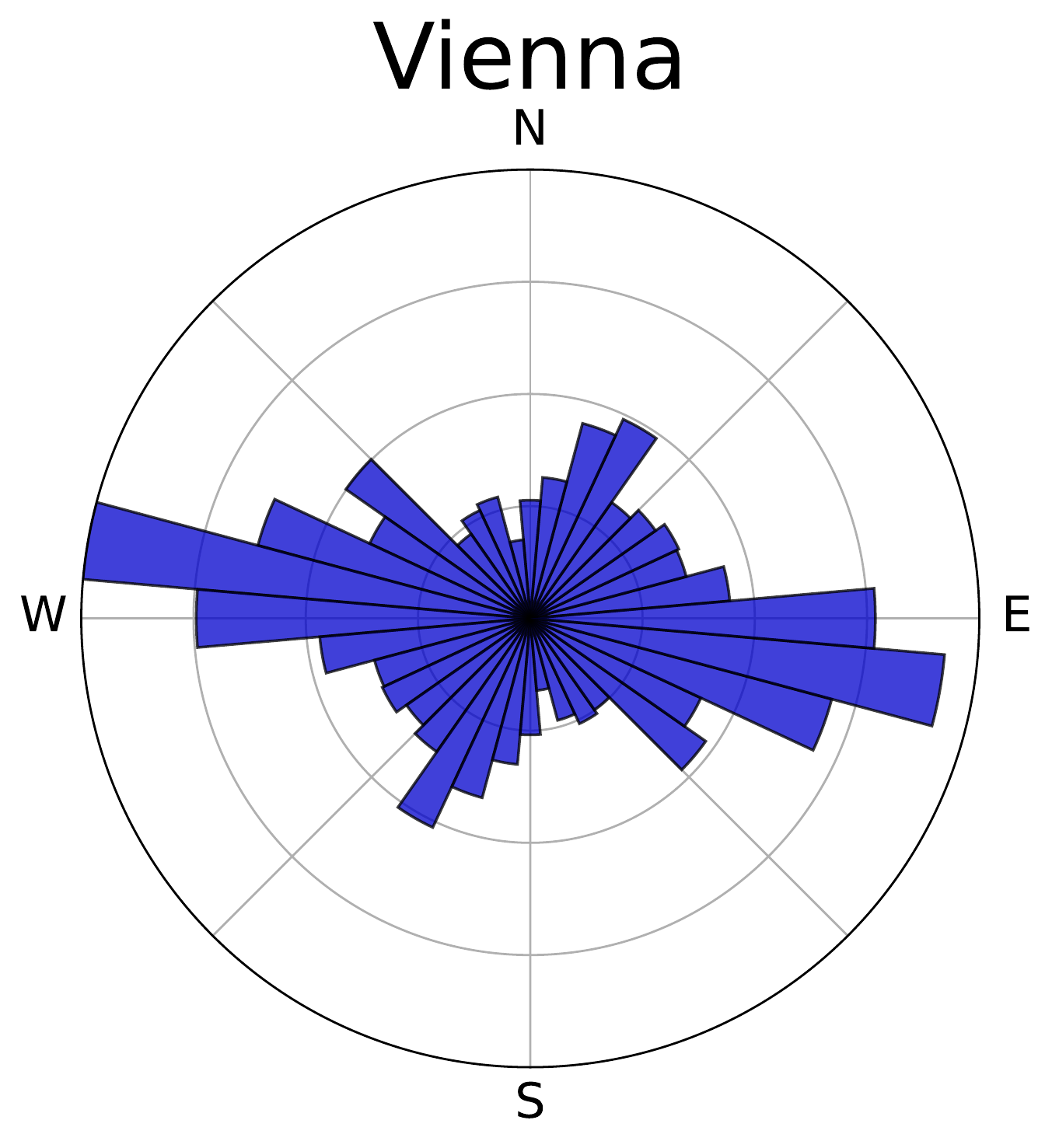}
	\includegraphics[width=.15\textwidth]{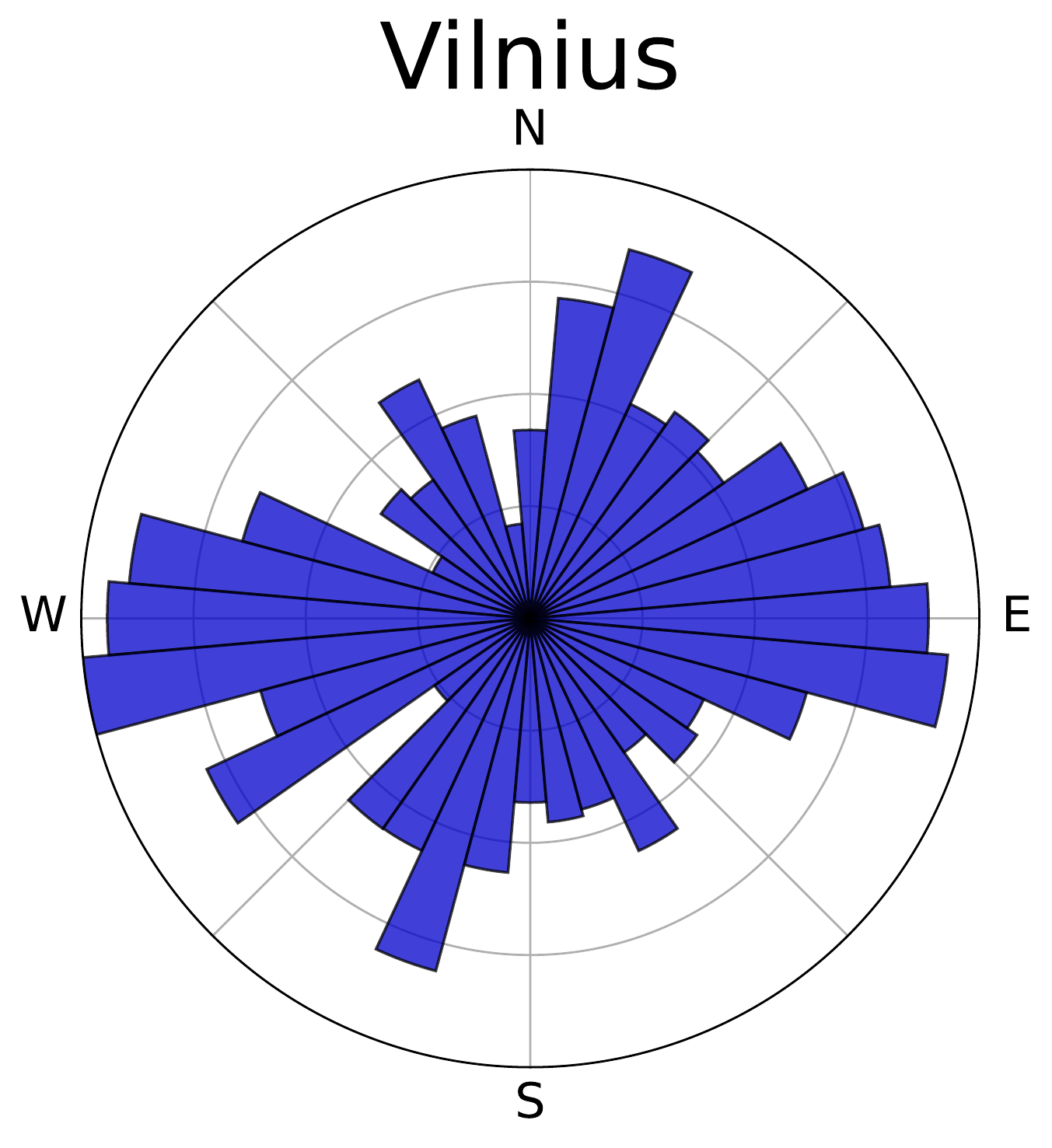}
	\includegraphics[width=.15\textwidth]{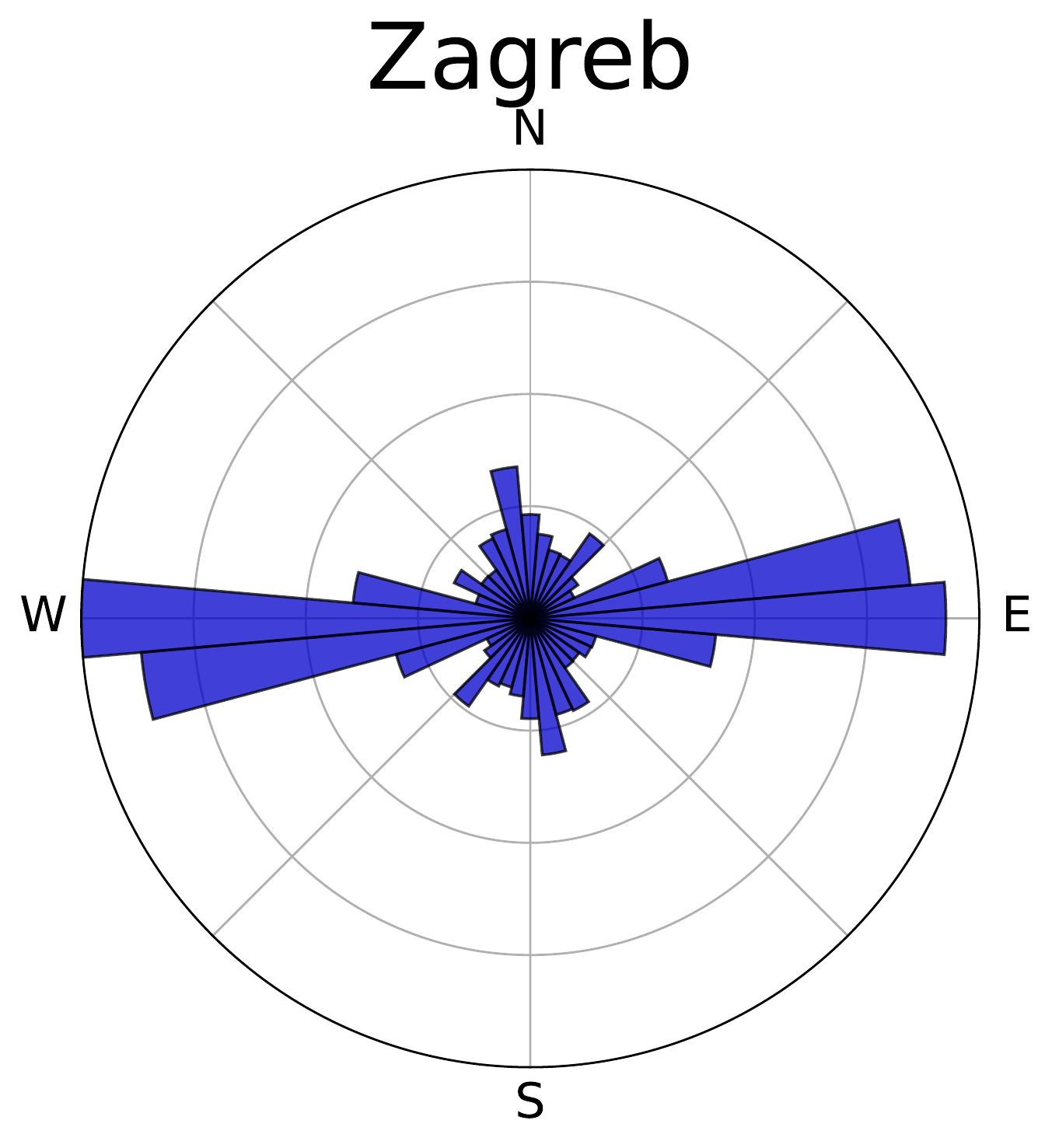}
	\caption{Public transport network orientations of $18$ selected European cities for which valid GTFS data could be obtained via \url{https://transitfeeds.com/l/60-europe}.}\label{fig:orientations_public_transport_europe}
	\end{center}
\end{figure}

For comparison, we also apply the methods described in section \nameref{sec:orientations_methodology} to $18$ selected major European cities.
\Cref{fig:orientations_streets_europe} shows their street orientation diagrams while \Cref{fig:orientations_public_transport_europe} illustrates the corresponding public transport orientations.
The key observations made for the German cities hold true for roughly half of the considered European cities:
street network orientations contain two orthogonal preferential orientations, which can sometimes be linked to geographical constraints, and main public transport network orientations tend to run along the east-west axis.
Deviating from the latter observation, merely Nice clearly possesses a more pronounced north-south orientation.
Conversely, street orientations of, e.g., Athens, Rome, and Stockholm are quite equally distributed, which could suggest that self-organizing dynamics dominated the spatial expansion of these cities \cite{batty2008size,barthelemy2008modeling,barthelemy2013self}.
The fact that Athens and Rome both have a history of being important European cultural centers for millennia might support that hypothesis.
Interestingly, some cities like Luxembourg City, Oslo, and Stockholm, whose street network orientations are quite equally distributed, still show a clearly pronounced (tilted) east-west axis in the public transport network orientation plots.

Finally, we comment on some European cities with unique properties in at least one of the two orientation plots.
The street network orientations of Brussels differ from other considered cities in the fact that the preferential directions are not orthogonal, but rather form an angle of $45$ degrees.
The public transport network orientations of Belgrade are remarkably nonsymmetric suggesting that a transport concept different from lines going back and forth along the same route with the same frequencies is in place.
Lastly, Barcelona is the only city considered in this paper with two equally pronounced orthogonal preferential directions in both orientation plots.
The latter observation is the result of a remarkable city planning effort by Cerd\`a in the year $1860$ to connect the medieval core of Barcelona with surrounding villages by an extensive road network in the form of an orthogonal grid
\cite{pallares2011cerda}, which is certainly only one of many stories behind the graphics in \Cref{fig:orientations_streets_germany,fig:orientations_public_transport_germany,fig:orientations_streets_europe,fig:orientations_public_transport_europe}.

All results presented in this section are reproducible with publicly available python codes\footnote{\url{https://github.com/KBergermann/Urban-multiplex-networks}}.
The code is designed according to the General Transit Feed Specification (GTFS) and should work on any valid GTFS data set making it easily utilizable for data of all regions around the world.

\section{Multiplex network model}\label{sec:multiplex_networks}

We now move to the second aspect of geometrical city modeling studied in this paper: the application of matrix function-based centrality measures to a multiplex network representation of the urban public transport systems.
To this end, we introduce the multiplex framework employed to formalize these systems.
We discuss the relation between network centralities and orientations in the section \nameref{sec:results_discussion}.

Network models have a long history in urban science:
going back almost $300$ years to Euler's K\"onigsberg bridge problem \cite{euler1741solutio} it could be argued that graph theory was motivated by an urban science problem.
Especially the abstraction of street networks to (almost) planar single-layer graphs has been the basis for many mathematical models of cities ever since \cite{barthelemy2011spatial,barthelemy2016structure}.
More recently, following the advent of multilayer graphs in various scientific disciplines \cite{kivela2014multilayer,boccaletti2014structure}, urban science profited from the flexibility provided by multi-layered network structures to construct more realistic models of complex interactions \cite{aleta2017multilayer,curado2021understanding,alessandretti2016user,zheng2018understanding,strano2015multiplex}.
Some of these works consider multi-layered public transport networks in which layers correspond to either lines or modes of transportation.

In this section, we introduce a primal \cite{porta2006network} multiplex representation of public transport networks in which layers represent lines.
We assume the layers to be node-aligned so that each layer contains an instance of each physical node representing a stop within the city limits.
We denote the number of physical nodes by $n$ and the number of layers by $L$.
Each copy of a physical node in any of the layers is called a node-layer pair.

We distinguish between two types of edges that can connect pairs of node-layer pairs: intra- and inter-layer edges.
Intra-layer edges connect node-layer pairs belonging to the same layer whenever the corresponding line directly connects the two stops, i.e., if a line approaches stops A-B-C, then A and B as well as B and C are connected by an edge, but A and C are not.
Inter-layer edges, on the other hand, connect instances of the same physical nodes belonging to different layers whenever both lines serve the corresponding stop, i.e., if the stop can be used to change between the lines.
We discuss modeling approaches to assign weights to both intra- and inter-layer connections later in this section.

In our multiplex networks we restrict ourselves to undirected edges although directed models taking travel directions into account would appear suitable for the application, cf.~our approach to network orientations in the section \nameref{sec:orientations}.
Our choice accounts for the structure of the GTFS data, which contains multiple different stop-ID's, i.e., physical nodes per stop.
This structure entails that, e.g., a bus stop with one stop on each side of a street but the same name is not represented by one but two physical nodes.
Employing a strategy to aggregate stops with the same name is error-prone as it provokes unexpected behavior when different data providers use inconsistent stop naming logics, e.g., in- and excluding track numbers in stop names.
In order to prevent in- and outbound traffic of a stop to be unequally distributed across two different physical nodes we use undirected edges, which equally represent in- and outbound traffic at the involved node-layer pair regardless of the current travel direction.
In the example scenario described above this leads to both stop-ID's carrying the same in- and outbound information.
This way of ``counting each connection twice'' appears preferable over each node-layer pair of the network carrying only half of the information available for the respective stop.

More formally, the multiplex networks described so far can be defined as the multilayer graph $\G=(\tilde{\V}, \E^{(1)}, \dots , \E^{(L)}, \tilde{\E})$ consisting of a common vertex set $\tilde{\V}$ for all layers, intra-layer edge sets $\E^{(l)}$ for all layers $l=1, 2, \dots , L$, and an inter-layer edge set $\tilde{\E}$.
Note that similar networks have been considered before, e.g., in \cite{taylor2017eigenvector,taylor2019supracentrality,taylor2021tunable,bergermann2021matrix} but in this paper we employ a different notion of inter-layer edges, which is determined by the data.

We choose a supra-adjacency matrix representation as the linear algebraic formulation of these multiplex networks \cite{kivela2014multilayer,taylor2017eigenvector,taylor2019supracentrality,taylor2021tunable,bergermann2021matrix}.
The two different types of edges are represented by two separate matrices.
The multilayer intra-layer adjacency matrix $\Aintra\in\R^{nL \times nL}$ contains the edges representing connections by public transport lines and is defined as the block-diagonal matrix
\begin{equation}\label{eq:def_aintra}
\Aintra = \text{blkdiag}[\A1, \dots , \AL] = 
\begin{bmatrix}
\A1 & \bm{0} & \dots & \bm{0}\\
\bm{0} & \bm{A}^{(2)} & \dots  & \bm{0}\\
\vdots & \vdots & \ddots & \vdots\\
\bm{0} & \bm{0} & \dots & \AL
\end{bmatrix},
\end{equation}
with $\bm{0} \in \R^{n \times n}$ the matrix of all zeros.
Each block-diagonal entry $\Al\in\R^{n \times n}$ corresponds to the adjacency matrix of layer $l$.
It contains the weight of the edge between the physical nodes $i$ and $j$ in layer $l$ in the entry $[\Al]_{ij}$ and zero if no edge is present for $i,j\in\{1, 2, \dots , n\}$ and $l\in\{1, 2, \dots , L\}$.

The inter-layer adjacency matrix $\Ainter\in\R^{nL \times nL}$ contains the edges representing possible changes between public transport lines and is defined as
\begin{equation}\label{eq:def_ainter}
\Ainter = 
\begin{bmatrix}
\bm{0} & \bm{O}^{(12)} & \dots & \bm{O}^{(1L)}\\
\bm{O}^{(21)} & \bm{0} & \dots  & \bm{O}^{(2L)}\\
\vdots & \vdots & \ddots & \vdots\\
\bm{O}^{(L1)} & \bm{O}^{(L2)} & \dots & \bm{0}
\end{bmatrix},
\end{equation}
where
\begin{equation*}
[\bm{O}^{(lk)}]_{ij} = 
\begin{cases}
1, & \text{if $i=j$ and physical node $x_i$ is connected between layers $l$ and $k$,}\\
0, & \text{otherwise,}
\end{cases}
\end{equation*}
with $l, k \in \{1, 2, \dots , L\}$ such that $\bm{O}^{(lk)}\in\R^{n \times n}$ contains ones only on a subset of its diagonal entries.
Thus, the presence of inter-layer edges depends on whether the lines represented by layers $l$ and $k$ both stop at a given physical node.
We set the block-diagonal of $\Ainter$ to zero matrices $\bm{0}\in\R^{n \times n}$ as these entries would not represent a change of lines.

On the one hand, the above definition of $\Ainter$ is more general than those in \cite{taylor2017eigenvector,taylor2019supracentrality,taylor2021tunable,bergermann2021matrix}, where each $\bm{O}^{(lk)}$ is a full identity matrix, as it includes all combinations of up to $n$ zero diagonal entries.
On the other hand, the construction in \cite{taylor2017eigenvector,taylor2019supracentrality,taylor2021tunable,bergermann2021matrix} allows each layer-layer pair to be coupled with a different weight, while $\Ainter$ defined in \Cref{eq:def_ainter} is unweighted.
We will introduce the parameter $\omega$ in \Cref{eq:supra_adjacency} to scale the inter-layer adjacency matrix, which can be interpreted as modeling a constant transfer time between all lines at all stops of the network.
Note that all numerical methods would be equally applicable if each inter-layer edge was weighted individually.

The definitions in \Cref{eq:def_aintra,eq:def_ainter} now allow us to define the supra-adjacency matrix $\bm{A}\in\R^{nL \times nL}$ of the multiplex network as
\begin{equation}\label{eq:supra_adjacency}
\bm{A} = \Aintra + \omega \Ainter,
\end{equation}
where $\omega\in\R_{\geq 0}$ is a scalar coupling parameter that trades off the relative importance of intra- and inter-layer weights.
Note that by our choice of undirected edges discussed earlier in this section we have $\bm{A}=\bm{A}^T$ throughout this paper.

\begin{figure}
	\begin{center}
	\includegraphics[width=.5\textwidth,valign=m]{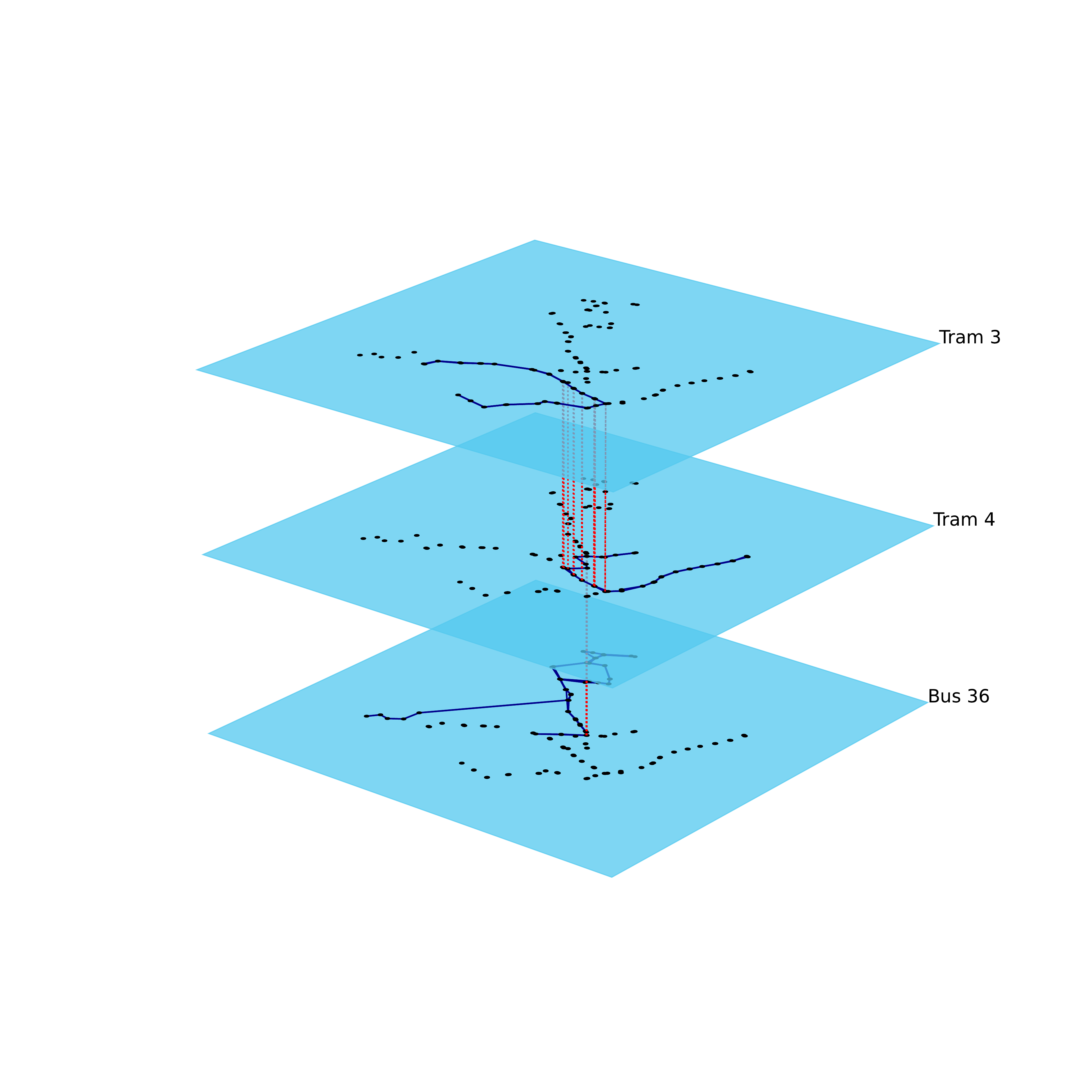}
	\includegraphics[width=.4\textwidth,valign=m]{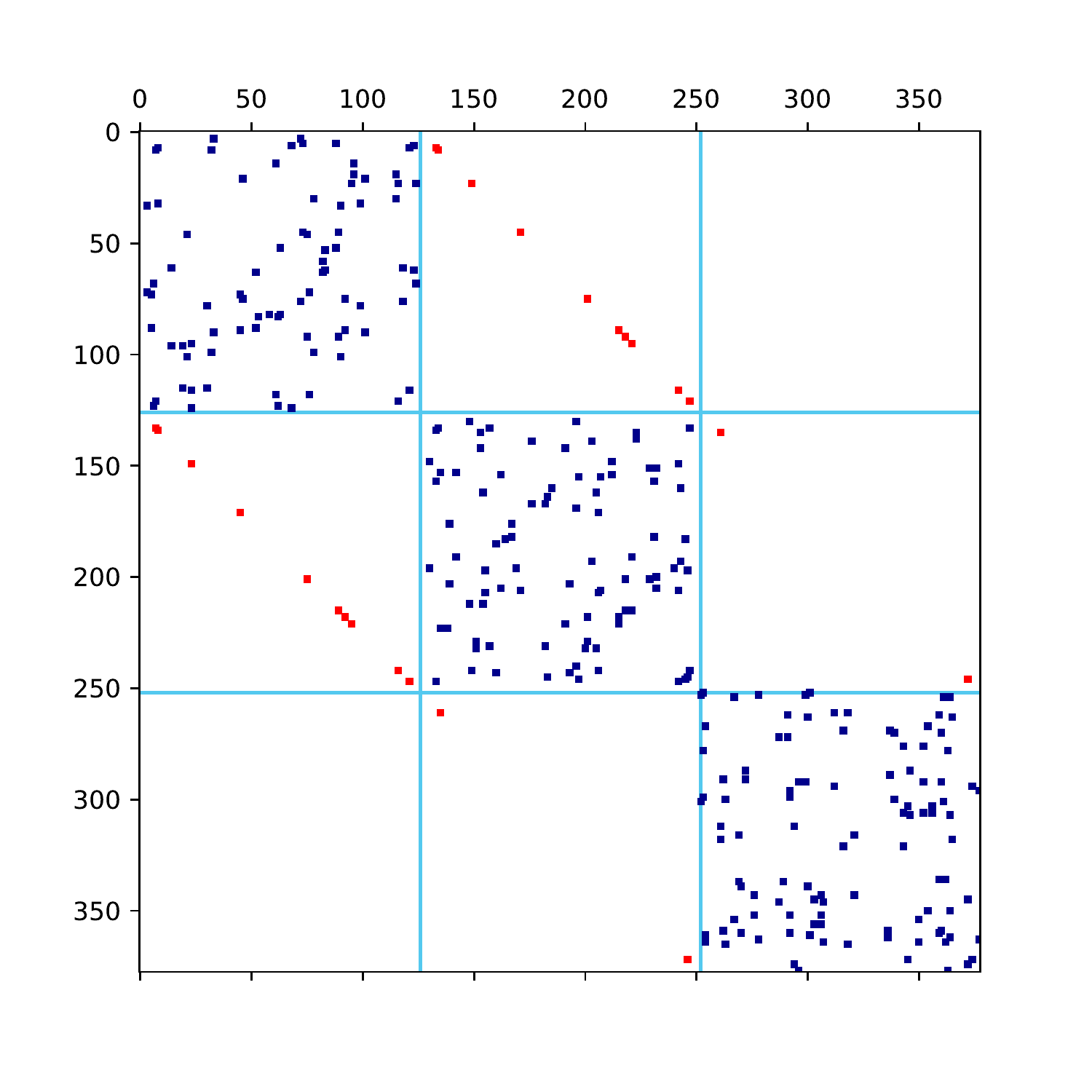}
	\caption{Example public transport multiplex network with $L=3$ selected layers of Freiburg (corresponding to two tram layers and one bus layer) and $n=126$ stops.
		The graphics were created with the python package pymnet \cite{kivela2017multilayer}.
		Left: Visualization of the multiplex network.
		Intra-layer edges are marked dark blue and inter-layer edges are marked red.
		Right: Sparsity structure of the corresponding supra-adjacency matrix.
		Dark blue entries represent intra-layer edges and red entries represent inter-layer edges.}\label{fig:multiplex_example}
	\end{center}
\end{figure}

We visualize the above definitions by a small example multiplex network in \Cref{fig:multiplex_example}.
The left plot shows a multiplex network that consists of three layers (two tram lines and one bus line) taken from the full multiplex network of the German city Freiburg.
Only the subset of physical nodes served by these lines is included in the figure.
The right plot shows the sparsity structure of the corresponding supra-adjacency matrix.
Both illustrations distinguish the two types of edges by different colors.

We remark that the supra-adjacency matrix $\bm{A}$ of our public transport networks is characterized by a high degree of sparsity.
This is true by the definitions in \Cref{eq:def_aintra,eq:def_ainter}, but our application is characterized by additional sparsity due to the facts that each line typically only serves a small subset of stops in the city and that most layers are approximately represented by chain graphs in which the number of edges is of the order of the number of non-isolated node-layer pairs.
\Cref{tab:germany_nodes_layers} illustrates the most important multiplex network properties of the $36$ German cities considered in the section \nameref{sec:orientations}.
It confirms the statements above: around $95-98\%$ of the node-layer pairs in the networks are isolated, i.e., possess no incident edge and the number of intra-layer edges is close to the number of non-isolated node-layer pairs for all networks.

Finally, we comment on our approach to assigning weights to the edges of the multiplex networks.
Many weighted urban network models rely on choosing either the geographical distance \cite{crucitti2006centrality,crucitti2006bcentrality,porta2006network} or travel times \cite{aleta2017multilayer,bast2016route} as weights, which is a sensible and straightforward choice for many problems like, e.g., routing algorithms. 
We, however, aim at applying matrix function-based centrality measures to the multiplex networks in which a node-layer pair is considered central if it is connected to many other node-layer pairs by short walks along edges with large weights.
To decide what it means to be ``close'' to many node-layer pairs in a public transport network, we rely on the combination of two concepts:
the frequency of the public transport line \cite{aleta2017multilayer} and a Gaussian kernel applied to travel times.
Gaussian kernels have become a well-established choice of similarity measures in many data-driven applications \cite{scholkopf2002learning,von2007tutorial,stoll2020literature,bergermann2021semi}.

We define the frequency $\phi^{(l)}_{ij}\in\mathbb{N}$ as the number of connections that line $l$ offers between stops $i$ and $j$ per day.
Furthermore, we denote the travel time between node-layer pairs $(i,l)$ and $(j,l)$ with $i,j\in\{1, 2, \dots , n\}$ and $l\in\{1, 2, \dots , L\}$ by $(\Delta t)_{ij}^{(l)}\in\R_{\geq 0}$ and use a Gaussian kernel to define the similarity of them as $\exp \left(-\frac{\left( (\Delta t)_{ij}^{(l)} \right)^2}{\sigma^2}\right)\in(0,1]$.
The scalar parameter $\sigma\in\R_{>0}$ determines the distribution of these similarities, e.g., $\sigma \ll (\Delta t)_{ij}^{(l)}$ leads to most weights being close to zero and $\sigma \gg (\Delta t)_{ij}^{(l)}$ leads to most weights being close to one.
As this parameter scales the travel times of public transport lines in the urban network it can be interpreted as a ``normalizing travel time''.
Numerical experiments on the role of the parameter $\sigma$ are presented in the section \nameref{sec:results_normalizing_travel_time}.
As described at the beginning of this section, edge weights are set to zero if the corresponding layer offers no direct connection.
Note, that information about both line frequencies and travel times can be extracted from GTFS data.

We allow different combinations of both weighting concepts by defining the intra-layer weights as
\begin{equation*}
[\Al]_{ij} = 
\begin{cases}
w^{(l)}_{ij}, & \text{if $(i,l)$ and $(j,l)$ are connected,}\\
0, & \text{otherwise,}
\end{cases}
\end{equation*}
with $w^{(l)}_{ij}$ equal to one of the following expressions:
\begin{equation}\label{eq:weighting_models}
\color{white}
\begin{cases}
\color{black}\bullet\quad 1, & \color{black}\text{using no travel times and no frequencies,}\\
\color{black}\bullet\quad \phi^{(l)}_{ij}, & \color{black}\text{using only frequencies,}\\
\color{black}\bullet\quad \exp \left(-\frac{\left( (\Delta t)_{ij}^{(l)} \right)^2}{\sigma^2}\right), & \color{black}\text{using only travel times,}\\
\color{black}\bullet\quad \phi^{(l)}_{ij} \exp \left(-\frac{\left( (\Delta t)_{ij}^{(l)} \right)^2}{\sigma^2}\right), & \color{black}\text{using both travel times and frequencies.}
\end{cases}
\end{equation}
With the fourth choice of weights we combine travel times and frequencies by multiplication of the two individual weights.
This leads to a (linearly) proportional dependence of the weight on the frequency and a (non-linearly) anti-proportional dependence of the weight on the travel time.
With this weighting approach, stops connected to many other stops via highly frequented routes with short travel times will be recognized as most central.
We compare the different weighting approaches in the section \nameref{sec:results_weight_models}.

As described in the section \nameref{sec:data}, files containing information on transfer options and transfer times between different public transport lines are not required but only optional in GTFS data.
Hence, we do not possess knowledge of realistic transfer times, which are represented by inter-layer edge weights in our multiplex network formulation.
We thus propose to utilize the unweighted inter-layer adjacency matrix defined in \Cref{eq:def_ainter} and to include the cost of changing lines into the coupling parameter $\omega$ from \Cref{eq:supra_adjacency} by defining a transfer time $\Delta t_{\mathrm{transfer}}\in\R_{\geq 0}$, which we assume to be constant across all pairs of lines and all stops.
For consistency, we also apply the same Gaussian kernel to $\Delta t_{\mathrm{transfer}}$ whenever we include travel times in the weights of intra-layer edges.
Thus, the inter-layer weights are given by
\begin{equation}\label{eq:transfer_time_weight}
\omega = \exp \left(-\frac{\left( \Delta t_{\mathrm{transfer}} \right)^2}{\sigma^2}\right)\in(0,1].
\end{equation}
Future public transport models should include realistic individual transfer times for all stops and all combinations of lines serving these stops.
From an implementation point of view these transfer times would be easy to include in our multiplex model without causing any issues for our algorithms.
We envision that more detailed knowledge of transfer times (including distances and walking times between different lines at the same stop) of practitioners can lead to more advanced city-specific multiplex public transport models in the future.

\section{Matrix function-based centralities}\label{sec:centralities}

Methods to identify and rank the most important nodes of complex networks have a long history in complex network science \cite{katz1953new,freeman1977set,freeman1978centrality,bonacich1987power,brin1998anatomy,kleinberg1999authoritative,page1999pagerank}.
Among many others, urban networks have been a key application of a variety of centrality measures including degree, closeness, betweenness, and different variants of eigenvector centralities \cite{crucitti2006centrality,crucitti2006bcentrality,porta2006network,scheurer2006centrality,to2015centrality,nourian2016spectral,wang2017research,agryzkov2019centrality,hellervik2019preferential,curado2021understanding}.

In this section, we consider matrix function-based centrality measures \cite{benzi2020matrix,estrada2010network,estrada2005subgraph,benzi2013total,katz1953new,bergermann2021matrix}, which interpolate between local degree centrality and global eigenvector centrality \cite{benzi2015limiting}.
An advantage of this class of centrality measures over eigenvector centrality is its more general applicability.\footnote{Eigenvector centrality is defined by the eigenvector belonging to the largest eigenvalue of a suitable matrix like, e.g., the graph's supra-adjacency matrix, cf.~\cite{taylor2017eigenvector,taylor2019supracentrality,taylor2021tunable}.
	The unique existence of this eigenvector can, however, only be guaranteed if the assumptions of the Perron--Frobenius theorem are satisfied \cite[Thm.~3.7]{taylor2021tunable}.
	This restriction does not apply to matrix function-based centrality measures, cf.~e.g.~\cite[Sec.~6.3]{bergermann2021matrix} for an example.
	Note, however, that variants of eigenvector centrality circumventing this shortcoming exist, cf.~e.g.~\cite{tudisco2018node}.}.
Recently, the class of matrix function-based centrality measures has been generalized to layer-coupled multiplex networks \cite{bergermann2021matrix}.
We illustrate that the same methods are also applicable to the more general multiplex network framework introduced in the section \nameref{sec:multiplex_networks}.
In the remainder of this section, we briefly motivate and introduce matrix function-based centrality measures and present efficient numerical methods for their computation as well as numerical results including a discussion of the impact of certain hyper-parameters.

\subsection{Definition}\label{sec:centralities_definition}

Matrix function-based centrality measures are based on the application of the matrix exponential or the matrix resolvent function to the adjacency matrix of a network.
We briefly motivate the definitions by considering walks on single-layer networks and refer the reader to \cite[Sec.~3]{bergermann2021matrix} for more details.
At the end of this subsection, we comment on the generalization of the definitions to the case of the multiplex networks introduced in the section \nameref{sec:multiplex_networks}.

In the case of an undirected and unweighted single-layer graph with $n$ nodes, the entry $[\bm{A}]_{ij}$ of the adjacency matrix $\bm{A}\in\R^{n \times n}$ is $1$ if an edge is present between nodes $i$ and $j$ and $0$ otherwise. 
Note that in the formalism introduced in the section \nameref{sec:multiplex_networks} this corresponds to $L=1$ and $\bm{A}=\Aintra=\A1$.
It is well-known from graph theory that an entry $[\bm{A}^p]_{ij}$ of the $p$th power of such an adjacency matrix contains the number of walks of length $p$ that exist between nodes $i$ and $j$ \cite{estrada2012structure}.
While (local) degree centrality can be formulated in terms of the (first power of the) adjacency matrix and (global) eigenvector centrality in terms of the limit of adjacency matrix powers, matrix function-based centralities consider walks of all lengths.
This approach can be represented by considering the adjacency matrix power series $\sum_{p=0}^{\infty} \bm{A}^p$.
Furthermore, one typically introduces a damping factor in this power series to control the magnitude of the entries of the matrix powers, which typically grows rapidly with $p$.
The following two choices of damping factors lead to the power series of the matrix exponential (left) and the matrix resolvent function (right)
\begin{equation}\label{eq:matrix_functions}
\sum_{p=0}^{\infty} \frac{\beta^p}{p!} \bm{A}^p = \exp \left( \beta\bm{A}\right), \qquad \sum_{p=0}^{\infty} \alpha^p \bm{A}^p = (\bm{I} - \alpha\bm{A})^{-1},
\end{equation}
where $\bm{I}\in\R^{n \times n}$ denotes the identity matrix and $\beta\in\R_{>0}$ and $0<\alpha<1/\lmax$\footnote{$\lmax$ denotes the largest eigenvalue of $\bm{A}$ and this choice of $\alpha$ ensures the existence of the inverse in \Cref{eq:matrix_functions} and determines the sign of all derivatives, cf.~\cite[Sec.~5.2.1]{bergermann2021matrix} and the references therein.} denote scalar hyper-parameters.
These parameters control the trade-off of locality and globality in the considered walks on the network by assigning more or less weight to longer walks.

The centrality of any given node $i$ in the network is then revealed by the following two expressions of these matrix functions:
$\bm{e}_i^T f(\bm{A}) \bm{e}_i$ denotes the diagonal element of $f(\bm{A})$, which contains the weighted sum of walks of all lengths, which start and end at node $i$;
$\bm{e}_i^T f(\bm{A})\bm{1}$ denotes the sum of the $i$th row of $f(\bm{A})$, which contains the weighted sum of all walks starting at node $i$ (regardless of where they end).
Here, $\bm{e}_i=[0, \dots , 0, 1, 0, \dots , 0]^T\in\R^n$ denotes the $i$th unit vector and $\bm{1}=[1, 1, \dots , 1]^T\in\R^n$ the vector of all ones.
This leads to the definitions of subgraph centrality \cite{estrada2005subgraph} (left) and resolvent-based subgraph centrality \cite{benzi2020matrix} (right)
\begin{equation}\label{eq:def_subgraph_resolvent-based_subgraph}
SC(i,\beta)=\bm{e}_i^T \exp \left(\beta \bm{A}\right) \bm{e}_i, \qquad SC_{\mathrm{res}}(i,\alpha)=\bm{e}_i^T (\bm{I}-\alpha \bm{A})^{-1} \bm{e}_i,
\end{equation}
as well as total communicability \cite{benzi2013total} (left) and Katz centrality \cite{katz1953new} (right)
\begin{equation}\label{eq:def_total_communicability_katz}
TC(i,\beta)=\bm{e}_i^T \exp \left( \beta \bm{A}\right) \bm{1}, \qquad KC(i,\alpha)=\bm{e}_i^T (\bm{I}-\alpha \bm{A})^{-1} \bm{1}.
\end{equation}
The same definitions hold true for weighted graphs and although extensions to directed networks exist \cite{bergermann2021matrix}, we restrict ourselves to undirected networks in this paper, which leads to symmetric adjacency matrices $\bm{A}^T=\bm{A}$.

It has recently been proposed to generalize the above definitions to the case of layer-coupled multiplex networks by replacing the graph adjacency matrix by the supra-adjacency matrix of the corresponding multiplex networks \cite[Sec.~4]{bergermann2021matrix}.
We will demonstrate that all methods from \cite{bergermann2021matrix} for the symmetric case $\bm{A}=\bm{A}^T$ still apply for the more general multiplex networks introduced in the section \nameref{sec:multiplex_networks}.

Note that by the construction of the supra-adjacency matrix defined in \Cref{eq:supra_adjacency} the above quantities yield centrality values for all node-layer pairs, which allows to identify and rank the most central node-layer pairs of the network.
Following \cite{taylor2017eigenvector}, we call these values joint centralities and denote the joint centrality of physical node $i$ in layer $l$ by $JC(i,l)$.
However, we remark that the numerical methods introduced in the following subsection are unable to compute subgraph and resolvent-based subgraph centralities of isolated node-layer pairs, i.e., node-layer pairs without any adjacent edge.
We set the centrality values of these node-layer pairs to $1$, which is consistent with \Cref{eq:matrix_functions}, in which the respective rows and columns of the supra-adjacency matrix power series are given by unit vectors.

Finally, we adapt the concept of marginal node ($MNC$) and marginal layer centralities ($MLC$) \cite{taylor2017eigenvector}.
These quantities are defined as
\begin{equation}\label{eq:marginal_centralities}
MNC(i) = \sum_{l=1}^L JC(i,l), \qquad MLC(l) = \sum_{i=1}^n JC(i,l),
\end{equation}
and can be used to assess the importance of all physical nodes and layers of the multiplex network, which will play a key role in our interpretation of the results.

\subsection{Numerical methods}\label{sec:centralities_methods}

We now discuss strategies for the numerical evaluation of the matrix function-based centrality measures defined in \Cref{eq:def_subgraph_resolvent-based_subgraph,eq:def_total_communicability_katz}.
For small network sizes, many software packages provide accurate algorithms for the explicit evaluation of the matrix exponential and the solution of a regular linear system (an efficient numerical implementation of, e.g., $(\bm{I} - \alpha \bm{A})^{-1} \bm{1}$ would solve the linear system $(\bm{I} - \alpha \bm{A}) \bm{x} = \bm{1}$ for the vector $\bm{x}$).
These methods, however, quickly become infeasible for medium to large network sizes and we briefly present efficient and scalable approximations based on Krylov subspace methods.
More details can be found in \cite{benzi2020matrix} for single-layer networks and in \cite[Sec.~5]{bergermann2021matrix} for multiplex networks.

As we only consider undirected graphs, i.e., symmetric supra-adjacency matrices in this paper we rely on the symmetric methods introduced in \cite[Sec.~5]{bergermann2021matrix}, which are based on the symmetric Lanczos method \cite{lanczos1950iteration,golub2013matrix}.
In its $k$th iteration, this method constructs the $k$th column of an orthogonal basis $\bm{Q}_k\in\R^{nL \times k}$ of the Krylov subspace
\begin{equation}\label{eq:krylov_subspace}
\mathcal{K}_k(\bm{A}, \bm{v}) = \text{span}\{ \bm{v} , \bm{A} \bm{v} , \bm{A}^2 \bm{v} , \dots , \bm{A}^{k-1} \bm{v} \}
\end{equation}
to a matrix $\bm{A}=\bm{A}^T\in\R^{nL \times nL}$ and a vector $\bm{v}\in\R^{nL}$.
This allows the decomposition of $\bm{A}$ into the form $\bm{A} \approx \bm{Q}_k \bm{T}_k \bm{Q}_k^T$, where $\bm{T}_k=\bm{T}_k^T\in\R^{k \times k}$ has tridiagonal form.
This approximation typically achieves a high accuracy for $k \ll nL$, which makes the eigendecomposition $\bm{T}_k = \bm{S}_k \bm{\Theta}_k \bm{S}_k^T$ easy to compute with standard methods.
These two matrix factorizations can then be combined to compute total communicability and Katz centrality by evaluating the quantity
\begin{equation}\label{eq:fAb}
f(\bm{A})\bm{1} \approx \bm{Q}_k \bm{S}_k f(\bm{\Theta}_k) \bm{S}_k^T \bm{Q}_k^T \bm{1},
\end{equation}
where $f$ is applied elementwise to the eigenvalues of $\bm{T}_k$  \cite{higham2008functions}.
The $i$th entry of the resulting vector then corresponds to the centrality value of the $i$th node-layer pair.

For subgraph and resolvent-based subgraph centrality we rely on an elegant relation between the symmetric Lanczos method, orthogonal polynomials, and Gauss quadrature discussed by Golub and Meurant \cite{golub2009matrices,golub1969calculation,golub1994matrices,golub1997matrices}, which can be used to compute lower and upper bounds on the sought quantities.
The final result of this approach yields a lower Gauss quadrature bound, which reads
\begin{equation}\label{eq:gauss_quadrature}
\bm{e}_i^T f(\bm{A}) \bm{e}_i \approx \bm{e}_1^T \bm{S}_k f(\bm{\Theta}_k) \bm{S}_k^T \bm{e}_1,
\end{equation}
where the computation of $\bm{T}_k=\bm{S}_k \bm{\Theta}_k \bm{S}_k^T$ relies on the basis of the Krylov subspace from \Cref{eq:krylov_subspace} with $\bm{v}=\bm{e}_i$.
We refer to \cite{golub2009matrices,golub1969calculation,golub1994matrices,golub1997matrices} and \cite[Sec.~5.2.1]{bergermann2021matrix} for theoretical background on this approximation as well as details on the construction of Gauss--Radau and Gauss--Lobatto rules, which yield an additional lower as well as two upper bounds on $\bm{e}_i^T f(\bm{A}) \bm{e}_i$.
Note that due to the high degree of sparsity in the supra-adjacency matrices of our multiplex public transport networks, Gauss quadrature rules can only be applied to the small subset of non-isolated node-layer pairs.
As described in the previous subsection, the centrality value of the remaining node-layer pairs is set to $1$.

All numerical methods introduced in this subsection scale linearly in the number of node-layer pairs under the assumption of sparsity in the supra-adjacency matrix, cf.~\cite[Sec.~5]{bergermann2021matrix}, and usually obtain highly accurate approximations in only around $10$ Lanczos iterations.
However, the quantities $\bm{e}_i^T f(\bm{A}) \bm{e}_i$ require the approximation of a separate quantity for each non-isolated node-layer pair, which makes them computationally more demanding than the quantities $f(\bm{A})\bm{1}$, which only need to be computed once.
As discussed in \cite[Sec.~6.1]{bergermann2021matrix}, the convergence of \Cref{eq:gauss_quadrature} is usually faster than that of \Cref{eq:fAb}.
A python implementation of the above methods is available at \url{https://github.com/KBergermann/Urban-multiplex-networks}.

\subsection{Results}

In this subsection, we present numerical results of the different matrix function-based centrality measures applied to the multiplex network representation of several German urban public transport networks.
For the interpretation of the resulting rankings of node-layer pairs (representing stop-line pairs), we mainly rely on marginal node centralities defined in \Cref{eq:marginal_centralities}.
This approach corresponds to the identification and ranking of the most central locations in urban networks, which has been the subject of many earlier studies with different centrality measures \cite{crucitti2006centrality,crucitti2006bcentrality,porta2006network,scheurer2006centrality,to2015centrality,nourian2016spectral,agryzkov2019centrality,curado2021understanding}.
We compare different weight models and matrix function-based centrality measures, which were introduced in the preceding sections.
Furthermore, we study the impact of all involved hyper-parameters and close with a discussion of the obtained results including their relation to public transport orientations.

Previous results from the literature on the application of centrality measures to urban transport networks indicate qualitative differences in the distribution of the most central nodes between the cases of public transport \cite{scheurer2006centrality,to2015centrality,curado2021understanding} and street networks \cite{crucitti2006centrality,crucitti2006bcentrality,porta2006network,nourian2016spectral,agryzkov2019centrality}.
As street networks are usually modeled as (almost) planar graphs in which the ``closeness'' of nodes is determined by their geographical distance the distribution of central nodes is usually characterized by a relatively smooth transition from more to less central nodes in terms of this geographical distance.
For instance, one often observes approximately circular shapes of descending centrality around the most central node of such a network \cite{crucitti2006centrality,crucitti2006bcentrality,porta2006network,nourian2016spectral,agryzkov2019centrality}.
Conversely, in public transport networks the ``closeness'' of nodes is typically tied to different indicators such as travel times or line frequencies.
This results in a less smooth distribution of node centralities with respect to the geographical position of the nodes.
For instance, stops, which are geographically very close to the most central stop of a public transport network can be classified as non-central if the two stops are not directly connected by any line \cite{scheurer2006centrality,to2015centrality,curado2021understanding}.
This general qualitative behavior is confirmed by the multiplex matrix function-based centrality measures and can be observed throughout the subsequent results.
Note that a combination of the different modeling approaches by, e.g., adding car or walking layers to public transport networks could mitigate this effect and constitute an interesting future research direction.

\begin{figure}
	\begin{center}
		\subfloat[Marginal layer centralities]{
			\includegraphics[width=.45\textwidth]{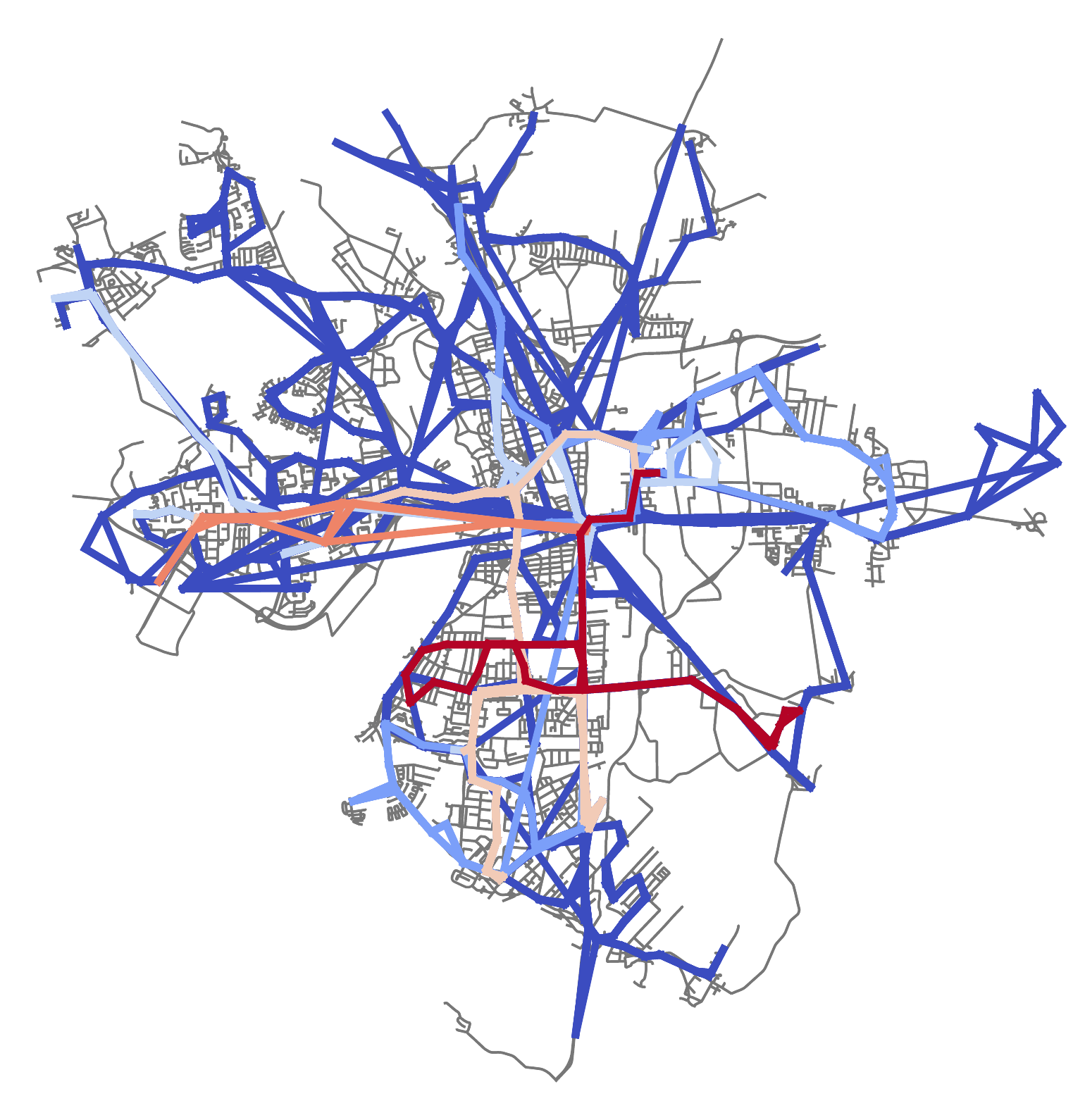}\label{fig:MLC_MNC_Halle_layer}
		}
		\hfill
		\subfloat[Marginal node centralities]{
			\includegraphics[width=.45\textwidth]{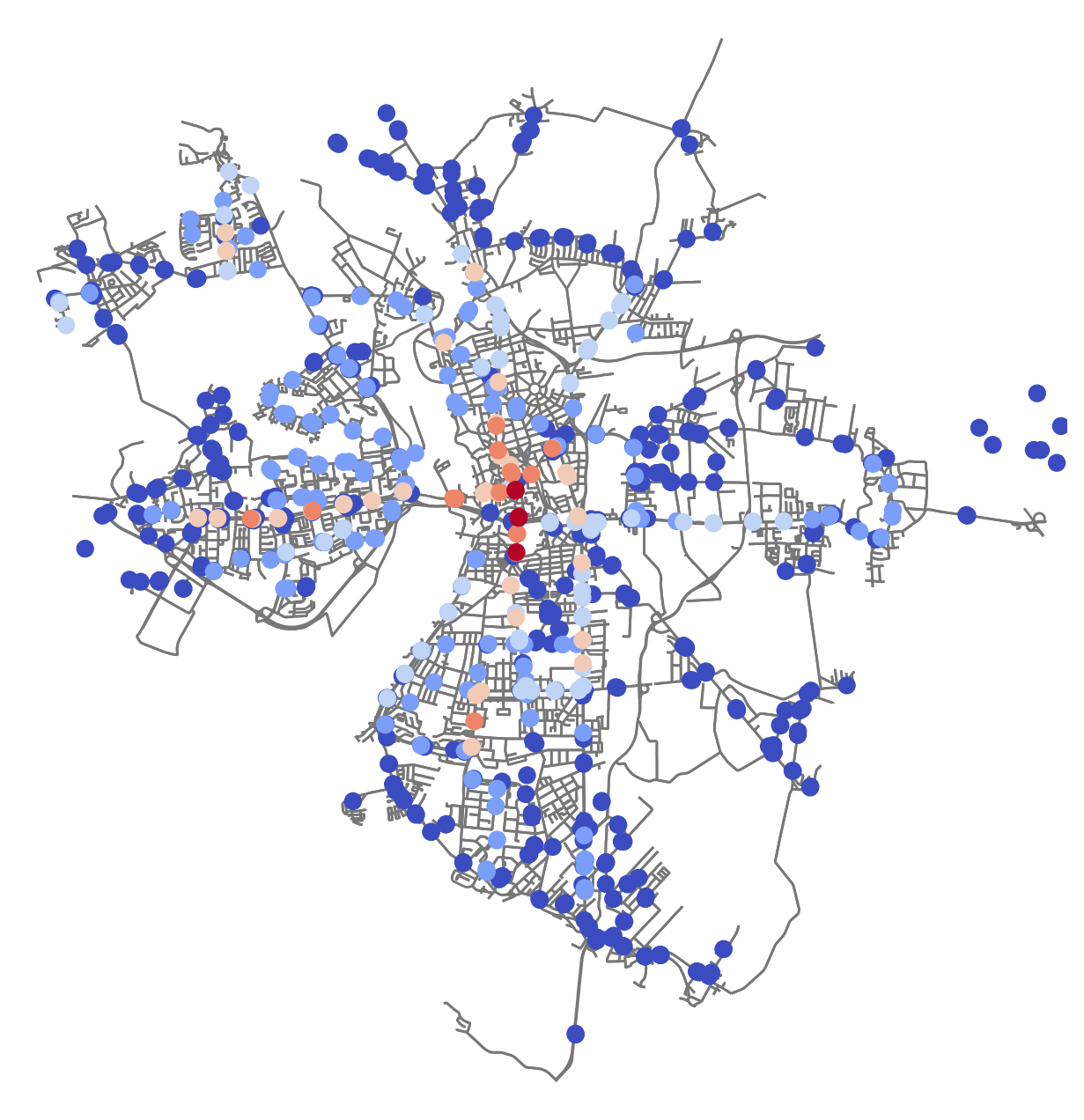}\label{fig:MLC_MNC_Halle_node}
		}
	\end{center}
	\caption{Example of marginal layer and marginal node subgraph centralities of Halle (Saale).
		Central lines and stops are marked red and non-central lines and stops are marked blue.
		The entries in the supra-adjacency matrix are weighted with travel times and frequencies.
		The parameters are chosen $\beta=0.5/\lmax$, $\sigma=5$, and $\Delta t_{\mathrm{transfer}}=5$.
		The street network in the background of the plots is created with the OSMnx python package \cite{boeing2017osmnx}.
		Left: Marginal layer centralities (corresponding to a ranking of the lines).
		Right: Marginal node centralities (corresponding to a ranking of the stops).}\label{fig:MLC_MNC_Halle}
\end{figure}

To illustrate this general behavior for our modeling approach we start by considering marginal layer and marginal node subgraph centralities of the German city Halle (Saale) as a first example.
\Cref{fig:MLC_MNC_Halle_layer} illustrates the ranking of lines while \Cref{fig:MLC_MNC_Halle_node} illustrates the ranking of stops in the corresponding multiplex public transport network.
All marginal centrality plots in this paper employ a heat map color scheme consisting of six categories representing centrality value sub-intervals of equal length.
While there is a tendency of geographically central stops to be classified as central public transport stops there are various exceptions in the form of dark blue (least central) stops in or around the city's geographical center, cf.~\Cref{fig:MLC_MNC_Halle_node}.
Instead, the comparison with \Cref{fig:MLC_MNC_Halle_layer} shows that there is a tendency of central stops to line up along important transport axes of the urban public transport system.
It is also interesting to note that the orientations of the most central lines in \Cref{fig:MLC_MNC_Halle_layer} coincide with the preferential directions of the public transport network identified in \Cref{fig:orientations_public_transport_germany}.

\subsubsection{Comparison of different measures}\label{sec:results_measures}

We continue our discussion with the comparison of the four matrix function-based centrality measures for the multiplex public transport networks defined in \Cref{eq:def_subgraph_resolvent-based_subgraph,eq:def_total_communicability_katz} at the example of marginal node centralities of Cologne.

\begin{figure}
	\begin{center}
		\subfloat[$SC$]{
			\includegraphics[width=.45\textwidth]{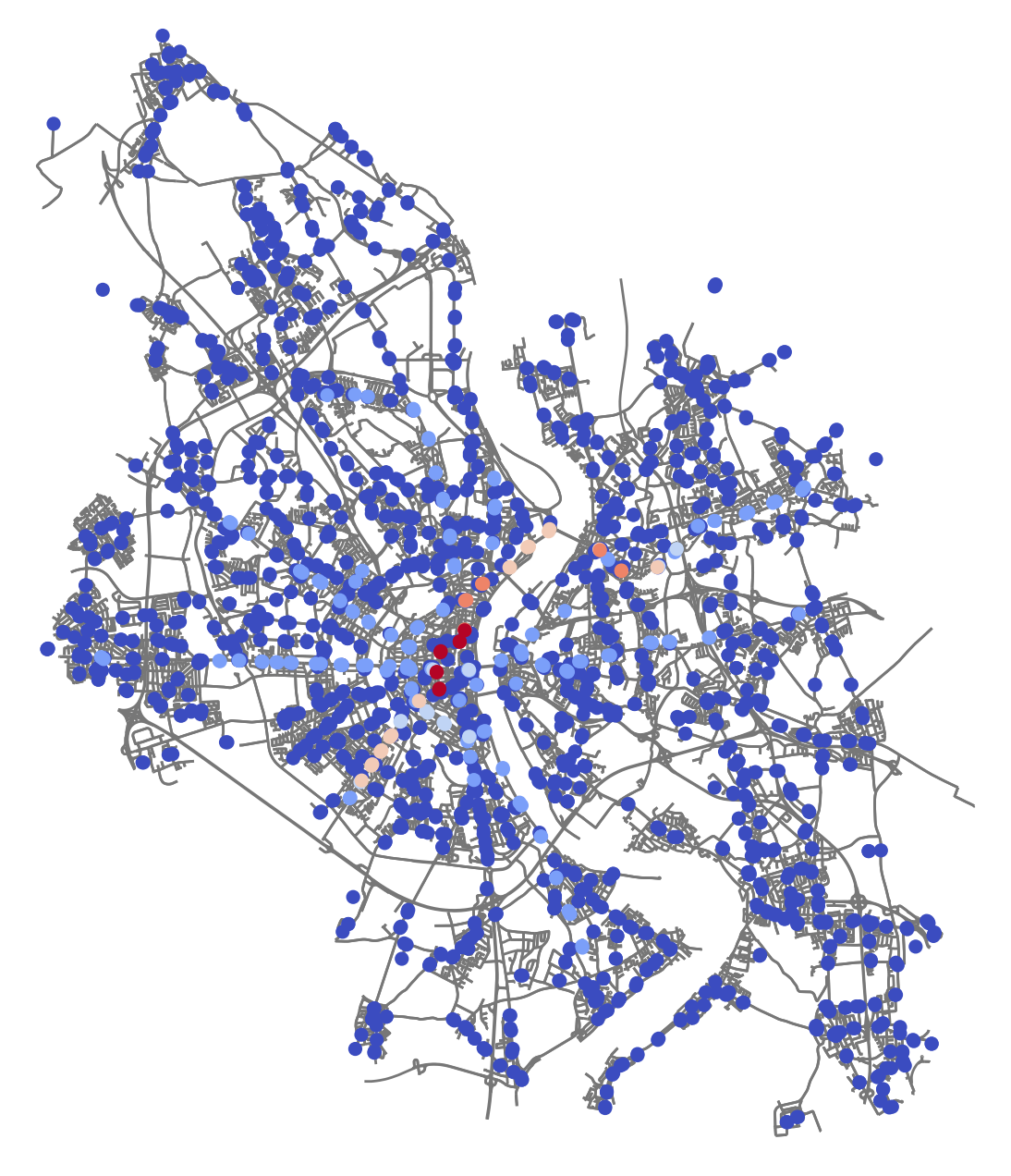}\label{fig:comparison_four_measures_SC}
		}
		\hfill
		\subfloat[$SC_{\mathrm{res}}$]{
			\includegraphics[width=.45\textwidth]{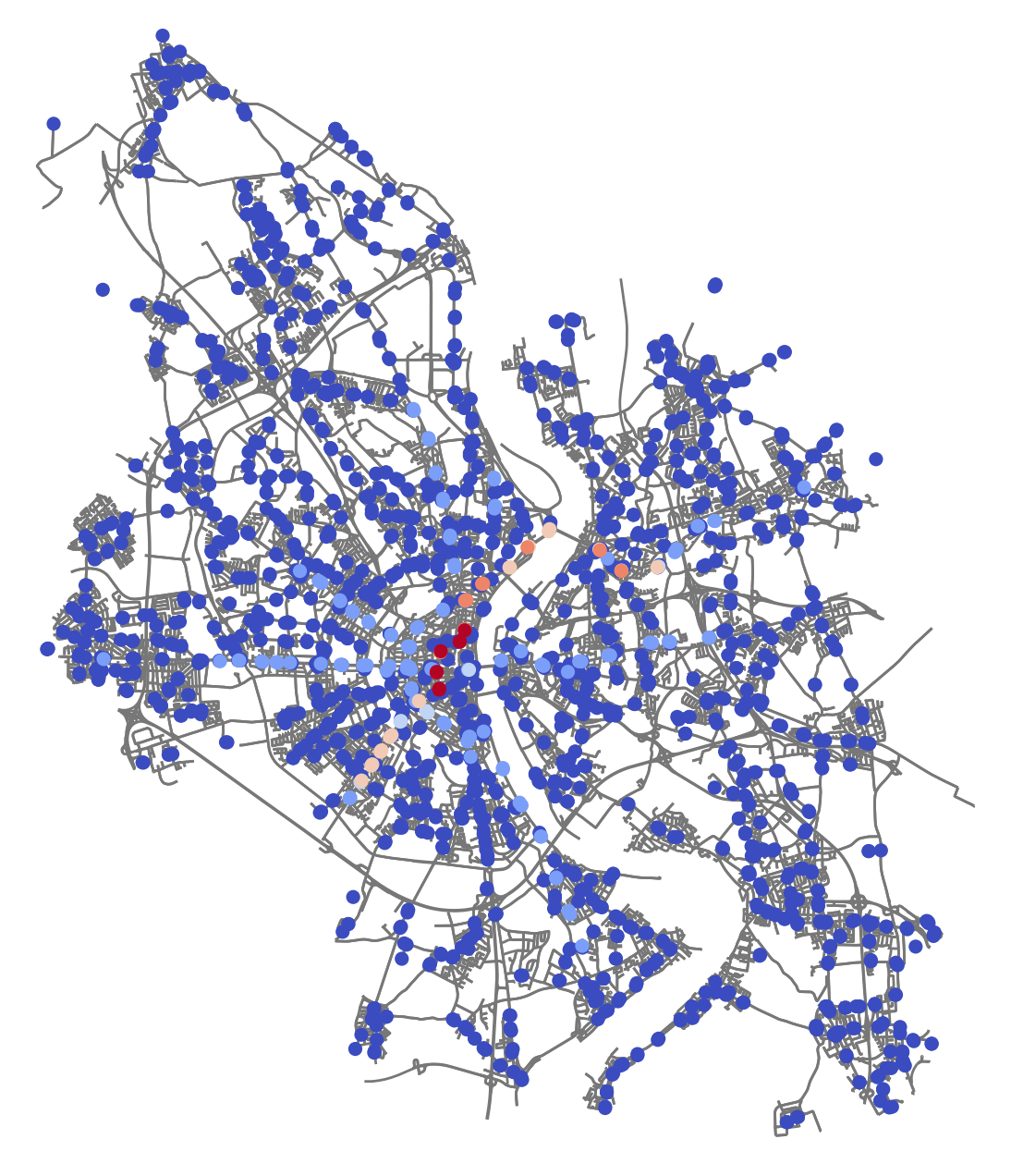}\label{fig:comparison_four_measures_SCres}
		}
		
		\subfloat[$TC$]{
			\includegraphics[width=.45\textwidth]{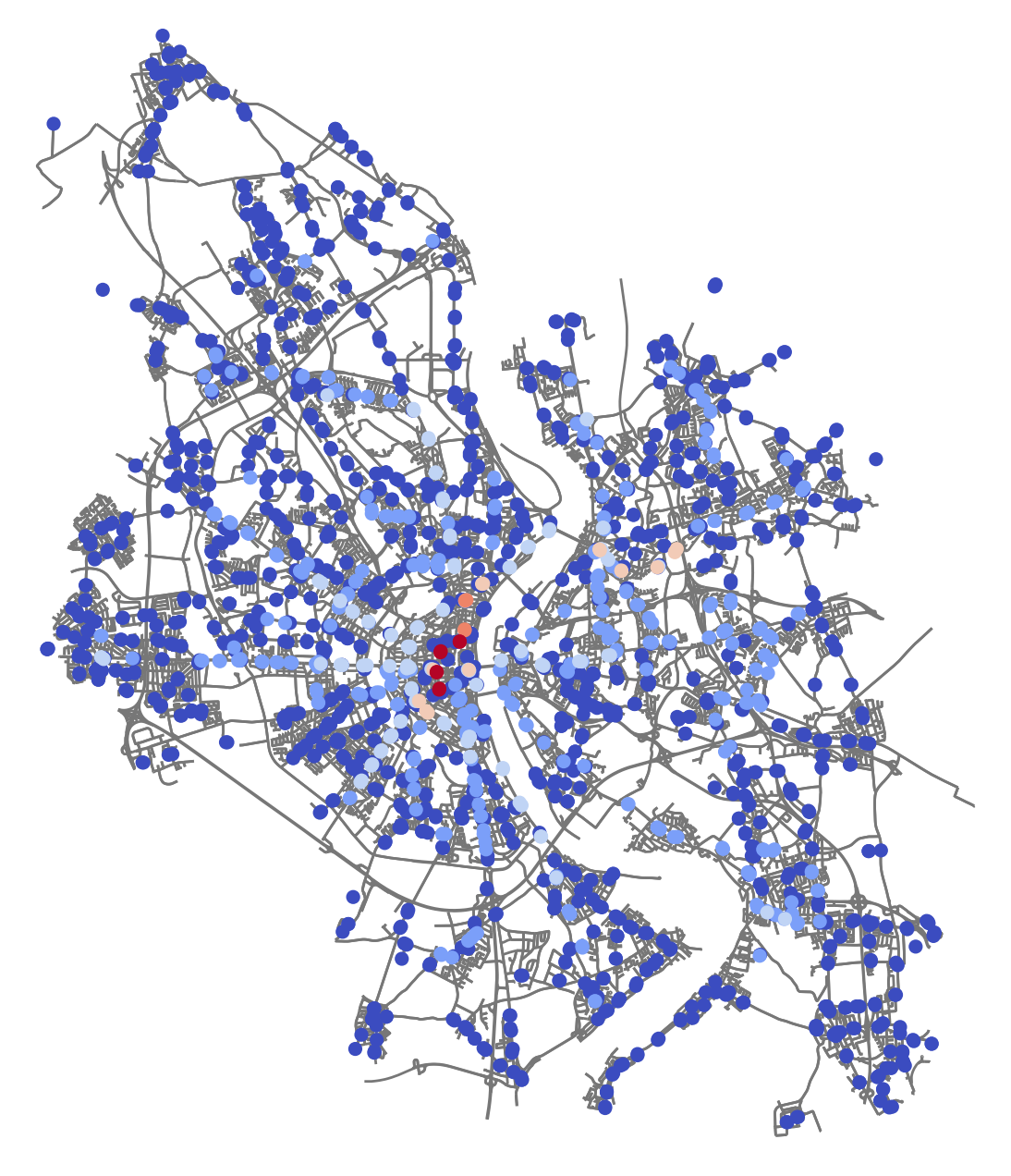}\label{fig:comparison_four_measures_TC}
		}
		\hfill
		\subfloat[$KC$]{
			\includegraphics[width=.45\textwidth]{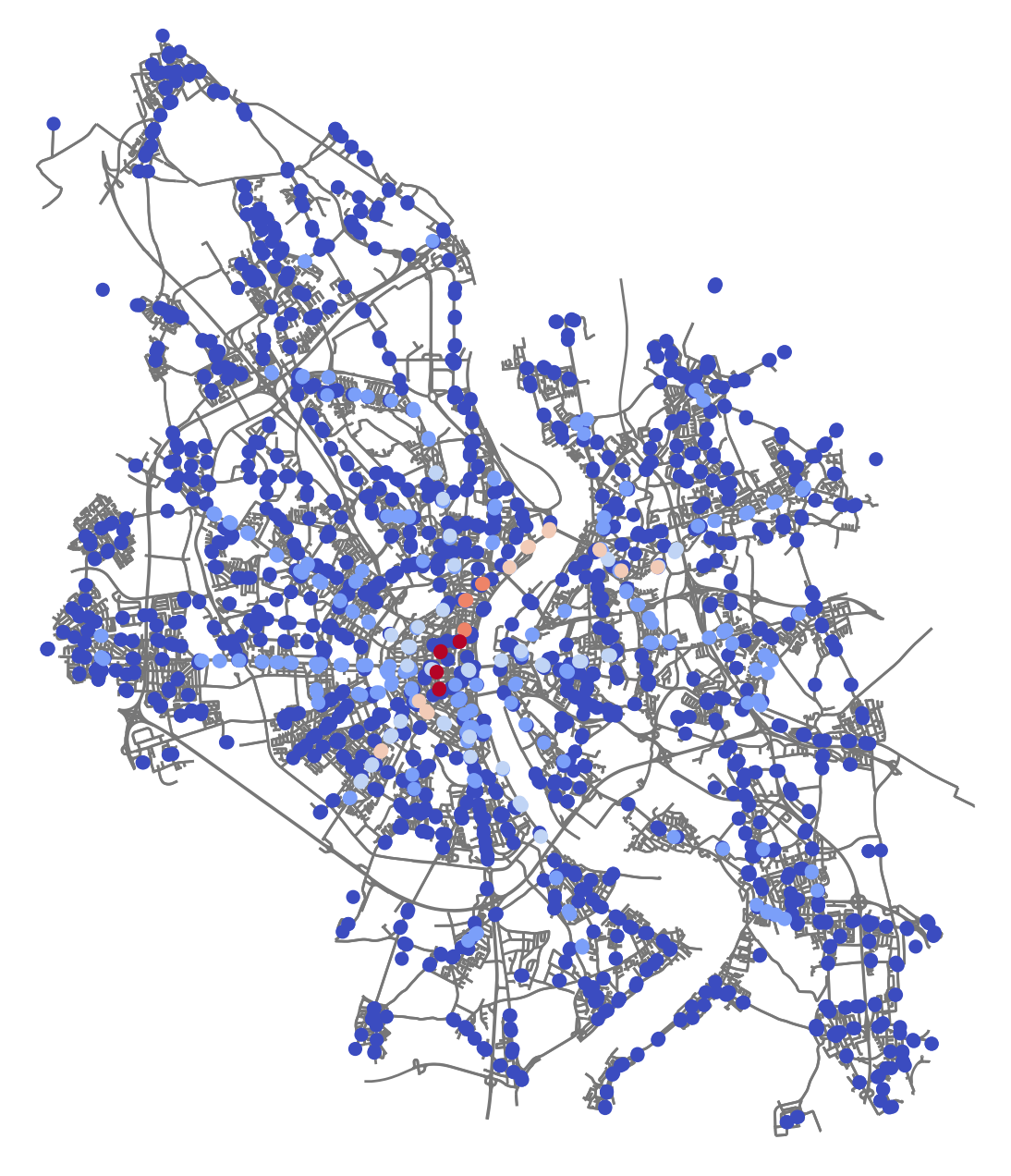}\label{fig:comparison_four_measures_KC}
		}
		\caption{Comparison of the four different matrix function-based centralities at the example of marginal node centralities of Cologne.
			Central stops are marked red and non-central stops are marked blue.
			The entries in the supra-adjacency matrix are weighted with travel times and frequencies.
			The parameters are set to $\alpha=\beta=0.5/\lmax$, $\sigma=5$, and $\Delta t_{\mathrm{transfer}}=5$.
			The street network in the background of the plots is created with the OSMnx python package \cite{boeing2017osmnx}.}\label{fig:comparison_four_measures}
	\end{center}
\end{figure}

\Cref{fig:comparison_four_measures} reflects the typical behavior of matrix function-based centrality measures that rankings produced by the four measures are similar but not equal \cite{benzi2020matrix,bergermann2021matrix}.
For the multiplex public transport networks considered in this paper one typically observes very similar results for $SC$ and $SC_{\mathrm{res}}$ as well as for $TC$ and $KC$.
However, $SC$ and $SC_{\mathrm{res}}$ typically show a centrality value distribution that is more uniformly distributed across the six centrality value categories, which results in more stops being classified into the top three categories (marked in red and orange) than it is the case for $TC$ and $KC$.
The difference between the two groups of measures is that $SC$ and $SC_{\mathrm{res}}$ are defined by the matrix functions' diagonal entries, while off-diagonal entries of the matrix functions are additionally included in $TC$ and $KC$.
This corresponds to $SC$ and $SC_{\mathrm{res}}$ only considering closed walks on the networks whereas $TC$ and $KC$ consider all walks.
It thus follows that the inclusion of the off-diagonals amplifies the quantitative ranking of the stops due to a steeper distribution of the off-diagonals in comparison to the diagonal entries of the matrix functions.

\subsubsection{Comparison of different weight models}\label{sec:results_weight_models}

We also compare the different weight models discussed in the section \nameref{sec:multiplex_networks}.
The modeling approaches include all combinations of in- and excluding travel times and line frequencies in the intra-layer weights $w^{(l)}_{ij}$ as specified in \Cref{eq:weighting_models}.
Whenever intra-layer travel times are included in $w^{(l)}_{ij}$ we also include the transfer time between lines in the coupling parameter $\omega$ as defined in \Cref{eq:transfer_time_weight}.

\begin{figure}
	\begin{center}
	\subfloat[{$[\Al]_{ij}=1$}]{
		\includegraphics[width=.4\textwidth]{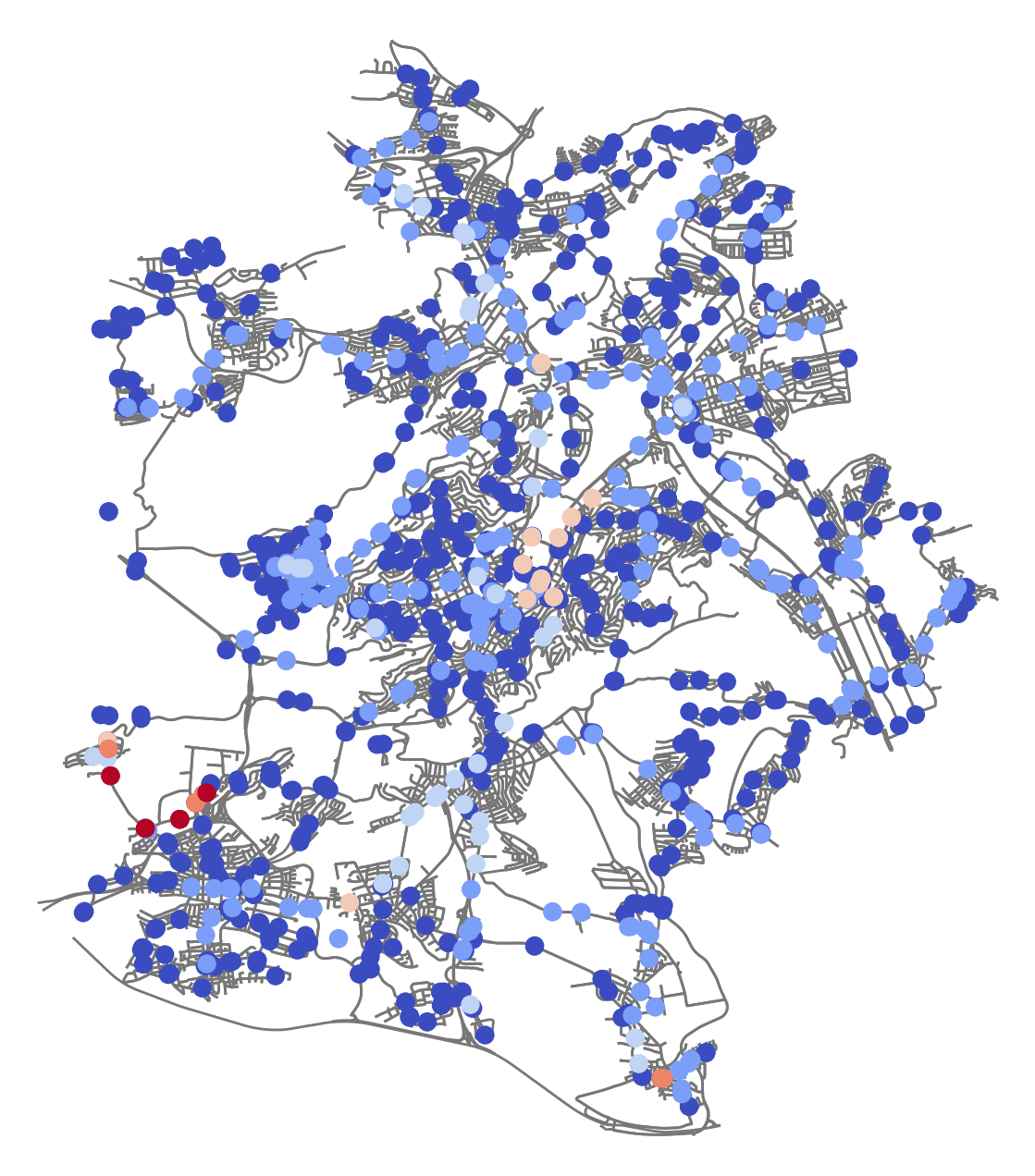}\label{fig:weight_models_unweighted_no_frequencies}
	}
	\hfill
	\subfloat[{$[\Al]_{ij}=\phi_{ij}^{(l)}$}]{
		\includegraphics[width=.4\textwidth]{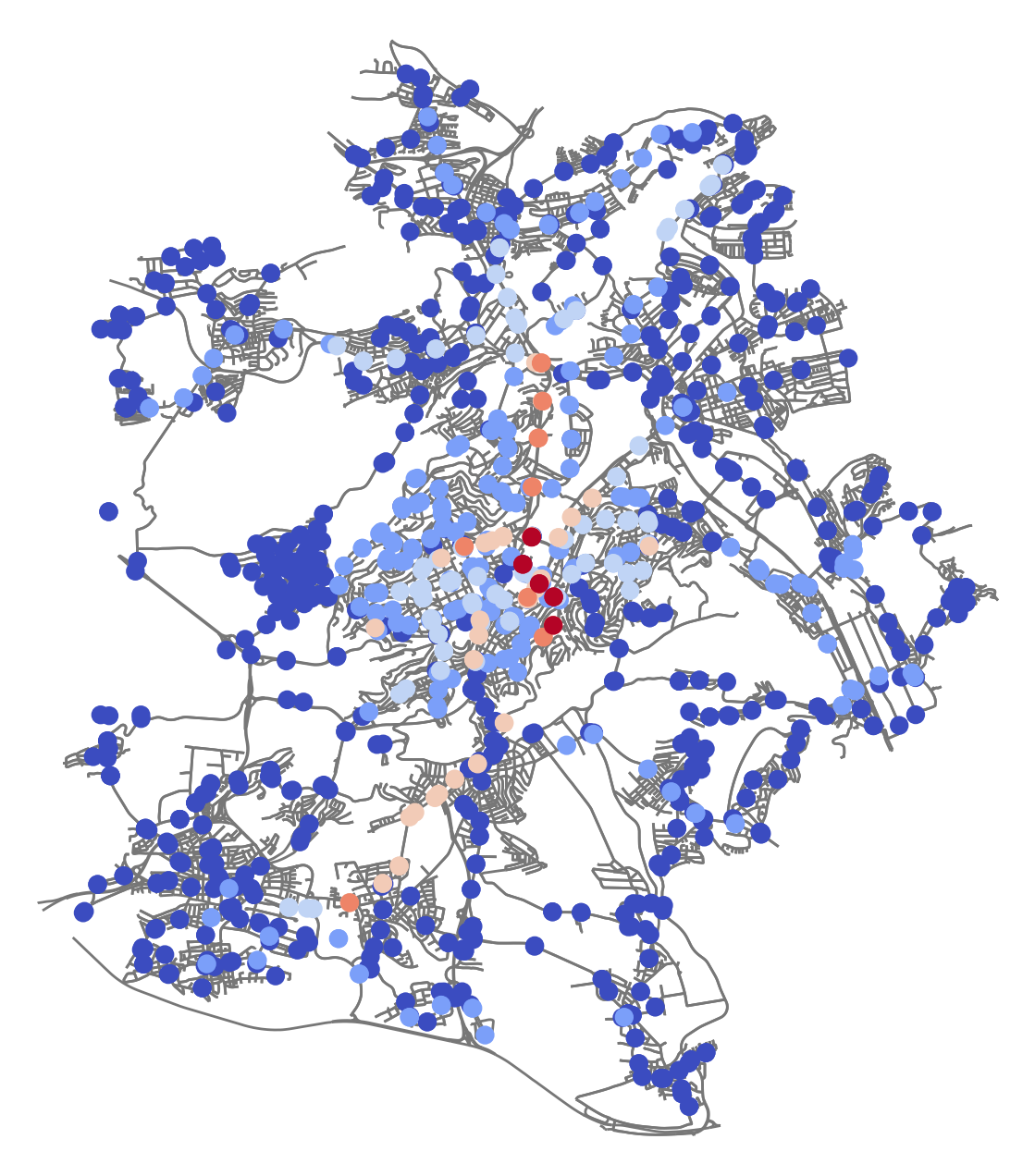}\label{fig:weight_models_unweighted_frequencies}
	}
	
	\subfloat[{$[\Al]_{ij}=\exp \left(-\frac{\left( (\Delta t)_{ij}^{(l)} \right)^2}{\sigma^2}\right)$}]{
		\includegraphics[width=.4\textwidth]{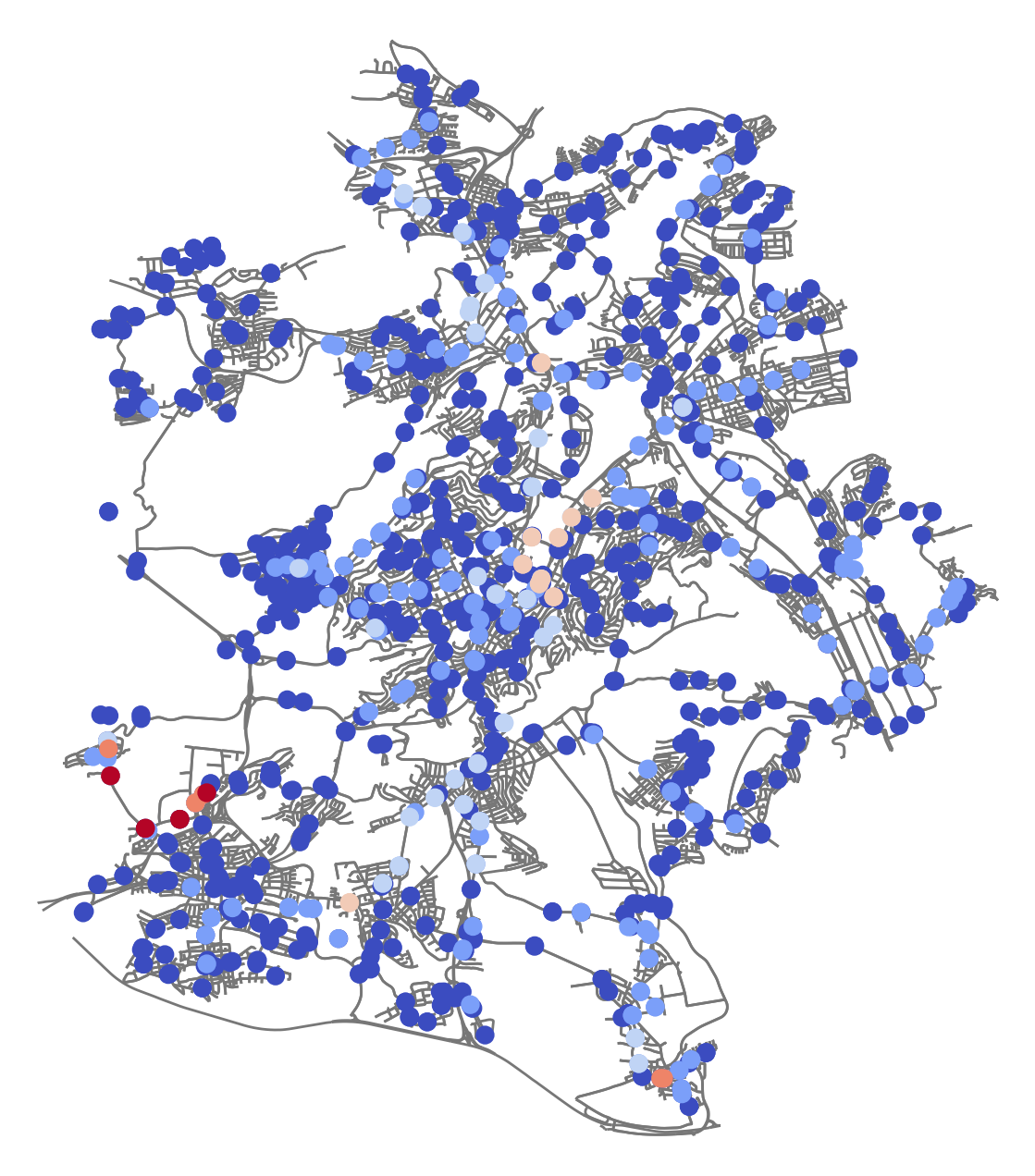}\label{fig:weight_models_weighted_no_frequencies}
	}
	\hfill
	\subfloat[{$[\Al]_{ij}=\phi_{ij}^{(l)} \exp \left(-\frac{\left( (\Delta t)_{ij}^{(l)} \right)^2}{\sigma^2}\right)$}]{
		\includegraphics[width=.4\textwidth]{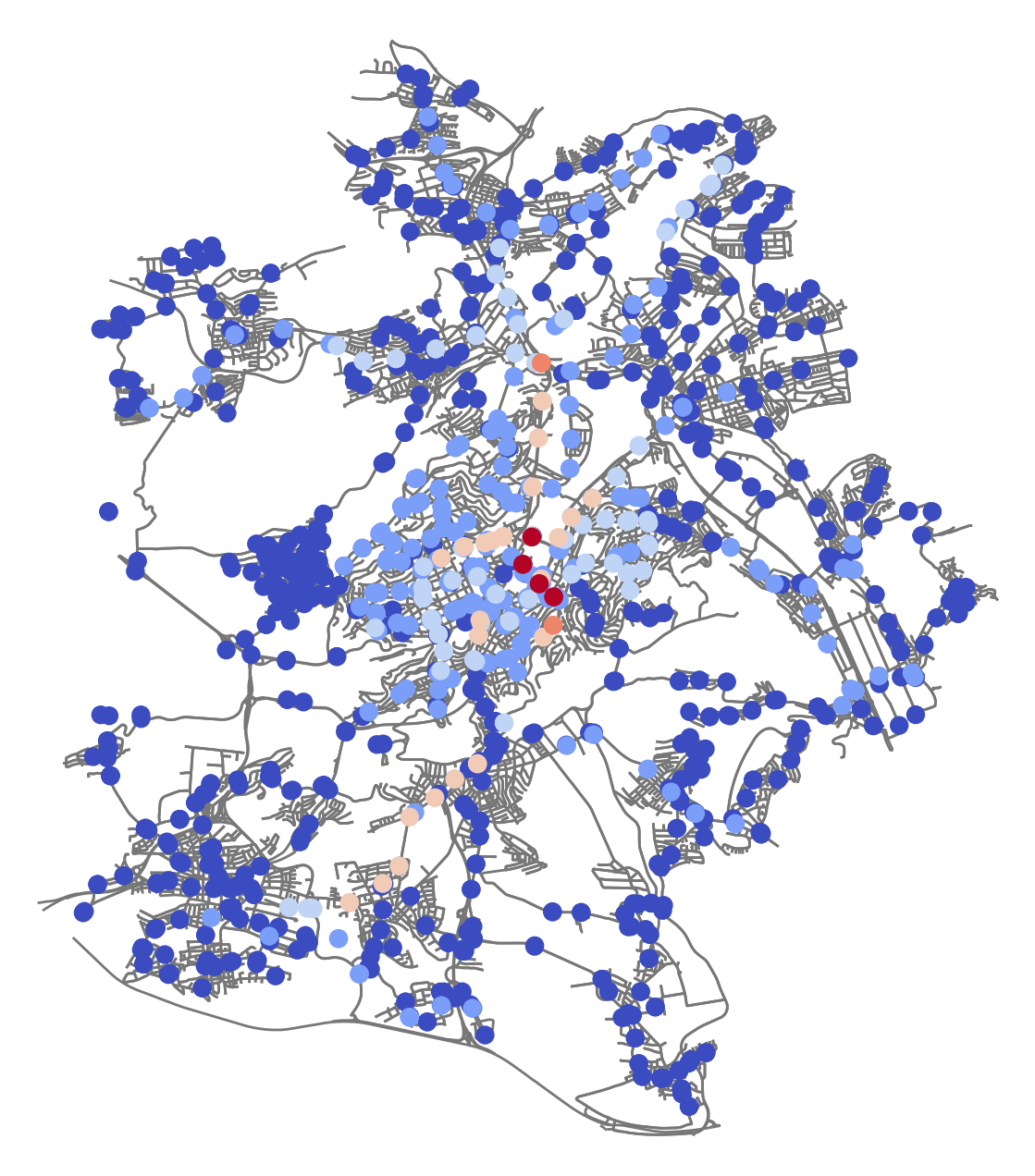}\label{fig:weight_models_weighted_frequencies}
	}
	\caption{Comparison of different weight models at the example of marginal node resolvent-based subgraph centralities of Stuttgart.
		Central stops are marked red and non-central stops are marked blue.
		The subcaptions denote the respective intra-layer weights $[\Al]_{ij}$ in the case that nodes $i$ and $j$ are connected in layer $l$.
		Otherwise, $[\Al]_{ij}$ is set to $0$ in all cases.
		The plots in the first column are obtained without frequencies, whereas the plots in the second column are obtained with frequencies.
		The plots in the first row are obtained without a Gaussian kernel applied to travel times, whereas the plots in the second row are obtained with a Gaussian kernel applied to travel times.
		Furthermore, the same Gaussian kernel is also applied to inter-layer transfer times in the second row.
		The parameters are chosen as $\alpha=0.5/\lmax$, $\sigma=5$, and $\Delta t_{\mathrm{transfer}}=5$.
		The street network in the background of the plots is created with the OSMnx python package \cite{boeing2017osmnx}.}\label{fig:weight_models}
	\end{center}
\end{figure}

\Cref{fig:weight_models} presents marginal node resolvent-based subgraph centralities of Stuttgart in the different scenarios.
It shows that the inclusion of line frequencies has a larger impact on the centralities than travel times do.
Interestingly, the exclusion of line frequencies has the effect of advantaging stops with large inter-layer degrees, i.e., stops with many transfer options.
In the case of \Cref{fig:weight_models_unweighted_no_frequencies,fig:weight_models_weighted_no_frequencies} this leads to ``Stuttgart Universit\"at'' (Stuttgart university) being classified as the most central stop of the city albeit being geographically located near the city limit.
Later in the section \nameref{sec:results_transfer_times} we discuss an interesting property of weight models without frequencies regarding the influence of transfer times on marginal node centralities.

\Cref{fig:weight_models_unweighted_frequencies,fig:weight_models_weighted_frequencies}, which include line frequencies, show what is likely to be a more expectable stop ranking for Stuttgart.
Here, the increased frequencies of lines serving the geographical center of the city shift the highest ranked stops to this central region.

Overall, \Cref{fig:weight_models} suggests that travel times do not have a large impact on the obtained stop rankings.
However, their inclusion adds an important ingredient for a realistic modeling of urban public transport networks as it allows the inclusion of transfer times between lines.

In the following three subsections, we discuss the influence of all involved hyper-parameters on the obtained stop rankings.
These parameters include the normalizing travel time $\sigma$, the matrix function parameters $\alpha$ and $\beta$, and the transfer time $\Delta t_{\mathrm{transfer}}$.

\subsubsection{Influence of the normalizing travel time}\label{sec:results_normalizing_travel_time}

Numerical experiments suggest that the normalizing travel time $\sigma$ only has a small impact on the obtained rankings in a small parameter range:
while increasing $\sigma$ from $0.2$ to $2$ (corresponding to $12$ seconds and $2$ minutes normalizing travel time, respectively) slightly increases the number of highly ranked stops in the city center, a further increase of $\sigma$ to $20$ or $200$ minutes has no noticeable effect.

\subsubsection{From local to global}

As proven for single-layer networks in \cite{benzi2015limiting} and as confirmed empirically for layer-coupled multiplex networks in \cite[Sec.~6]{bergermann2021matrix}, all four matrix function-based centrality measures defined in \Cref{eq:def_subgraph_resolvent-based_subgraph,eq:def_total_communicability_katz} contain degree and eigenvector centrality as limit cases of the parameters $\alpha$ or $\beta$.
In this subsection, we empirically confirm this behavior for marginal node centralities of our more general multiplex network models.

\begin{figure}
	\begin{center}
		\subfloat[$\alpha=0.01 / \lmax$]{
			\includegraphics[width=.3\textwidth]{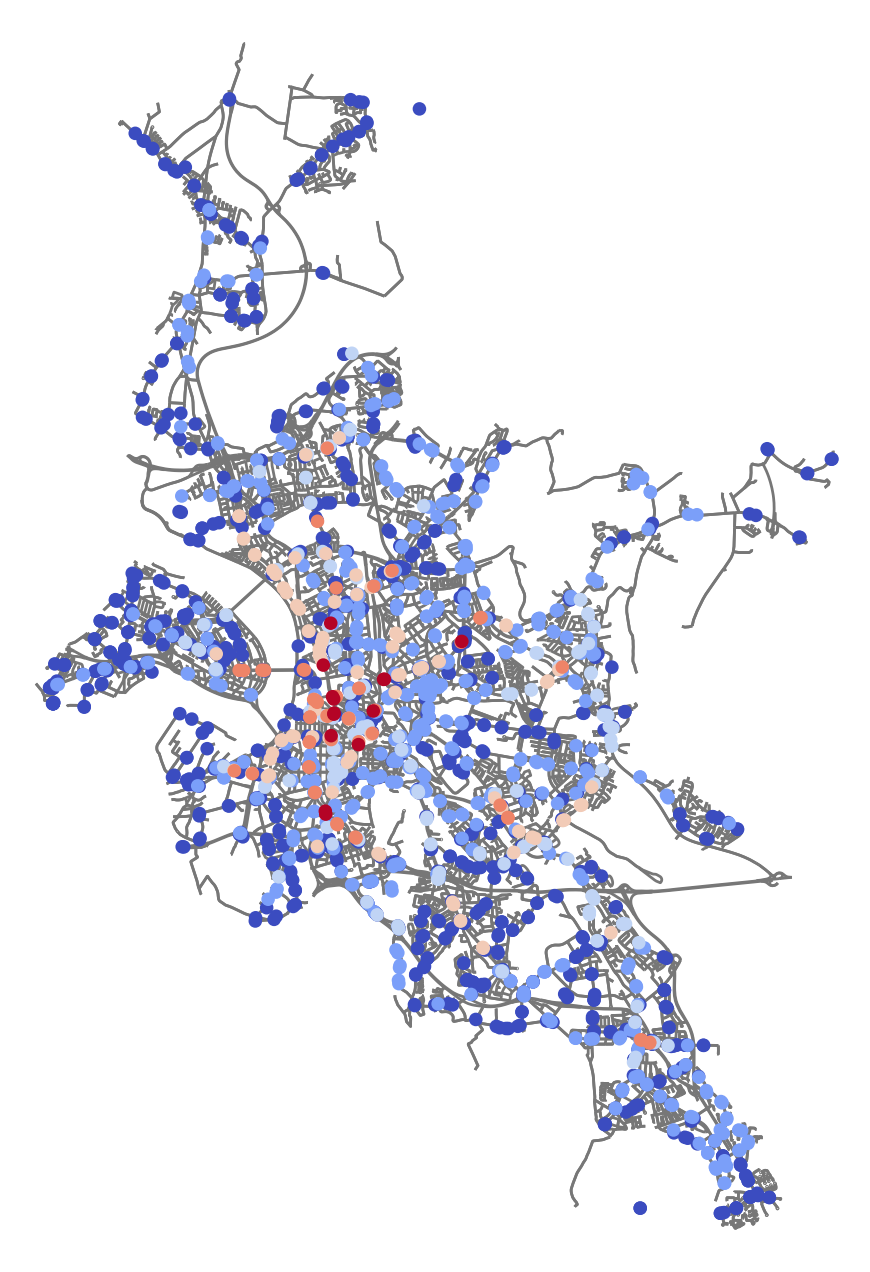}\label{fig:local_to_global_duesseldorf_0_01}
		}
		\hfill
		\subfloat[$\alpha=0.75 / \lmax$]{
			\includegraphics[width=.3\textwidth]{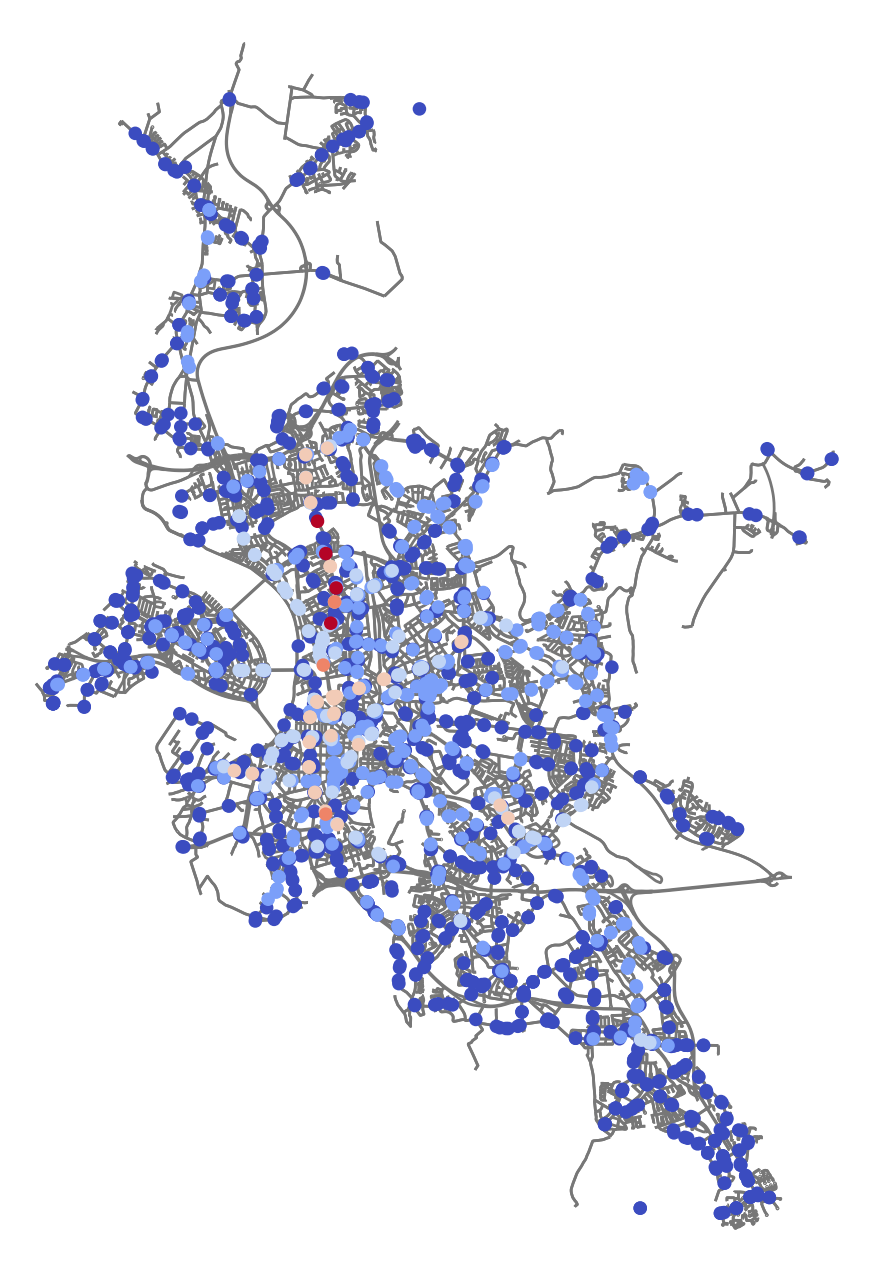}\label{fig:local_to_global_duesseldorf_0_75}
		}
		\hfill
		\subfloat[$\alpha=0.99 / \lmax$]{
			\includegraphics[width=.3\textwidth]{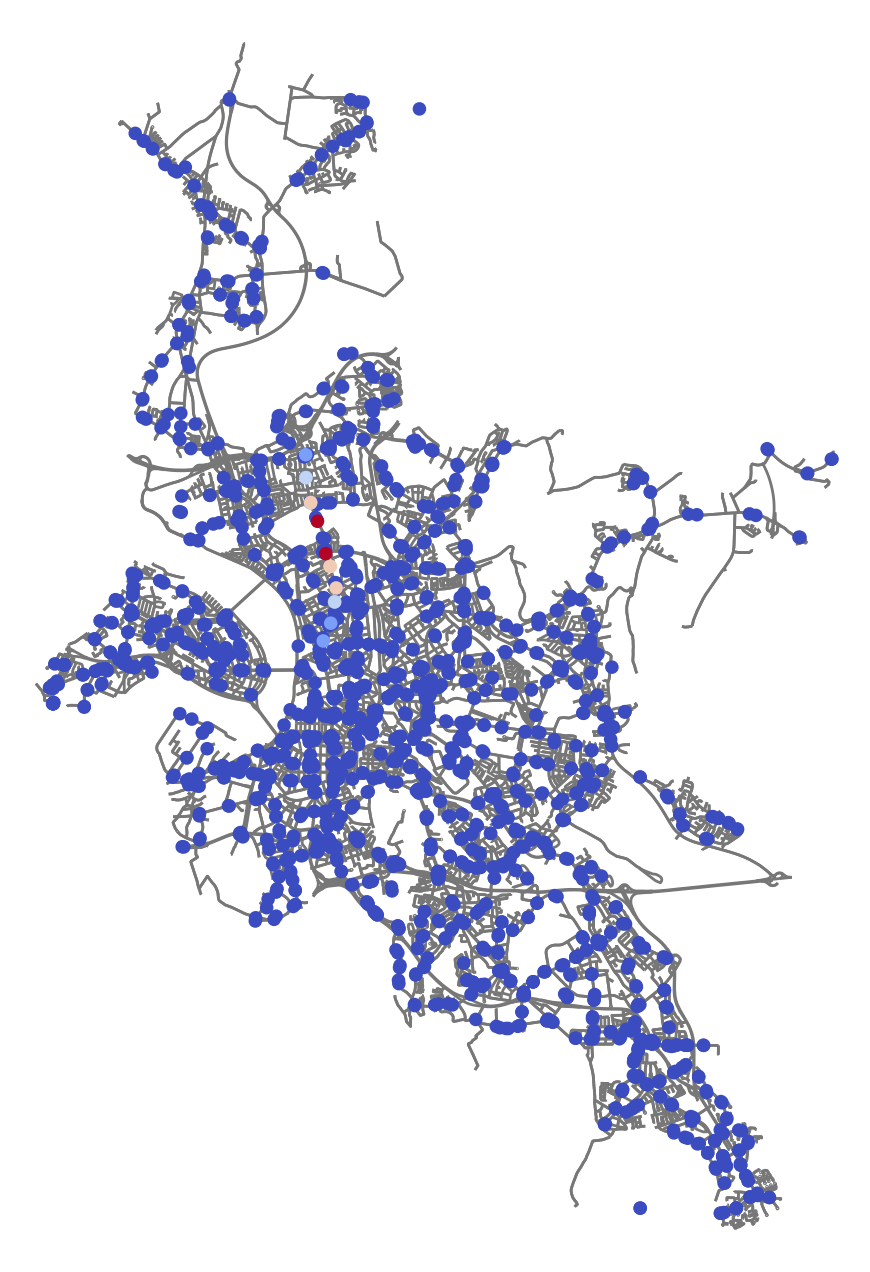}\label{fig:local_to_global_duesseldorf_0_99}
		}
	\end{center}
	\caption{Marginal node Katz centrality plots of D\"usseldorf with varying parameter $\alpha$.
		The left figure corresponds to a local measure close to degree centrality, the right figure corresponds to a global measure close to eigenvector centrality, and the center figure corresponds to an interpolated intermediate situation.
		Central stops are marked red and non-central stops are marked blue.
		The entries in the supra-adjacency matrix are weighted with travel times and frequencies.
		The remaining hyper-parameters are chosen as $\sigma=5$ and $\Delta t_{\mathrm{transfer}}=15$.
		The street network in the background of the plots is created with the OSMnx python package \cite{boeing2017osmnx}.}\label{fig:local_to_global_duesseldorf}
\end{figure}

We illustrate this at the example of marginal node Katz centralities of D\"usseldorf in \Cref{fig:local_to_global_duesseldorf}.
As discussed before, the admissible parameter range for $\alpha$ for this measure is given by $(0, 1/\lmax)$, where $\lmax$ is the largest eigenvalue of the supra-adjacency matrix.
In \Cref{fig:local_to_global_duesseldorf_0_01}, $\alpha=0.01/\lmax$ is chosen to be close to the lower end of that interval, which corresponds to being close to degree centrality in which only direct neighbors of each node-layer pair are considered.
The other extreme is represented by \Cref{fig:local_to_global_duesseldorf_0_99} in which $\alpha=0.99/\lmax$ is chosen to be close to the upper end of the admissible interval, which corresponds to being close to eigenvector centrality in which nodes are considered central if their closest neighbors are also central.

It can be seen in \Cref{fig:local_to_global_duesseldorf_0_01} that the local case is characterized by a relatively uniform distribution of marginal node centralities into the six centrality categories.
Furthermore, relatively central stops are geographically distributed across the whole city.
In this situation, each local neighborhood can possess its own central stops, regardless of their relative importance for the whole city.
Conversely, eigenvector centrality in \Cref{fig:local_to_global_duesseldorf_0_99} considers the stationary distribution of any initial distribution of walkers on the node-layer pairs of the network.
This often leads to localization effects \cite{martin2014localization}, which entail a centrality distribution that is almost uniform for all but a few very central stops.
The strength of matrix function-based centrality measures now lies in the fact that the choice of $\alpha$ allows to continuously interpolate between these two established concepts of centrality measures.
In \Cref{fig:local_to_global_duesseldorf_0_75}, the choice of $\alpha=0.75/\lmax$ illustrates this property by being ``visually in between'' the two extreme cases.

\subsubsection{Impact of transfer times}\label{sec:results_transfer_times}

In \Cref{eq:transfer_time_weight} we specified our approach to modeling a constant transfer time across all lines and stops of the network by means of the coupling parameter $\omega$, which is defined by a Gaussian kernel applied to the transfer time $\Delta t_{\mathrm{transfer}}$.
We start by presenting an example of a multiplex network of Chemnitz in which intra-layer edges are weighted with travel times and line frequencies.
However, an interesting property more frequently emerges when line frequencies are excluded from the intra-layer weight model.
We conclude the results section with a parameter study of the coupling parameter $\omega$ in the situation without frequencies.

\begin{figure}
	\begin{center}
		\subfloat[$\Delta t_{\mathrm{transfer}}=0$]{
			\includegraphics[width=.3\textwidth]{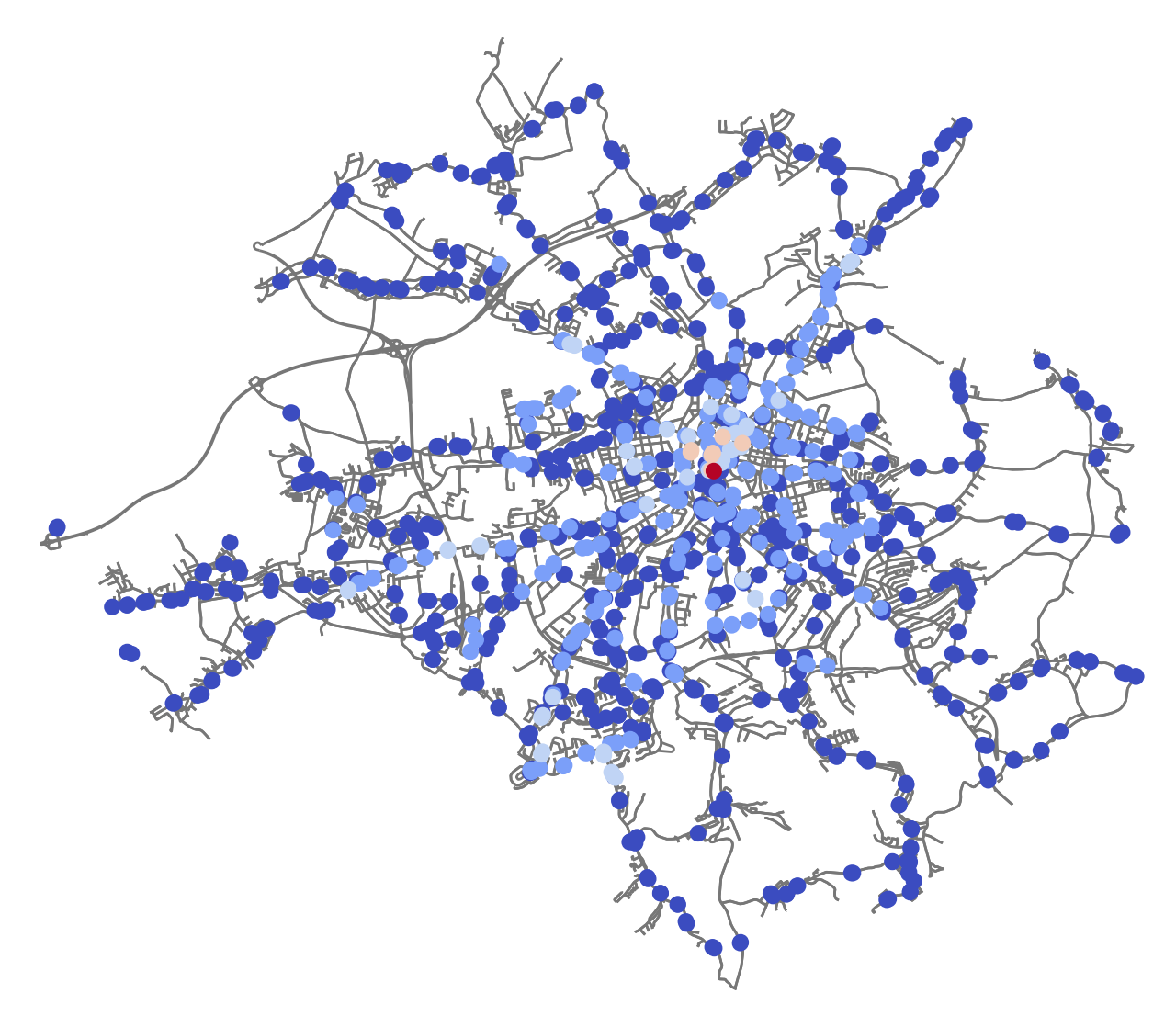}\label{fig:transfer_time_chemnitz_0}
		}
		\hfill
		\subfloat[$\Delta t_{\mathrm{transfer}}=5$]{
			\includegraphics[width=.3\textwidth]{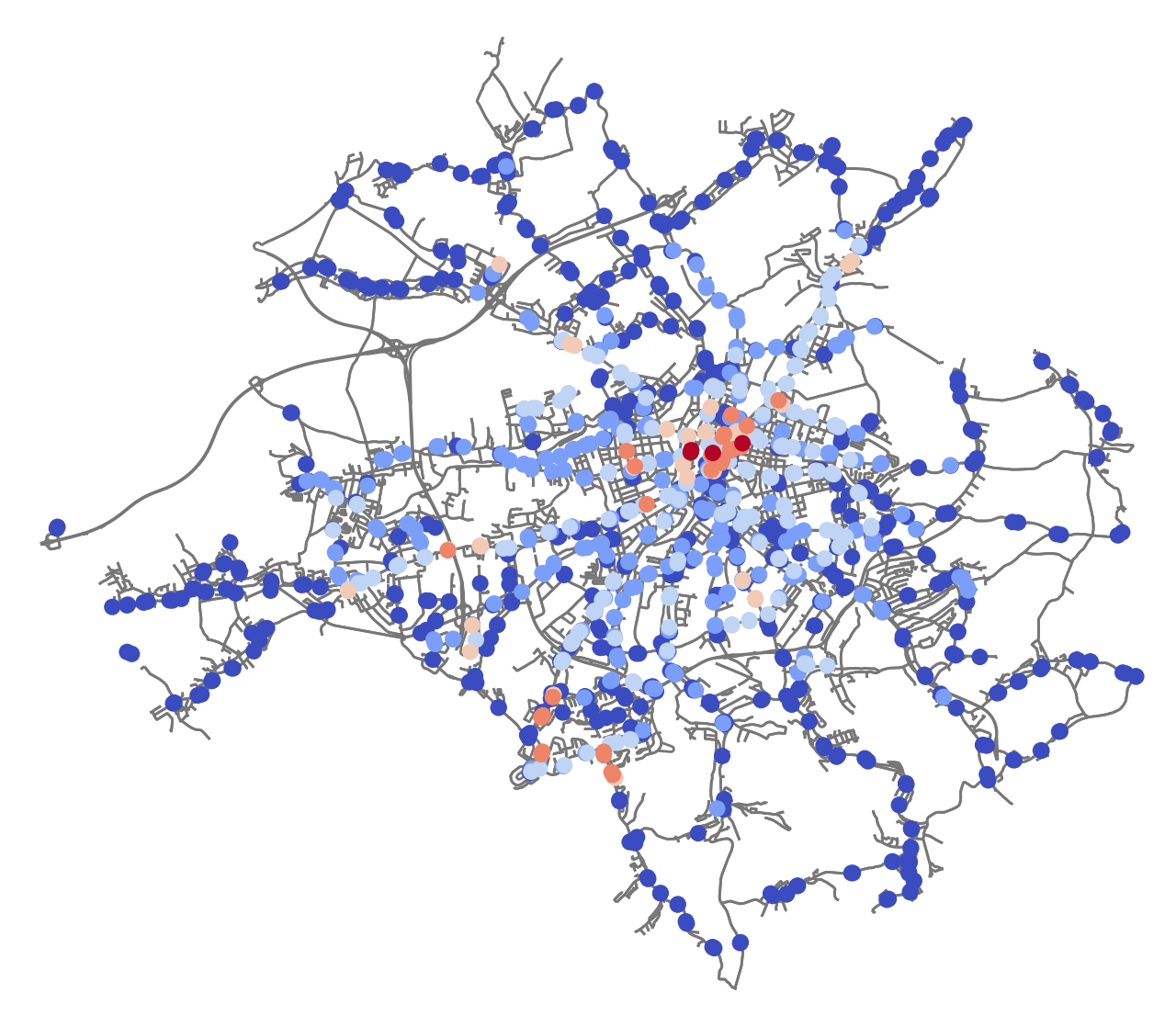}\label{fig:transfer_time_chemnitz_5}
		}
		\hfill
		\subfloat[$\Delta t_{\mathrm{transfer}}=15$]{
			\includegraphics[width=.3\textwidth]{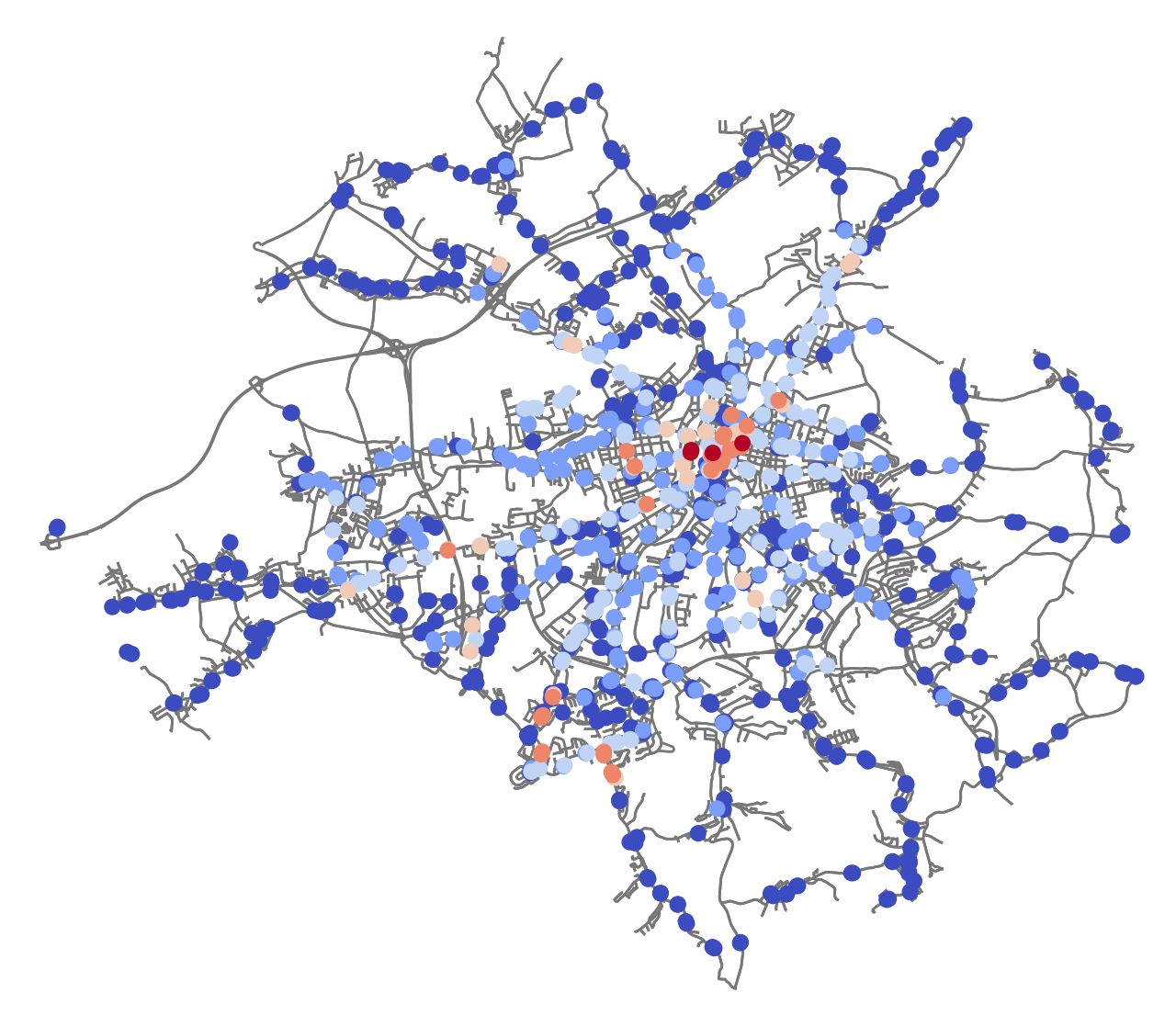}\label{fig:transfer_time_chemnitz_15}
		}
	\end{center}
	\caption{Marginal node total communicability plots of Chemnitz with varying transfer time $\Delta t_{\mathrm{transfer}}$.
		Central stops are marked red and non-central stops are marked blue.
		The entries in the supra-adjacency matrix are weighted with travel times and frequencies.
		The remaining hyper-parameters are chosen as $\beta=0.5/\lmax$ and $\sigma=1$.
		The street network in the background of the plots is created with the OSMnx python package \cite{boeing2017osmnx}.}\label{fig:transfer_time_chemnitz}
\end{figure}

\Cref{fig:transfer_time_chemnitz} illustrates marginal node total communicabilities of Chemnitz with the constant transfer time varying between $0$ and $15$ minutes.
Assigning no cost for changing lines in \Cref{fig:transfer_time_chemnitz_0} leads to Chemnitz's stop ``Zentralhaltestelle'' to be ranked as the most central stop of the city by a large margin.
The role of this stop is encoded in its name, which literally translates to ``central stop''.
The public transport system of Chemnitz is organized such that most of the city's lines stop at this geographically central stop, i.e., it offers by far the most opportunities to change lines.
However, this characteristic forfeits its significance as a cost for changing lines is introduced.
\Cref{fig:transfer_time_chemnitz_5} shows that many other stops in- and outside of the geographical city center are classified as central when the transfer time is increased to $5$ minutes.
\Cref{fig:transfer_time_chemnitz_15} illustrates that a further increase from $5$ to $15$ minutes (and above) has no significant additional impact.
This behavior, however, emerges only for a relatively small normalizing travel time of $\sigma=1$.
For larger parameters $\sigma$ or other cities with a more uniform distribution of inter-layer degrees this behavior is less pronounced when line frequencies are included in the weight model.

\begin{figure}
	\begin{center}
		\subfloat[Marginal layer centralities]{
			\includegraphics[width=.47\textwidth]{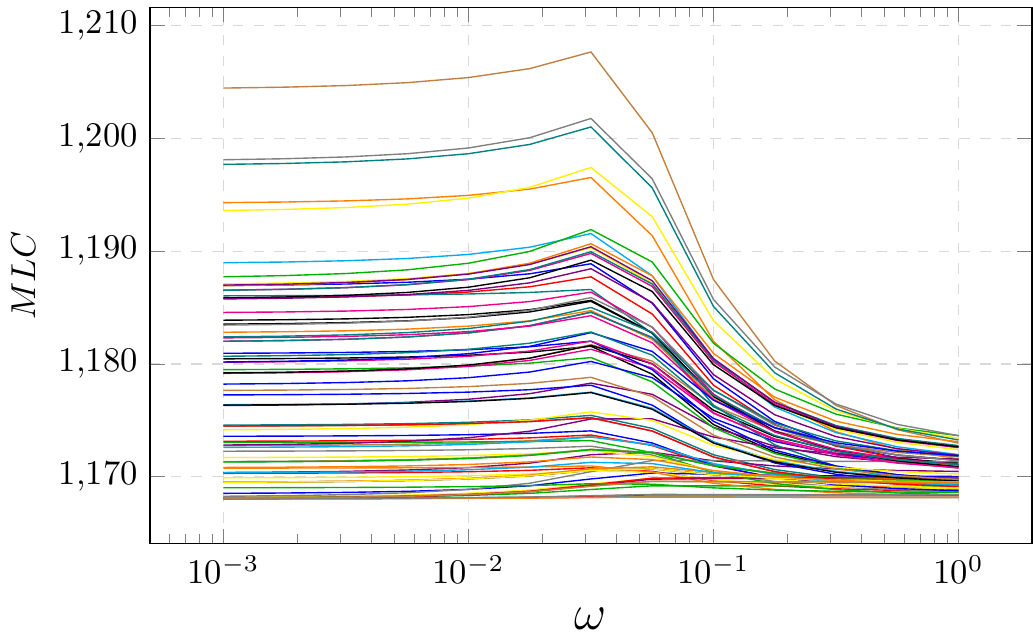}\label{fig:varying_omega_MLC}
		}
		\hfill
		\subfloat[Marginal node centralities]{
			\includegraphics[width=.45\textwidth]{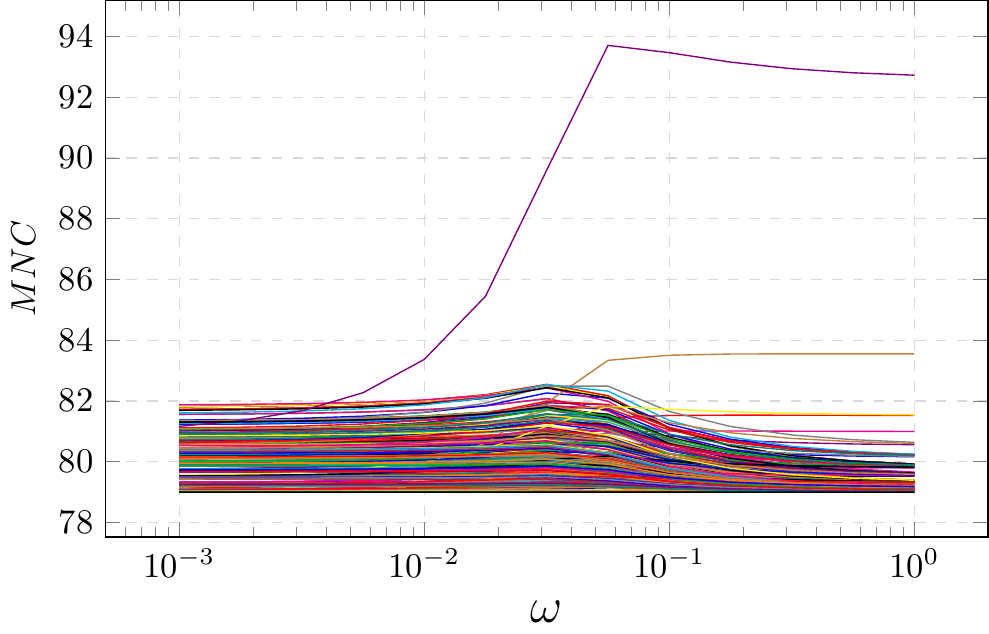}\label{fig:varying_omega_MNC}
		}
	\end{center}
	\caption{Marginal layer and marginal node total communicabilities of Chemnitz with varying coupling parameter $\omega$.
		The entries in the supra-adjacency matrix are weighted with travel times only.
		The remaining hyper-parameters are chosen as $\beta=0.5/\lmax$ and $\sigma=1$.
		The largest MNC for $\omega>5\cdot 10^{-3}$ corresponds to the stop ``Zentralhaltestelle'' (engl.: ``central stop'').}\label{fig:varying_omega}
\end{figure}

\begin{figure}
	\captionsetup[subfigure]{labelformat=empty, justification=centering}
	\begin{center}
		\subfloat[Aachen]{
			\includegraphics[width=.13\textwidth]{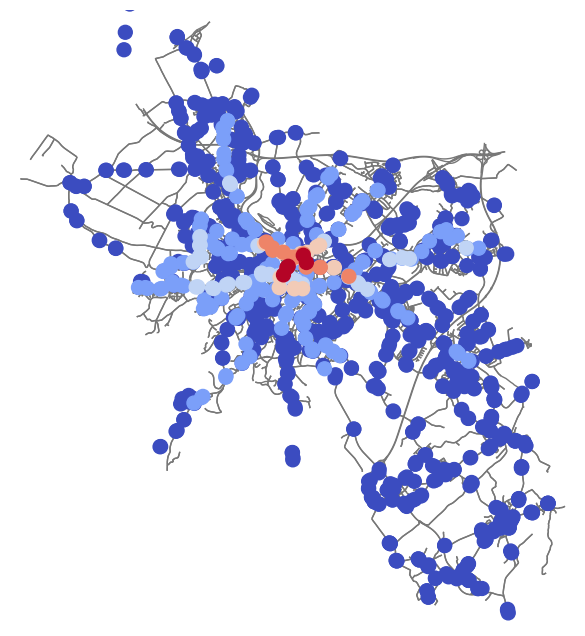}}\hfill
		\subfloat[Augsburg]{
			\includegraphics[width=.13\textwidth]{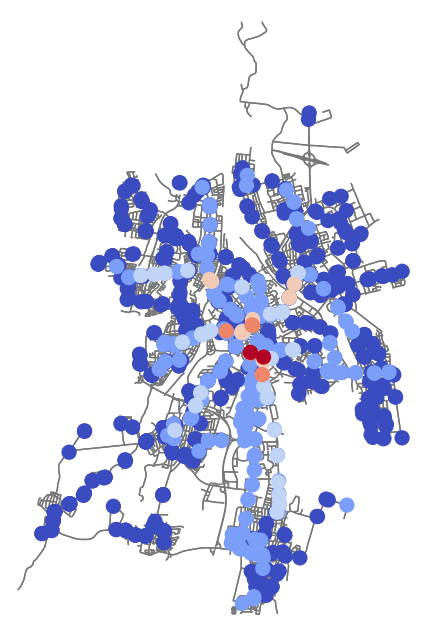}}\hfill
		\subfloat[Berlin]{
			\includegraphics[width=.13\textwidth]{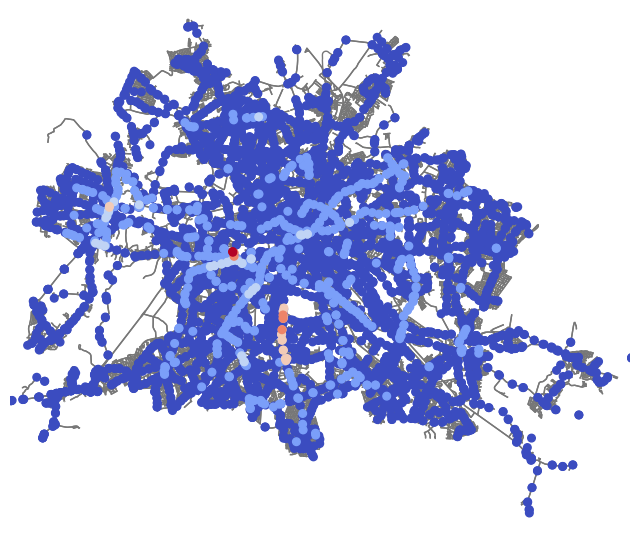}}\hfill
		\subfloat[Bielefeld]{
			\includegraphics[width=.13\textwidth]{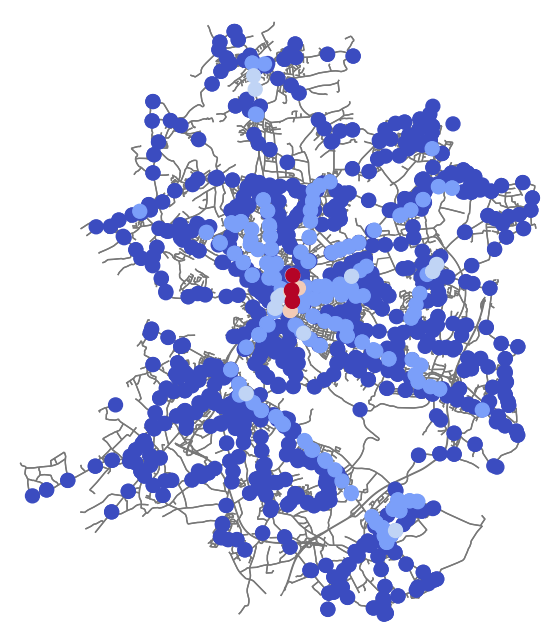}}\hfill
		\subfloat[Bonn]{
			\includegraphics[width=.13\textwidth]{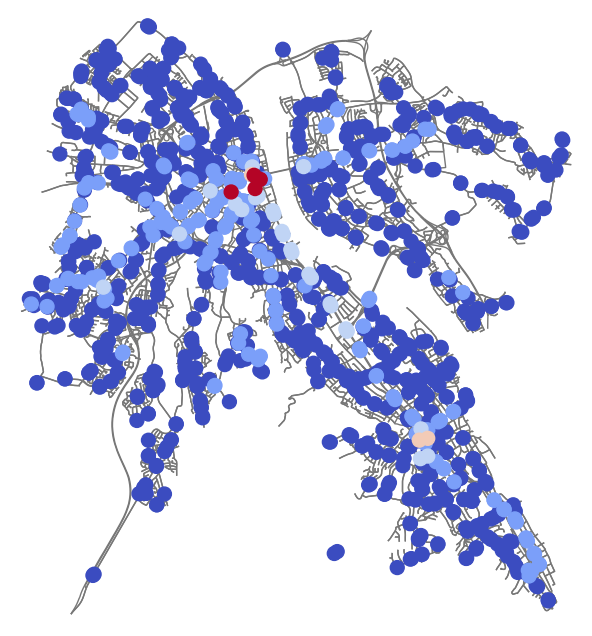}}\hfill
		\subfloat[Braunschweig]{
			\includegraphics[width=.13\textwidth]{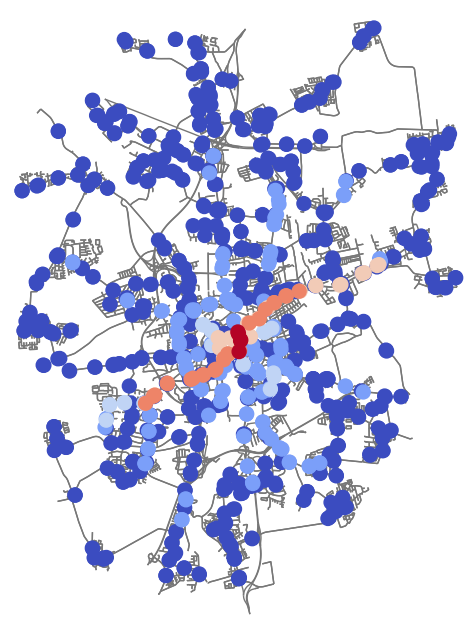}}
		
		\subfloat[Bremen]{
			\includegraphics[width=.13\textwidth]{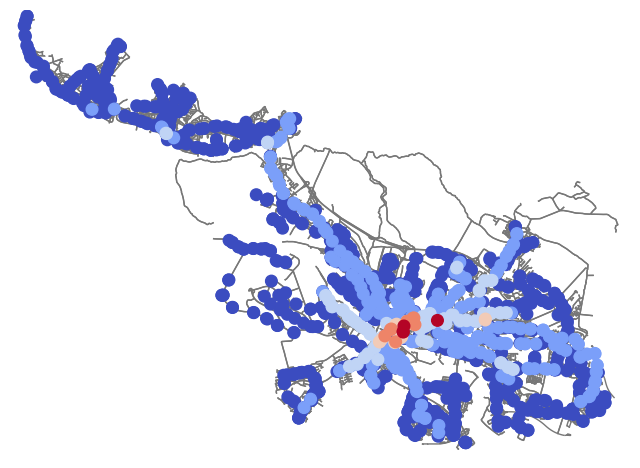}}\hfill
		\subfloat[Chemnitz]{
			\includegraphics[width=.13\textwidth]{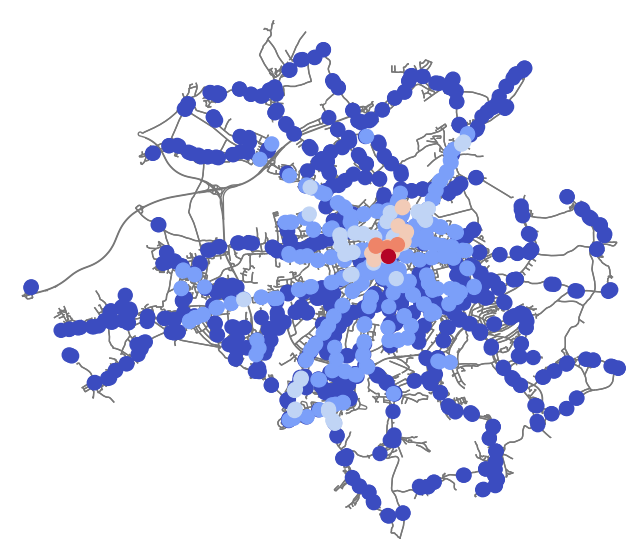}}\hfill
		\subfloat[Cologne]{
			\includegraphics[width=.13\textwidth]{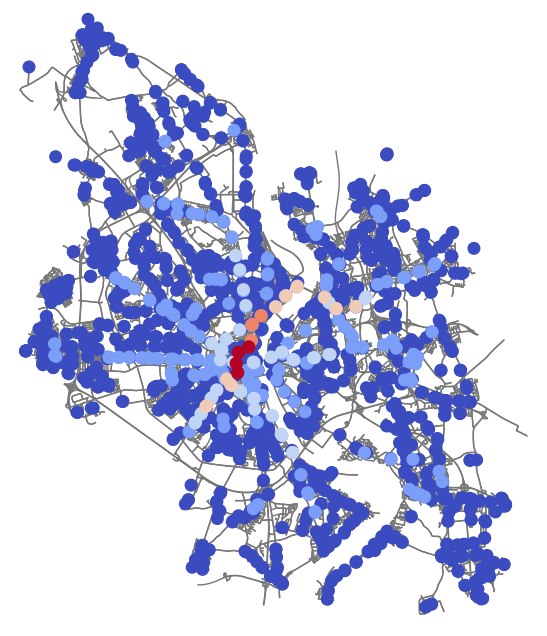}}\hfill
		\subfloat[Dortmund]{
			\includegraphics[width=.13\textwidth]{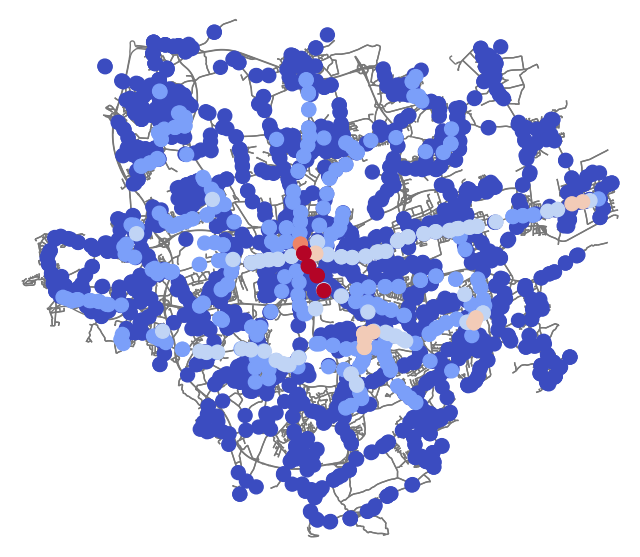}}\hfill
		\subfloat[Dresden]{
			\includegraphics[width=.13\textwidth]{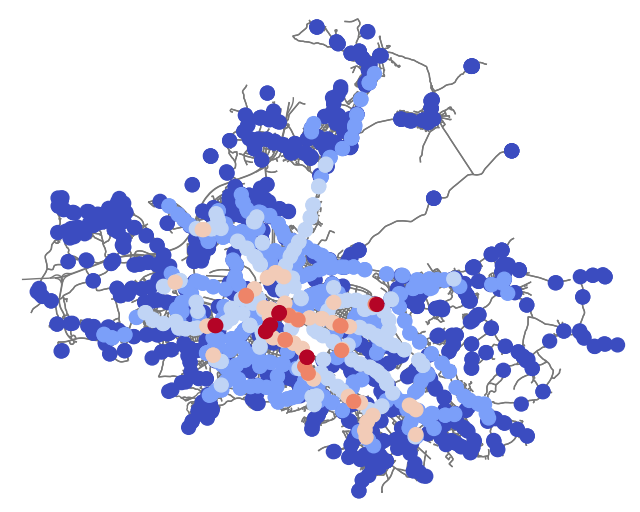}}\hfill
		\subfloat[Duisburg]{
			\includegraphics[width=.11\textwidth]{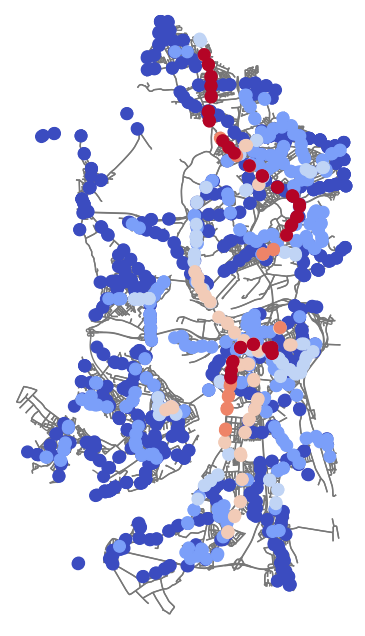}}
		
		\subfloat[Düsseldorf]{
			\includegraphics[width=.13\textwidth]{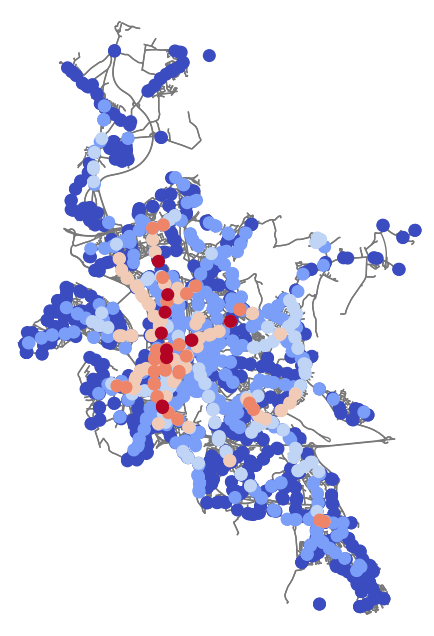}}\hfill
		\subfloat[Erfurt]{
			\includegraphics[width=.13\textwidth]{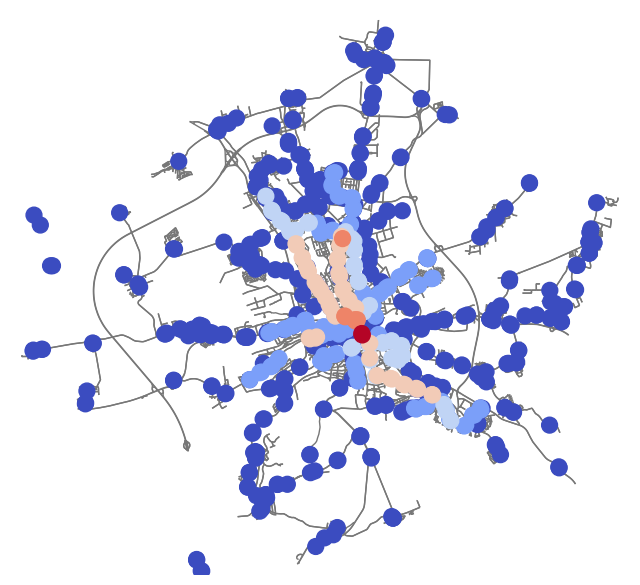}}\hfill
		\subfloat[Essen]{
			\includegraphics[width=.13\textwidth]{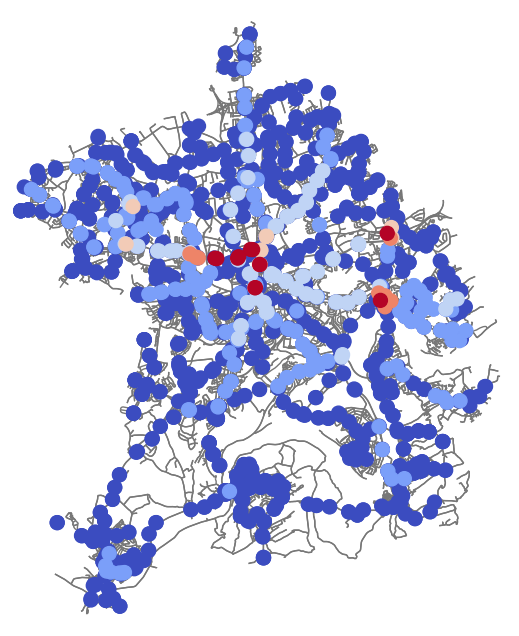}}\hfill
		\subfloat[Frankfurt am Main]{
			\includegraphics[width=.13\textwidth]{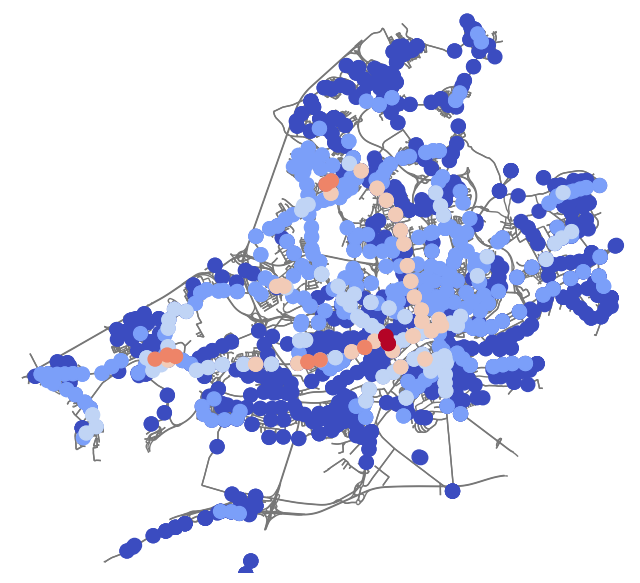}}\hfill
		\subfloat[Freiburg]{
			\includegraphics[width=.13\textwidth]{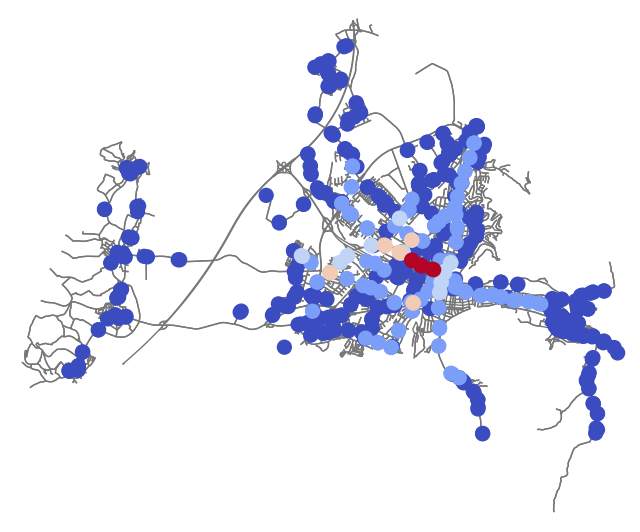}}\hfill
		\subfloat[Halle (Saale)]{
			\includegraphics[width=.13\textwidth]{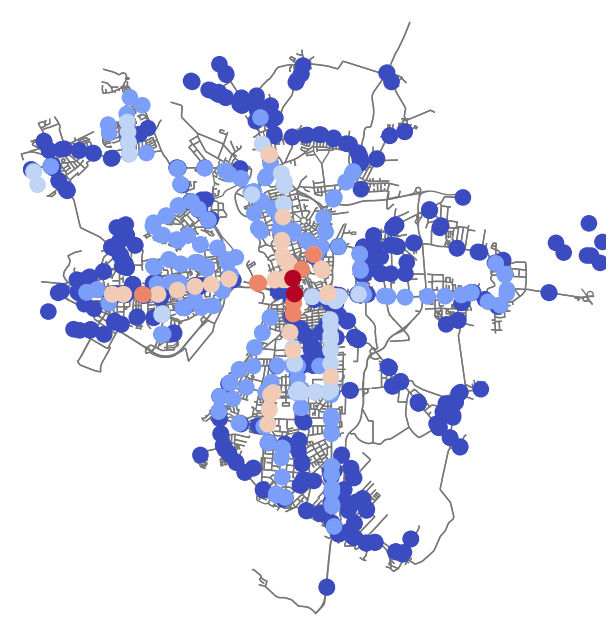}}
		
		\subfloat[Hamburg]{
			\includegraphics[width=.13\textwidth]{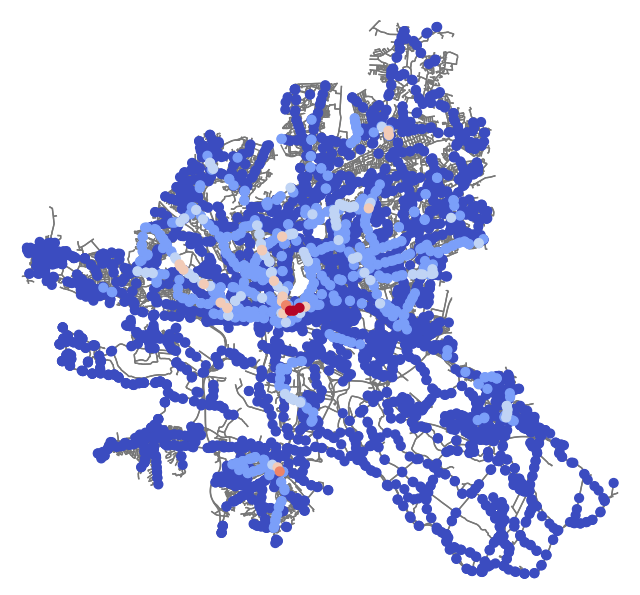}}\hfill
		\subfloat[Hanover]{
			\includegraphics[width=.13\textwidth]{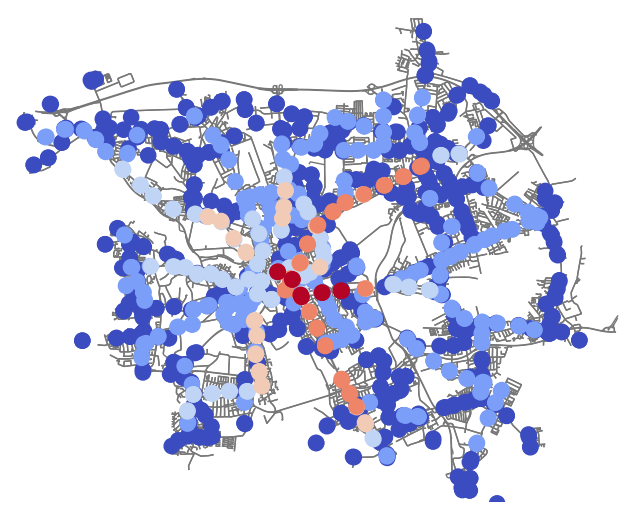}}\hfill
		\subfloat[Karlsruhe]{
			\includegraphics[width=.13\textwidth]{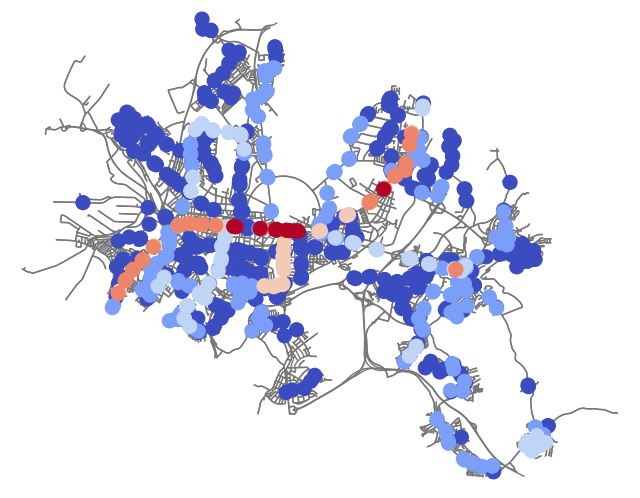}}\hfill
		\subfloat[Kiel]{
			\includegraphics[width=.11\textwidth]{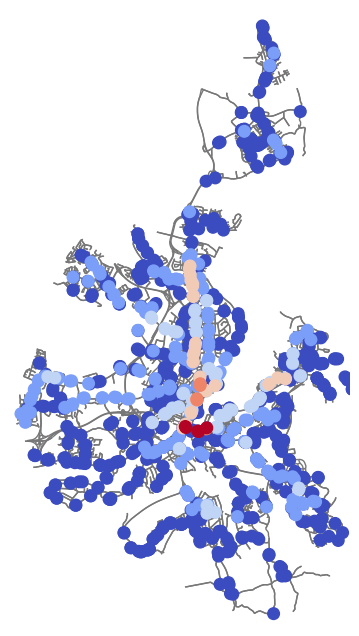}}\hfill
		\subfloat[Krefeld]{
			\includegraphics[width=.13\textwidth]{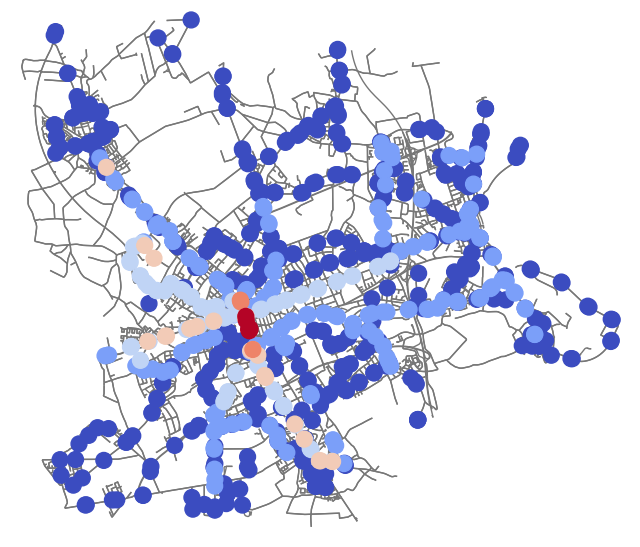}}\hfill
		\subfloat[Leipzig]{
			\includegraphics[width=.13\textwidth]{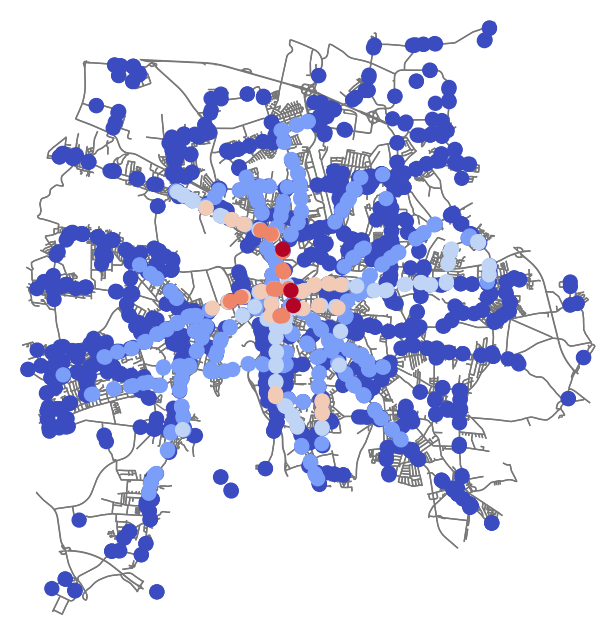}}
		
		\subfloat[Lübeck]{
			\includegraphics[width=.13\textwidth]{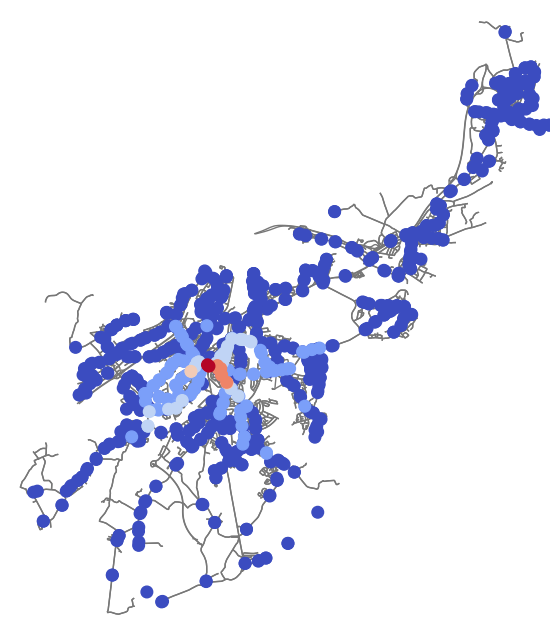}}\hfill
		\subfloat[Mainz]{
			\includegraphics[width=.13\textwidth]{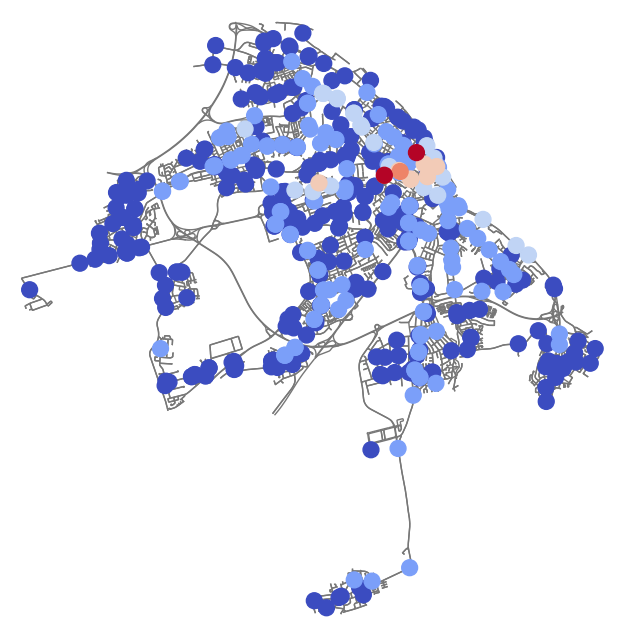}}\hfill
		\subfloat[Mannheim]{
			\includegraphics[width=.13\textwidth]{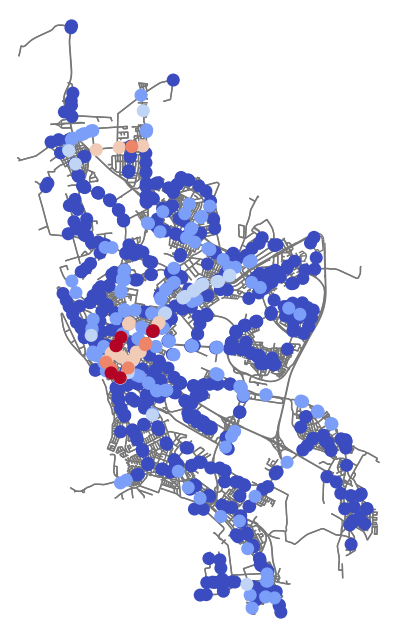}}\hfill
		\subfloat[Munich]{
			\includegraphics[width=.13\textwidth]{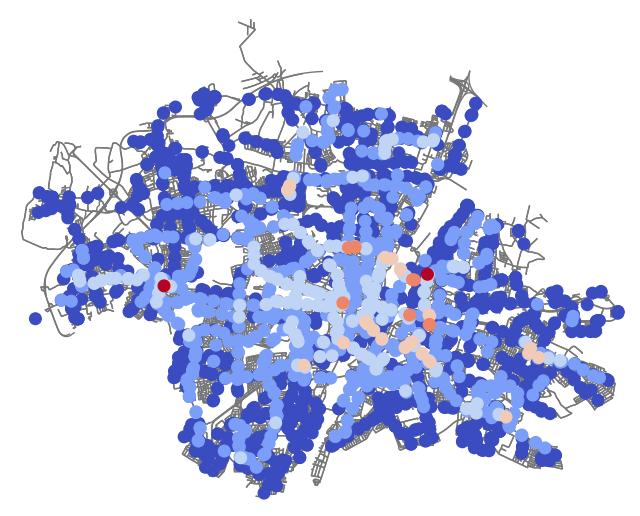}}\hfill
		\subfloat[Mönchengladbach]{
			\includegraphics[width=.13\textwidth]{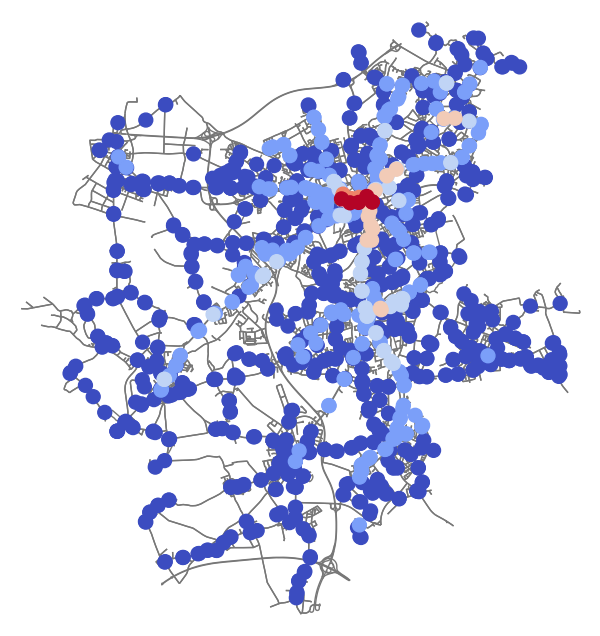}}\hfill
		\subfloat[Münster]{
			\includegraphics[width=.13\textwidth]{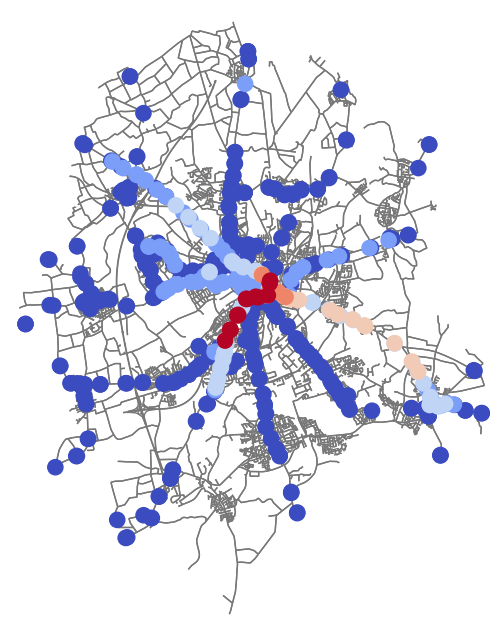}}
		
		\subfloat[Nuremberg]{
			\includegraphics[width=.13\textwidth]{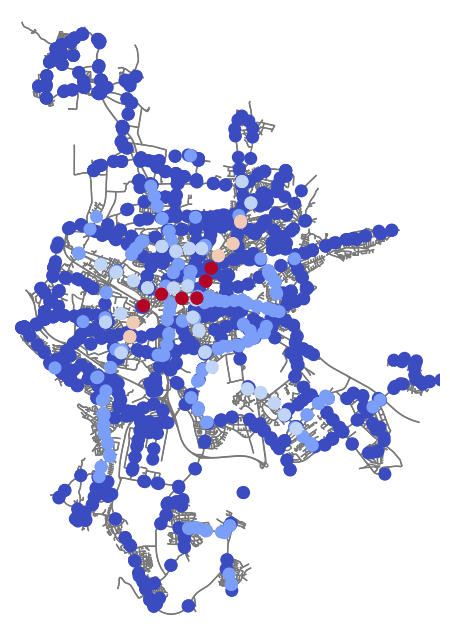}}\hfill
		\subfloat[Oberhausen]{
			\includegraphics[width=.13\textwidth]{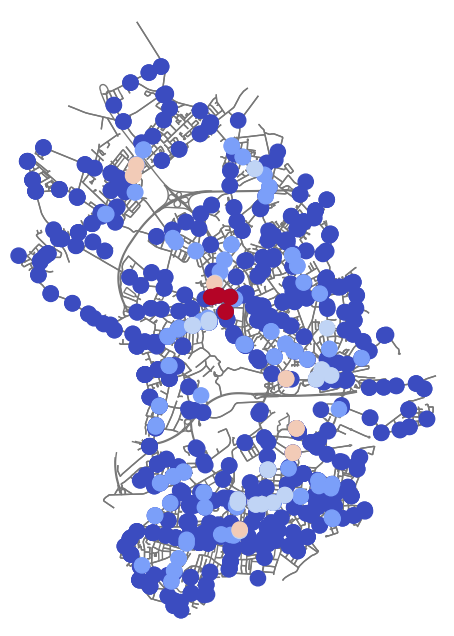}}\hfill
		\subfloat[Rostock]{
			\includegraphics[width=.13\textwidth]{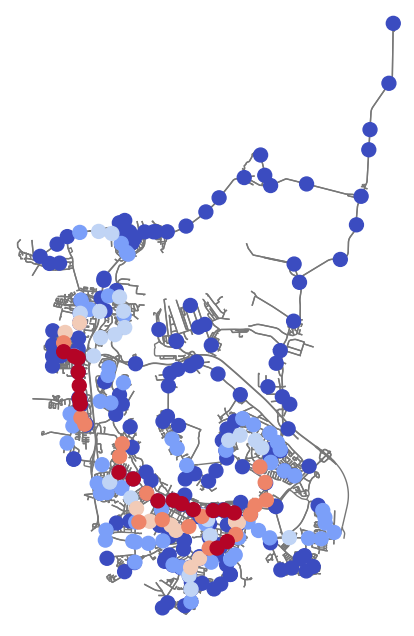}}\hfill
		\subfloat[Stuttgart]{
			\includegraphics[width=.13\textwidth]{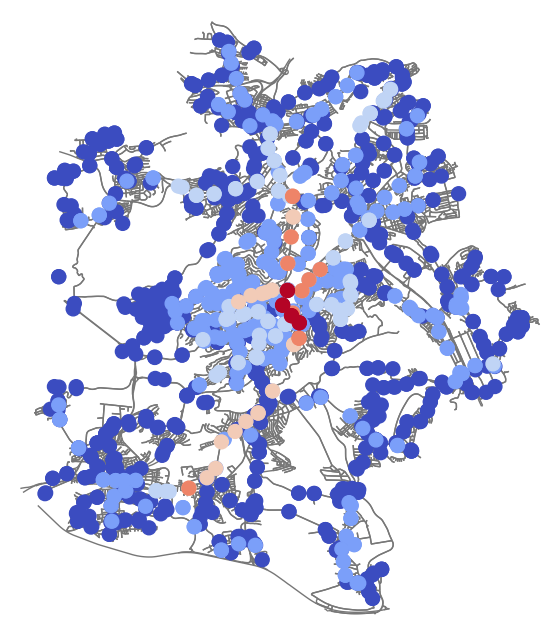}}\hfill
		\subfloat[Wiesbaden]{
			\includegraphics[width=.13\textwidth]{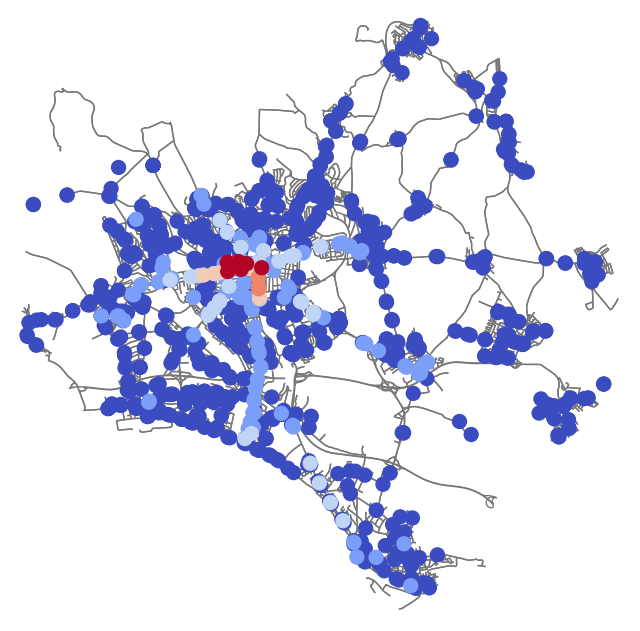}}\hfill
		\subfloat[Wuppertal]{
			\includegraphics[width=.13\textwidth]{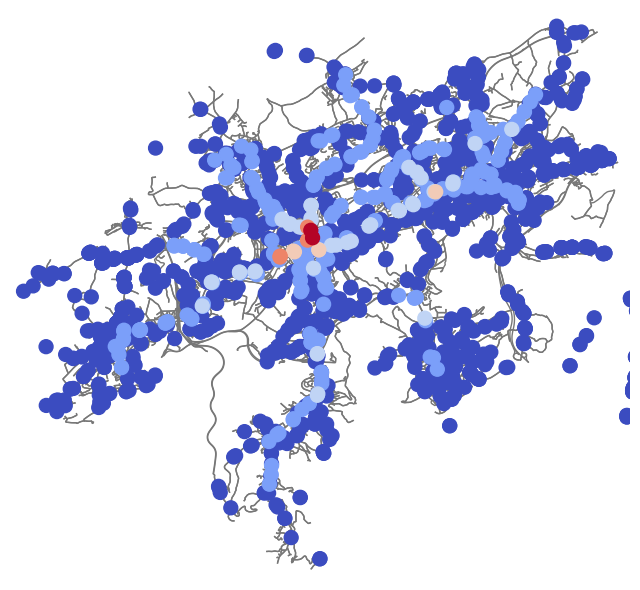}}
	\end{center}
	\caption{Marginal node Katz centrality plots of the $36$ largest German cities with valid GTFS data.
		Central stops are marked red and non-central stops are marked blue.
		The entries in all supra-adjacency matrices are weighted with travel times and frequencies.
		The hyper-parameters are chosen as $\alpha=0.5/\lmax$, $\sigma=5$, and $\Delta t_{\mathrm{transfer}}=5$ for all cities.
		The street networks in the background of the plots are created with the OSMnx python package \cite{boeing2017osmnx}.}\label{fig:Germany_KC_MNC}
\end{figure}

Excluding line frequencies from the weight model leads to a stronger localization of marginal node centralities in the situation $\Delta t_{\mathrm{transfer}} \rightarrow 0$ across different networks and a larger range of $\sigma$.
The effect of the variation of $\Delta t_{\mathrm{transfer}}$ and subsequently $\omega$ on marginal layer and marginal node centralities is illustrated in \Cref{fig:varying_omega}.
Here, we observe an interesting clustering behavior of marginal node centralities in the strong coupling limit in which $\omega$ approaches its maximum value of $1$ corresponding to $\Delta t_{\mathrm{transfer}}=0$.
These clusters are determined by the stops' inter-layer degrees, i.e., by the number of lines stopping at the corresponding stop.
This behavior reflects the accessibility of a larger part of the network when a larger number of lines can be used within a constant total travel time (defined as the sum of intra- and inter-layer travel times).

Another interesting observation in \Cref{fig:varying_omega} concerns the peak in most marginal layer and marginal node centralities between $\omega=0.01$ and $\omega=0.1$.
This phenomenon has been similarly encountered in other applications before \cite[Sec.~6.3]{bergermann2021matrix} and is an interesting question for future research.

\subsubsection{Discussion}\label{sec:results_discussion}

The preceding subsections gave a detailed account of the differences between the introduced centrality measures and weight models as well as the influence of all involved hyper-parameters.
Earlier in this paper, we commented on conclusions obtained from public transport network orientations and their relation to the corresponding street network orientations.
We now discuss the relation between these two aspects of urban public transport networks.

\begin{figure}
	\begin{center}
		\subfloat[Bielefeld]{
			\includegraphics[width=.3\textwidth]{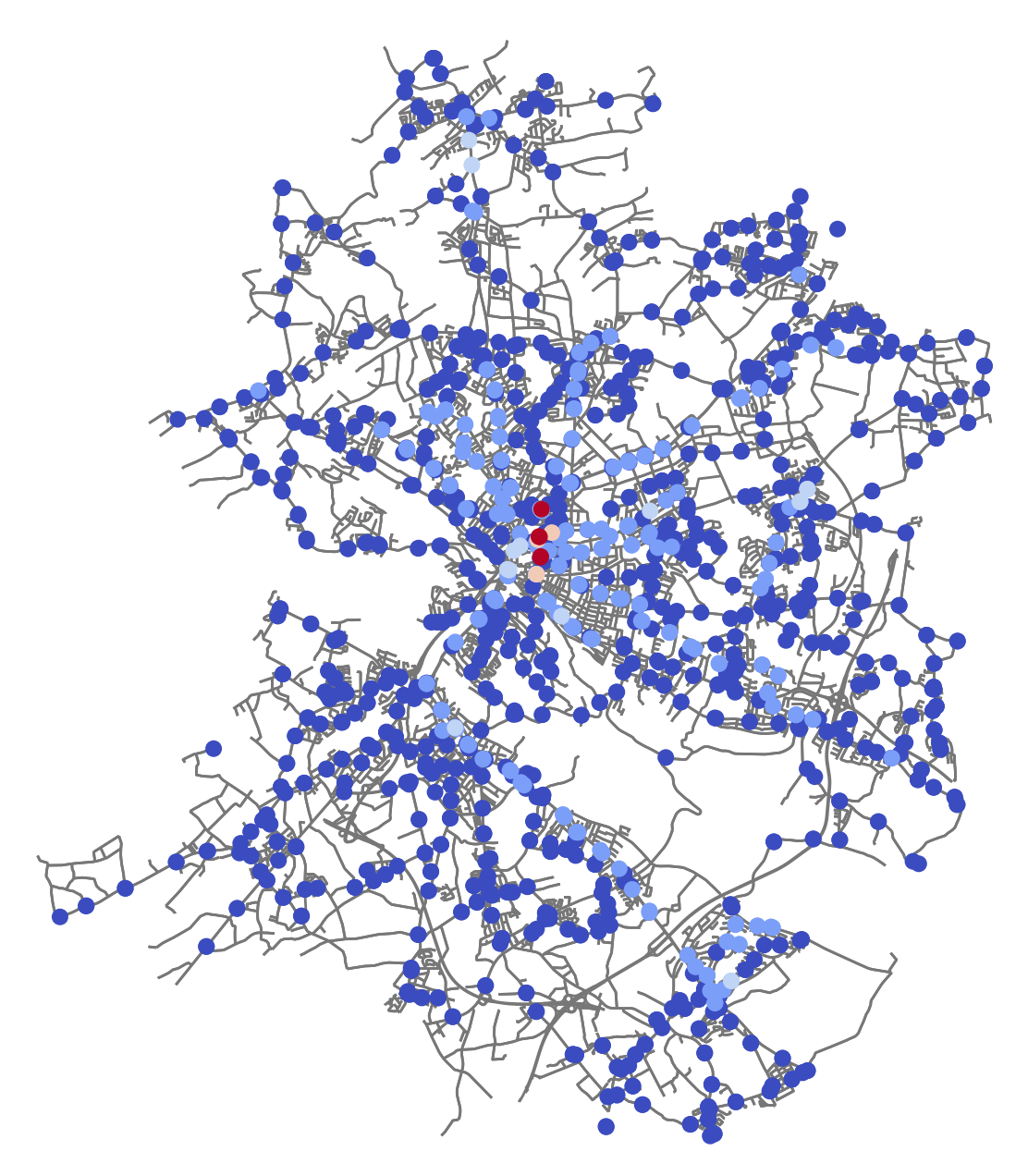}\includegraphics[width=.15\textwidth]{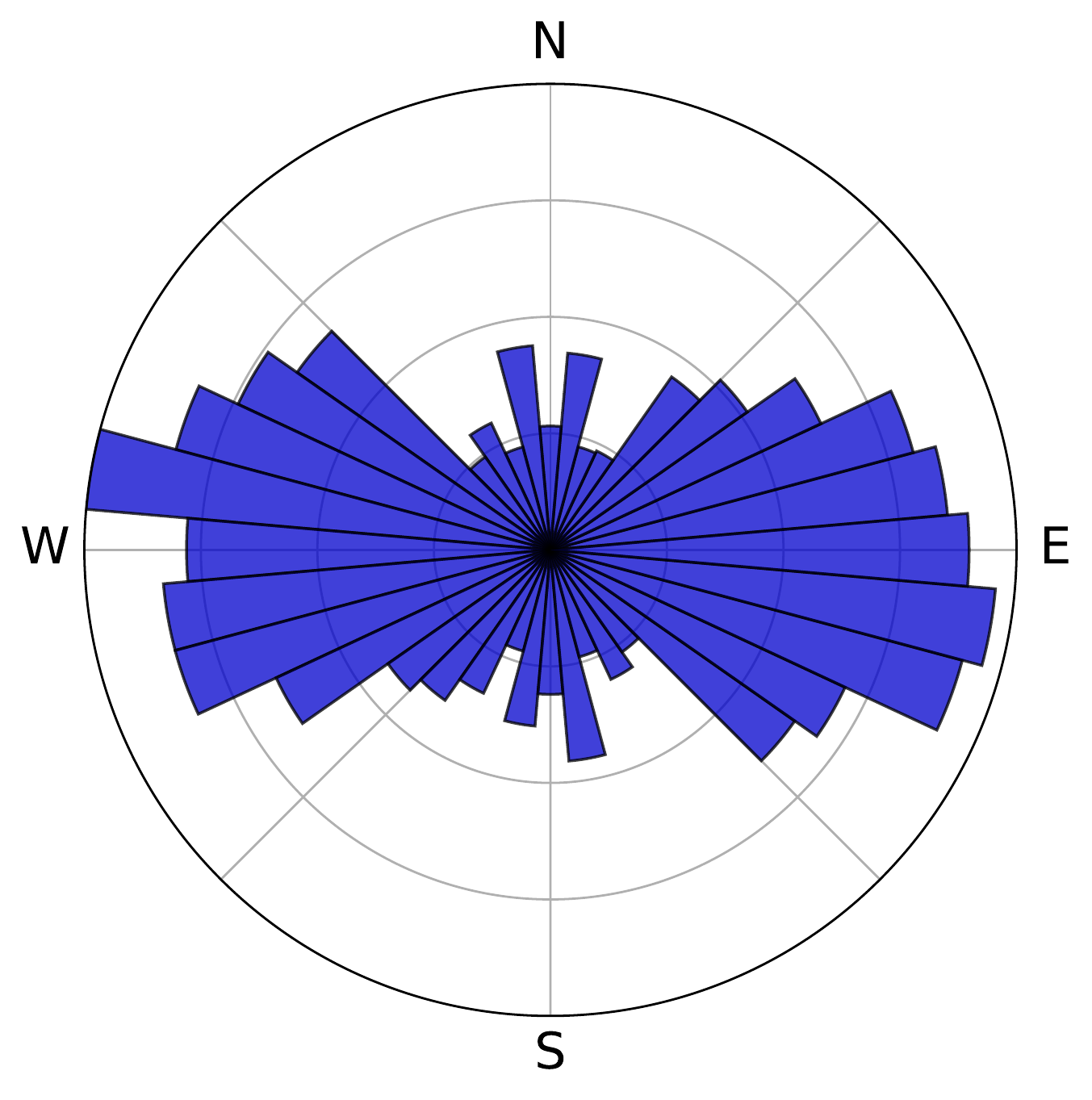}\label{fig:comparison_centralities_orientations_4_examples_bielefeld}}\hfill
		\subfloat[Munich]{
			\includegraphics[width=.3\textwidth]{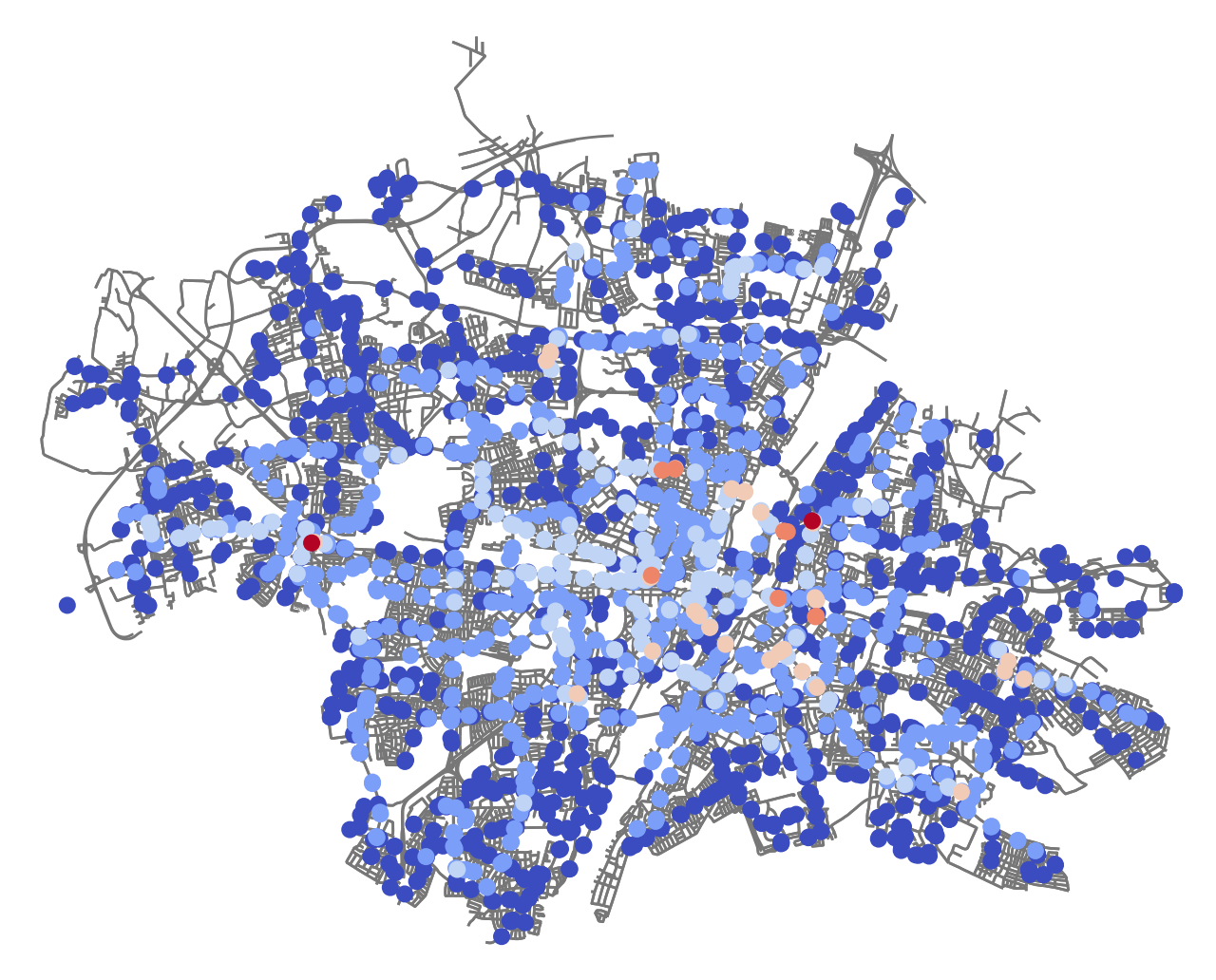}\includegraphics[width=.15\textwidth]{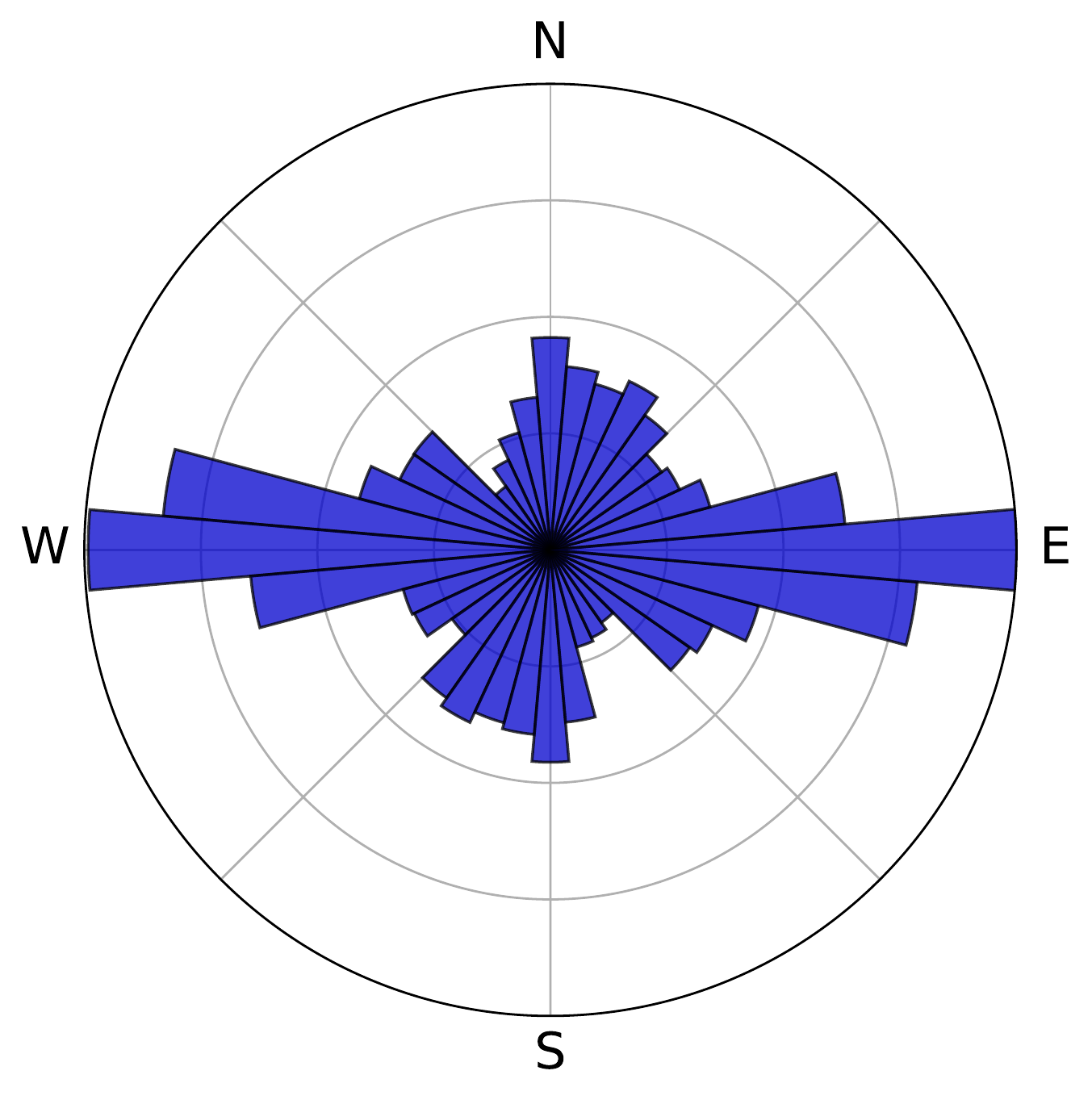}\label{fig:comparison_centralities_orientations_4_examples_munich}}
		
		\subfloat[Duisburg]{
			\hspace{.05\textwidth}\includegraphics[width=.25\textwidth]{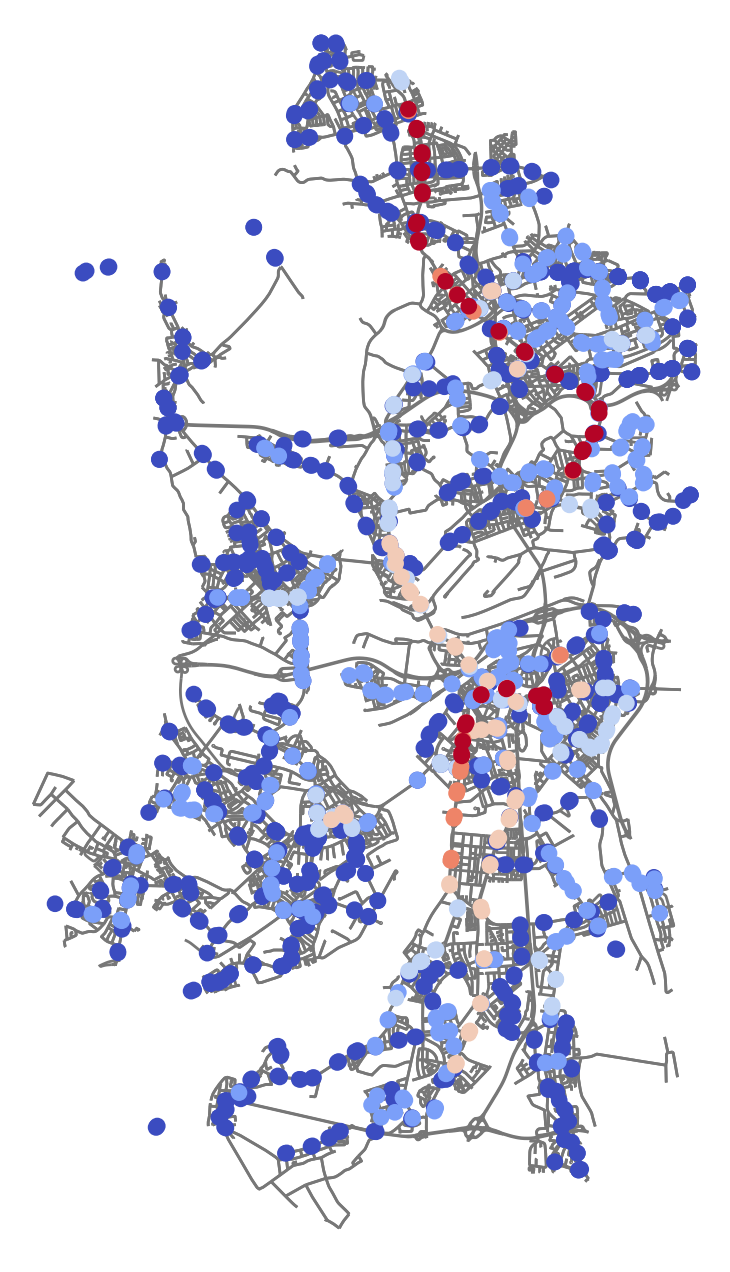}\includegraphics[width=.15\textwidth]{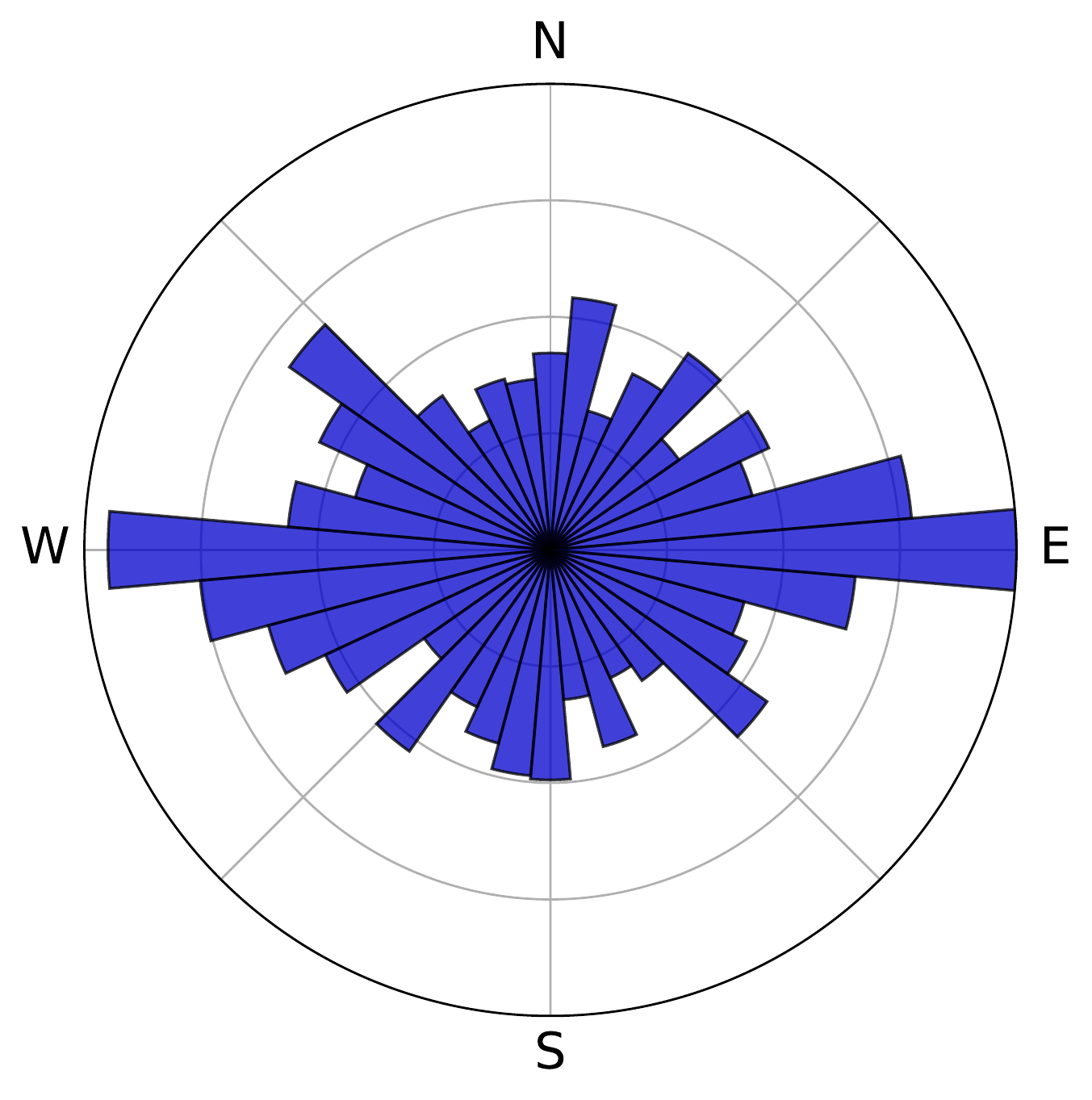}\label{fig:comparison_centralities_orientations_4_examples_duisburg}}\hfill
		\subfloat[Karlsruhe]{
			\includegraphics[width=.3\textwidth]{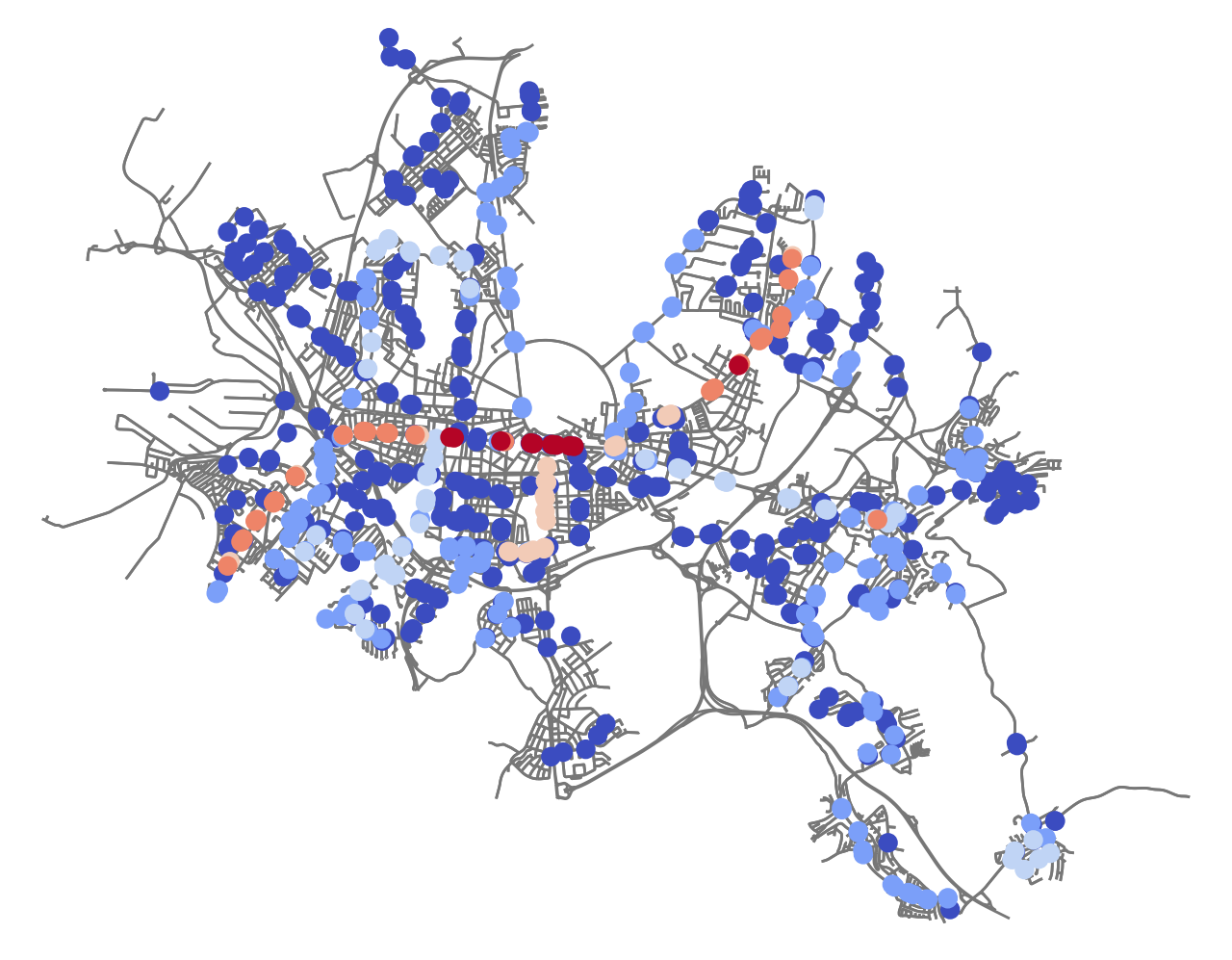}\includegraphics[width=.15\textwidth]{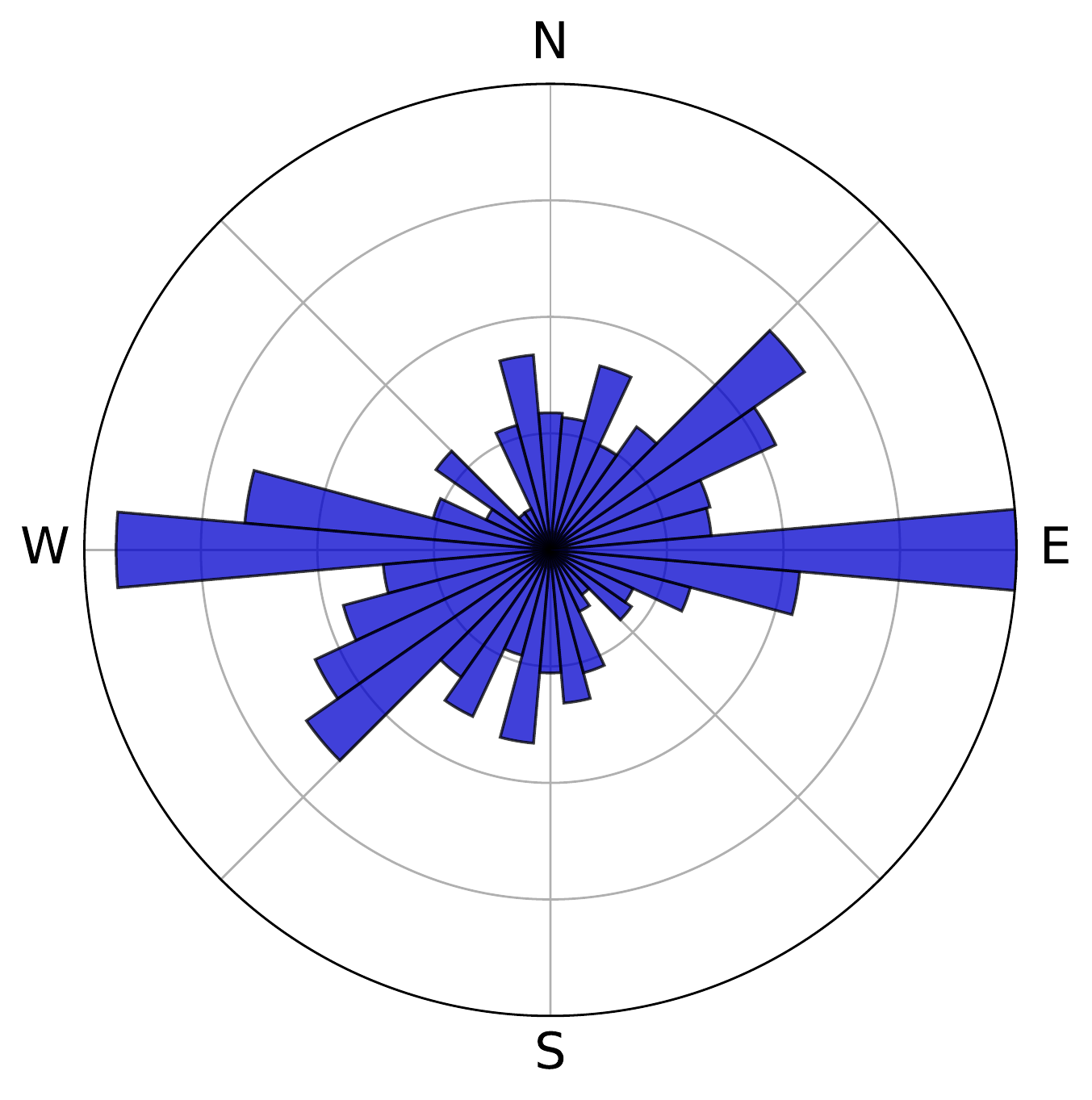}\label{fig:comparison_centralities_orientations_4_examples_karlsruhe}}
	\end{center}
	\caption{Comparison of selected German cities' margial node Katz centralities and public transport orientations.
		The entries in all supra-adjacency matrices are weighted with travel times and frequencies.
		The hyper-parameters are chosen as $\alpha=0.5/\lmax$, $\sigma=5$, and $\Delta t_{\mathrm{transfer}}=5$ for the centrality plots of all cities.
		The street networks in the background of the centrality plots are created with the OSMnx python package \cite{boeing2017osmnx}.}\label{fig:comparison_centralities_orientations_4_examples}
\end{figure}

\Cref{fig:Germany_KC_MNC} provides an overview of marginal node Katz centralities for all $36$ German cities from \Cref{fig:orientations_public_transport_germany} using a common set of hyper-parameters, which allows a direct comparison of the results.
While we identified a clear pattern in the public transport network orientations across all German cities the centrality plots in \Cref{fig:Germany_KC_MNC} give a mixed picture.
Here, we observe various different distributions of central stops both in terms of the relative number of central stops per city as well as their geographical arrangement.

In \Cref{fig:comparison_centralities_orientations_4_examples} we focus on four example cities with rather different centrality distributions, which are not necessarily reflected in the corresponding public transport orientation plots.
\Cref{fig:comparison_centralities_orientations_4_examples_bielefeld} shows that Bielefeld has only a small number of central stops located at the geographical city center.
This center is the origin of various sequences of light blue stops (indicating the second to last central category), which seem to equally spread out into all directions and not only into the preferential directions of the corresponding orientation plot.
\Cref{fig:comparison_centralities_orientations_4_examples_munich} reveals that several central stops are spread out over the city area of Munich.
However, the connections between these central stops can not clearly be linked to the preferential direction of the corresponding orientation plot.
In \Cref{fig:comparison_centralities_orientations_4_examples_duisburg,fig:comparison_centralities_orientations_4_examples_karlsruhe} and several other examples we observe highly central sequences of stops lining up along major public transport axes.
In some cases, the directions of these axes can be linked to the most pronounced bearings of the corresponding orientation plots, cf.\ Karlsruhe in \Cref{fig:comparison_centralities_orientations_4_examples_karlsruhe} but also Braunschweig, Erfurt, Halle (Saale), Hanover, M\"unster, and Rostock.
This pattern, however, can not be observed in other examples like Duisburg in \Cref{fig:comparison_centralities_orientations_4_examples_duisburg}, Cologne, or Stuttgart.

We conclude that orientation plots of street and public transport networks are capable of revealing interesting high-level properties of urban systems.
For a more detailed analysis of the geometrical properties of cities like, e.g., the spatial distribution of major transport axes additional measures must be taken into account.
This manuscript proposes multiplex matrix function-based centralities as one such measure.

\section{Conclusion}

We studied two geometrical aspects of urban public transport networks:
orientations and centralities.
We determined orientations of directed public transport networks of the $36$ largest German as well as $18$ major European cities and compared them to orientations of undirected street networks.
All considered German cities revealed two more or less pronounced orthogonal preferential street network directions, which can often be linked to geographical constraints.
We found that most considered German public transport network orientations concentrate around the one of the two preferential street network directions, which is closer to the cardinal east-west axis.
The same qualitative behavior could only be observed for a small subset of the considered European cities.
However, north-south-like preferential public transport directions remained rare.

Furthermore, we formally introduced urban public transport multiplex networks in which nodes correspond to stops and layers to lines and applied multiplex matrix function-based centrality measures in order to identify and rank the most central lines and stops of the considered German cities.
These measures generate rankings, which are consistent with previous investigations.
In addition, they offer the benefit of being able to flexibly choose the desired degree of locality, possessing efficient and scalable numerical implementations, and being applicable to a wide range of problems.
The influence of different hyper-parameters, which all have meaningful interpretations in terms of the urban science application, was thoroughly studied.
Our study showed that matrix function-based centralities are capable of revealing insights into geometrical aspects of urban systems on a more granular level than orientation plots are.

We believe that multiplex models of cities, ideally combining various aspects like, e.g., various modes of transportation or additional aspects of urban life can contribute to a better understanding of urban systems.
We hope that the presented methodology can add to urban scientists' toolkits and that city-specific modeling as well as parameter-tuning of domain experts yield further contributions to the economic, environmental, and social challenges lying ahead.

\section*{Acknowledgments}
We thank Peter Bernd Oehme for his support in processing the GTFS data and programming basic python routines.
We thank Geoff Boeing for pointing us into the right direction for reproducing street network orientations using OSMnx.


\end{document}